\documentclass[oneside, 12pt]{article}
\usepackage[spanish]{babel}
\usepackage[mathscr]{euscript}
\usepackage[utf8]{inputenc}
\usepackage[T1]{fontenc}
\usepackage{lmodern}
\usepackage{amssymb}
\usepackage[square,numbers]{natbib}
\usepackage{enumerate} 
\usepackage{makeidx}
\makeindex
\usepackage{graphicx}
\usepackage{float}
\floatstyle{boxed} 
\restylefloat{figure}

\newcommand{\keyword}[1]{\textbf{#1}}

\begin{document}

\pagestyle{empty}
\renewcommand{\indexname}{Índice de nombres}
\setcounter{page}{2}
\pagestyle{plain} 
\date{}
\bibliographystyle{plain}
\begin{titlepage}
\title{INFLUENCIAS ASTRONÓMICAS SOBRE LA EVOLUCIÓN GEOLÓGICA Y BIOLÓGICA DE LA TIERRA (Parte II).}
\author{Carlos A. Olano\\ \small{Facultad de Ciencias Astron\'{o}micas y Geof\'{i}sicas,}\\ \small{Universidad Nacional de La Plata, Argentina}\\ \small{(E-mail: colano@fcaglp.fcaglp.unlp.edu.ar)}}
\maketitle
\end{titlepage} 
\maketitle
\begin{abstract}
 With the same general purposes as Part I of this monograph, we analyze here major events in the history of the Earth, such as the formation of the Earth itself, the origin of life, the great glaciations and the mass extinctions of species, and we also analyze the astronomical context in which they occurred. 
 
 The Earth was formed by the accretion of planetesimals, where each impact on the Earth's surface  created a crater of molten rock. Using simple models, we estimate the cooling times of the craters by radiation and the average temperature of each layer of rock added to the Earth's surface. We conclude that the major source of the Earth's internal heat distribution was the Earth's gravitational contraction.
 
 Comets have played an important role in different stages of the evolution of the Solar System and of the Earth in particular. We present a brief history of the main theories on the origin and nature of comets. We propose the hypothesis that embryonic comets are gestated in dense clouds of plasma ejected by stars into the circumstellar and interstellar medium and that they complete their formation within cold and dense interstellar clouds of dust and molecular gas. Using simple models, we show that asteroids, precursors of minicomets or comets, with radii of up to  kilometers can be formed within dense plasma clouds. According to our hypothesis, the Sun's protoplanetary disk was formed within a molecular cloud containing interstellar comets.
The protoplanetary disk captured comets from the parent molecular cloud.
 A part of the captured comets was integrated into the protoplanetary disk as planetesimals; and the rest of the comets formed the comet cloud, called the Oort cloud, that  surrounds the Solar System.

 The great glaciations produced profound geological and biological consequences. In particular, the glaciations of the Cryogenic period, that lasted $\approx 200$ million years (Myr), turned the Earth into an ice ball and  meant a turning point in the history of life. The causes of global ice ages are not yet well known. First, we refer to Milankovitch's theory that explains the Quaternary glaciations as a consequence of the secular variations in the Earth's orbital parameters. Second, we refer to Arrhenius's climatic theory that explains glaciations by the reduction of greenhouse gases in the atmosphere (inverse greenhouse effect). We extended the Arrhenius model, taking into account the possibility of an atmosphere impregnated with volcanic ash and/or cosmic dust. Using our models, we calculate the evolution of the Earth's surface temperature in different scenarios, which include deposition of dust in the atmosphere due to comet collisions and/or immersion in dense interstellar clouds. Our models predict global land surface temperatures as low as -59 degrees Celsius.

 The Sun on its galactic path probably approached many times  interstellar clouds  and stars. Assuming a penetrating encounter between the Sun and a giant molecular cloud (GMC), we analyze the non-gravitational effects on the Solar System. The passage of the Sun through a dense clump of the GMC could trigger an ice age on Earth lasting  thousands of  years. In order to estimate gravitational effects of such encounters on the Solar System, we simulate a grazing encounter between the Sun and a GMC of $10^{4}$ M$_{\odot}$, where the relative velocity between the Sun and the GMC is 20 km s$^{-1}$ and the impact parameter is 20 pc. As a result of the encounter, tidal forces act on the Oort cloud, and trigger an intense shower of comets over the interior of the Solar System, with a high probability that  one comet  impacts the Earth.
   Close encounters with interstellar clouds, as well as stars, remove comets from the Solar System, and may have devastated the primordial comet cloud. Therefore,  we wonder whether  the Oort cloud was acquired in relatively recent times by the Solar System.

In previous studies by this author, it was found that the Sun could  be captured by a supercloud around 500-700  Myr ago. This explains the present position and kinematics of the Sun in the local system of gas and stars. The approximate  temporal coincidence of the capture of the Sun and the beginning of the Cryogenic period is very suggestive. This could indicate that the Sun was immersed for 200  Myr in a dense cloud of dust, molecular gas and interstellar comets. In this situation, we calculate that there was a massive capture of interstellar comets by the Sun, giving rise to a new Oort cloud. In addition, we estimate a copious fall of dust, molecular gas and comets on the Solar System, and in particular on the Earth.

We assume that, during the Cryogenic period, the Earth received abundant cosmic material, and we wonder whether there was also a panspermia event of complex life. An episode of complex life panspermia could explain the astonishing emergence of multicellular life at the beginning of the Cambrian period, a phenomenon known as ``the Cambrian explosion of life''. Since then, life suffered mass extinctions of species every $\approx 26$ Myr. This period of extinctions agrees with half the period of vertical oscillation of the Sun around the median plane of the supercloud, to which the Sun is gravitationally bound. The greatest disturbances to the Oort cloud, and the subsequent comet showers on Earth, occur in the encounters of the Sun with supercloud's substructures that lie near the extremes of the Sun's vertical oscillation.

\end{abstract}
\keyword{\textbf{Keywords}: Solar System--Earth--Comets--Oort's cloud--Comet impacts
--Interstellar clouds--Ice ages--Origin of life--Mass extinctions.}

\newpage

\section{Introducción}

\begin{quote}
\small \it{El drama de la vida no es en realidad sino una expresión particular y tal vez efímera de las fuerzas encontradas que han señalado el paso de la Tierra a través del tiempo y del espacio. Criaturas vivientes surgieron de algún modo del suelo y soportaron las vicisitudes de muchos largos periodos hasta llegar a la época presente. Su suerte y la suerte de la superficie de la Tierra ha sido una sola.} \rm 

John Hodgdon Bradley\index{Bradley, J.H.} (1898-1962) en el libro ``Autobiografía de la Tierra'' (Autobiography of Earth) \footnote{Libro publicado originalmente en el año 1935  pero que es  aún hoy un placer leerlo. Conceptos científicos  difíciles  son bellamente explicados  con  un lenguaje un tanto poético y filosófico.}
\end{quote}

El origen de la vida es un gran misterio, que quizás nunca  descubramos. Sin embargo, nuestros conocimientos actuales nos permiten entrever las fascinantes circunstancias en que la vida apareció y evolucionó en la Tierra. Al respecto, hay tres hechos intrigantes. Primero, la vida surgió en  la Tierra primitiva  apenas ésta estuvo en condiciones de albergarla, hace aproximadamente 3500 millones años. Segundo,  durante  los primeros 3000 millones años, es decir durante la mayor parte de la historia de la Tierra, la vida solo existió en forma bacteriana y unicelular. Tercero, la vida multicelular compleja emergió abruptamente hace apenas $\approx 550$ millones de años en un corto periodo geológico conocido como  ``la explosión del Cámbrico\index{Cámbrico, la explosión del}''. En esta monografía estudiaremos,  entre otras cosas, el contexto astronómico y geológico en el cual se desarrolló la vida  primitiva. Ello podría ayudarnos a entender los hechos arriba puntualizados.

Aquí tratamos un intervalo de tiempo geológico mucho mayor que el considerado en la parte I de esta monografía \citep{Olano5}, el cual se extiende desde la formación de la Tierra primitiva  (Sección \ref{TierraPrimitiva1}) hasta las últimas glaciaciones del Cuaternario (Sección \ref{ErasHielo}). Como parte de la formación de la Tierra, nos referimos brevemente a la formación del Sistema Tierra-Luna (Sección \ref{TierraLuna}), a la determinación de la edad de la Tierra como planeta (Sección \ref{EdadTierra}) y a la configuración del perfil térmico del interior de la Tierra, esencialmente  como producto de su contracción gravitatoria (Sección \ref{ContracciónTierra}).

Los enigmáticos cometas, que  son uno de los focos de este estudio,
han jugado probablemente un papel central en el origen de la vida y en las recurrentes catástrofes terrestres que provocaron extinciones masivas de plantas y animales.  En la Sección \ref{HistoriaCometas}, presentamos una breve historia sobre las teorías del origen y naturaleza de los cometas. También tratamos el medio interestelar como el substrato fundamental   donde nacen y mueren las estrellas (Sección \ref{MedioInterestelar}),  donde  se forman y diseminan los cometas interestelares (Sección \ref{CometasInterestelares}) y donde se forman las nubes moleculares y los sistemas protoplanetarios (Sección \ref{DiscoProtoplanetario}).

En numerosas ocasiones, la vida estuvo expuesta a circunstancias dramáticas, como aquellas ocasionadas por las prolongadas eras de hielo que convirtieron a la Tierra en una bola de hielo. En la Sección \ref{ErasHielo} y sus subsecciones, analizamos las posibles causas, tanto internas como externas a la Tierra, que ocasionaron las grandes glaciaciones. En primer lugar,  analizamos con algún detalle la teoría de Milankovitch\index{Milankovitch, M.}, que atribuye las variaciones climáticas cíclicas y las consecuentes glaciaciones del Cuaternario  a las variaciones seculares de los parámetros orbitales de la Tierra. En segundo lugar, analizamos el modelo climatológico de Arrhenius\index{Arrhenius, S.}, en el cual la concentración de gases de efecto invernadero determina la temperatura superficial de la Tierra, y extendemos el modelo  de Arrhenius\index{Arrhenius, S.} para estudiar el comportamiento climático con  una atmósfera impregnada de polvo volcánico y/o polvo cósmico. Para explicar las grandes eras de hielo, consideramos varios escenarios que incluyen el efecto invernadero inverso, actividad volcánica masiva, choques de grandes cometas sobre la Tierra y la inmersión de la Tierra en densas nubes oscuras interestelares.

En la Sección \ref{Penetrante1}, estudiamos los efectos del  encuentro del Sistema Solar con una densa nube interestelar. Primeramente, tratamos  un encuentro penetrante, donde los efectos no-gravitatorios como la deposición de gas y polvo en la alta atmósfera de la Tierra pueden ser importantes. Finalmente, tratamos  un encuentro rasante con una nube molecular gigante y  mostramos que los efectos gravitatorios sobre la nube de Oort\index{Oort, nube de} producen  una intensa lluvia de cometas sobre el interior del Sistema Solar y sobre la Tierra en particular.

 Por encuentros con estrellas y nubes interestelares, el Sistema Solar habría sido despojado de gran parte de  los cometas primordiales que forman la llamada nube de Oort. Por ello, en la Sección \ref{Reposición}, proponemos la hipótesis  de que gran parte de la nube de Oort se adquirió  durante el período geológico Criogénico, en el cual la Tierra sufrió una era de  hielo global que duró $\approx 200$ millones de años. En la Sección \ref{CapturaCometas}, demostramos que en  un lento encuentro penetrante   con una densa nube interestelar de polvo, gas molecular y cometas, el Sol puede  capturar una plétora de cometas de la nube interestelar.   
 
 El Sol y grupos de estrellas y nubes interestelares vecinos constituyen, según este autor, una supernube. En el encuentro circunstancial con la supernube en proceso de formación, el Sol quedó cautivo de ella. En la Sección \ref{TioVivo}, sugerimos que  la captura del Sol por parte de la supernube y el subsecuente encuentro penetrante del Sol con una condensación de la supernube condujeron a la captura masiva de cometas por parte del Sol y a las grandes glaciaciones del Criogénico. Sugerimos también que las consecuencias de esos hechos serían  notables: 1) Evento de panspermia de vida evolucionada y la explosión de vida del Cámbrico (Sección\ref{panspermia}); 2)  Encuentros cuasi periódicos  del Sol con subestructuras de la supernube y extinciones masivas de  especies de plantas y animales (Sección \ref{TioVivo}).
 
\section{Formación de la Tierra primitiva \label{TierraPrimitiva1}}

En el frío disco proto planetario con temperaturas cercanas al cero absoluto, se formaron copos de nieve a los que se les fueron pegando granos de polvo, átomos y moléculas gaseosas  creándose cuerpos llamados  planetesimales  \footnote{Este es el mecanismo clásico invocado para explicar la formación de los planetesimales. Sin embargo,  nosotros mostraremos en esta monografía que dicho mecanismo puede ser mucho más complejo (Ver Sección \ref{CometasInterestelares}).}. Los planetesimales fueron los ladrillos con los cuales se construyó el  sistema planetario. Se piensa que los planetas, y en particular la Tierra, se formaron  acrecentando   
planetesimales. Las colisiones inelásticas entre planetesimales formaron cuerpos mayores que finalmente constituyeron los planetas con órbitas definidas. 
Hoy se sabe que los casi 10.000 asteroides que se encuentran entre Marte y Júpiter son en realidad los planetesimales que no lograron fundirse para formar un planeta, debido a las intensas fuerzas de marea gravitacionales ejercidas principalmente por Júpiter.

\subsection{Modelo de una Tierra ígnea}

Se podría pensar que como consecuencia de la cuantiosa energía liberada en las múltiples colisiones, la Tierra fue en un principio un cuerpo ígneo de muy alta temperatura, o  en otras palabras, {\it una bola de roca fundida incandescente}. A fin de discutir esta posibilidad,
haremos algunos cálculos elementales. Supongamos que la proto-Tierra incorporó $n$ planetesimales, y que  queremos saber en cuanto se incrementa la temperatura de la proto-Tierra después de incorporar un nuevo planetesimal.
La relación entre el calor o energía  $\Delta Q$ que ingresa o escapa de un cuerpo de masa $m$ y la variación de la temperatura $ \Delta T$ de dicho cuerpo, que se produce como consecuencia,  es dada por la ecuación
\begin{equation}
\Delta Q=m\, c_{m}\, \Delta T, 
\label{calor}
\end{equation}
donde $c_{m}$  es el calor especifico del material del cuerpo \footnote{El calor específico ($c_{m}$) es la energía necesaria para elevar en un 1 grado la temperatura de 1 kg de masa. Sus unidades en el Sistema Internacional son J/(kg K). En rigor $c_{m}$ depende de la temperatura del cuerpo, pero para nuestro propósito puede tomarse como una constante.  A temperaturas muy bajas $c_{m}$ tiende a cero, un efecto cuántico que fue explicado por Einstein\index{Einstein, A.} en el año 1907. En nuestro caso tampoco es necesario diferenciar entre las capacidades calóricas a volumen constante $c_{v}$ y a presión constante $c_{p}$.}.

Si, al momento del choque, la velocidad del planetesimal con respecto la proto-Tierra es $v_{P}$, la energía que se inyecta en la proto-Tierra es $\Delta Q=\frac{1}{2} m_{P}\, v_{P}^{2}$, es decir la energía cinética del planetesimal, donde $m_{P}$ es la masa del planetesimal. Si todos los planetesimales tienen la misma masa $m_{P}$, entonces la masa de la proto-Tierra es $m=n\,  m_{P}$. 

A partir de la ecuación (\ref{calor}), obtenemos que el incremento de la temperatura de la proto-Tierra debido al choque con el planetesimal es $\Delta T=\frac{\Delta Q}{m \, c_{m}}=\frac{\frac{1}{2} m_{P}\, v_{P}^{2}}{n\,  m_{P}\, c_{m}}=\frac{\frac{1}{2} v_{P}^{2}}{n\,  c_{m}}$. Cuando los planetesimales se fueron haciendo escasos, el proceso de acrecimiento se detuvo. En la última colisión,  $n=N-1$, donde $N$ es el número total de planetesimales que colisionaron para formar la Tierra, y  $m=M_{T}-m_{P}$, donde $M_{T}$ es la masa de la Tierra. Dado que  $m=n\,  m_{P}$,   $N=\frac{M_{T}}{ m_{P}}$. A fin de obtener una estimación de $N$, supondremos  que la densidad de los  planetesimales fue igual a la de la Tierra. Entonces $N=\left(\frac{R_{T}}{R_{P}}\right)^{3} \approx 600^{3}$, adoptando $R_{T}= 6000 $ km para el radio de la   Tierra y $R_{P}=10$ km para el radio del planetesimal. 

Si despreciamos las pérdidas de calor y consideramos que todos  los planetesimales que incorporó la Tierra tenían la misma velocidad $v_{P}$, la temperatura de la Tierra al finalizar el proceso de acrecimiento es dada por 
\begin{equation}
T=\sum_{n=1}^{N-1} \Delta T=\frac{v_{P}^{2}}{2 \, c_{m}} \sum_{n=1}^{N-1} \frac{1}{n}.
\label{Tierra1}
\end{equation}

Dado que $N=216 \times 10^{6}$, $\sum_{n=1}^{N-1} \,\,(\frac{1}{n})=1+\frac{1}{2}+\frac{1}{3}+\frac{1}{4}+\frac{1}{5}+...=19.768$. Por lo tanto, la ecuación (\ref{Tierra1}) puede escribirse del siguiente modo:
\begin{equation}
T=19.768 \frac{v_{P}^{2}}{2 \, c_{m}}.
\label{Tierra2}
\end{equation} 
Reemplazando en la ecuación (\ref{Tierra2})  $c_{m}$ por el calor específico de la roca fundida (o magma),  el cual es del orden de 1135 J/(kg K), y $v_{P}$ por 1000 m/s (=1 km/s), una velocidad relativamente pequeña comparada con la velocidad orbital de la Tierra ($\approx 30$ km/s), encontramos que la Tierra primitiva tenía una temperatura $T \approx  8800$ K. Valor muy superior  a la temperatura de fusión  de la roca a la presión atmosférica. El resultado de $T$ depende fuertemente del valor adoptado para $v_{P}$. Si usamos la velocidad cósmica con la que chocan actualmente los meteoritos, $v_{P}\approx 20$ km s$^{-1}$, debemos multiplicar por 400 el valor que obtuvimos arriba para la temperatura de la Tierra primitiva.

Este problema también puede resolverse, a partir de (\ref{calor}),  con la siguiente ecuación diferencial: $m c_{m}dT= dQ$, donde $dQ=\frac{1}{2} m_{P}\, v_{P}^{2} dn$ y $m=n m_{p}$. Separando las variables, a un  lado del signo igual la variable T y al otro lado la variable n,  e integrando, obtenemos 
$T= \frac{v_{P}^{2}}{2 \, c_{m}} \int_{1}^{N} \frac{1}{n} dn$. 
Es decir, $T= (ln N) \frac{v_{P}^{2}}{2 \, c_{m}} $ y reemplazando N por el valor hallado arriba, obtenemos  casi la misma ecuación que (\ref{Tierra2}).

El resultado sería que, terminada la etapa de fusión de planetesimales, la tierra fue un cuerpo ígneo de alta temperatura. El cálculo de la temperatura se basó en la suposición de que las pérdidas energéticas no fueron importantes y de que el calor se repartió uniformemente en la masa de la Tierra. ¿Pero realmente fue así?

 \subsection{Modelo de una Tierra congelada o tibia \label{TierraTibia}}
Para contestar la pregunta formulada en el última frase del sección anterior, a continuación consideraremos la generación y pérdida de calor en un proceso elemental de acreción  de masa ; es decir, la adición de un planetesimal a la Tierra primitiva. La Tierra se formó ``choque a choque''. En cada  choque de un planetesimal con la superficie solidificada de la Tierra, se forma un cráter con una capa de lava de una temperatura del orden de 1200 K. Si esta capa de lava es suficientemente delgada, la pérdida de calor por irradiación de su superficie  expuesta al vacío puede enfriar uniformemente a dicha capa de lava. Con el propósito de estimar el tiempo que demanda el enfriamiento de la capa de lava, utilizaremos la ley de Stefan-Boltzmann\index{Stefan-Boltzmann, ley de}. Esta ley establece  que un cuerpo con temperatura $T$ y superficie $S$ irradia por segundo la energía dada por $S \sigma T^{4}$, donde $\sigma (=5.670 \times \rm 10^{-8} J\, m^{-2}\, s^{-1} \,K^{-4})$  es la constante de Stefan-Boltzmann. Por lo tanto, la pérdida de calor de dicho cuerpo por unidad de tiempo ($\frac{dQ}{dt}$) debe ser igual la energía irradiada, es decir 
 $\frac{dQ}{dt}=-S \sigma T^{4}$. Introdujimos el signo menos porque se trata de una pérdida de energía. Reemplazando $dQ$ por la expresión (\ref{calor}),  obtenemos
 \begin{equation}
 m\, c_{m}\,  \frac{dT}{dt}=-S \sigma T^{4}.
 \label{corteza}
 \end{equation}
 Si $\rho_{m}$ representa la densidad del magma, $S$ la superficie externa del cráter y $E$ espesor de la capa de lava del cráter, $m = \rho_{m} S E$. Así, la ecuación diferencial (\ref{corteza}) se convierte en $\rho_{m}  E c_{m}\,  \frac{dT}{dt}=-\sigma T^{4}$, cuyos términos  pueden también ser ordenados del siguiente modo:
 \begin{equation}
 dt=-\frac{\rho_{m}\, E\, c_{m}}{\sigma} \frac{dT}{T^{4}},
 \label{corteza2}
 \end{equation}
  donde el miembro izquierdo solo depende del tiempo y el derecho solo de la temperatura, permitiendo integrarlos separadamente. Entonces, $\int_{t_{0}}^{t} dt= -\frac{\rho_{m}\,  E\, c_{m}}{\sigma} \int_{T_{0}}^{T} \frac{1}{T^{4}} dT$, donde $t_{0}$ y $T_{0}$ son el tiempo y la temperatura inicial, respectivamente. Con lo cual obtenemos, 
  \begin{equation}
  t-t_{0}=-\frac{\rho_{m}\,  E\, c_{m}}{\sigma} \left(\frac{1}{3 T_{0}^{3}}-\frac{1}{3 T^{3}}\right).
 \label{tiempoE}
  \end{equation}

A partir de la ecuación (\ref{tiempoE}) podemos obtener una estimación de nuestra incógnita: t,  el tiempo de enfriamiento del cráter. Al formarse el cráter, $t_{0}=0$, suponemos que  $T_{0}=1200$ K (temperatura a la cual se funde el granito) y  además que la etapa de enfriamiento del cráter termina cuando la temperatura del magma solidificado alcanza T=100 K (173 grados centígrados bajo cero, similar a la temperatura de la cara oscura de la Luna). Si adoptamos   $E=10$ metros y  $\rho_{m}= 3 \rm \,g \,cm^{-3}$ para la densidad del magma, resulta que $t= 6$ años, un instante para los tiempos geológicos.

Si la capa de lava del cráter  es muy gruesa, el enfriamiento de la misma no solo depende del calor que pierde por radiación térmica  una delgada capa externa, sino también del  transporte  del calor por conducción a través de todo su interior.  Por ello, planteamos ahora una aproximación mejor que la dada arriba para el cálculo del enfriamiento de un cráter. Dividimos la capa de lava de espesor $E$ en $N$ subcapas paralelas al suelo del cráter y por lo tanto el espesor de cada subcapa es $\Delta x= \frac{E}{N}$. Con el índice $i=1$, identificamos a la subcapa que tiene  la superficie superior expuesta al vacío e irradia térmicamente según la ley de Stefan-Boltzmann\index{Stefan-Boltzmann, ley de} de cuerpo negro. Con  $i=N$, denotamos la última subcapa, asentada sobre la superficie fría de la Tierra. Como suponemos una estructura de capas planas paralelas, en la cual su parte superior pierde calor por irradiación y la inferior por contacto con el suelo de la Tierra a una  temperatura de solo 100 K, se produce un flujo unidimensional de calor en  la  dirección $x$ perpendicular a la capa de lava.

  La ley de Fourier\index{Fourier, ley de} sobre la conducción del calor aplicada a un flujo unidimensional puede escribirse
  
 \begin{equation}
  \frac{\Delta Q}{\Delta t}=- \lambda_{0} S \frac{\Delta T(x,t)}{\Delta x},
  \label{LeyFourier}
  \end{equation}
  donde $S$ es la superficie de la capa de lava y $\lambda_{0}$ es el coeficiente de conductividad  de las rocas ígneas. La ley (\ref{LeyFourier}) establece que, en un cierto tiempo $t$,  el flujo  de calor en la posición $x$ es proporcional al gradiente de temperatura en dicha posición. Nuestro problema es encontrar, a partir de una especificada distribución de temperatura inicial $T(x,0)$ de la capa de lava, la distribución $T(x,t)$ en tiempos posteriores.
  
Denotaremos la posición de la subcapa  $i$ por $x_{i}$ y su temperatura $T(x_{i},t)$ por $T_{i}(t)$ para abreviar. El flujo de calor que entra o sale de la subcapa $i$ de o hacia  las subcapas  adyacentes es, según (\ref{LeyFourier}),  $- \lambda_{0} S \frac{ T_{i+1}(t) -T_{i}(t)}{\Delta x}$ con la que yace debajo ($i+1$), y $- \lambda_{0} S \frac{ T_{i}(t) -T_{i-1}(t)}{\Delta x}$ con la que yace encima ($i-1$). Por lo tanto, el flujo neto que pasa por la subcapa $i$ es
\begin{equation}
\frac{\Delta Q_{i}}{\Delta t}= \lambda_{0} S \left(\frac{T_{i+1}(t)-T_{i}(t)}{\Delta x}-\frac{T_{i}(t)-T_{i-1}(t)}{\Delta x}\right).
\label{FlujoCalor1}
\end{equation}
  El calor ganado o perdido  por la subcapa i produce un cambio de su temperatura $\Delta T _{i}(t)$ de acuerdo con  (\ref{calor}): $\Delta Q_{i}= c_{m} m_{i} \Delta T _{i}(t)$. Suponiendo que la capa de lava es homogénea y de densidad $\rho_{m}$, todas las subcapas tienen la misma masa y por lo  tanto $m= \rho_{m} S \Delta x$ y $\Delta Q_{i}= c_{m}  \rho_{m} S \Delta x \Delta T _{i}(t)$. Reemplazando esta expresión para  
  $\Delta Q_{i}$ en  (\ref{FlujoCalor1}),  obtenemos:
  \begin{equation}
  \frac{\Delta T_{i}(t)}{\Delta t}= \lambda_{1}  \frac{1}{\Delta x}\left(\frac{T_{i+1}(t)-T_{i}(t)}{\Delta x}-\frac{T_{i}(t)-T_{i-1}(t)}{\Delta x}\right),
  \label{FlujoCalor2}
  \end{equation}
donde $\lambda_{1}=\frac{\lambda_{0}}{c_{m} \rho_{m}}$. Si hacemos tender $\Delta x$ y $\Delta t$ a valores infinitamente pequeños, (\ref{FlujoCalor2}) se convierte en la famosa ecuación de conducción térmica de Fourier\index{Fourier, J.B.}
\begin{equation}
\frac{\partial T(x,t)}{\partial t}= \lambda_{1} \frac{\partial^{2} T(x,t)}{\partial x^{2}}.
\label{Fourier}
\end{equation} 
Esta ecuación es de interés histórico \footnote{A fin de  resolver analíticamente la ecuación (\ref{Fourier}), Fourier\index{Fourier, J.B.} desarrolló en 1810-20 la teoría de  series infinitas de  senos y cosenos, hoy conocidas como series de Fourier\index{Fourier, J.B.}.}.

La ecuación (\ref{FlujoCalor2}) gobierna el flujo de calor en las subcapas $i\geq 2$. En cambio, la superficie superior de la  subcapa 1 ($i=1$) pierde calor por radiación de acuerdo con  ley de Stefan-Boltzmann\index{Stefan-Boltzmann, ley de} de cuerpo negro :$\frac{\Delta Q_{-}}{\Delta t}=- \sigma S T_{1}(t)^{4}$. Mientras, el flujo de calor que ingresa a la subcapa 1 por su superficie inferior, en contacto con la subcapa 2, es $\frac{\Delta Q_{+}}{\Delta t}= \lambda_{0} S (\frac{T_{2}(t)-T_{1}(t)}{\Delta x})$. Entonces, a partir  del flujo neto, $\frac{\Delta Q_{+}}{\Delta t}+\frac{\Delta Q_{-}}{\Delta t}$, y la fórmula (\ref{calor}), obtenemos

\begin{equation}
\frac{\Delta T_{1}(t)}{\Delta t}=  \frac{1}{\Delta x} \left( \lambda_{1} \frac{T_{2}(t)-T_{1}(t)}{\Delta x}-  \lambda_{2} T_{1}(t)^{4}\right),
\label{FlujoCalor3}
\end{equation} 
donde $\lambda_{2}=\frac{\sigma}{c_{m} \rho_{m}}$.

Ahora,  mediante las ecuaciones (\ref{FlujoCalor2}) y (\ref{FlujoCalor3}),  calculamos el enfriamiento de una capa de lava, o variación de su temperatura $T(x,t)$, que tiene inicialmente una temperatura uniforme  de $1200$ K. Es decir, suponemos que todas las subcapas tienen  la misma temperatura inicial, lo cual  en términos matemáticos equivale a que, en $t=0$, $T_{i}(0)= 1200$ K para $i=1,2,3...N$. A fin de calcular el gradiente de temperatura de la subcapa $N$, necesitamos conocer la temperatura $T_{N+1}$ correspondiente a la temperatura del suelo sobre el cual se apoya la subcapa $N$. Suponemos que $T_{N+1}(t)=100$ K para todo tiempo $t$. Adoptamos para $\Delta t$ y $\Delta x$ un valor constante y suficientemente pequeño. Las ecuaciones (\ref{FlujoCalor2}) y (\ref{FlujoCalor3})  nos permiten determinar $\Delta T_{i}(0)$ conociendo $T_{i}(0)$, y obtener las temperaturas de las subcapas en el primer paso, $t=\Delta t$, $T_{i}(\Delta t)= T_{i}(0)+ \Delta T_{i}(0)$. En este primer paso en $t$, solo las temperaturas de la subcapa 1 y $N$ varían, luego por un efecto tipo ``dominó'' variaran las temperaturas de las subcapas del interior. Nuevamente usando  (\ref{FlujoCalor2}) y (\ref{FlujoCalor3})  con $T_{i}(\Delta t)$ determinamos $\Delta T_{i}(\Delta t)$ y con ello las temperaturas del siguiente paso, $T_{i}(2 \Delta t)= T_{i}(\Delta t)+ \Delta T_{i}(\Delta t)$. Asi, siguiendo, podemos calcular las  temperaturas de las diferentes  subcapas en un instante posterior a $t=n \Delta t$   usando (\ref{FlujoCalor2}) y (\ref{FlujoCalor3}) y la fórmula recursiva $T_{i}(t+\Delta t)= T_{i}(t)+ \Delta T_{i}(t)$.

\begin{figure}
\includegraphics[scale=0.9]{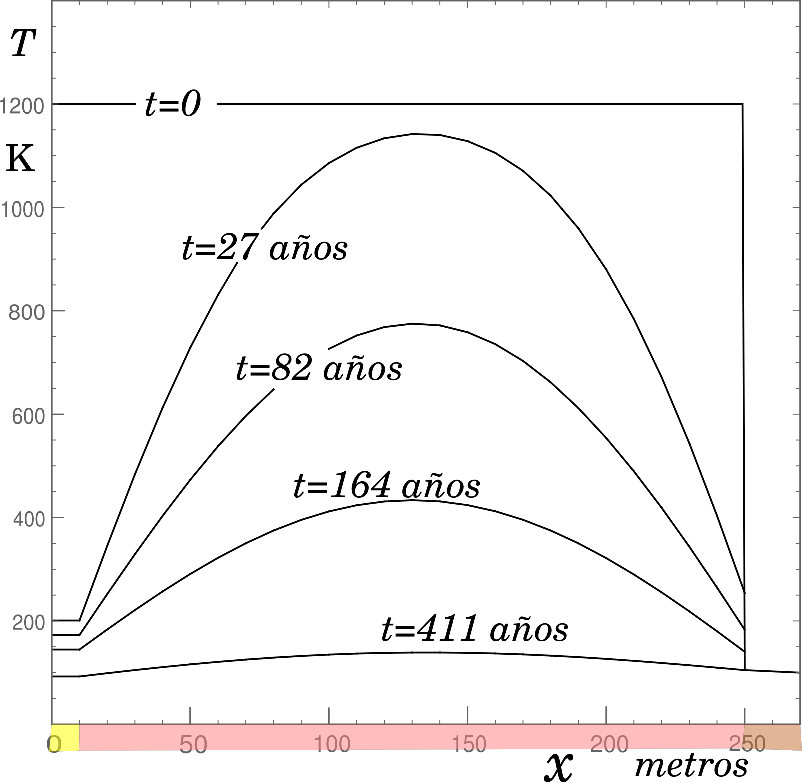} 
\caption{Evolución de la distribución de temperatura de la capa de lava de un cráter como el de  Sudbury\index{Sudbury, cráter}. El tiempo indicado sobre las curvas es el tiempo transcurrido desde el origen del cráter. En la escala vertical se representa la temperatura en grados Kelvin y en la horizontal la profundidad de la capa en metros. La zona amarilla es el estrato cuya superficie expuesta al vacío irradia térmicamente. La zona marrón en el extremo derecho representa el suelo terrestre.}
\label{PerfilTemperatura}
\end{figure}

Como ejemplo, estimaremos el tiempo de enfriamiento de uno de los cráteres de impacto más grandes y antiguos de la Tierra: el cráter Sudbury\index{Sudbury, cráter} (Canadá) (ver Tabla 1 de \citep{Olano5}). Se estima que  este cráter con un diámetro $D_{c}= 200$ km  tuvo un volumen de lava $V_{c}$= 8000 km$^{3}$ \footnote{ver Tabla 1 de R.A.F. Grieve\index{Grieve, R.A.F.} y M.J. Cintala\index{Cintala, M.J.} 1992}. Por lo tanto, el espesor medio de la capa de lava es $E=\frac{V_{c}}{\pi(0.5 D_{c})^{2}} \approx 250$ metros. Adoptamos $\Delta t= 1$ día, $\Delta x=10$ m y la constante $k= 1.86 \times 10^{-6}$ m$^{2}$ s$^{-1}$. La Fig. \ref{PerfilTemperatura} representa la distribución de la temperatura del cráter para distintos tiempos transcurridos desde el momento de su formación (t=0). La Fig. \ref{PerfilTemperatura} muestra que el enfriamiento de la superficie y la base del cráter fue muy rápido. A partir de las distribuciones representadas en la Fig. \ref{PerfilTemperatura}, podemos considerar que el tiempo de enfriamiento fue de  411 años, dado que en ese tiempo todo el cráter alcanzó una temperatura mínima y uniforme de 100 K.  Si hubo en la capa de lava corrientes convectivas que transportaron lava desde el interior a la superficie,  el tiempo de enfriamiento pudo  reducirse significativamente. En efecto, la lava en la superficie se enfría por radiación muy eficientemente y puede  agrietarse. A través de esas grietas, lava más caliente puede   emerger desde el interior  y también enfriarse rápidamente por radiación.
 
 Consideremos que la Tierra en formación incorporó una capa de planetesimales  cuando los cráteres de impactos de estos cubrieron  la superficie total de la Tierra. Si todos los cráteres tuvieron el mismo radio $r_{c}=\frac{D_{c}}{2} =100$ km,  entonces el número de cráteres de la capa $n_{c}$ debe cumplir la siguiente relación: $\epsilon \pi r_{c}^{2} n_{c}=4 \pi R_{T}(t)^{2}$, donde el radio de la tierra $R_{T}$ depende del tiempo y $\epsilon$ tiene en cuenta el porcentaje  de superposición de los cráteres. Por lo tanto, $n_{c}=\frac{4 \pi R_{T}(t)^{2}}{\epsilon \pi r_{c}^{2}}$. El espesor de la capa de planetesimales, que es igual al incremento del radio de la Tierra $\Delta R_{T}$, debe cumplir la relación: $4 \pi R_{T}(t)^{2} \Delta R_{T}= \frac{4}{3} R_{p}^{3} n_{c}$, donde $R_{p}$ es el radio de los planetesimales. Por lo tanto, $\Delta R_{T}=\frac{4 R_{p}^{3}}{3 \epsilon r_{c}^{2}}$. Note que $\Delta R_{T}$ es una constante y entonces el número de capas de planetesimales que se necesitó  para formar la Tierra es $\cal N$= $\frac{R_{T}}{\Delta R_{T}}=\frac{3 \epsilon}{4} \frac{R_{T} r_{c}^{2}}{R_{p}^{3}}$, donde $R_{T}$ es el radio final de la Tierra ($\approx 6400$ km). Si designamos $\Delta t_{c}$ al tiempo que  se tardó en formar cada capa de planetesimales, el tiempo total que demandó la formación de Tierra es $t_{f}= \cal N$ $\Delta t_{c}$. Si $\Delta t_{c}$ fue mayor que el tiempo de enfriamiento $t_{e}$ de los cráteres, cada nueva capa  se formó sobre la capa anterior que ya se encontraba en su mayor parte solidificada y fría. De acuerdo con dicha posibilidad, adoptaremos   $\Delta t_{c}=2 t_{e}$, donde  $t_{e}=411$ años según el resultado arriba obtenido. Si el porcentaje de superposición de los cráteres fue del 50 por ciento,  $\epsilon=0.5$. Si además suponemos que   $R_{p}=5$ km y  $r_{c}=100$ km,  encontramos que  $t_{f}= 1.6 \times 10^{8}$ años. A pesar de lo rudimentario de nuestro modelo, el tiempo de formación de la Tierra de 160 millones de años que hemos  estimado está dentro del rango de valores dados por  modelos detallados de acreción de planetesimales.

   En base a esos resultados, podemos decir que cada nueva capa de planetesimales que la Tierra acumuló se formó sobre las capas subyacentes ya casi totalmente frías. Por otra parte, cuando la Tierra adquirió un tamaño considerable, la superficie de ésta afectada por el impacto de un planetesimal es pequeña: el cráter derretido es de sólo  unos pocos  cientos kilómetros de radio (ejemplo el cráter de Chicxulub\index{Chicxulub, cráter}, ver Tabla 1 de \citep{Olano5}).   Por lo tanto, despreciar la pérdida de calor, como lo hemos hecho más arriba, es incorrecto. Entonces, los procesos de ganancia y pérdida  de energía deben ser tratados  simultáneamente.
  
Esto nos muestra que  la Tierra  pudo formarse a temperaturas muy bajas, en contra de la idea tradicional de que la Tierra tuvo un origen ígneo. La hipótesis  del origen ígneo tiene el atractivo de que explica el actual interior caliente de la Tierra (así lo prueba la  erupción volcánica de magma) como un remanente del calor  original de la Tierra. En cambio, si la  Tierra primitiva fue fría, debemos explicar por qué  en el presente la Tierra tiene tan altas temperaturas en su interior (ver Sección \ref{ContracciónTierra}).

\section{Teorías sobre la formación del sistema Tierra Luna \index{La Luna} \label{TierraLuna}}

\begin{quote}
\small \it{Estas son conjeturas sobre la formación de las estrellas y el sistema solar, conjeturas, que yo plasmo con toda la duda que debe inspirarme todo aquello que no es resultado de las observaciones o cálculos.} \rm 

Pedro Simón Laplace \index{Laplace, P.S.} (1749-1827)
\end{quote}

Según el modelo estándar,  los planetas crecieron por aglomeración de planetesimales que se formaron en la nebulosa solar. A la luz de dicho modelo,
hemos encontrado muchas piezas sueltas del rompecabezas pero aún nos falta mucho para completar el armado. En verdad,  nuestro conocimiento de la formación de la Tierra es rudimentario. Ahora está a nuestro alcance el estudio de otras estrellas con discos protoplanetarios en distintos estados evolutivos, y ello sin dudas nos permitirá obtener conocimientos más asentados sobre la formación de los planetas y lunas.   Los eventos que hemos descripto  son en realidad  hipótesis de trabajo que no están totalmente corroboradas, pero aun así son muy útiles porque nos permiten avanzar dentro de un programa de investigación. Por ejemplo, hemos visto en la sección anterior que la colisión de los planetesimales produce un calentamiento superficial del planeta en formación, que rápidamente se enfría radiativamente. Esto es lógico dado que el contenido de calor es proporcional al volumen de la masa  fundida por el impacto y la pérdida de calor por radiación es proporcional a la superficie expuesta de esa masa; y la relación de dicha superficie a su volumen es comparativamente grande.
Sin embargo, si grandes cuerpos del tamaño de la Luna (``embriones planetarios'')  chocan, la razón superficie-volumen de la parte fusionada es relativamente pequeña, y una parte importante del calor generado queda atrapado en el interior del protoplaneta y no puede fácilmente escapar al exterior. Los embriones planetarios se formaron por acumulación de planetesimales y a su vez  algunos planetas mayores podrían haberse formado por la colisión y fusión  de  dos o más embriones planetarios.

En la sección siguiente tratamos otros procesos que generan calor en las profundidades del protoplaneta, quedando allí confinado por largo tiempo. Como veremos, la autogravedad de la Tierra produce  acomodamientos y compactación de su masa interior, generando calor.  

Otro fenómeno que nos intriga es el alto contenido de hierro del núcleo de la Tierra. ¿Se debe esto a la decantación por gravedad de los elementos pesados hacia el centro de la Tierra? También podría explicarse si los primeros planetesimales acumulados por la Tierra eran ricos en hierro. En efecto, si la Tierra migró  a la órbita actual desde  una órbita  más alejada del Sol,  la composición del  material con el cual la Tierra creció en la etapa inicial  puede ser distinta a la del material que incorporó en la etapa final. Por último, nos referiremos  brevemente a otro hecho intrigante: la formación de la Luna. Hay varias hipótesis o teorías sobre el origen de la Luna que pueden dividirse del siguiente modo:

 a) {\bf Hipótesis de la fisión}. \rm  Una porción de la Tierra se desprendió, debido a su originalmente  alta velocidad de rotación, formando la Luna. La idea de la fisión de la Tierra fue propuesta por  George Darwin \index{Darwin, G.} (hijo del famoso biólogo Charles Darwin) \index{Darwin, C.} y mantuvo su popularidad hasta bien entrado el siglo veinte. El geólogo austríaco Otto Ampherer  \index{Ampherer, O.} vio en  1925 que el desprendimiento de una parte de la Tierra podría ser la causa de la deriva de los continentes \footnote{Hoy se sabe que los continentes son arrastrados por las corrientes convectivas de magma del manto terrestre.}, fenómeno descubierto unos años antes por Alfred Wegener\index{Wegener, A.} \footnote{Su nombre se asocia con toda justicia a este descubrimiento, si bien también otros se habían percatado de la coincidencia del contorno de los continentes que bordean el Atlántico. Uno de  ellos fue el explorador alemán Alejandro von Humboldt,\index{von Humboldt, A.} pero no sugirió la ruptura y desplazamiento  de los continentes como explicación.}. Se pensó también que el Océano Pacífico era la cicatriz de tal suceso;  pero  hoy sabemos que la corteza oceánica es mucho más joven que la Luna. Sin embargo, ese hecho no invalida la teoría. Un  punto a favor de la teoría de fisión terrestre es que puede explicar el hecho de que  la composición de las rocas lunares (recogidas por las misiones Apolo\index{Apolo, misión espacial})  es muy similar a la composición del manto terrestre. El lado débil de la teoría es que no puede explicar el momento angular del sistema Tierra-Luna.
 
 b) {\bf Hipótesis de la co-formación}. \rm La Luna se formó junto  a la Tierra formando un sistema binario. Esta hipótesis no puede dar cuenta de la composición lunar ni del momento angular del sistema. Si bien esta hipótesis no concuerda con los hechos del sistema Tierra-Luna,  las lunas de Júpiter sí  parecen haberse formado junto a Júpiter en un disco circum-planetario de gas y polvo, de manera análoga a la formación de los planetas en la nebulosa solar.
 Cuando Galileo\index{Galileo} descubrió los cuatro principales satélites de Júpiter, conocidos ahora como las lunas Galileanas, vio en ellos un sistema análogo al heliocéntrico en donde los cuerpos menores giraban en torno al  principal. Los planetas exteriores se diferencian de los interiores en que los primeros tienen en general muchas lunas.

 c) {\bf {Hipótesis de la captura}. \rm  La Luna se formó en otro lugar del sistema solar y fue capturada por la Tierra. La dificultad de esta hipótesis es que no puede explicar la composición química  y la dinámica orbital de la Luna. Además para que se produzca la captura, la luna debe interactuar antes con al menos un tercer cuerpo cediendo a éste su exceso de energía para quedar atrapada en el campo gravitatorio de la Tierra. El geólogo Norteamericano  F.B. Taylor\index{Taylor, F.B.}, un antecesor de Wegener\index{Wegener, A.}, propuso en 1910 que los continentes fueron arrastrados en gran escala hacia el ecuador como consecuencia de las fuerzas de mareas originadas por la hipotética captura de la Luna. Según esta propuesta, la captura habría ocurrido hacia el final del Cretácico, \index{Cretácico} es decir apenas 65 millones atrás (en el fin de la era de los dinosaurios). A pesar de que el mecanismo propuesto  por Taylor\index{Taylor, F.B.} no es correcto, su atrevida idea de la deriva continental ha sido ampliamente corroborada.
 
 d) {\bf Hipótesis de la gran colisión} \rm. Esta hipótesis, formulada en el año 1975 por  William K. Hartmann\index{Hartmann, W.K} y  Donald R. Davis\index{Davis, D.R.}, plantea que  la colisión de un gran cuerpo del tamaño de Marte con la Tierra eyectó material
 fundido  entorno a la Tierra, material  que luego se reagrupó formando la Luna. Esta violenta colisión contra la Tierra, ya casi formada,  debió ser una colisión excéntrica u oblicua, con lo cual se puede explicar el momento angular actual del sistema Tierra-Luna, como  así también la deficiencia lunar en hierro. El choque le imprimió a la Tierra su rápida  velocidad inicial de rotación \footnote{Las fuerzas de marea de la Luna desaceleraron desde entonces la rotación de La Tierra. En ese entonces, hace $\approx$ 4500 millones de años, la duración del día era mucho menor que la actual. Si la desaceleración ha sido siempre como la actual, en el comienzo el día fue de sólo 3 horas.}. Parte del material fundido del cuerpo que impactó y del arrancado de la Tierra circundó la Tierra, moviéndose en el mismo sentido que la rotación de  la Tierra; rotación que el mismo impacto produjo. Por ello la Luna gira en torno a la Tierra en el mismo sentido que la Tierra gira sobre su eje \footnote{El lector puede comprobar que la Luna se desplaza hacia el Este con respecto a las estrella fijas ($\approx 12$ grados por día) y como es bien sabido la Tierra también gira hacia el Este, lo cual hace que los astros se pongan por el Oeste.}. Como la Luna es pobre en hierro, se piensa que  gran parte del material caliente expelido con el cual se formó la Luna provino del manto de ambos planetas.
 
 Las hipótesis a, b y c no superan las pruebas y la única que quedaría en pie es la hipótesis de la gran colisión. Sin embargo, aún ésta última no esta exenta de dificultades. Un hecho inesperado en el contexto de esta hipótesis es que  las abundancias isotópicas del oxígeno de la Tierra y de la Luna son indistinguibles con un alto nivel de precisión,  cuando pequeñas diferencias en  la composición de la proto Tierra y la del cuerpo que impactó deberían observarse. La explicación estándar  es que el cuerpo que chocó contra la Tierra se formó en la misma región del Sistema Solar que se formó la Tierra. La hipótesis del gran impacto es aplicable, en principio, sólo al  origen  de nuestra Luna, como un caso singular. Esto es en apariencia un punto en  contra de la teoría. Sin embargo, hay un hecho que habla en favor del origen contingente de la Luna. En efecto, el tamaño de nuestra Luna es excepcionalmente grande  en relación al tamaño del planeta al cual está gravitacionalmente ligada.

 \section{El calor interno de la Tierra \label{ContracciónTierra}}
 \begin{quote}
\small \it{Es como si debiéramos recomponer los fragmentos rotos de un periódico comparando primero la coincidencia de los bordes y luego las de las líneas de texto impreso. Si coinciden, sólo podemos concluir que los trozos estaban originalmente en esa posición. Si solo dispusiéramos de una línea escrita, aún tendríamos una importante probabilidad de que el ajuste fuera exacto: pero si tenemos $n$ líneas, entonces esa probabilidad se elevará a la enésima potencia.}\footnote{Celebrada analogía del padre de la deriva continental. Un ejemplo de una ``línea impresa'' es la cadena de montañas del Cabo, Sudáfrica\index{Sudáfrica}, que se extiende de este a oeste y se formó en el Pérmico\index{Pérmico}. Su prolongación occidental se encuentra en un pliegue similar constituido por las sierras de la ventana y de Tandil\index{Tandil}, Argentina\index{Argentina}.} \rm
Alfred Wegener\index{Wegener, A.} (1880-1930) 
\end{quote}
 
\begin{quote}
\small \it{A la luz del conocimiento moderno, el estudioso debe considerar las tremendas presiones en el interior profundo como la causa principal de la contracción. Bajo estas presiones, es posible que las materias sufran redistribuciones moleculares progresivas que se traducen en nuevos compuestos de mayor densidad  y menor volumen. Es así como, a medida que pasaban los siglos, la tierra puede haberse vuelto más densa y más pequeña.}
\rm 

John Hodgdon Bradley \index{Bradley, J.H.} (1898-1962)  
\end{quote}

Hemos visto que el proceso de acrecentamiento por impactos de planetesimales no puede explicar por sí solo las condiciones térmicas  actuales de la Tierra. En consecuencia,  otras fuentes de energía deben ser las responsables del calentamiento interior de la Tierra. Si la Tierra en la etapa final de su formación sufrió el impacto de un embrión planetario, el volumen de la masa derretida de ambos cuerpos debió ser enorme. De hecho  se piensa que la formación de la Luna surgió como consecuencia de un gran impacto contra la Tierra. La energía involucrada en ese episodio fue 100 millones de veces mayor que aquella que creó el cráter de Chicxulub\index{Chicxulub, cráter} y que extinguió a los dinosaurios. Como consecuencia  del gran impacto, se formó en la Tierra un profundo océano de magma. El rápido enfriamiento radiativo de la superficie del océano de magma formó una corteza aislante que puede mantener la alta temperatura  del magma interior por mucho tiempo \footnote{Esto se debe a la baja conductibilidad de las rocas y a las grandes distancias que debe atravesar el calor antes de salir a la superficie de la Tierra. El tiempo de enfriamiento de la Tierra por conducción térmica es del orden de $10^{12}$ años, tiempo mucho mayor que la edad de la Tierra que es $\approx 4.5 \times 10^{9}$ años.}.

Otra de las fuentes de energía que puede calentar la Tierra, bien conocida por los físicos y astrónomos desde el siglo 18,  es la energía gravitacional o potencial. La otra fuente de energía que mantiene el horno de la Tierra es la radioactividad de las rocas, un fenómeno de la materia   que descubrió Becquerel\index{Becquerel, A.H.} en el año 1896.

A continuación mostraremos como a través de su propia fuerza de gravedad la Tierra puede contraerse y generar calor en su interior. Dividamos la Tierra en capas concéntricas como las capas de una cebolla, con respecto al centro de la Tierra. Por lo tanto, cada delgada capa esférica de la Tierra puede  caracterizarse por su radio $r$ y su espesor $dr$. Si $M(r)$  es la masa de la Tierra que está dentro de la esfera de radio $r$ y $dm(r)$ es una porción de la masa de la capa de radio $r$, la porción de la capa es atraída gravitatoriamente como si $M(r)$ estuviera concentrada en el centro de la Tierra. Es decir, por la ley de gravedad $dF(r)= \frac{G\, M(r)\, dm(r)}{r^{2}}$. Las capas exteriores (de radio $>r$) no ejercen una atracción  neta sobre la capa $r$.
Si tomamos  una pequeña superficie de área $dS$ de la esfera que define la capa r y denominamos $dm(r)$ a la masa que contiene el  cilindro  de base $dS$ y altura $dr$, $dm(r)=\rho(r)\, dS\, dr$.
 Dado que $M(r)=\frac{3}{4} \pi \rho(r) r^{3}$  la fuerza gravitatoria $dF(r)$ es 
\begin{equation}
dF(r)=\frac{4}{3}  \pi G \rho(r)^{2} r dr dS.
\label{dF}
\end{equation}

Consideremos ahora el peso que soporta la capa de radio $r_{1}$, debido al peso conjunto  de  las capas que están por encima de la capa $r_{1}$, es decir aquellas con radios mayores que $r_{1}$ y menores que $R_{0}$ (el radio de  Tierra al inicio de su autocontracción). Sobre una superficie $dS$ de la capa $r_{1}$ actúa una fuerza $W(r_{1})$ debido al peso de la materia contenida en la columna de base $dS$ y altura $R_{0}-r_{1}$. Teniendo en cuenta (\ref{dF}), $W(r_{1})= \int_{r_{1}}^{R_{0}} dF(r)= \frac{4}{3}  \pi G (\int_{r_{1}}^{R_{0}} \rho(r)^{2} r dr)  dS$. Por definición,  $\frac{W(r_{1})}{dS}$ es la presión sobre la capa $r_{1}$. Por lo tanto, denominando a esta presión $P(r_{1})$, y con la simplificación de que todas las capas tienen la misma densidad ($\rho(r)=\rho_{0}=$ constante),  obtenemos  que 
\begin{equation}
P(r_{1})=\frac{2}{3}  \pi G \rho_{0}^{2}R_{0}^{2} \left (1-\left(\frac{r_{1}}{R_{0}}\right) ^{2}\right).
\label{P(r1)}
\end{equation}
Obsérvese que $r_{1}=0$ corresponde al centro de la Tierra, donde según (\ref{P(r1)}) la presión $P(0)=\frac{2}{3}  \pi G \rho_{0}^{2}R_{0}^{2}$ es máxima. En la superficie, $r_{1}=R_{0}$, la presión es naturalmente cero.

En condiciones de equilibrio, la presión $P(r_{1})$ es contrarrestada  por la presión interna del material de la capa $r_{1}$. Pero supongamos que éste no es el caso y que $P(r_{1})$ supera a la presión interna que se contrapone. Entonces,  $P(r_{1})$ empuja como un pistón contrayendo el espesor de la capa $r_{1}$, que originalmente es $dr_{1}$.  Supongamos que alcanza un nuevo equilibrio cuando el espesor de la capa es $dr_{1}-dh$. Por lo tanto, se realizó un  trabajo sobre la capa $dr_{1}$ dado por $P(r_{1})\, dS\, dh$. Si este trabajo se convierte totalmente en energía dentro de la capa en cuestión, tenemos
\begin{equation}
dQ=P(0) \left (1-\left(\frac{r_{1}}{R_{0}}\right) ^{2} \right) dS dh.
\label{EP}
\end{equation}
Utilizando la ecuación (\ref{calor}), y  teniendo en cuenta que  $m=\rho_{0} \, dS\, dr_{1}$ dado que aquí se toma para $m$ la masa contenida en el diferencial de volumen $dS \, dr_{1}$, obtenemos a partir de (\ref{EP}) que
\begin{equation}
\Delta T= \frac{2}{3}  \frac{\pi G \rho_{0} R_{0}^{2}}{c_{m}} \mu \left (1-\left(\frac{r_{1}}{R_{0}}\right) ^{2}\right) \\,
\label{DeltaT2}
\end{equation}
donde $\mu =\frac{dh}{dr_{1}}$. Dado que consideramos que la masa de la Tierra permanece constante en el proceso de contracción, vamos a introducirla en (\ref{DeltaT2})  teniendo en cuenta que $M_{T}=\frac{4}{3} \rho_{0} R_{0}^{3}$ y además vamos  a expresar (\ref{DeltaT2})  en función del radio adimensional  $x= \frac{r_{1}}{R_{0}}$  
\begin{equation}
\Delta T(x)_{1}= \frac{G M_{T}}{2 c_{m}}  \frac{\mu  }{R_{0}}  (1-x ^{2}).
\label{DeltaT3}
\end{equation}
El subíndice de $\Delta T(x)$ indica el incremento de la temperatura de la capa en el primer paso o episodio de contracción. Terminado el primer paso de una contracción pequeña ($\mu \ll 1$), y dado que la masa se conserva $\rho_{1} (dr_{1}-dh) dS =\rho_{0} dr_{1} dS$, la nueva densidad de la capa es $\rho_{1}= \frac{\rho_{0}}{1-\mu}$. 
Si todas las capas se contraen simultáneamente con el mismo coeficiente $\mu$, supuesto muy pequeño, la ecuación (\ref{DeltaT3}) nos da sus incrementos de temperaturas variando $x$ entre 0 y 1. Dado que el incremento de la densidad de todas las capas es el mismo, la distribución de la densidad sigue siendo homogénea y por lo tanto se trata de una contracción homóloga. Como $dh=\mu dr_{1}$, $h=\mu  R_{0}$, y  el nuevo radio de la Tierra es $R_{1}= R_{0}-h=(1-\mu) R_{0}$. Después de $n$ pasos, $\rho_{n}= \frac{\rho_{0}}{(1-\mu)^{n}}$,  $R_{n}=(1-\mu)^{n} R_{0}$ y por lo tanto

\begin{equation}
\Delta T(x)_{n}= \frac{G M_{T}}{2 c_{m}}  \frac{\mu}{R_{0}} \frac{1}{(1-\mu)^{n -1}} (1-x ^{2})
\label{DeltaT4}
\end{equation}
y
\begin{equation}
T(x)=\sum_{n=1}^{N} \Delta T(x)_{n}=\Delta T(x)_{1} \sum_{n=1}^{N} \frac{1}{(1-\mu)^{n-1}} \,\,,
\label{Temp}
\end{equation}
donde $N$ es el número total de pasos discretos que adoptamos  para simular el proceso que llevó a la Tierra contraerse desde el radio inicial $R_{0}$ al presente radio de la Tierra $R_{T}$. Entonces, $R_{N}=R_{T}=(1-\mu)^{N} R_{0}$ y por lo tanto  $\mu=1-\left(\frac{R_{T}}{R_{0}} \right)^{(1/N)}$.

Reemplazando la fórmula (\ref{DeltaT3}) en (\ref{Temp}), obtenemos que  la temperatura  en función del radio adimensional $x$ está dada por 
\begin{equation}
T(x)=T(0) \, (1-x ^{2}),
\label{DdeT}
\end{equation}
donde $T(0)=\frac{G M_{T}}{2 c_{m}}  \frac{\mu  }{R_{0}} \sum_{n=1}^{N} \frac{1}{(1-\mu)^{n-1}}$. Si adoptamos $R_{0}=7400$ km y $N\geq 20$ y tenemos en cuenta que $\mu$ es función de $N$, es decir $\mu(N)$, obtenemos 
$T(0)\approx 3740$ grados Kelvin. El valor de $T(0)$ es poco sensible al valor de $N$ adoptado para el cálculo. En efecto, entre   $N=20$ y  $N\rightarrow \infty$, el valor de $\sum_{n=1}^{N} \frac{\mu(N)}{(1-\mu(N))^{n-1}}$ se encuentra entre 0.159 y 0.160. La expresión (\ref{DdeT}) es representada en  la Fig. \ref{PerfilTermico}. Nuestro tratamiento semi-cuantitativo tiene sólo el propósito de mostrar la capacidad de la energía gravitacional de la Tierra de dar cuenta de gran parte de la estructura térmica de la Tierra. Para el cálculo de modelos evolutivos de la Tierra, se necesita conocer la ecuación de estado $P=P(\rho, T)$ \footnote{A diferencia de la ecuación de estado para los gases ideales, la ecuación de estado de los líquidos y sólidos es compleja y no disponemos de ecuaciones analíticas completas para ella. La ecuación de estado para las rocas  se expresa corrientemente como una suma $P=P_{1}(\rho) + P_{2} (\rho, T)$, donde $P_{1}$ es la presión de Fermi\index{Fermi, presión de} debida a los electrones libres y $P_{2}$ es la presión térmica.} que gobierna la presión del interior de la Tierra que equilibra la presión gravitatoria, junto a  condiciones iniciales realistas. 
\begin{figure}
\includegraphics[scale=0.9]{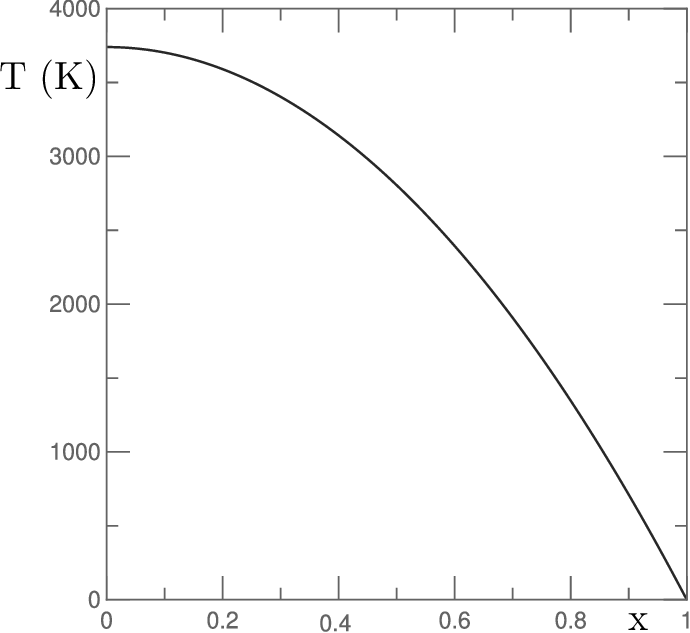} 
\caption{Perfil térmico de la Tierra creado  por la energía potencial liberada en el proceso de contracción terrestre. En el eje vertical,  se  representa la temperatura $T$ en grados Kelvin  en función del radio adimensional $x$, representado en el eje horizontal.}
\label{PerfilTermico}
\end{figure}

Los datos sismológicos hacen pensar que el núcleo interno de la Tierra es sólido\footnote{La estructura y dinámica del núcleo terrestre son muy complejas. El núcleo interno, que está rodeado por una capa de metales fundidos,  genera  corrientes eléctricas y con ello el campo magnético de la Tierra.}. Por lo tanto, el punto de fusión del hierro será el límite superior de la temperatura posible del núcleo interno.
Según algunas estimaciones,  el punto de fusión del hierro a la presión del centro de la Tierra $P(0)$ es del orden de $4500 \pm 1500$ grados Kelvin. Note que la temperatura $T(0)$ que hemos  calculado en  nuestra simulación es levemente inferior al límite superior estimado  para la temperatura del núcleo terrestre (ver Fig. \ref{PerfilTermico}). 

  A continuación evaluaremos el calentamiento producido por los elementos radioactivos como el uranio, el torio y el potasio contenidos en las rocas. La concentración de las fuentes radiógenas de calor en los granitos y los basaltos está bien estudiada. Esto permite calcular la cantidad de calor generado por las rocas graníticas, $\approx$ 17.4  cal m$^{-3}$ por año,  y de las basálticas, $\approx$ 3.5 cal m$^{-3}$  por año \footnote{Una caloría (cal)=\, 4.184 Joules.}. Si consideramos que la corteza terrestre está compuesta por una capa de granito de 15 km de espesor descansando sobre una capa de basalto de también 15 km, la cantidad de calor producido en una columna de 30 km de altura y 1 metro cuadrado de base es $ 313 \times 10^{3}$   calorías  por año.
 Es interesante comparar este número con las mediciones  del flujo de calor  realizadas en numerosos lugares de la superficie terrestre. La pérdida media mundial de calor  es igual  $467 \times 10^{3}$ cal m$^{-2}$  por año. Por lo tanto el 70 por ciento del calor que se desprende de la Tierra corresponde al calor generado en la corteza terrestre por los elementos radioactivos. Esto indicaría  que en primera aproximación la generación y pérdida de calor están equilibradas. Consecuentemente, el 30 por ciento  restante del calor que se pierde proviene del calor generado por  las fuentes radioactivas del interior de la Tierra. La proporción de elementos radiactivos del núcleo  de la Tierra sería  muy pequeña, comparable a la que existe en los meteoritos férreos; y
curiosamente,  la proporción de componentes  radiactivos de la corteza terrestre es muy similar a la de los meteoritos pétreos.

\section{La edad de la Tierra \label{EdadTierra}}
Se sabe que la edad de la Tierra es de unos 4.500 millones de años. A continuación esbozamos el método que se empleó para determinar dicha cifra. La edad exacta de la Tierra se obtiene mediante el uso de la composición isotópica del plomo \footnote{También hay otros isótopos como el rubidio ($Rb^{87}$) y el estroncio ($Sr^{87}$) que pueden usarse como relojes naturales para datar las rocas. Los elementos radioactivos de larga vida son los fundamentos de los relojes geológicos.}. Los isótopos del uranio (U$^{238}$ y U$^{235}$) y el torio (Th$^{232}$) se desintegran espontáneamente perdiendo partículas subatómicas y convirtiéndose en diferentes isótopos estables del plomo (Pb):

\begin{eqnarray}
U^{238} & \longrightarrow & Pb^{206}, \nonumber\\
U^{235} & \longrightarrow & Pb^{207}, \nonumber\\
Th^{232} & \longrightarrow & Pb^{208} .
\end{eqnarray}
A los isótopos listados a la izquierda se les llama ``progenitores'' $(P)$ y los de la derecha ``descendientes'' $(D)$.
Muchos procesos de la naturaleza, entre ellos la desintegración radioactiva,  siguen lo que se conoce como ley exponencial. Un ejemplo típico es el crecimiento de la población humana. El proceso de desintegración atómica puede compararse al siguiente símil biológico. Supongamos que dentro de una caja introducimos   un número de moscas estériles, de modo que ellas puedan vivir en la caja en condiciones naturales. Si volvemos a abrir la caja pasado un tiempo mayor que  tres veces la vida media de las moscas, las encontraremos a casi todas muertas. La relación entre  el número de moscas vivas y el número  de moscas muertas puede darnos una estimación del tiempo que trascurrió desde que introdujimos las moscas en la caja. Del mismo modo,  la proporción  entre el número de átomos progenitores y  de átomos descendientes de una especie, que existe en una muestra de una roca, puede dar la edad de la roca.

La probabilidad de que un átomo radioactivo se desintegre durante  un intervalo pequeño de tiempo $\Delta t$ es dado por $\lambda \, \Delta t$, donde $\lambda$ es la constante de desintegración del elemento. Por lo tanto, si en el momento $t_{0}$ hay $N$  átomos, en una muestra de roca, que aún no se han desintegrado, en el intervalo   $(t_{0},\, t_{0}+\Delta t)$ se desintegraran $\lambda\, N \, \Delta t$. Como $N$ es la cantidad  del elemento progenitor (en nuestro caso el uranio), lo denotaremos por  $P (t_{0})$, es decir $P (t_{0})=N$. Entonces, la disminución de la cantidad de progenitores en el tiempo $t_{1}=t_{0}+\Delta t$ es $\Delta P (t_{1})= P (t_{1})-P (t_{0})=-\lambda \,P (t_{0}) \, \Delta t$. Esto  también puede escribirse como $P (t_{1})=P (t_{0})-\lambda \,P (t_{0}) \, \Delta t=P (t_{0})(1-\lambda \, \Delta t)$. Para un tiempo posterior $t_{2}=t_{1}+\Delta t=t_{0}+ 2 \Delta t$,  $P (t_{2})=P (t_{1})(1-\lambda  \, \Delta t)$ y reemplazando  la expresión para  $P (t_{1})$ obtenemos  $P (t_{2})=P (t_{0})(1-\lambda  \, \Delta t)^
{2}$. Avanzando otro paso en el tiempo, obtenemos que  $P(t_{3})=P (t_{0})(1-\lambda  \, \Delta t)^{3}$.
Generalizando, 
\begin{eqnarray}
P (t_{n+1})&=&P (t_{0}) (1-\lambda \, \Delta t)^{n+1},\nonumber\\
t_{n+1}&=&t_{0}+ (n+1)\, \Delta t, \, \, \, \,(n=0,1,2,...).
\label{uranio}
\end{eqnarray}
De la segunda ecuación de (\ref{uranio}), $\Delta t=\frac{t_{n+1}-t_{0}}{n+1}$. Si adoptamos $t_{0}=0$ y $t_{n+1}=t$, la primera ecuación de (\ref{uranio}) resulta igual a 
\begin{equation}
P(t)= P(0)(1- \frac{\lambda \, t}{n+1})^{n+1}.
\label{uranio2}
\end{equation}
Se puede demostrar que 
\begin{equation}
\lim_{n \rightarrow \infty} (1-  \frac{\lambda \, t}{n+1})^{n+1}=e^{-\lambda \,t}.  \nonumber\\
\label{uranio3}
\end{equation}
Reemplazando la identidad (\ref{uranio3}) \footnote{Esta identidad se puede demostrar desarrollando en series de Taylor\index{Taylor, series de } el primer miembro y llevándolo al límite, con lo cual se obtiene una serie de potencias que coincide con la de $e^{-\lambda t}$ ($=\sum_{n=0}^{\infty} \frac{(-\lambda t)^{n}}{n!}$).} en (\ref{uranio2}) obtenemos 
\begin{equation}
P(t)= P(0)\,\, e^{-\lambda \,t};
\label{uranio4}
\end{equation}
ecuación que nos da el número de átomos progenitores $P(t)$ en el presente, partiendo del número de átomos progenitores $P(0)$ que existían en el tiempo que tomamos como origen ($t=0)$.
La solución (\ref{uranio4}) se puede también obtener  planteando la ecuación diferencial $dP=-\lambda P dt$, e integrándola por el método de separación de variables: $\int \frac{dP}{P}= -\lambda \int \,dt$.

La ecuación (\ref{uranio4}) nos permite definir el concepto de vida media de un átomo radioactivo. La vida media $t_{1/2}$ de un elemento radioactivo es el intervalo de tiempo que se necesita para que el número de átomos progenitores  decrezca a la mitad de su número original; es decir $\frac{P(t)}{P(0)}= \frac{1}{2}$. Reemplazando está relación en (\ref{uranio4}), y despejando $t$, que aquí llamamos  $t_{1/2}$, obtenemos
\begin{equation}
t_{1/2}=\frac{\ln 2}{\lambda}.
\label{Half-time}
\end{equation}
Las vidas medias del $U^{238}$,  $U^{235}$ y  $Th^{232}$ son $4.51 \times 10^{9}$, $7.04 \times 10^{8}$ y $1.40 \times 10^{10}$ años, respectivamente. La larga vida media de estos elementos los hace apropiados como relojes geológicos.

Denominando  $D(t)$ y $D(0)$ a la  cantidad del elemento descendiente en el presente y en $t=0$, respectivamente,
$D(t)=D(0) + (P(0)-P(t))$, donde $(P(0)-P(t))$ es la cantidad del elemento progenitor que se convirtió en descendiente. De (\ref{uranio4}), obtenemos que $P(0)= P(t)\,\, e^{\lambda \,t}$, con lo cual la ecuación para $D(t)$ se puede escribir: $D(t)=D(0) + P(t) (e^{\lambda \,t}-1).$ Es conveniente usar la cantidad de un isótopo no-radigénico  $R$, perteneciente a la especie del descendiente, como referencia \footnote{En el caso del plomo se usa el isótopo $Pb^{204}$  que no es radiogénico; es decir se trata de plomo común que no se generó por la desintegración de un progenitor.}. Para lo cual, dividimos ambos miembros de la última ecuación por R, obteniendo
\begin{equation}
D_{r}(t)=D_{r}(0) + P_{r}(t) (e^{\lambda \,t}-1),
\label{uranio5}
\end{equation}
donde $D_{r}(t)=\frac{D(t)}{R}$, $D_{r}(0)=\frac{D(0)}{R}$ y $P_{r}(t)=\frac{P(t)}{R}$. 
Mediante los espectrómetros de masa podemos medir muestras de la roca que deseamos datar, y  obtener  los valores de $D_{r}(t)$ y $P_{r}(t)$. De modo que la ecuación (\ref{uranio5}) tiene dos incógnitas:  $t$ y $D_{r}(0)$. Si la roca en cuestión es un sistema cerrado desde su cristalización, es decir que no hubo intercambios de progenitores y descendientes con su entorno, todas las muestras de distintos minerales de la roca deben tener la misma edad $t$ y  el mismo valor para $D_{r}(0)$. Dado que $R$ y $D(0)$ no son radioactivas,  sus cantidades permanecieron constantes e igualmente distribuidas dentro de la roca puesto que $R$ y $D(0)$ tienen propiedades físico-químicas similares. Por lo tanto  $D_{r}(0)$ es constante para todos los minerales que constituyen la roca. Denotemos  con $a=D_{r}(0)$ y $b=(e^{\lambda \,t}-1)$,  que son constantes e iguales para todos los minerales de la roca, y con $y=D_{r}(t)$ y $x=P_{r}(t)$ los valores que pueden determinarse experimentalmente para distintos minerales de la roca. Con ello, la ecuación (\ref{uranio5}) se escribe $y= a + b \, x$, ecuación que representa una recta de pendiente $b$ y ordenada al origen $a$. Para cada mineral de la roca podemos tomar una muestra y medir el par de datos $(x,y)_{i}=(P_{r}(t), D_{r}(t))_{i}$, el cual representado  en un sistema de coordenadas Cartesianas es el punto correspondiente a la muestra $i$. Con las muestras  $i=1,2,3,..$,  obtenemos un número de puntos con lo cual se puede  determinar la recta de mejor ajuste, y con ello el valor de $a$ y el de $b$. Entonces, despejando $t$ de $b=(e^{\lambda \,t}-1)$, obtenemos la edad absoluta de la roca:
\begin{equation}
t({\rm edad\, de\, la\, roca})\, =\, \frac{1}{\lambda} \ln (b+1).
\end{equation}

Las edades de las rocas más antiguas de la corteza terrestre son $\approx  3.75 \, 10^{9}$ años. Es claro que la edad de la Tierra tiene que ser mayor que esa cifra. El primero en determinar la edad de la Tierra como planeta fue Claire Patterson \index{Patterson, C.}
en el trabajo ``La edad de los meteoritos y la Tierra''(año 1956).
A partir de la datación de sedimentos oceánicos de alta profundidad, Patterson\index{Patterson, C.} obtuvo una edad de  $4.55 \pm 0.07 \times 10^{9}$ años para la Tierra. Él también analizó un conjunto de meteoritos y obtuvo para ellos la misma edad que la Tierra, corroborando que la Tierra y los meteoritos tuvieron un origen común.

\section{Origen y naturaleza de los cometas \label{HistoriaCometas}}

\begin{quote}
\small \it{La nube de Oort es una fábula en ambos sentidos de la palabra: asombrosa, por una  parte, y totalmente hipotética, por otra. Como se apresuran a señalar los críticos, es por definición invisible, cual nube de mosquitos a millones de kilómetros de distancia y no hay ninguna posibilidad de observarla directamente, quizás hasta que la primera astronave salga del Sistema Solar y la atraviese en su camino. Pero si los científicos tuvieran que atenerse a lo que ven, no tendrían nada que decir sobre el interior del Sol ni sobre la  vida familiar de los dinosaurios, y tampoco tendríamos ninguna clase de electrodomésticos, puesto que los electrones son absolutamente invisibles.} \rm 
 
Nigel Calder\index{Calder, N.} (1931-2014)  en el libro ``The comet is coming'' 

\end{quote}

Uno o dos cometas se acercan  por año al Sol dentro de la zona de visibilidad \footnote{Esfera de  $\approx 2$ Unidades Astronómicas de radio, centrada en el Sol, dentro de la cual el cometa recibe la iluminación solar  suficiente para la visualizacion del cometa con telescopios o a simple vista.}. Estos cuerpos  que aparecen con tanta frecuencia en el cielo han intrigado al hombre desde la antigüedad. Hay una rica historia de las ideas sobre el origen y naturaleza de los cometas que se han propuesto desde la época de los primeros filósofos de la naturaleza hasta  la actualidad (ver \cite{Bailey} \cite{Clube1} para  un raconto histórico). Esta historia milenaria nos enseña en particular  las dificultades que los grandes científicos tuvieron en dilucidar  el papel que  los cometas juegan en  la trama de los cielos.

En la antigüedad y la edad media, a los cometas se los relacionaba con fenómenos atmosféricos o celestes. En la cosmología geocéntrica de Aristóteles\index{Aristóteles}, a los fenómenos temporarios como los cometas se los ubicaba en el mundo sublunar, pues la región que se encuentra más allá de la Luna debe ser, según Aristóteles\index{Aristóteles},  inmutable y eterna. A partir del desarrollo de la teoría heliocéntrica de Copérnico\index{Copérnico}, se descarta la hipótesis atmosférica y se piensa en general que los cometas son cuerpos que transitan el Sistema Solar. En 1577 apareció un gran cometa y el astrónomo danés Tycho Brahe\index{Tycho Brahe} realizó cuidadosas mediciones que mostraron que dicho cometa se encontraba más allá de la Luna. Johannes Kepler\index{Kepler}, discípulo de  Tycho Brahe\index{Tycho Brahe} \footnote{Tycho Brahe\index{Tycho Brahe} llevó a cabo durante muchos años mediciones sistemáticas de las posiciones de los planetas, particularmente de Marte, con una  precisión muy alta para la época. A partir del análisis de todo ese material observacional, Kepler\index{Kepler} descubrió las tres famosas leyes que rigen el movimiento de los planetas.}, sugirió  que los cometas atravesaban en línea recta el Sistema Solar.

Con el desarrollo de la mecánica Newtoniana, se aclaró que los cometas son cuerpos que responden a la fuerza gravitatoria, y que hay  dos alternativas para explicar el origen de ellos: a) son cuerpos ligados al Sistema Solar desde su formación  como los planetas o; b) son cuerpos que provienen del espacio entre las estrellas, es decir del medio interestelar, y que circunstancialmente atraviesan el Sistema Solar tal como lo sugirió Kepler\index{Kepler}. La existencia de cometas que periódicamente se acercan al Sol, como el cometa Halley,  parece hablar en favor de que los cometas pertenecen al Sistema Solar y que se formaron en la misma nebulosa primigenia que dio  origen  a los planetas. En efecto, el filósofo Inmanuel Kant\index{Kant, I.}, que había creado  la teoría nebular para explicar la formación del Sistema Solar, creía que la formación de los cometas podía incorporarse a esa teoría.  Sin embargo, el gran mecánico celeste Laplace\index{Laplace, P.S.},  que propuso también en forma independiente la teoría nebular  \footnote{La teoría nebular de Laplace\index{Laplace, P.S.} data del año 1796 (Ver Sección 3 de la parte I de esta monografía \citep{Olano5}) y la teoría interestelar de los cometas del año 1805.}, sostenía el origen interestelar de los cometas. Él encontraba difícil conciliar la existencia   de cometas con órbitas casi parabólicas o hiperbólicas, que indicarían distancias interestelares, con el origen de los cometas en el Sistema Solar. Los cometas periódicos se explicarían como cometas interestelares capturados por el Sistema Solar. 

Hay una consecuencia que se deduce de la  teoría interestelar de Laplace\index{Laplace, P.S.}: el Sol debe estar en reposo con respecto al medio interestelar local, tal como el mismo Laplace\index{Laplace, P.S.} lo destacó. De otro modo,  los cometas tenderían a aparecer en torno a  la posición del cielo hacia la cual se dirige el Sol. Hecho que no ocurre, pues los cometas aparecen en direcciones distribuidas isotrópicamente en el cielo. Sin embargo, ¿está en realidad el Sol  en reposo con respecto al medio interestelar local?  En principio, dando respuesta a esa pregunta, podemos refutar o corroborar la teoría interestelar de Laplace\index{Laplace, P.S.}. Ya en 1733, W. Herschel\index{Herschel, W.} determinó por primera vez el movimiento del Sol sobre  la base de movimientos propios \footnote{Los movimientos propios de las estrellas fueron descubiertos en 1718 por E. Halley\index{Halley, E.}, e indican los cambios de posición de las estrellas en el cielo. Debido al  movimiento del Sistema Solar, la posición en el plano celeste  de una estrella relativamente cercana con respecto a las posiciones de las estrellas de fondo que se encuentran muy lejanas (estrellas fijas) cambia con el tiempo. El movimiento propio  se mide en segundos de arco por año. La separación angular entre  dos posiciones de una estrella, expresada en segundos de arco, se la  divide por el tiempo en años que demandó a la estrella desplazarse entre las dos posiciones. Si conocemos la distancia a la estrella podemos determinar su velocidad tangencial, componente de la velocidad total perpendicular a la línea de la visual. Como estas velocidades son relativas al Sol, ellas reflejan  el movimiento del Sistema Solar como un todo a través del medio interestelar local.} de unas pocas estrellas, pero ello fue desestimado por Laplace\index{Laplace, P.S.}. Con el desarrollo de la astrofísica y los análisis espectroscópicos de las estrellas, se pudieron medir velocidades radiales de las estrellas \footnote{componente de la velocidad total de una  estrella en la dirección de la línea de la visual, o dicho de otro modo, velocidad con la cual la estrella  se aleja o se  acerca al  observador.} a partir del efecto Doppler\index{Doppler, efecto}. Con esta mejora en los estudios de los movimientos estelares, se estableció definitivamente  que el Sol se mueve con respecto al entorno estelar e interestelar con una velocidad peculiar de $\approx20$ km s$^{-1}$ hacía la constelación de Hércules.

El astrónomo italiano Giovanni V. Schiaparelli\index{Schiaparelli, G.V.} (1835-1910) fue el primero en destacar la falla de la hipótesis interestelar de Laplace\index{Laplace, P.S.} en cuanto a que  el movimiento  del Sol, es decir de todo el Sistema Solar, con respecto al medio interestelar local  es significativo y no puede despreciarse. Hacia fines del siglo diecinueve el número de cometas con órbitas elípticas que se habían descubierto  era considerable. El astrónomo inglés  Richard A. Proctor\index{Proctor, R.A.} (1837–1888)  señaló  alrededor del año 1870 que la hipótesis de Laplace\index{Laplace, P.S.} sobre la captura de cometas interestelares no puede explicar cuantitativamente la proporción observada de cometas con  órbitas elípticas. Sin embargo, la mayoría de los astrónomos del siglo diecinueve tendieron a seguir la teoría interestelar de Laplace. Recién bien entrado el siglo veinte,  nuevas investigaciones mostraron claramente las debilidades de la teoría de Laplace\index{Laplace, P.S.} y en consecuencia llevaron a un cambio de paradigma, es decir al regreso a la vieja idea de que los cometas se formaron en el disco protoplanetario del Sistema Solar.
 
 Al examinar una muestra de cometas de largo periodo con precisas determinaciones de sus  parámetros orbitales, Jan H. Oort\index{Oort, J.H.}  halló  en el año 1950 que los sitios de origen a partir de los cuales esos   cometas ``caían'' hacia   el interior del Sistema Solar se encontraban  a grandes distancias del Sol. Oort\index{Oort, J.H.} dedujo que la distribución de los cometas formaría una nube esférica con un  radio del orden 100000 UA  \footnote{La Unidad Astronómica es igual a la distancia media Tierra-Sol.} (la nube de Oort), en cuyo núcleo yace  el Sistema Solar. El borde de la nube de Oort\index{Oort, nube de} se encontraría a  media distancia entre el Sol y la estrella más cercana al Sol (Alfa Centauri). Se estima que la nube de Oort\index{Oort, nube de} contiene $10^{12}$ cometas. 
 
 En la teoría nebular o planetesimal, los cometas se formaron en el disco protoplanetario y por lo tanto se esperaría que la actual distribución de los cometas  conserve su distribución original en forma de disco. Sin embargo la nube de Oort\index{Oort, nube de}   es esférica. La explicación de Oort\index{Oort, J.H.} fue que una gran explosión en el seno del disco protoplanetario eyectó en todas direcciones fragmentos cometarios. Dieciocho  años antes del  estudio de Oort\index{Oort, J.H.}, Ernst Öpik\index{Oepik, E.} entrevió que  cometas del Sistema Solar que orbitan a grandes distancias  del Sol pueden  ser dispersados por fuerzas gravitatorías debidas a encuentros lejanos con las estrellas. Por otra parte, a fin de explicar las características de la nube de Oort\index{Oort, nube de}, el astrónomo uruguayo Julio Fernández\index{Fernández, J.} \footnote{También Julio Fernández  y el astrónomo argentino Adrián Brunini\index{Brunini, A.} propusieron que las estructuras de la nube de Oort y del resto del Sistema Solar primigenio se configuraron dentro de un cúmulo de estrellas en formación a partir de una misma densa nube de polvo y gas molecular\cite{Fernandez}.} y otros propusieron que  los cometas fueron eyectados desde el disco protoplanetario por perturbaciones gravitacionales de los planetas  gigantes en formación. Además, demuestran  que una fracción de los cometas  pueden ser  eyectados con velocidades que les permiten escapar del campo gravitatorio del Sistema Solar  y convertirse en cometas interestelares. Por lo tanto, así como nuestro Sistema Solar puede ser una fuente de cometas interestelares, sistemas planetarios de otras estrellas también lo pueden ser.

El hecho de que los cometas de la nube de Oort\index{Oort, nube de} estén ligados gravitacionalmente al Sistema Solar no implica necesariamente que ellos son cometas primordiales, es decir originados en la nube protoplanetaria. A lo largo de la larga vida del Sistema Solar, la nube de Oort\index{Oort, nube de} fue en numerosas veces fuertemente perturbada por los encuentros del Sistema Solar con masivas nubes interestelares y con estrellas de fondo, con lo cual muy probablemente se la despojó  de la mayoría de los cometas primordiales. Los astrónomos ingleses Victor Clube\index{Clube, V.} y Bill Napier\index{Napier, B.} han sugerido que  el Sistema Solar, al  atravesar  un denso enjambre de cometas interestelares contenidos en un masiva nube interestelar (nube molecular, ver Sección \ref{MedioInterestelar}),  puede capturar cometas interestelares y así reponer los cometas perdidos \cite{Clube2} (ver también \cite{Bailey} y referencias que allí se encuentran sobre antecedentes de ideas similares). En los artículos \citep{Olano10},\citep{Olano1} y \citep{Olano9} del autor,  se muestra que el Sistema Solar se encontró con masivas nubes moleculares hace $\sim 10$ y $40$ millones de años en coincidencia con las dataciones de las extinciones masivas del Mioceno\index{Mioceno} y Eoceno\index{Eoceno}. 

Ambas teorías sobre el origen de los cometas, capturados del medio interestelar versus formados con el Sistema Solar, pueden tener una parte de la verdad. A menudo se cumple en las ciencias la fórmula epistemológica del filósofo alemán Hegel\index{Hegel, G.W.F.}, según la cual a una \bf{tesis} \rm se contrapone una \bf{antítesis} \rm  y finalmente se compone  una \bf{síntesis}\rm.  El hecho de que la mayor parte de los cometas que ingresan a la zona de visibilidad  pertenezca al Sistema Solar, es decir a la nube de Oort\index{Oort, nube de}, no excluye la existencia de una gran población de cometas interestelares. Al contrario, se refuerza la idea de que ambos  tipos de cometas tuvieron un mecanismo común de formación: se gestaron en nubes de plasma estelar y crecieron en nubes moleculares interestelares, de acuerdo con las hipótesis que desarrollaremos en la Sección \ref{CometasInterestelares}. 

\newpage

\section{El medio interestelar y su relación con las estrellas\label{MedioInterestelar}}

\begin{quote} \it{Dadme materia y os construiré un mundo con ella.} \rm

 Inmanuel Kant (1724-1804) \index{Kant, I.} \footnote{Célebre filósofo alemán que hizo importantes aportes filosóficos y científicos, como por ejemplo su teoría del conocimiento y en el campo de la astronomía su teoría nebular sobre la formación del Sistema Solar. Él también anticipó que muchas de las llamadas nebulosas no formarían parte de nuestro sistema estelar (la Vía Láctea), sino que serían ``universos islas'': es decir galaxias como la Vía Láctea pero muy lejanas. Otros de sus anticipos científicos fue su duda acerca de que la velocidad de rotación de la Tierra permanece constante (ver Sección 3) y por lo tanto de su uso como un reloj preciso. Actualmente, el tiempo astronómico, que es el que básicamente utilizamos, se corrige con gran precisión mediante relojes atómicos.} 
 
\end{quote}

Cuando observamos el cielo estrellado, la primera impresión que tenemos es que el espacio entre las estrellas (el medio interestelar) está vacío. Sin embargo, está lleno de materia invisible al ojo humano \footnote{El astrónomo alemán Johannes Franz Hartmann\index{Hartmann, J.F.}, director del Observatorio Astronómico de La Plata (Argentina) entre los años 1921 y 1934, encontró las primeras evidencias de la existencia de la materia interestelar. Descubrió, en los espectros estelares, líneas de absorción que no correspondían a las estrellas, sino a material ubicado entre el observador  y las estrellas.}. En efecto, el material interestelar se encuentra en general a temperaturas cercanas al cero absoluto y por lo tanto no emite luz en el espectro visible. Sobre la franja de estrellas débiles de la Vía Láctea, hay regiones que se observan como agujeros o ventanas donde casi no hay estrellas. En realidad, ello no se debe a la ausencia de estrellas en esas regiones, sino a extensas nubes oscuras o de polvo que se encuentran en el medio interestelar y ocultan a  las estrellas que se encuentran detrás. Un ejemplo de ello es la nube oscura conocida como el Saco de Carbón, la cual  se encuentra en la constelación de la Cruz del Sur. Las nubes oscuras contienen gas y polvo y son el hogar de  moléculas interestelares complejas, pues los átomos se apoyan    en las superficies de los granos de polvo para unirse y formar moléculas. Por otra parte, el polvo de las nubes oscuras  apantalla a las moléculas contra la radiación de fondo que pueda disociarlas. Por ello, las nubes oscuras o de polvo son casi sinónimos de nubes moleculares interestelares. 

Las abundancias de los elementos químicos del  gas interestelar son similares a aquellas observadas en las atmósferas estelares. El $\approx 70 $ por ciento de la masa del material interestelar está constituido por hidrógeno (atómico y molecular),  el $\approx 28 \%$ por helio y el restante por elementos más pesados. La masa del polvo interestelar es de solo el 1 por ciento de la masa del  gas interestelar ($\frac{m_{p}}{m_{g}}=\frac{1}{100})$. Nuestra Galaxia (la Vía Láctea) es un enorme disco en rotación de estrellas, gas y polvo interestelares. Gran parte del disco galáctico es llenado por un substrato templado de hidrógeno atómico parcialmente ionizado y  de muy baja densidad ($0.1$ átomos/cm $ ^{3})$, dentro del cual hay extensas nubes frías de hidrógeno atómico de baja densidad (1-100 átomos/cm $^{3})$ y densas nubes moleculares ($10^3-10^{6}$ moléculas por cm$^{3})$
 \footnote{Las densidades del medio interestelar son mucho menores que las que pueden lograrse con las mejores bombas de vacío de los laboratorios terrestres. La densidad de la atmósfera sobre la superficie de la Tierra es del orden de $10^{19}$ moléculas por cm $^{3}$.}. Puesto que la distancia media  entre las estrellas es muy grande 
 y por lo tanto  el volumen que ocupa el medio interestelar es muy  grande, la cantidad de materia que  el medio interestelar contiene   es inmensa,  aún con las bajas   densidades  del gas interestelar. Por ejemplo, la distancia a Alfa Centauri, que es la estrella más cercana al Sol, es de $\approx 1,3$ pc \footnote{Los astrónomos usan como unidad de distancia  el parsec (pc) que equivale a  $3.086 \times 10^{13}$ km= $3.26$ años luz. Una estrella está a una distancia de un pc, si su paralaje es de un segundo de arco, de allí la abreviación  parsec. La paralaje de una estrella se obtiene a partir del ángulo $\alpha$ con que se desplaza aparentemente la posición celeste de una  estrella cercana mirada desde dos puntos diametralmente opuestos de la órbita de la Tierra en torno al Sol, con respecto a las estrellas lejanas de fondo que se encuentran en la misma región del cielo. La paralaje $p$ de una estrella es definida por la mitad de $\alpha$ expresada en segundos de arco. Por lo tanto la distancia a una estrella en km es $\frac{D_{TS}}{p}$, donde $D_{TS}\approx 150 \times 10^{6}$ km es  el radio de la órbita terrestre 
 y $p$ se expresa en radianes ($p$ es un número muy pequeño, por lo tanto $tan \, \, p \approx p$).} \footnote{Arquímides\index{Arquímides} (287-212 a. C.), el Einstein\index{Einstein, A.} de la antigüedad, intuyó  que,  en el marco de  la teoría heliocéntrica de Aristarco de Samos\index{Aristarco de Samos} (310-230 a. C.),  las estrellas deberían estar muy lejos pues ellas no mostraban paralajes apreciables. El primero en vislumbrar la naturaleza de las estrellas fue Giordano Bruno\index{Bruno, G.} (1548-1600) quien enseñaba que las estrellas son soles o que el Sol es una estrella.}.  Supongamos un volumen esférico centrado en el Sol con un radio $R_{\star \star}$ igual a la distancia entre Alfa Centauri y el Sol,  
 es decir $R_{\star \star}=1,3$ pc. Si dicho volumen contiene gas interestelar de una densidad numérica $\rho = 10$ átomos de hidrógeno por cm $^{3}$, típico de las nubes interestelares de hidrógeno atómico, la masa contenida dentro del volumen ($=\frac{4}{3} \pi R_{\star \star}^{3}\rho$) es igual aproximadamente a dos masas solares ($=4 \times 10^{33}$ gramos). La concentración gravitatoria de una nube de similares dimensiones y masa puede haber originado el Sistema Solar. 
 
 Dada la gran abundancia del hidrógeno (H) en la materia interestelar, las nubes moleculares están esencialmente  compuestas de moléculas de hidrógeno ($H_{2}$).  Las nubes moleculares tienen muy bajas temperaturas (10-20 K) y en consecuencia no emiten radiación en la parte  visible del espectro electromagnético. Sin embargo,  varias especies moleculares, como la molécula de  CO, pueden  emitir   ondas de radio en dichas condiciones físicas y por lo tanto las nubes moleculares pueden observarse mediante el uso de radiotelescopios. 
 
 ¿De dónde provino el material interestelar? Para contestar esta pregunta, debemos remontarnos al origen del Universo. Hace aproximadamente 15000 millones de años  todo el Universo estuvo concentrado en un ``punto'', al que el
sacerdote belga, matemático y  astrónomo  Georges Lemaître\index{Lemaître, G.} (1894-1966) \footnote{Lemaître\index{Lemaître, G.} fue el primero en proponer la expansión del Universo, y se lo puede considerar con justicia  el padre de la teoría del ``Big Bang''. A partir del corrimiento al rojo de los espectros de las  galaxias, el astrónomo Norteamericano  Edwin Hubble\index{Hubble, E.} (1889-1953) comprobó observacionalmente la expansión del Universo.}   llamó el átomo primigenio o ``huevo cósmico''. A los pocos minutos \footnote{Hay un interesante libro de divulgación titulado ``los tres primeros minutos del Universo''  en el cual su autor, el premio Nobel de física Steven Weinberg\index{Weinberg, S.} (1933-2021), describe muy didácticamente el origen del Universo sin recurrir a fórmulas matemáticas.} de la Gran Explosión (Big Bang) del átomo primitivo se formaron las partículas elementales. Pasados 500.000 años de la explosión, el Universo se expandió tanto  que la energía radiante que no se había convertido en materia permaneció como fotones desacoplados de la materia. Esta radiación remanente de la Gran Explosión se observa actualmente como la radiación de fondo de microondas de 3 grados Kelvin \footnote{El descubrimiento de esta radiación cósmica de fondo por Penzias\index{Penzias, A.A.}  y Wilson\index{Wilson, R.W.} en el año 1965, que les valió el premio Nobel de física, fue un gran espaldarazo para la teoría de la Gran Explosión.}. Poco después del desacople de la radiación de la materia, la temperatura del Universo descendió a 3000 K y permitió la creación de los átomos, mayoritariamente hidrógeno, a partir de  la unión de núcleos y electrones \footnote{La nucleosíntesis de la Teoría de la Gran Explosión predice que   el $\approx 75$   por ciento  de la masa del Universo consiste en Hidrógeno (H) y  el $\approx  25$ por ciento en Helio (H$_{e})$. La formación de átomos más pesados fue muy pequeña. Ello concuerda con las abundancias de los elementos observados en el Universo y es una fuerte evidencia en favor de la teoría de la Gran Expansión.}. A este tiempo se lo llama era de la materia, en el cual la fuerza de gravedad Newtoniana juega un papel importante. Aun con una distribución inicial uniforme de la materia en todo el Universo, se formaron  grandes  nubes o condensaciones gaseosas debido a la inestabilidad gravitatoria llamada inestabilidad de Jeans\index{Jeans, inestabilidad de} } \footnote{James Jeans\index{Jeans, J.} (1877-1946) fue un famoso astrónomo y físico británico que demostró la inestabilidad gravitatoria que lleva su nombre.}. Esas grandes nubes de gas se enfriaron y formaron protogalaxias, sin ninguna estrella aún. Con el tiempo, la inestabilidad de Jeans\index{Jeans, inestabilidad de} hizo también su trabajo en las protogalaxias formando nubes en contracción gravitatoria que crearon las primeras estrellas. Es decir, el gas primitivo  que  dio origen a la materia interestelar de una galaxia fue el gas de la protogalaxia misma, sintetizado en la Gran Explosión. 

Las primeras estrellas se formaron del prístino gas de casi puramente H y H$_{e}$, creados en la Gran Explosión. A medida que una protoestrella se contrae por su propia fuerza de  gravedad \footnote{El proceso de contracción gravitatoria de una estrella es muy similar al analizado en la Sección 4 para la Tierra, claro que las masas de las estrellas son muy superiores a la masa terrestre.}, su núcleo se calienta y alcanza densidades y temperaturas que inician los procesos termonucleares que hacen que una estrella pueda brillar por millones de años. Las enormes  energías que liberan las estrellas se generan por el proceso de fusión nuclear, en el cual los núcleos de átomos livianos se unen para formar núcleos de átomos más pesados. En esa compactación o empaquetamiento de núcleos atómicos, se pierde cierta masa que se convierte directamente en energía  tal como lo prescribe la famosa fórmula de Einstein\index{Einstein, A.} sobre la equivalencia entre masa y energía, a saber $E=m c^{2}$. En la etapa final de la vida de las estrellas, cuando queman todo su combustible nuclear, esparcen sus cenizas en el medio interestelar, enriqueciéndolo con los átomos pesados cocinados en sus hornos nucleares. El medio interestelar, como el suelo a la vida, es el substrato fundamental en el cual se forman los sistemas estelares y planetarios y al cual vuelven cuando se desintegran. El Sistema Solar se formó en una región del medio interestelar abonada por  desechos de estrellas extintas, y por lo tanto rica en elementos pesados tales como el carbono, el oxigeno, el calcio y  el hierro que son esenciales para la vida. Literalmente, estamos hechos de polvo de estrellas \footnote{Somos ``Hijos de las estrellas'' tal como se titula el interesante libro de divulgación de  Daniel Altschuler\index{Altschuler, D.}.}.

El polvo interestelar, tal como lo hemos descripto arriba, es  uno de los componentes fundamentales del medio interestelar. Esas pequeñas partículas sólidas son los ladrillos con los cuales se construyeron los cometas y el conocimiento del origen de las mismas puede ayudarnos a investigar la naturaleza de los cometas. Como veremos debajo, el origen del polvo interestelar  está también relacionado con la muerte de las estrellas. Las estrellas de baja masa  tienen una larga vida y mueren  apaciblemente. Por ejemplo el Sol, que es   una estrella de relativamente baja masa,  ha vivido $\approx$ 5000 millones de años y se encuentra en la mitad de su vida. La longevidad de esas estrellas se explica porque  son estrellas ``ahorrativas'' que consumen lentamente el combustible nuclear que poseen. En cambio, las estrellas de alta masa, y por lo tanto con mayor cantidad de combustible nuclear que las estrellas de baja masa, gastan su combustible nuclear dispendiosamente y en consecuencia son muy luminosas y sus vidas relativamente cortas. Una estrella de 50 masas solares vive solo unos pocos millones de años. Estas estrellas masivas mueren como han vivido, con un fogonazo final que las hace brillar como miles de millones de soles. Ello se debe a que  explotan como \it{supernovas} \rm  \footnote{La razón histórica de  la  raíz  \it{nova}  \rm de la palabra  es porque  los astrónomos  antiguos veían aparentemente una estrella nueva en un lugar donde antes no había nada. En realidad estaba la misma estrella pero su brillo era tan débil que no era percibido por el ojo desnudo. Se descubrió que hay dos tipos de novas de naturalezas muy distintas: una corresponde a una estrella que se destruye (supernova) y otra a una estrella que tiene un aumento episódico  de luminosidad (nova).} eyectando al espacio interestelar gran parte de sus masas estelares con velocidades  de miles de kilómetros por segundo \footnote{El flujo de rayos cósmicos, y de radiación ionizante en general, que recibiría la Tierra desde una supernova situada a una distancia del Sol menor que 20 pc  sería tan intenso  que  destruiría la vida terrestre o gran parte de ella. La probabilidad de un encuentro cercano del  Sol con una estrella que explota como supernova es afortunadamente muy baja. Si bien no podemos excluir que dicho evento haya ocurrido en la historia de la Tierra y que explique alguna extinción masiva en particular,  la periodicidad  con que se sucedieron las extinciones en masa indica que otras causas fueron en general las responsables de las extinciones masivas (ver Sección \ref{TioVivo}).}

Las fuentes y mecanismos de formación de los granos de polvo no están totalmente esclarecidos. Es aún un tema de debate entre los especialistas. Una posibilidad es que las partículas sólidas (polvo)  se formen 
en el medio interestelar vía la aglomeración  de átomos de sílice, hierro y otros elementos químicos que se adhieran al chocar entre ellos.  Sin embargo, el tiempo que demanda la formación de un grano típico de polvo por dicho mecanismo es demasiado alto. Los astrónomos aceptan en general que el polvo puede formarse en envolturas circunestelares producidas por la  expulsión  de masa de las estrellas centrales  hacia el medio interestelar. En las condiciones de densidades y temperaturas no muy altas, del orden de 1000 K,  que existen  en las envolturas circunestelares se podrían sintetizar los silicatos,  grafitos y minerales ferrosos que componen  los granos de polvo.

Las  estrellas empiezan a generar energía nuclear  transmutando el hidrógeno del núcleo estelar  en helio. El 90 por ciento de la vida de una estrella transcurre en este estado ``normal'', pero luego sobreviene una rápida evolución. Cuando se agota el hidrógeno de su núcleo, la parte central de la estrella se contrae y calienta mientras que las capas más externas se expanden y enfrían hasta restablecerse  el equilibrio hidroestático total. La fusión nuclear no se detiene puesto que la estrella comienza a ``quemar'' el hidrógeno de la capa que rodea al núcleo de helio. Como consecuencia, la luminosidad y el tamaño de la estrella aumentan y a la estrella se la llama \it{gigante roja} \rm  \footnote{La estrella se enrojece porque la temperatura de su superficie disminuye a valores de 2000-3000 K, con las cuales la estrella emite la mayor parte de la energía lumínica en longitudes de onda que corresponden a la luz roja. A pesar de que la energía emitida por unidad de superficie de la estrella disminuye, pues depende de la temperatura superficial de la estrella de acuerdo con la ley de Stefan-Boltzmann\index{Stefan-Boltzmann, ley de}, la luminosidad de la estrella aumenta por el gran aumento del tamaño y superficie de la estrella.}. Dentro de 5000 millones de años, el Sol ingresará en la etapa de gigante roja y envolverá a los planetas interiores y a la misma Tierra. Naturalmente, será el fin de la Tierra. Las estrellas de masas bajas y medias que alcancen en sus núcleos temperaturas de cientos de millones de grados Kelvin pueden quemar helio y a partir de lo cual producir carbono (C) y oxígeno (O). Durante la evolución como gigantes, las estrellas sufren inestabilidades que provocan desprendimientos de sus  capas externas, formando envolturas circunestelares ricas en C y O. Por lo tanto, esas estrellas serían fuentes importantes del polvo cósmico, en particular de  granos de polvo carbonados. Las envolturas circunestelares de polvo  emiten  radiación infrarroja, la cual es consistente
con la emisión infrarroja debida a granos de polvo. Se  observan detalles en los espectros infrarrojos de las envolturas circunestelares que indican la presencia de granos de polvo con temperaturas cercanas a 1000 K. Esta es una indicación de las condiciones en que el polvo puede formarse.

Las estrellas de gran masa y sus estampidos finales como supernovas son unas de las fuentes más poderosas de energía de la naturaleza que desencadenan procesos astrofísicos fundamentales. Los núcleos de las estrellas masivas, con masas mayores que ocho masas solares, están sujetos a presiones y temperaturas extremadamente  altas. Con esas condiciones físicas extremas,  se produce  la  núcleo-síntesis de elementos pesados,  como por ejemplo el aluminio,   el silicio y  el hierro. A las estrellas, podemos describirlas muy simplificadamente como inmensos globos de gas incandescente. Las estrellas se mantienen  en equilibrio hidrostático,  si  la fuerza de gravedad, que tiende a contraer la estrella hacia su centro, es contrarrestada por la presión térmica y la presión de radiación. En las estrellas masivas, la generación de energía por segundo es muy grande y por lo tanto la presión de radiación es también muy grande. La gran fuerza de gravedad que poseen las estrellas masivas es esencialmente contrarrestada por la presión de radiación. Cuando se agota el combustible nuclear, la presión de radiación cesa y   la estrella se desploma hacia su centro.

 La materia contiene el mismo número de protones y electrones y en consecuencia tiende a ser  eléctricamente neutra en una escala macroscópica. Cuando la estrella masiva moribunda se derrumba sobre su núcleo central, este núcleo central se comprime de tal modo que los protones y los electrones se acoplan físicamente formando neutrones. El colapso se detiene cuando el núcleo adquiere densidades supra-nucleares y el material que continua cayendo sobre la superficie del núcleo revota generando una onda de choque que se expande hacia
 afuera y detiene en parte la caída del resto de la estrella. Sin embargo, los astrofísicos se dieron cuenta de que la onda expansiva que se genera en el revote contra el núcleo duro neutrónico no puede invertir el movimiento total de caída de la estrella, convirtiendo una implosión en una explosión. Algo más debía ocurrir.
 
 Con la fusión de un protón y un electrón,  se crea un neutrón, tal como lo hemos dicho,  y se  emite un neutrino. Como consecuencia, el núcleo colapsado de la estrella genera un flujo colosal de neutrinos que  se dirige hacia el espacio exterior, acarreando una enorme cantidad de energía. El neutrino es una partícula, sin carga eléctrica,   que tiene  una masa  extremadamente pequeña y  que  se propaga  casi a la velocidad de la luz. La mayor parte del flujo de neutrinos escapa de la estrella, puesto que los neutrinos casi no interaccionan con la materia. Sin embargo, dado que  el número de neutrinos emitidos es  altísimo, los neutrinos absorbidos por la estrella   inyectan, en un instante, ingentes cantidades de energía sobre todo el volumen de la estrella.  En particular, la onda expansiva generada por el revote del material en caída sobre el núcleo duro, se revitaliza.  Con todo ello, la estrella se expande en forma explosiva. En el lugar de la estrella solo queda el núcleo superdenso que se convierte en una estrella de neutrones \footnote{Las estrellas de neutrones se observan como pulsáres, pues estas estrellas emiten pulsos de ondas  electromagnéticas que se repiten con un periodo de tiempo muy corto y regular. El primer pulsar  fue detectado en forma casual  con un radiotelescopio especial. Las observaciones e interpretaciones del fenómeno fueron realizadas  por un equipo inglés  dirigido por Anthony Hewish\index{Hewish, A.}, a quien le valió  en 1974 el Premio Nobel de Física. Las estrellas de neutrones tienen tamaños muy pequeños, de 10 a 20 km, campos magnéticos intensos y velocidades de rotación muy altas. A través de los polos magnéticos de la estrella, se emiten sendos haces de radiación electromagnética, luz y ondas de radio. Como   el eje de rotación de la estrella no coincide en general con su eje magnético,  la estrella de neutrones se comporta como un faro cósmico.}, si las fuerzas nucleares detienen el colapso gravitatorio, o del otro modo en un agujero negro. En coincidencia con la aparición en el cielo de la supernova 1987a  \footnote{Las supernovas son fenómenos relativamente raros y se designan con el año de  aparición en el cielo. La  supernova 1987a se encuentra en la Nube Mayor de Magallanes, a 160000 años-luz desde la Tierra,  y se podía ver a simple vista. Esta supernova es la más cercana que apareció desde el año 1604 en que apareció la supernova descubierta por Kepler\index{Kepler}, ubicada dentro de la Vía Láctea.}, dos detectores de neutrinos, uno ubicado en Japón y el otro en Estados Unidos, registraron el paso del flujo de neutrinos. Ello corroboró las ideas que se habían propuesto para explicar el fenómeno supernova. 
 
Se ha sostenido clásicamente, tal como lo hemos arriba mencionado,   que los granos de polvo se forman principalmente en las envolturas estelares que  se desprenden de las estrellas en la etapa de gigantes rojas. Sin embargo, investigaciones modernas revelan sorprendentemente que el polvo puede también formarse en el corazón de las supernovas. La evolución de  la famosa supernova 1987a ha sido estudiada en detalle desde casi el momento de la explosión en el año 1987 hasta el presente con instrumental observacional de alta sensibilidad y resolución.  Los astrofísicos teóricos deberían afilar sus lápices para dar cuenta de los complejos mecanismos de formación y destrucción de granos de polvo que operan en la materia expulsada  por las supernovas.

\section{Formación de los cometas en el medio interestelar \label{CometasInterestelares}}

\begin{quote}
\small \it{Comets remain one of the greatest enigmas of all.} \rm

 Henry Norris Russell\index{Russell, H.N.} (1877-1957) en el libro ``The solar system and its origin'' (1935) \footnote{Russell fue un astrónomo estadounidense, muy conocido por  el diagrama de Hertzsprung-Russell que desarrolló junto a  Ejnar Hertzsprung\index{Hertzsprung, E.} y que tiene gran importancia para los estudios de evolución estelar y la determinación de distancias a las estrellas.}
\end{quote}

La frase citada de Henry Norris Russell\index{Russell, H.N.}  es aún cierta hoy día. Sin pretender resolver el enigma, esbozaremos algunas ideas básicas sobre los posibles mecanismos que pueden operar en la formación de los cometas. Se piensa que los granos de polvo son los ladrillos con los cuales se formaron los cometas y, por lo tanto, deberíamos explicar primero el origen y formación de los granos de polvo. Hemos visto que las fuentes de  los granos de polvo pueden ser varias. Sin embargo, el mecanismo de formación de los granos sería esencialmente el mismo en todos los casos: un pequeño agregado de partículas atómicas crece con la incorporación de las partículas que chocan  contra su superficie y quedan pegadas a ella.
 
 Consideremos un grano embrionario de  radio $a$  que se encuentra dentro de una nube rica  en núcleos  atómicos pesados,  como por ejemplo el carbono, de densidad numérica $n$.  Supongamos que, en el marco de un sistema de coordenadas solidario con la nube, los núcleos atómicos pesados de la nube tienen todos las mismas velocidades $v$ en cuanto a su magnitud, pero sus direcciones de movimiento están distribuidas al azar. Dentro de un diferencial de tiempo $dt$, todas las partículas atómicas que se encuentran dentro de una esfera de radio $R=v dt$ centrada en el grano de polvo y que sus velocidades  se dirigen hacia el grano de polvo chocarán contra la superficie del grano de polvo.
 
  En particular, calculemos cuantas partículas atómicas que yacen en el cascarón esférico de radio  $R$ y espesor $dR$  chocan y se adhieren al grano de polvo. Para tal fin, tomemos un diferencial de volumen $dV= dR dS$ dentro del cascarón. Consideremos una esfera que está centrada en $dV$ y que contiene la sección $\pi a^{2}$ del grano de polvo.  Dado que la distribución de velocidades de las partículas  es  considerada isotrópica,  la proporción de partículas que se dirigen hacia la superficie $\pi a^{2}$ con respecto a la totalidad de la esfera, $4 \pi R^{2}$, es naturalmente $\frac{\pi a^{2}}{4 \pi R^{2}}$. Por lo tanto la cantidad de partículas de $dV$ que chocan el grano de polvo es $\frac{ a^{2}}{4 R^{2}} n dV$. Como el volumen del cascarón es $4 \pi R^{2} dR$, el número de $dV=dR dS$  contenido en el cascarón es $N=\frac{4 \pi R^{2} dR}{dR dS}$ y el número total de partículas del cascarón que chocan contra el grano es $ N \frac{ a^{2}}{4 R^{2}} n\, dV = \pi a^{2} n \, dR$.
  
  Si dividimos la esfera de radio $R$ en $N_{c}$ cascarones y definimos el espesor de cada cascarón por $dR=\frac{R}{N_{c}}$, el aporte que cada cascarón hace al número de las partículas que chocan contra el grano de polvo es igual a $\pi a^{2} n \, dR= \pi a^{2} n \,\frac{R}{N_{c}}$. Por lo tanto, la cantidad total de partículas, $\cal N$, que chocan con el grano de polvo es $\cal N \it = N_{c} \pi a^{2} n \,\frac{R}{N_{c}}= \pi a^{2} n \, R$ y si tenemos en cuenta que $R = v dt$:
  
\begin{equation}
  \cal N \it = \pi a^{2} n \, v \, dt.
  \label{Nchoques}
\end{equation}
 
Si $m$ es la masa media de las partículas pesadas y $\rho$ la densidad de ellas, el volumen de cada partícula es $\frac{m}{\rho}$ y por lo tanto el volumen agregado al grano de polvo, $\Delta \cal V$,  es $\Delta \cal V \it=  \cal N \it \frac{m}{\rho}$, con lo cual resulta:
\begin{equation}
\Delta \cal V \it = \pi a^{2} n   \frac{m}{\rho} \, v \, dt .
\label{Vpolvo}
\end{equation}
Por otro lado, si  las partículas incorporadas al grano se reparten uniformente sobre la superficie del grano inicial forman una capa esférica de radio $a$ y espesor $da$ y por lo tanto $\Delta \cal V \it= \pi a^{2} da$, e igualando con (\ref{Vpolvo})  obtenemos el aumento del radio del grano de polvo:
\begin{equation}
da= n   \frac{m}{\rho} \, v \, dt .
\label{Vpolvo2}
\end{equation}
La ecuación (\ref{Vpolvo2}) muestra que la variación de $a$ con respecto al tiempo es una constante: $\frac{da}{dt}=\frac{a(t)-a_{0}}{t}=n   \frac{m}{\rho} \, v =const.$ En consecuencia, la ecuación (\ref{Vpolvo2}) implica que
\begin{equation}
a(t)=a_{0}+  n \frac{m}{\rho} \, v \, t,
\label{acrecion}
\end{equation}
donde $a_{0}$ es el radio inicial del grano de polvo. La fórmula (\ref{acrecion}) es aplicable  tanto para los granos de polvo formados en una nube gaseosa  fría como para aquellos formados en una nube caliente. A continuación estudiamos el caso en el cual los granos de polvo se forman y aumentan de tamaño en una nube de gas totalmente ionizado, debido a su alta temperatura.

\subsection{Tiempo de enfriamiento de una densa nube de plasma y formación de  granos de polvo, de meteoroides y de asteroides en ella}

\begin{quote}
\small \it{Así, el helio y todos los elementos  más pesados son el resultado de la evolución cosmológica. Su historia, y en especial la historia de los elementos más pesados, es,  según las concepciones cosmológicas actuales, una historia extraña. Los elementos más pesados  se consideran actualmente productos de explosiones de supernovas..., todos los núcleos más pesados parecen ser extremadamente raros. Así, la Tierra y, presumiblemente, los demás planetas de nuestro sistema solar están hechos principalmente de materiales muy raros (y yo diría que preciosos).} \rm   
 
 Karl R. Popper\index{Popper, K.R.} (1902-1994) en el libro ``El universo abierto: Un argumento en favor del indeterminismo'' (Post Scriptum a La lógica de la investigación científica. Vol II, 1982)
\end{quote}
 
 Los propulsores de las teorías catastrofistas sobre el origen del Sistema Solar (Ver parte I de esta monografía \citep{Olano5}) intuyeron correctamente que al enfriarse el material  incandescente desprendido de las estrellas  puede formar cuerpos metálicos y rocosos, tales  como asteroides, cometas y planetas. Sin embargo, la naturaleza es muy sutil \footnote{Einstein\index{Einstein, A.} decía que Dios es sutil, pero no malicioso. Aunque a veces dudaba.} y no lo hace de la manera directa en que la imaginaron los catastrofistas para el origen de nuestro sistema planetario. Tal como lo hemos mencionado, las estrellas de gran masa explotan al final de sus vidas como supernovas y esparcen violentamente al medio interestelar la materia procesada en sus núcleos. Los remanentes de supernovas se comportan  como mezcladores del  material interestelar, compuesto de desechos estelares  y de gas primigenio, favoreciendo el reciclaje del material interestelar para formar nuevas estrellas y sistemas planetarios. En general,  las estrellas emiten a lo largo sus vidas,  además de luz, un flujo de partículas eléctricamente cargadas  que se denomina   viento estelar. El viento estelar es particularmente  intenso en las estrellas de gran masa. En las etapas evolutivas finales, las estrellas sufren  inestabilidades y episodios de expulsión de masa, con lo cual pueden formarse  envolturas circunestelares. Si el viento  de la estrella central es suficientemente fuerte, el material circunestelar es impulsado hacia el medio interestelar, como un velero  empujado por el viento. Este es otro mecanismo por el cual las estrellas  pueden inyectar su material inicialmente ígneo  en el medio interestelar.

La materia expulsada por las estrellas, ya sea que pase a formar parte del medio circunestelar o del medio interestelar, se encuentra originalmente en estado de plasma  debido a la alta temperatura de dicha materia. El plasma es  gas ionizado de alta conductividad eléctrica  formado esencialmente por  protones y electrones libres. Las estrellas poseen fuertes campos magnéticos (de 300  a 30000 Gauss) y el plasma eyectado por una estrella lleva consigo las líneas del campo magnético que contenía mientras estaba en el interior de la estrella. Este fenómeno de ``congelamiento'' del campo magnético en el interior del plasma  fue descubierto por el físico sueco Hannes Alfvén\index{Alfvén, H.} (1908-1995), Premio Nobel de Física en el año 1970. Hemos visto en la Parte I de esta monografía \citep{Olano5} que las líneas de fuerza magnéticas frenan a las partículas cargadas que intentan atravesarlas perpendicularmente.  Por lo tanto, una nube de plasma adquiere una  cierta rigidez gracias a su magnetismo y se comporta en cierto modo como un cuerpo sólido. A continuación,  demostraremos que las nubes de plasma estelar pueden ser sitios de formación del polvo cósmico.

Los campos magnéticos, densidades numéricas y temperaturas, denotados aquí por  $B$, $n$ y $T$  respectivamente, que poseen las nubes de plasma al ser eyectadas por una estrella, deben ser similares a las de las capas exteriores de la estrella, incluyendo su fotosfera, a partir de las cuales  dichas nubes de plasma emergen. En una situación de equilibrio, la presión térmica $P_{th}$ y  la presión magnética $P_{m}$ (= $\frac{B^{2}}{8 \pi})$, a las que están sujetas las capas exteriores de una estrella, deben igualarse. Según la clásica fórmula de la presión de un gas perfecto, $P_{th}= n k T$, donde $k$ es la constante de Boltzmann, y como hemos dicho $n$ es el número de partículas por cm$^{3}$. Por lo tanto, $n k T=\frac{B^{2}}{8 \pi}$  y,  a partir de esta igualdad,  podemos estimar $n$  si poseemos las magnitudes observacionales de  $B$ y $T$:
\begin{equation}
n= \frac{B^{2}}{8 \pi k T} \, .
\label{densidadfotosferica}
\end{equation}
Si adoptamos $B=300$ Gauss,  cota inferior  estimada a través de la medición  del efecto Zeeman\index{Zeeman, efecto} sobre las líneas espectrales estelares, y $T=10000$ K  como una temperatura  media de las fotosferas de estrellas masivas evolucionadas, obtenemos mediante la fórmula(\ref{densidadfotosferica})  que $n \sim 10^{15}$ cm$^{-3}$. Por cierto, ese número para la densidad fotosférica es una mera estimación y puede variar dentro de unos pocos  ordenes de magnitud. Pues,  los movimientos convectivos y de expansión del plasma determinan también  las condiciones físicas de las fotósferas estelares. Nosotros usaremos densidades numéricas para el plasma de las envolturas circunestelares entre $10^{12}$ y $10^{15}$ cm$^{-3}$.

El auto confinamiento magnético de una nube de plasma de tamaño $L$ y conductividad eléctrica $\Gamma$ tiene una vida media que puede estimarse por medio  del tiempo de decaimiento magnético $\tau_{m}$ dado por

\begin{equation}
\tau_{m}= \Gamma \mu_{0} L^{2},
\end{equation} 
donde $\mu_{0}= 4 \pi \times 10^{-7} \rm T.m.A^{-1}$ es la permeabilidad magnética. La conductividad eléctrica de un plasma depende de su temperatura $T$ y es dada por la fórmula 
$\Gamma \approx 10^{-2} T^{3/2} \rm (ohm\,  m)^{-1}$. Por ejemplo,  para una nube de plasma de $T=10000$ K y $L=60000$ km (cinco veces el diámetro de la Tierra),  $\tau_{m}\sim 10^{6}$ años. Es decir, las vidas de las nubes magnéticamente confinadas  son  largas aun para nubes de plasma relativamente pequeñas.

 Se hace difícil concebir la idea de que, con temperaturas del plasma del orden de 10000 K, las partículas del plasma puedan amalgamarse para formar granos de polvo. Sin embargo, debemos tener en cuenta que la temperatura de los granos de polvo no necesariamente es igual a la temperatura del plasma en el cual los granos de polvo se encuentran inmersos. Por ejemplo, las órbitas de los satélites y estaciones espaciales  se ubican generalmente en  la termósfera de la Tierra, que tiene temperaturas de $\approx 3000$ K  y, sin embargo, esas naves espaciales y sus astronautas no se vaporizan. Para evitar por ahora el problema del inicio de la formación de los granos de polvo,  supongamos que granos de polvo de muy baja temperatura penetran desde el exterior a una nube de plasma de 10000 K. En dichas condiciones, trataremos de estimar cuál es la temperatura que adquiere un grano de polvo.

La energía cinética  de cada partícula del plasma que choca contra un grano de polvo puede ser, al menos en parte,  absorbida  por el grano de polvo y así contribuir al calentamiento del mismo. Si  el hidrógeno,  que es el elemento más abundante del Universo, se encuentra ionizado en el plasma, el número $n_{e}$ de  electrones libres y el número $n_{p}$ de protones libres por $cm^{3}$ son aproximadamente iguales: $n_{e}\approx n_{p}\approx n/2$. Las abundancias relativas al hidrógeno de los elementos pesados, C, Mg, Al, Si, Fe, O, Ni y S, todas sumadas dan $9.5 \times 10^{-4}$. Por lo tanto, la densidad numérica de los núcleos atómicos pesados es $n_{np}\approx 9.5 \times 10^{-4} n/2$. Mientras los elementos livianos del plasma están totalmente despojados de electrones,  los núcleos pesados, es decir con muchos protones y neutrones, conservan un gran séquito  de electrones en su entorno. Los tamaños de las zonas en torno a las cuales se mueven los electrones ligados al núcleo pesado son cientos de miles de veces mayores que los núcleos desnudos. Por lo tanto, los núcleos pesados ofrecen un gran blanco y, con gran probabilidad, pueden ser detenidos al ingresar al interior de los granos de polvo. 
Denotamos con los símbolos $\zeta_{e}$,  $\zeta_{p}$ y  $\zeta_{np}$, la fracción de la energía cinética de los electrones, los protones y núcleos pesados, respectivamente, que es depositada al chocar dichas partículas contra un grano de polvo. 
Sin embargo, los valores de  $\zeta_{e}$,  $\zeta_{p}$ y  $\zeta_{np}$ son difíciles de determinar tanto por el lado teórico, que requiere la aplicación de complejos conceptos de la mecánica cuántica, como por el lado experimental.

La velocidad de calentamiento de un grano de polvo debido a la energía inyectada por las  partículas que chocan con el grano de polvo (ver expresión \ref{Nchoques}) es :

\begin{equation}
\frac{\Delta Q_{+}}{\Delta t}= \zeta_{e} \pi a^{2} n_{e} v_{e} E_{e} + \zeta_{p} \pi a^{2} n_{p} v_{p} E_{p} +\zeta_{np} \pi a^{2} n_{np} v_{np} E_{np},
\label{EnergiaAbsorbida1}
\end{equation}
donde $a$ es el radio del grano de polvo, $E_{e}$, $E_{p}$
y $E_{np}$ son las energías cinéticas medias de los electrones, protones y núcleos pesados, respectivamente, y  $v_{e}$, $v_{p}$ y $v_{np}$ son las correspondientes velocidades medias. Por lo tanto, $E_{e}=\frac{1}{2} m_{e} v_{e}^{2}$,  $E_{p}=\frac{1}{2} m_{p} v_{p}^{2}$ y $E_{np}=\frac{1}{2} m_{np} v_{np}^{2}$, donde $m_{e}$,   $m_{p}$ y 
 $m_{np}$ son las masas de los electrones, los protones y los núcleos pesados, respectivamente. Por otro lado, de acuerdo con la ley de equipartición de la energía y la teoría cinética de los gases, $E_{e}=E_{p}=E_{np}=\frac{3}{2} k T$, donde $k$ es  
la constante de Boltzmann y $T$ la temperatura del plasma. Dado que $E_{e}=\frac{1}{2} m_{e} v_{e}^{2}=\frac{3}{2} k T$, obtenemos 

\begin{equation}
v_{e}=\sqrt{\frac{3 k T}{m_{e}}},
\label{Ve}
\end{equation}
y procediendo de forma similar,  
\begin{equation}
v_{p} = \sqrt{\frac{3 k T}{m_{p}}}
\label{Vp}
\end{equation}
y
\begin{equation}
v_{np}  =  \sqrt{\frac{3 k T}{16\, m_{p}}} .
\label{Vnp}
\end{equation}
En la fórmula (\ref{Vnp}), hemos utilizado como masa media de los elementos pesados $m_{np}= 16\,  m_{p}$.

Además, el grano de polvo se enfría irradiando luz como un cuerpo negro según la siguiente expresión,
\begin{equation}
\frac{\Delta Q_{-}}{\Delta t}= 4 \pi a^{2} \sigma T_{d}^{4},
\label{EnergiaIrradiada}
\end{equation}
donde $T_{d}$ es la temperatura del grano de polvo. En un estado estacionario, las ganancias y pérdidas de energía del grano de polvo se igualan y por lo tanto   las expresiónes (\ref{EnergiaAbsorbida1})  y (\ref{EnergiaIrradiada})
deben ser iguales, condición que nos permite calcular la temperatura $T_{d}$ de equilibrio:
\begin{equation}
T_{d}= \left(\zeta_{e} \frac{3 k T}{16\, \sigma} n \sqrt{\frac{3 k T}{m_{e}}} + \zeta_{p} \frac{3 k T}{16\, \sigma} n \sqrt{\frac{3 k T}{m_{p}}}+ \zeta_{np} \frac{3 k T}{16 \, \sigma} 9.5 \times 10^{-4} n \sqrt{\frac{3 k T}{16 \, m_{p}}}\right)^{1/4} .
\label{Tdust}
\end{equation}
Para obtener (\ref{Tdust}), hemos reemplazado previamente en (\ref{EnergiaAbsorbida1}) las velocidades $v_{e}$, $v_{p}$ y $v_{np}$ por sus expresiones (\ref{Ve}), (\ref{Vp}) y (\ref{Vnp}). Obsérvese que  el parámetro $a$, radio del grano de polvo,   se canceló en la fórmula (\ref{Tdust}). Sin embargo, implícitamente depende de $a$ pues $\zeta_{e}$,  $\zeta_{p}$ y $\zeta_{np}$ son funciones de  $a$ y de $T$. En efecto, cuanto más grande es el tamaño del grano  de polvo, mayor es la probabilidad de que las partículas que penetran el grano de polvo queden atrapadas en su interior. Por otra parte, al aumentar  la temperatura $T$ del plasma,  aumenta la energía cinética $E$ de las partículas y así la probabilidad de que las partículas traspasen el grano de polvo. 

Supongamos que un electrón traspasa un grano de polvo. Si $E$ y $E'$ son  la energía cinética del electrón al ingresar  y su energía al emerger de dicho grano de polvo, respectivamente,  la energía transferida al grano de polvo es $E-E'= E(1-\frac{E'}{E})= \zeta_{e} E$. Es decir, $\zeta_{e} (a,E)= (1-\frac{ E'}{E})$. Si el electrón es detenido en el interior del grano de polvo, toda la energía  $E$ fue absorbida por el grano de polvo y, por lo tanto, $E'=0$ y $\zeta_{e} (a,E)=1$. En cambio, si $E'=E$, $\zeta_{e} (a,E)=0$, lo cual significa que  el electrón traspasó el grano de polvo sin transferir parte de su energía al grano de polvo. Basado en datos experimentales \cite{Dwek},
\begin{eqnarray}
\zeta_{e} & \approx & 0.875 \, ,  \nonumber \\
\zeta_{p} & \approx & 1,   \nonumber \\ 
\zeta_{np} & \approx & 1,  \nonumber \\  
\label{Zeta}
\end{eqnarray}
si $E<E_{u}$, donde $E_{u}$ es una energía umbral de $\approx$ 1 kev para $a=0.1 \mu m$. Como para $T\le$15000 K, $E\le 2$ ev, podemos aplicar los valores dados por  (\ref{Zeta}) en la expresión (\ref{Tdust}). Comparando las relaciones (\ref{Ve}) y (\ref{Vp}), $v_{e}=\sqrt{\frac{m_{p}}{m_{e}}}\, v_{p}\approx 43\, v_{p}$. Es decir, la velocidad de los electrones es mucho mayor que la velocidad de los protones y de los núcleos pesados. Entonces, en un dado intervalo de tiempo, el número de choques de los electrones contra un grano de polvo es mucho mayor que el número de choques de las otras partículas. En consecuencia, los granos de polvo son calentados básicamente por los choques con electrones. Hecho que podemos comprobar con la ecuación (\ref{EnergiaAbsorbida1}), donde debemos considerar que las energías de las diferentes partículas son iguales y que los valores de los parámetros $\zeta$ son dados por las relaciones (\ref{Zeta}).

\begin{figure} 
\includegraphics[scale=0.65]{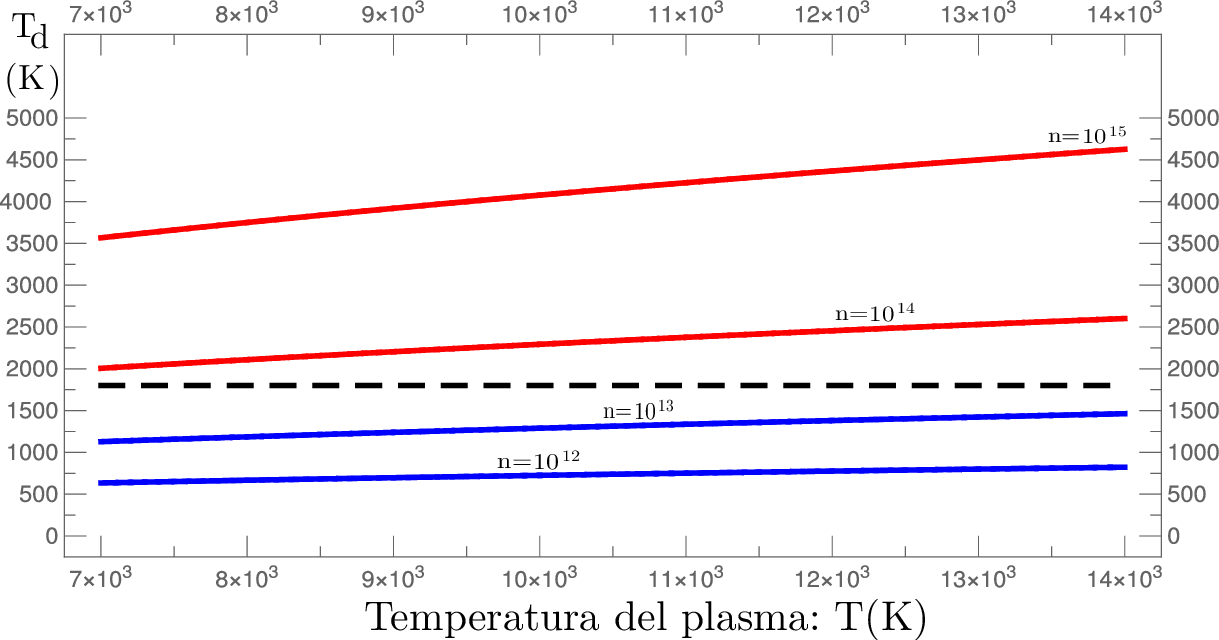} 
\caption{Temperatura $T_{d}$ del polvo en grados Kelvin (K), eje vertical, en función de la temperatura del plasma, eje horizontal, para diferentes valores de la densidad  $n$  de partículas del plasma por $cm^{3}$. La línea horizontal de trazos indica la temperatura de fusión del hierro en grados Kelvin. Sobre cada curva se anota el valor de $n$ dado para el plasma.}
\label{TemperaturaPolvo}
\end{figure}

Consideremos una nube de plasma cuya densidad $n$ se conserva en el tiempo, pero su temperatura $T$ disminuye debido al enfriamiento del plasma. Para calcular las temperaturas $T_{d}$ de los granos de polvo que se encuentran inmersos en dicha nube de plasma,  podemos emplear la fórmula (\ref{Tdust}) con el valor de $n$ dado  y las distintas temperaturas $T$ que el plasma puede tomar a lo largo del tiempo. En la Fig. \ref{TemperaturaPolvo}, representamos curvas  de $T_{d}$ en función de $T$ para cuatro  densidades del plasma: $n= 10^{12}, 10^{13}, 10^{14}$ y $10^{15}$ $cm^{-3}$, las cuales son  indicadas sobre las respectivas curvas. Cada curva muestra la variación de $T_{d}$ en el rango representado de $T$, entre 7000 y 14000 K. En el caso de la nube con densidad $n= 10^{12}$ $cm^{-3}$, la temperatura del polvo es de solo 822 K cuando la  temperatura del plasma es de 14000 K y de 634 K cuando $T=7000$ K.  En el caso de la nube con densidad $n= 10^{13}$ $cm^{-3}$, $T_{d}=1128-1463$ K. Las temperaturas del polvo de ambas nubes (representadas por las curvas azules en la Fig. \ref{TemperaturaPolvo}) se encuentran por debajo de la temperatura de fusión del hierro ($\approx 1500$ K)  y, por lo tanto,  el polvo puede coexistir con dichas nubes de plasma sin destruirse. En cambio, $T_{d}>1500$ K para $7000<T<14000$ en las nubes de plasma con $n= 10^{14}$ y  $10^{15}$ $cm^{-3}$, como lo muestran las curvas en rojo de la Fig. \ref{TemperaturaPolvo}. Es decir, los granos de polvo que ingresen en nubes de plasma con densidades $n \geq 10^{14}$ $cm^{-3}$ son calentados a temperaturas mayores que la temperatura de fusión del hierro y, en consecuencia, se evaporarán. La Fig. \ref{TemperaturaPolvo} muestra que la densidad del plasma es un factor importante en la determinación de la temperatura del polvo.

Arriba,  hemos supuesto que el número de electrones y protones libres del plasma es dado por la ionización casi completa del hidrógeno atómico. Por lo tanto, la cantidad de hidrógeno atómico neutro, generalmente denotado por $H_{I}$, es muy pequeña en el plasma considerado. En lenguaje matemático,  $\frac{n_{H_{I}}}{n} \ll 1$. A medida que la temperatura $T$ del plasma disminuye, aumenta la probabilidad de que electrones y protones se recombinen y formen $H_{I}$. La pregunta que deseamos contestar es: ¿por debajo de qué temperatura $T$, la densidad  $n_{H_{I}}$ del plasma comienza a aumentar significativamente? Para responder a ello, debemos resolver la ecuación de equilibrio de ionización de Saha \footnote{Meghnad Saha\index{Saha, M.} (1893-1956) fue un astrónomo indio, quien calculó el estado de ionización de los diferentes elementos de la atmósfera del Sol y, a partir de las intensidades de las rayas de Fraunhofer\index{Fraunhofer, rayas de} en el espectro solar, obtuvo la correcta composición química del Sol. En fase gaseosa, los átomos y moléculas emiten o absorben luz en un conjunto discreto  de frecuencias (``rayas de  Fraunhofer\index{Fraunhofer, rayas de}'') que caracterizan a cada especie atómica y permiten identificarla. En los espectroscopios clásicos, la luz de la estrella a analizar ingresa a través de una delgada ranura, de modo que cada raya de Fraunhofer\index{Fraunhofer, rayas de} corresponde a la imagen de la ranura iluminada por luz de una determinada frecuencia o longitud de onda.} que relaciona, para una especie atómica dada,  el número $n_{i+1}$ de átomos en el estado de ionización $(i+1)$ con respecto al número $n_{i}$ en el estado de ionización $i$:
 
\begin{equation}
\frac{n_{i+1}}{n_{i}}= 2 \left(\frac{2 \pi m_{e} k T}{h^{2}}\right)^{3/2} \frac{g_{i+1}}{g_{i}} \frac{e^{-I_{i}/(k T)}}{n_{e}},
\label{ESaha}
\end{equation}
donde $I_{i}$ es el potencial de ionización del estado de ionización $i$. La aparición de la constante $h$ de Planck en la fórmula de Saha (\ref{ESaha}) indica que ésta  contiene  conceptos de la física cuántica.

 Nuestro propósito es aplicar (\ref{ESaha}) al hidrógeno, el cual  solo se puede encontrar en estado neutro, $H_{I}$, o totalmente ionizado, $H_{II}$. Por lo tanto, $n_{1}=n_{H_{I}}$ y $n_{2}=n_{H_{II}}$. Como el hidrógeno es el elemento más abundante, podemos considerar que $n_{H_{II}}$ es igual al número  de protones libres, el cual a su vez es igual al número de electrones libres $n_{e}$. En consecuencia, $n_{2}=n_{H_{II}}=n_{e}$. Por otra parte, para el átomo de hidrógeno, $I_{1}=13.8$ ev,  $g_{1}=2$ y $g_{2}=1$. Haciendo los reemplazos correspondientes en (\ref{ESaha}) obtenemos
 
 \begin{equation}
 \frac{n_{e}^{2}}{n_{H_{I}}}= \left(\frac{2 \pi m_{e} k T}{h^{2}}\right)^{3/2}  e^{-I_{1}/(k T)}.
\label{ESaha2}
 \end{equation}
 
 La densidad total $n$ de partículas (electrones y protones, tanto libres como formando $H_{I}$) por cm$^{-3}$ puede considerarse una constante para un plasma dado. En consecuencia,  $n=2 n_{e}+2 n_{H_{I}}$, de donde obtenemos que  $n_{e}=\frac{1}{2}(n-2 n_{H_{I}})$; y reemplazando $n_{e}$ por su igualdad en (\ref{ESaha2}), obtenemos la siguiente ecuación cuadrática de nuestra incógnita, $X=\frac{n_{H_{I}}}{n}$:
 \begin{equation}
 X^{2}+ b(T,n)\, X +1/4=0,
 \end{equation}
 donde  $b(T,n)$ es una función de $n$ y $T$ dada por la siguiente fórmula
 
 $b(T,n)= -1 - \frac{1}{n} \left(\frac{2 \pi m_{e} k T}{h^{2}}\right)^{3/2} e^{-I_{1}/(k T)}$. 
 Por lo tanto,
 \begin{equation}
 X=\frac{-b(T,n) \pm \sqrt{b(T,n)^{2}-1}}{2}.
 \label{SolSaha}
 \end{equation}
 
\begin{figure} 
\includegraphics[scale=0.55]{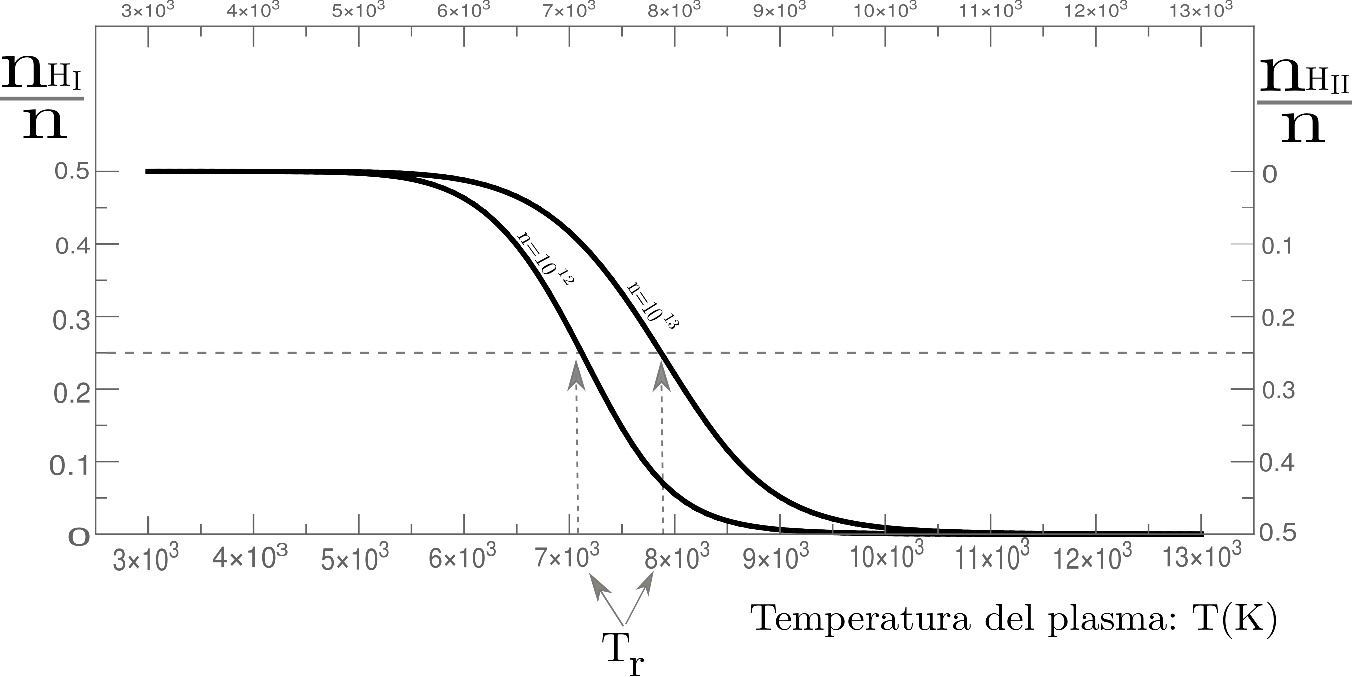} 
\caption{Proporción del número  de átomos de $H_{I}$ con respecto al número total de partículas $n$, $X(=\frac{n_{H_{I}}}{n})$ indicado sobre el eje vertical izquierdo, en función de la temperatura del plasma  para $n=10^{12} $ y $n=10^{13}$ cm$^{-3}$, valores anotados sobre sus respectivas curvas. Sobre el eje vertical derecho se indica $\frac{n_{H_{II}}}{n}$.}
\label{Saha}
\end{figure}

 La Fig. \ref{TemperaturaPolvo} muestra que el polvo puede sobrevivir  en plasmas de densidades $n=10^{12}$ y $n=10^{13}$  cm$^{-3}$. Si fijamos $n$ en uno de dichos valores,  la ecuación (\ref{SolSaha}) nos permite dibujar $X(=\frac{n_{H_{I}}}{n})$ en función de $T$ (ver Fig. \ref{Saha}). Cuando el plasma llega a temperaturas  menores de 5500 K,     $\frac{n_{H_{I}}}{n}=0.5$ para ambas curvas representadas en la Fig. \ref{Saha}. Ello significa que todos los protones y electrones se unieron formando $H_{I}$  y, por lo tanto, el número de parejas formadas, $n_{H_{I}}$,  es la mitad del número de partículas individuales: $n_{H_{I}}=0.5\, n$. Es decir, el plasma se convierte casi enteramente en átomos de hidrógeno  y, en consecuencia, $n_{H_{II}}=0$ tal como se indica en el eje vertical derecho de la Fig. \ref{Saha}. Por otro lado, para $T\ge 10000$ K, $\frac{n_{H_{I}}}{n}=0$, lo cual significa que el plasma se encuentra totalmente ionizado. En consecuencia, $\frac{n_{H_{II}}}{n}=\frac{1}{2}$, lo cual significa que la mitad de las partículas consiste en protones libres.

La longevidad de una nube de plasma está relacionada con su tiempo de enfriamiento $\tau_{e}$, superado este tiempo, el plasma se convierte en hidrógeno neutro. Por otra parte, a medida que el plasma se enfría, la densidad de electrones libres  y con ello  las corrientes eléctricas que mantienen a la nube magnetizada  disminuyen. Por lo tanto, la nube de plasma tendería a dispersarse. Si la temperatura inicial del plasma es $T_{0}$, digamos igual a 15000 K,  después de un cierto tiempo,  el plasma se enfría y alcanza la temperatura $T_{re}$ (temperatura de recombinación), con la cual las cantidades de $H_{I}$ y de $H_{II}$ son iguales.  La Fig. \ref{Saha} muestra que $T_{re}\approx 7000$ K y 8000 K para  plasmas de densidades de $10^{12}$ y  de $10^{13}$ cm$^{-3}$, respectivamente. Para nuestros fines es conveniente definir $\tau_{e}$ como el tiempo que tarda la nube de plasma  en pasar de $T_{0}$ a $T_{re}$. A continuación, exponemos un procedimiento para estimar $\tau_{e}$ de una nube de plasma, con lo cual podremos determinar el engrosamiento de los granos de polvo que se desarrollan en la nube de plasma.
 
En una  densa nube esférica de plasma, de radio $R_{c} \,(=\frac{L}{2})$, el calor $Q$ de la nube fluye hacia el exterior  esencialmente por  conducción térmica \footnote{Las nubes de  plasma de baja densidad son transparentes a la radiación de frenado, a veces referida con la palabra alemana ``bremsstrahlung'', que se produce en el interior del plasma por la desaceleración entre partículas eléctricamente cargadas. Por lo tanto, en estos casos, la  pérdida de calor  es esencialmente radiativa.},  de acuerdo con la ley clásica de Spitzer-Härm\index{Spitzer-Härm, ley de}:
\begin{equation}
\frac{\Delta Q}{\Delta t}=-k_{0} T^{5/2} \frac{\Delta T}{\Delta r},
\label{Spitzer-Härm}
\end{equation}
donde la temperatura $T$ es expresada en grados Kelvin ($\rm K$) y $k_{0} \approx 6 \times  10^{-12} \rm J \,m^{-1} s^{-1} K^{-7/2}$. Note la similitud de  (\ref{Spitzer-Härm}) con la ley de Fourier\index{Fourier, ley de} (\ref{LeyFourier}), y por lo tanto usaremos un procedimiento similar al utilizado  en la Sección \ref{TierraTibia}, pero en este caso  para calcular el tiempo de enfriamiento $\tau_{e}$ de una nube densa de plasma. Si dividimos la nube en $N$ cascarones de espesor $\Delta r= \frac{R_{c}}{N}$, tal como las capas de una cebolla, e identificamos a cada capa con  el índice $i$, el flujo neto de calor que pasa por el cascarón $i$ en el instante $t$ es
\begin{equation}
\frac{\Delta Q_{i}}{\Delta t}=4 \pi r_{i}^{2} k_{0}   T_{i} (t)^{5/2}  \left(\frac{T_{i+1}(t)-T_{i}(t)}{\Delta r}-\frac{T_{i}(t)-T_{i-1}(t)}{\Delta r}\right),
\label{FlujoCalor4}
\end{equation}
donde $r_{i}$ y $T_{i}(t)$ son  el radio medio y temperatura del cascarón $i$, mientras $T_{i+1}(t)$ y $T_{i-1}(t)$ son las temperaturas de los cascarones adyacentes en el instante $t$. La cantidad de calor que contiene el cascarón $i$, $Q_{i} (t)$,  es dado por $ n V_{i} E_{m} $, donde $n$ es la densidad numérica de partículas,  $V_{i}=4 \pi r_{i}^{2} \Delta r$ es el volumen del cascarón y  $ E_{m}$ es la energía media de cada partícula. De acuerdo con la teoría cinética de los gases,   $E_{m}=\frac{3}{2} k T_{i}(t)$. Por lo tanto,

\begin{equation}
Q_{i} (t)= 6 \pi k \,  n  \, r_{i}^{2}\,  \Delta r \, T_{i}(t)
\end{equation}
y
\begin{equation}
\frac{\Delta Q_{i}}{\Delta t}=6 \pi k \,  n  \, r_{i}^{2}\,  \Delta r \,\frac{\Delta T_{i}(t)}{\Delta t} .
\label{FlujoCalor5}
\end{equation}
Igualando (\ref{FlujoCalor4}) y (\ref{FlujoCalor5}), obtenemos

\begin{equation}
\frac{\Delta T_{i}(t)}{\Delta t}=\frac{2 k_{0}}{3 k} \frac{ T_{i} (t)^{5/2}}{n} \frac{1}{\Delta r} \left(\frac{T_{i+1}(t)-T_{i}(t)}{\Delta r}-\frac{T_{i}(t)-T_{i-1}(t)}{\Delta r}\right).
\label{FlujoCalor6}
\end{equation}

\begin{figure} 
\includegraphics[scale=0.9]{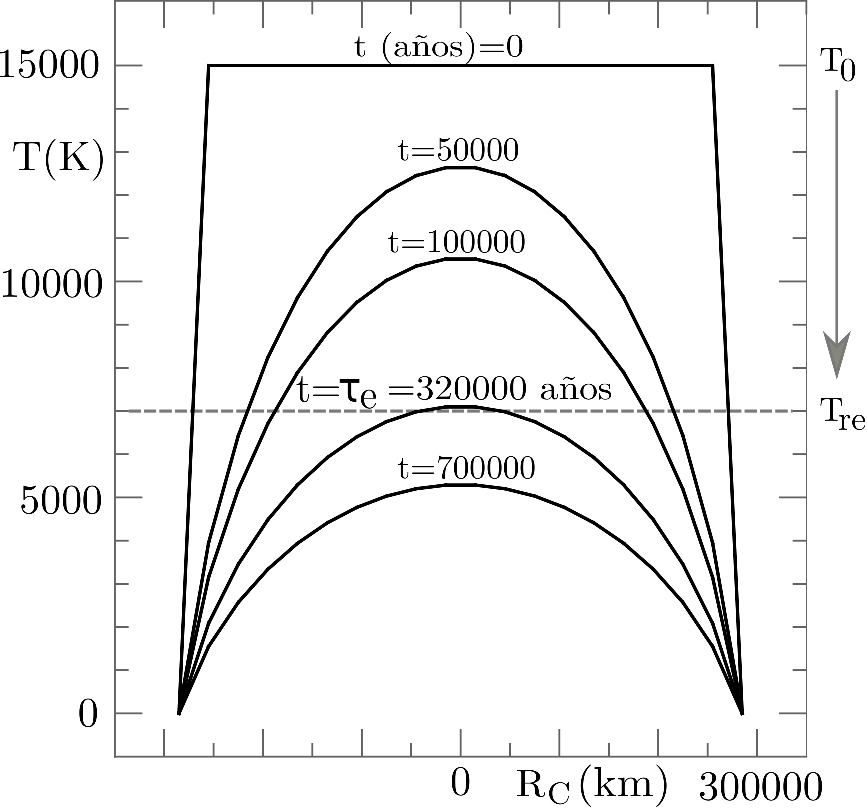} 
\caption{Perfil de temperatura de una nube de plasma de densidad $n=10^{12}$ cm$^{-3}$ y radio $R_{c}=300000$ km para diferentes tiempos $t$. Sobre cada curva se anota el tiempo expresado en años. 
El tiempo $t$ que tarda el centro de la nube en enfriarse a la temperatura $T_{re}=$ 7000 K (indicada por la línea de trazos)  es aquí llamado tiempo de enfriamiento $ \tau_{e}$ y en este caso $\tau_{e}=$ 320000 años.}
\label{EnfriamientoPlasma}
\end{figure}

El cascarón $i=1$ es la piel de la nube de plasma, donde su superficie externa  está en contacto con el vacío, y solo puede perder energía por radiación. Por lo tanto, el cascarón $i=1$ se enfría rápidamente y su temperatura permanece siempre muy baja. En consecuencia, supondremos por simplicidad que $T_{1} (t)=0$ para todo tiempo $t$. Además adoptamos como condición inicial, es decir para $t=0$,  que $T_{i} (0)=T_{0}=$constante para $i\geq 2$. Con dichas condiciones y pasos $\Delta t$ y $\Delta r$  suficientemente pequeños, podemos resolver de manera iterativa  la ecuación (\ref{FlujoCalor6}), mediante un procedimiento  similar  al empleado para resolver la ecuación (\ref{FlujoCalor2}) (Sección \ref{TierraTibia}).

 Con el procedimiento arriba delineado, analizaremos, como  ejemplos conservadores, nubes de plasma con radios $R_{c}$ entre 100000 y 300000 km (la distancia Tierra-Luna), muy pequeñas con respecto al tamaño de las estrellas gigantes y supergigantes. Consideramos que la temperatura inicial $T_{0}$ es uniforme e  igual a 15000 K para todas las nubes de plasma. Es decir, en el tiempo $t=0$,  $T_{0}=15000$ K. En la Fig. \ref{EnfriamientoPlasma}, representamos  la evolución de la distribución de temperatura para una nube de $R_{c}=300000$ km y $n=10^{12}$ cm$^{-3}$. Las curvas correspondientes a  los diferentes tiempos $t$ representados muestran que las temperaturas de la nube decaen abruptamente hacia los bordes de la nube y  que el centro de la nube tiene siempre la temperatura máxima. Definimos el tiempo de enfriamiento $\tau_{e}$ como el tiempo en el cual la temperatura del centro de la nube es igual a $T_{re}=7000$ K. En el caso particular mostrado en la Fig.  \ref{EnfriamientoPlasma},  encontramos que $\tau_{e}=320000$ años. En ese momento, la región central de la nube está parcialmente ionizado, pero hacia los bordes predomina el hidrógeno neutro ($H_{I}$).
 
  \begin{figure} 
\includegraphics[scale=0.9]{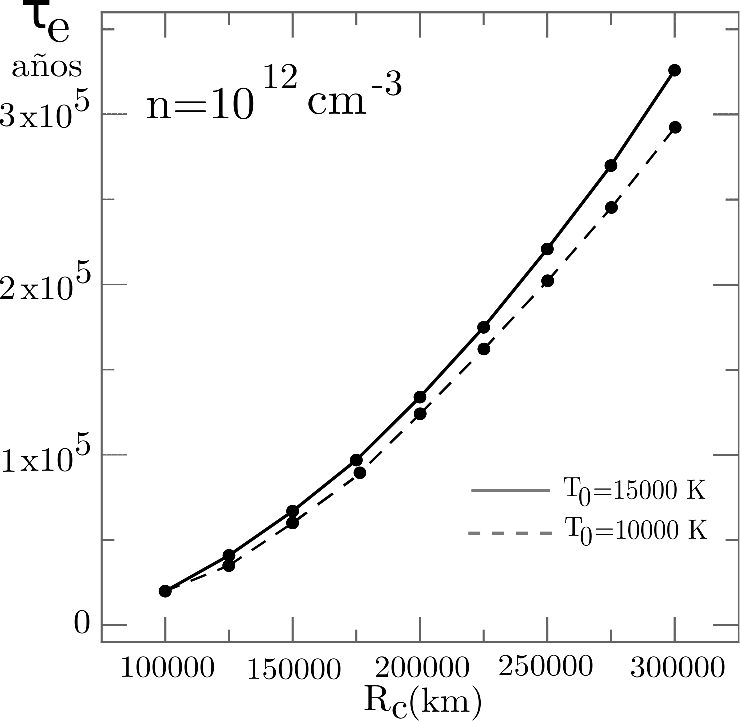} 
\caption{Tiempos de enfriamiento $\tau_{e}$ (eje vertical) en años para nubes de plasma de radios $R_{c}$ (eje horizontal)  entre 100000 y 300000 km. Todas las nubes tienen la misma  densidad  $n=10^{12}$ cm$^{-3}$. La curva de trazo lleno se calculó con $T_{0}=15000$ K y la de trazos con $T_{0}=10000$ K.}
\label{Enfriamiento1}
\end{figure}

\begin{figure} 
\includegraphics[scale=0.9]{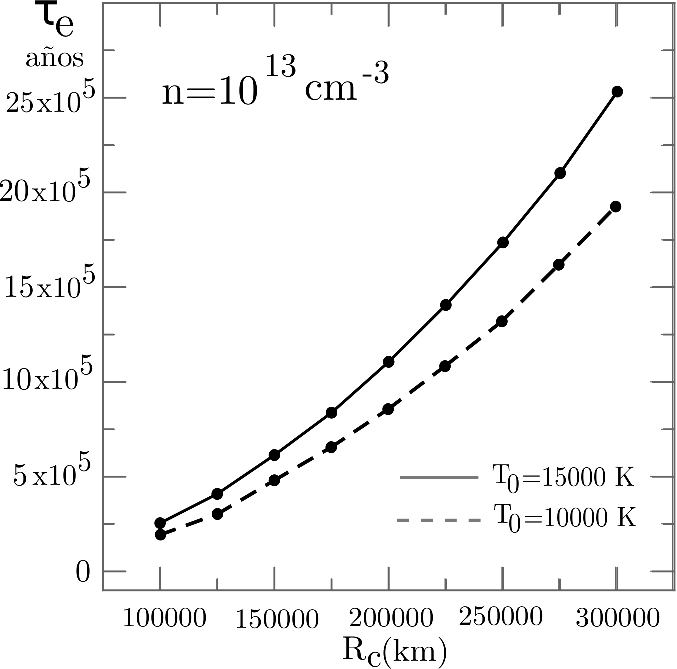} 
\caption{Lo mismo que la Fig. \ref{Enfriamiento1} pero para nubes de $n=10^{13}$ cm$^{-3}$.}
\label{Enfriamiento2}
\end{figure}

 Repitiendo los cálculos con los cuales obtuvimos  la Fig.  \ref{EnfriamientoPlasma}, pero para nubes de diferentes radios y densidades, podemos determinar como varía 
 $\tau_{e}$ en función del radio $R_{c}$ y de la densidad $n$ de las nubes. En la Fig. \ref{Enfriamiento1}, representamos $\tau_{e}$ para nubes con densidades  $n=10^{12}$ cm$^{-3}$ y radios $R_{c}$ entre 100000 y 300000 km.  En la Fig. \ref{Enfriamiento2}, representamos lo mismo que en la Fig.  \ref{EnfriamientoPlasma}, pero para  nubes con densidades  $n=10^{13}$ cm$^{-3}$ y con $T_{re}=8000$ K. Las dos figuras de marras nos enseñan que, para $n=$constante, las nubes más pequeñas se enfrían más rápidamente y que, para $R_{c}=$constante,  las nubes más densas tardan mucho más en enfriarse. Para las nubes de máximo radio que hemos tratado ($R_{c}=300000$), $\tau_{e}\approx$ 320 mil años si $n=10^{12}$ cm$^{-3}$, mientras  que  $\tau_{e}\approx 2.5$ millones años  si  $n=10^{13}$ cm$^{-3}$. 

A fin de averiguar cuán sensible es el valor resultante de $\tau_{e}$ al valor adoptado para  la temperatura inicial $T_{0}$,  repetimos los cálculos usando $T_{0}=10000$ K. Los resultados son  representados en sus respectivas figuras, Fig. \ref{Enfriamiento1} y \ref{Enfriamiento2}, por líneas de trazos. La comparación de la curva llena, correspondiente $T_{0}=15000$ K, con aquella a trazos de  $T_{0}=10000$ muestra que las diferencias  en ambas figuras son relativamente pequeñas.

En los bordes  de las nubes de plasma, que  se enfrían rápidamente a temperaturas inferiores  a 1500-2000 K (ver Fig. \ref{EnfriamientoPlasma}), las moléculas  pueden formarse rápida y eficientemente en fase gaseosa y asociarse en  grandes agregados. Estos  conglomerados de moléculas, a los cuales podemos denominar precursores de los granos de polvo, ingresan al interior de su nube de plasma y engrosan sus tamaños incorporando núcleos atómicos pesados. Las nubes grandes de plasma pueden también ser alimentadas de granos embrionarios de polvo  proporcionados por  nubes más pequeñas del entorno.  Si bien el número de protones libres que penetran los granos de polvo inmersos en el plasma es muy grande, probablemente la mayor parte de ellos no puede ser retenida y  escapa  a través de la superficie de los granos. En cambio, al chocar con un grano de polvo, los núcleos atómicos pesados del plasma se adhieren y se unen en el grano de polvo. A pesar de que la abundancia de los elementos pesados es baja, el crecimiento de la masa y del tamaño de los granos de polvo se debe básicamente a la recolección de dichos núcleos pesados. Los granos de polvo se comportan como redes de pescadores que atrapan a los peces grandes (núcleos pesados) y dejan escapar a los peces pequeños (electrones y protones libres). Los electrones que quedan atrapados dentro de un grano de polvo pueden neutralizar eléctricamente las cargas positivas de los núcleos pesados que forman el grano de polvo. Este mecanismo de la formación de los granos de polvo explica la composición mineralógica de los meteoritos y del polvo cósmico \footnote{La misión espacial  ``Stardust'' recolectó muestras de polvo cometario para su análisis químico en laboratorios terrestres. Uno de los grupos de minerales encontrados es el que constituye  el Olivino $(Mg_{2} Si O_{4}$, y $Fe_{2} Si O_{4})$, muy común en los meteoritos y en las rocas terrestres.}.

Ahora nuestro propósito es determinar los tamaños que pueden alcanzar los granos de polvo que evolucionan dentro de una nube de plasma. Para ello, reescribimos la fórmula (\ref{acrecion}) especificando que nos referimos exclusivamente a la acreción de núcleos atómicos pesados:

\begin{equation}
a(t)=a_{0}+  n_{np} \frac{m_{np}}{\rho} \, v_{np} \, (t-t_{0}).
\label{acrecion2}
\end{equation}
Teniendo en cuenta que $v_{np}$ es dada por la fórmula (\ref{Vnp}), la masa media de los elementos pesados por  $m_{np}= 16\,  m_{p}$ y $n_{np}\approx 9.5 \times 10^{-4} n/2$ (ver más arriba), la fórmula (\ref{acrecion2}) puede escribirse

\begin{equation}
a(\tau_{e})= a_{0} +  9.5 \times 10^{-4} \frac{n}{2}   \,  \frac{16\, m_{p}}{\rho}  \sqrt{\frac{3 k T}{16\, m_{p}} } \, (\tau_{e}-t_{0}),
\label{acrecion3}
\end{equation}
donde $\rho= 3$ gramos por cm$^{3}$, $m_{p}$ es la masa del protón y $n$ y $T$ son  el número de partículas  por cm$^{-3}$ y la temperatura del plasma, respectivamente. Como $T$ varía con el tiempo, lo remplazaremos  por el valor medio $\frac{T_{0}+T_{re}}{2}$. Note también que la variable  temporal de la fórmula (\ref{acrecion2}) es, en la fórmula  (\ref{acrecion3}),  dada por  $\tau_{e}$ que representa  el tiempo de vida de la nube de plasma considerada,  y  que $t_{0}$  es aquí el tiempo que transcurrió desde el origen de la nube hasta el momento en que  el grano precursor de radio $a_{0}$ empieza a interaccionar  con el plasma.  Por lo tanto,  el intervalo de  tiempo en que el grano de polvo acumula partículas del plasma es $\tau_{e}-t_{0}$.

\begin{figure} 
\includegraphics[scale=1.1]{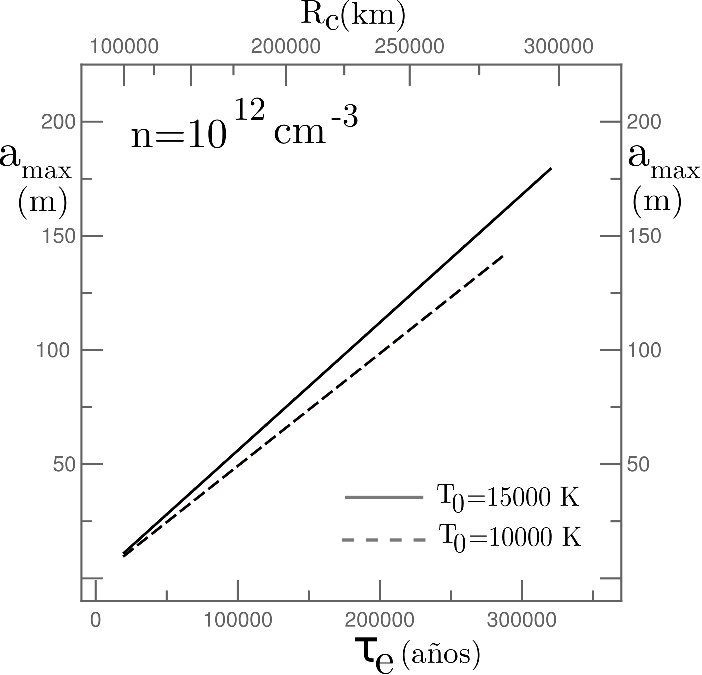} 
\caption{Radios máximos ($a_{max}$) de cuerpos formados por acreción dentro de nubes de plasma de densidad $n=10^{12}$ partículas por cm$^3$. Ambas curvas representan $a_{max}$ en función de $\tau_{e}$, de nubes de plasma con radios $R_{C}$ entre 100000 y 300000 km. 
En los ejes cartesianos horizontales inferior y superior, se indican los tiempos de enfriamiento $\tau_{e}$, expresados en años, y los correspondientes radios $R_{c}$, expresados en km, respectivamente. En los ejes verticales, se leen los valores de $a_{max}$ expresados en metros. La curva de trazo lleno se calculó con $T_{0}=15000$ K, en cambio, la curva de trazos con $T_{0}=10000$ K. Note que la escala de $R_{c}$ no es lineal.}
\label{TamañoPolvoMax}
\end{figure}

\begin{figure} 
\includegraphics[scale=1.1]{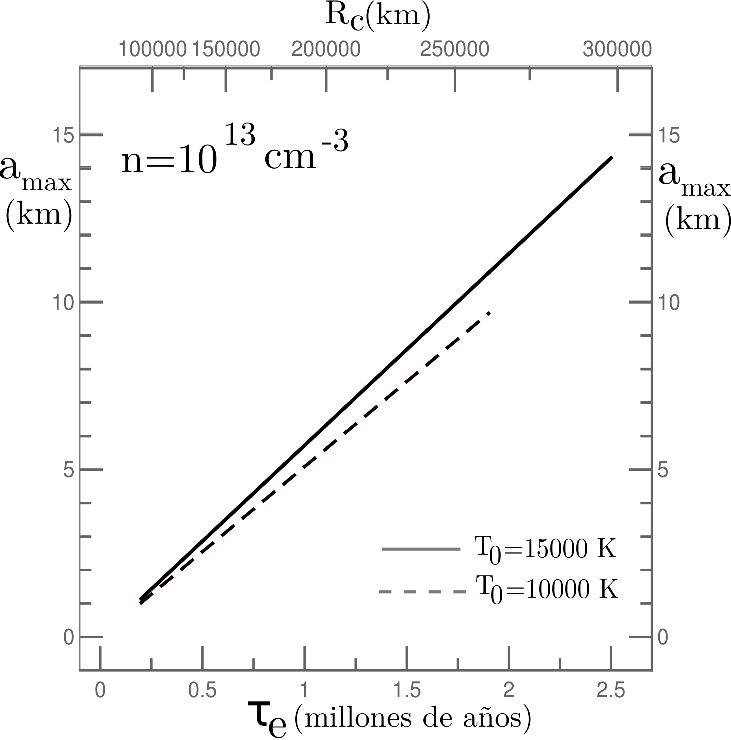} 
\caption{lo mismo que la Fig. \ref{TamañoPolvoMax}, pero para  $n=10^{13}$ partículas por cm$^3$. Además,  $a_{max}$ es expresado en kilómetros y $\tau_{e}$ en millones de años.}
\label{TamañoPolvoMax2}
\end{figure}

Si consideramos que hay un ingreso continuo de granos embrionarios a las nubes de plasma, entonces $t_{0}$ toma valores entre 0 y $\tau_{e}$. Por lo tanto, el radio final $a(\tau_{e})$ de los cuerpos,  que inicialmente son pequeños granos de polvo, es máximo  cuando $t_{0}=0$. Por otra parte, cuando $t_{0}=\tau_{e}$,  $a(\tau_{e})=a_{0}$, donde el radio  el grano precursor $a_{0}$ suponemos que no supera el micrón (ver fórmula \ref{acrecion3}).
En consecuencia,  una nube de plasma puede generar un conjunto de cuerpos sólidos que van desde granos submicrométricos a cuerpos con radios máximos $a_{max}(\tau_{e})$. El radio máximo $a_{max}(\tau_{e})$ se obtiene haciendo $t_{0}=0$ en la fórmula (\ref{acrecion3}).
La Fig. \ref{TamañoPolvoMax} representa $a_{max}(\tau_{e})$ en función de $\tau_{e}$ para nubes de plasma con radios $R_{c}$ entre 100000 y 300000 km y densidad $n$ igual para todas. Lo mismo se representa en la Fig.  \ref{TamañoPolvoMax2}, pero para otra densidad $n$. La correspondencia entre  el valor de $R_{c}$, indicado en la parte superior de ambas figuras,  y el valor de $\tau_{e}$, se obtuvieron usando los datos de las respectivas Figuras \ref{Enfriamiento1} y \ref{Enfriamiento2}.

La ordenada de la Fig. \ref{TamañoPolvoMax} muestra  que se pueden formar, en las nubes tratadas,  cuerpos sólidos con radios entre 10 y 170 metros y, en las nubes más densas representadas en la Fig. \ref{TamañoPolvoMax2}, se pueden formar cuerpos de radios entre 1 y 14 km. Cabe advertir que el tiempo de vida de las nubes de plasma puede ser inferior al dado por $\tau_{e}$.  Ello puede ocurrir  si la nube colisiona antes con otras nubes de plasma o con nubes interestelares. Aun así, en un tiempo de vida de solo medio millón de años, $a_{max}\approx 3$ km (ver  Fig. \ref{TamañoPolvoMax2}). Estos resultados nos indican la posibilidad de que las nubes circunestelares de plasma sean fuentes de meteoroides \footnote{Con el término genérico \it{meteoroides}, \rm llamados meteoritos  cuando han caído sobre  la superficie de la Tierra,    designamos a los cuerpos de naturaleza similar a la  de los meteoritos o pequeños asteroides que flotan en el espacio.} y asteroides interestelares.

\subsection{Diseminación de meteoroides y asteroides formados en el material expulsado por estrellas en la etapa final de sus vidas y convertidos, en el medio interestelar, en minicometas y cometas \label{Diseminación}}

Hemos visto que densas nubes de plasma expulsadas por  estrellas pueden formar desde granos de polvo sub-micrométricos hasta pequeños asteroides, incluso hasta asteroides no tan pequeños,  los cuales al internarse en el medio interestelar inician un largo viaje a través del cual experimentan diversas contingencias y transformaciones. Las nubes de polvo formadas en las nubes de plasma estelar pueden tardar algunos cientos de miles de años en encontrarse con las nubes interestelares del entorno y, en consecuencia, en  mezclarse con la materia interestelar. Si una tenue nube de esencialmente gas atómico y polvo aumenta  su densidad por compresión debido, por ejemplo, al choque con otra nube interestelar o con un remanente de supernova,  se pueden formar extensas nubes moleculares, donde la molécula de hidrógeno, $H_{2}$,  es naturalmente  la más abundante. En el medio interestelar con temperaturas muy bajas, $\le 100$ K, los átomos que chocan contra un grano de polvo se pegan  en su superficie y pueden unirse químicamente entre ellos. En la superficie de los granos de polvo, además de la molécula  $H_{2}$, se sintetizan  muchas otras moléculas como la del  agua, con lo cual los granos de polvo se cubren de un manto de hielo. Las colisiones entre los granos de polvo remueven parte  de sus mantos helados, con lo cual se liberan  moléculas que pasan  a la fase gaseosa. Las nubes moleculares, que  son sitios  muy fríos ($10-20$ K) y relativamente densos, pueden contraerse gravitacionalmente  y formar estrellas y sistemas planetarios. 

 Los granos de polvo, meteoroides y asteroides, formados dentro de una misma nube de plasma, tienen todos la misma velocidad media,  ``heredada''  de la nube madre. Si ellos penetran juntos una nube interestelar, los pequeños granos de polvo  son frenados y capturados por la nube interestelar y, en cambio, los meteoroides y asteroides la atraviesan completamente sin casi disminuir sus velocidades. Ello se debe naturalmente a que las masas de los meteoroides y asteroides son mucho mayores que las de los granos de polvo. Por lo tanto, esos cuerpos mayores pueden recorrer enormes distancias, atravesar densas y extensas nubes atómicas y moleculares y, al igual que los granos de polvo, cubrirse de espesos mantos de hielo. Es decir, los meteoroides y asteroides se convertirían  en mini-cometas y cometas interestelares y se distribuirían ampliamente en la Galaxia.  
 
En un estudio de  la  dinámica de las estrellas en una dirección perpendicular al plano  Galáctico en la región del Sol,  Oort\index{Oort, J.H.} \footnote{Jan H. Oort\index{Oort, J.H.} (1900-1992) fue un astrofísico holandés que realizó importantes aportes en la investigación  de la rotación y estructura de nuestra Galaxia y fue también pionero en la utilización de técnicas radioastronómicas para estudiar la distribución Galáctica  del hidrógeno neutro interestelar mediante su emisión en la longitud de onda de 21 cm. Él también propuso la existencia de la nube cometaría que envuelve al Sistema Solar y que, en su honor, lleva su nombre.}  estimó, en el año 1932,  la densidad del disco Galáctico $\rho_{0}$ en la posición del Sol. Dicha densidad $\rho_{0}$ es a veces  llamada ``límite de Oort'' en su honor. Es un límite en el sentido de que la suma de las masas estelares y gaseosas (materia visible) contenidas en una unidad de volumen del  disco Galáctico local no debe superar el valor de  $\rho_{0}$. La densidad de la materia visible, $\rho_{mv}$,  no solo no supera el valor de $\rho_{0}$, sino que está significativamente por debajo. La diferencia es llamada densidad de la materia oscura $\rho_{mo}$ $(=\rho_{0}-\rho_{mv})$. Según algunas estimaciones modernas, 
$\rho_{mo}\simeq 0.01$ $M_{\odot}$ pc$^{-3}$ (masas solares por parsec cúbico). La naturaleza de la materia oscura aún se desconoce y es posible  que, al menos en parte,  ni siquiera esté formada por la materia ordinaria (bariónica). Una posibilidad es que parte de la materia oscura esté constituida por una numerosa población de cometas interestelares, dado que ellos son muy pequeños y fríos como para detectarlos en el oscuro y frío espacio interestelar. Aún  queda mucho por aprender sobre los relativamente pequeños cuerpos que constituyen la materia interestelar.

Si la materia oscura está compuesta enteramente de cometas de radio $a=2$ km y densidad $\rho_{c}=$ 3 g cm$^{-3}$, la densidad de cometas por parsec cúbico, $n_{ci}$, es dada por $n_{ci}=\frac{\rho_{mo}}{\cal M \it}$, donde $\cal M \it=\frac{4}{3} \pi a^{3} \rho_{c}=1\times 10^{17}$ g es la masa de cada cometa. Unificando las unidades  para la masa solar ($M_{\odot}=2 \times10^{33}$ g), y el parsec (pc$=3.086\times 10^{18}$ cm), obtenemos
\begin{equation}
n_{ci} \simeq 2 \times 10^{14} \rm cometas\,\, por \,\,pc^{3}.
\label{MateriaOscura}
\end{equation}

Hasta el año 2017, en que se descubre un genuino objeto interestelar (cometa o asteroide)
llamado \it{Oumuamua}\index{Oumuamua 1I, cometa o asteroide interestelar}, \rm sólo se podía inferir,  sobre la base de la ausencia de detecciones de cometas interestelares, un límite superior para $n_{ci}$ del orden $6  \times 10^{12}$ cometas por parsec cúbico. En el año 2019, se descubre otro objeto interestelar en las cercanías del Sistema Solar, el cometa \it{2I/Borisov\index{Borisov 2I, cometa interestelar}} \footnote{Pocos días antes del envío de la versión original de este artículo a arXiv, fue descubierto el tercer cometa interestelar: 3I/ATLAS\index{Atlas 3I, cometa interestelar}. Ello abona nuestra hipótesis de una gran población de cometas interestelares.}\rm. Con todo ello, se ha estimado que $n_{ci} \approx 7  \times 10^{14}$ cometas por parsec cúbico \cite{Zwart}, número del orden del dado por la expresión (\ref{MateriaOscura}). En la concepción clásica, los cometas se forman en los discos proto-planetarios y, por razones dinámicas,  pueden ser expulsados al medio interestelar. Si ello es así, la alta densidad de cometas interestelares hallada implicaría una alta tasa de formación de exo-planetas. Por el contrario, nosotros hemos argumentado que  los cometas se forman en desechos estelares, originalmente calientes, que se mezclan con la fría materia interestelar. Y por lo tanto, el aporte de los cometas planetarios al valor de $n_{ci}$ sería pequeño.

\section{Captura de cometas interestelares por nubes moleculares y la formación de sistemas proto-planetarios \label{DiscoProtoplanetario}}

Vivimos dentro de un sistema de estrellas achatado con forma de disco. Cuando miramos hacia el plano del disco vemos un mayor número de estrellas que se proyectan sobre el cielo, formando una banda luminosa que llamamos Vía Láctea. Por lo tanto, nuestra Galaxia, la Vía Láctea, es una galaxia de disco. Sabemos además que el cosmos está poblado de galaxias, islas de estrellas o universos islas como las llamaba el filósofo Inmanuel Kant\index{Kant, I.}, muy alejadas unas de otras. La mayoría de las galaxias de disco contienen brazos espirales, delineados por estrellas brillantes,   que conforman una estructura similar a la de un vórtice o  remolino. Entonces, se sospechaba que nuestra Galaxia   podría ser  también una galaxia  espiral. El hecho de que  nos encontramos dentro de ella, y no podemos tener una vista externa como en el caso de las otras galaxias, requirió de laboriosas investigaciones  para comprobar que la Vía Láctea es efectivamente  una galaxia espiral. 

Para visualizar el fenómeno, supongamos una galaxia de disco con dos brazos espirales. Los dos brazos se extienden, dentro del disco,   a partir del centro galáctico pero en direcciones opuestas y se curvan también en sentidos opuestos, formando una estructura con la forma de la letra S centrada en el  núcleo  galáctico. Este tipo de galaxia espiral es la más simple, dentro de la clasificación morfológica  de las galaxias espirales. La mayoría de los astrónomos coinciden en que la Vía Láctea tiene cuatro largos brazos espirales.
Las galaxias no rotan como un cuerpo sólido, sino que su velocidad angular es función de la distancia al centro galáctico. Se dice que las galaxias tienen rotación diferencial. Ello plantea  el problema  de cómo puede mantenerse la estructura espiral sin deformarse por la rotación diferencial y, en consecuencia, desaparecer en unas pocas rotaciones de la galaxia.  A ello se lo denomina el dilema del enroscamiento. Un fenómeno similar ocurre cuando se revuelve el azúcar en el café: se genera una estructura espiral que rota más rápido en el centro que en el borde de la taza, pero pronto desaparece el patrón espiral.

El astrónomo sueco Bertil Lindblad\index{Lindblad, B.} (1895-1965) fue quien sentó las bases de la moderna teoría de la estructura espiral. Él reconoció que  el dilema del retorcimiento de los brazos espirales se resuelve si la materia que integra los brazos no es siempre la misma. Es decir no son brazos materiales, sino una perturbación gravitatoria u onda de densidad que se propaga a través del disco de estrellas y gas. La onda de densidad rota en torno al centro galáctico como un cuerpo rígido. El gas y las estrellas rotan con mayor  velocidad que la onda de densidad y, por lo tanto,  atraviesan los brazos espirales,  donde son transitoriamente demorados por la perturbación gravitatoria. Es como un flujo de automóviles que transitan por una autopista, cuando están por cruzar un paso a nivel  son  obligados a disminuir la velocidad y allí se produce un amontonamiento de vehículos. Es decir, en los brazos espirales, aumenta la concentración de estrellas, gas y polvo interestelares, lo cual permite que la onda de densidad se auto-mantenga.

La aplicación de técnicas matemáticas desarrolladas para la física de plasmas permitió  tratar exitosamente inestabilidades colectivas como las que  generan  los brazos espirales y demostrar el mantenimiento del fenómeno por un tiempo considerable de la vida de las galaxias. Sin embargo,  aún ignoramos el origen (u origenes) de la inestabilidad espiral. Una predicción importante de la teoría es que a lo largo de los brazos espirales se desarrolla un frente de choque que desencadena el colapso de nubes de gas y polvo y la consecuente formación de estrellas. Las estrellas más brillantes, que  tienen una corta vida y en consecuencia no se alejan  mucho del lugar de origen, hacen que los brazos espirales resplandezcan. La abrupta concentración y variación cinemática  del gas y polvo interestelares, provocadas por  la onda de densidad espiral, conllevan a la formación de nubes gigantes de gas, polvo y estrellas (o supernubes). El autor de esta monografía demostró que, en dichas condiciones, las supernubes pueden capturar el 4 por ciento de las estrellas viejas (de campo) que circunstancialmente circulan   dentro de las supernubes \cite{Olano4}. Aquí agregamos que, en forma similar, las supernubes pueden capturar el 4 por ciento de los meteoroides, asteroides y cometas interestelares que transitan dentro del volumen de las supernubes en formación.

 Hemos visto que las nubes interestelares de gas y polvo que pasan a través de las ondas de densidad Galáctica  se comprimen, desaceleran y,  en el proceso,  capturan cometas interestelares. También como resultado de ello,  en las más densas y frías concentraciones  de gas y polvo, se forman rápida y eficientemente nubes moleculares. Dentro de este
 escenario es posible imaginar  la formación de densas y extensas capas de gas molecular, dentro de las cuales se gesten subestructuras como discos proto-planetarios. Las ondas de choque de gran escala, como las que  se generan en las ondas de densidad, o aquellas de escala menor que generan las explosiones de supernovas, se comportan como topadoras que acumulan capas de materiales en el frente. 
 
 Desde el punto de vista clásico, los cometas del Sistema Solar son los planetesimales que se formaron  en el disco protoplanetario y que, debido a  perturbaciones gravitatorias de los planetas mayores, se alejaron del disco, adquirieron un manto de hielo y  pasaron a formar parte de la nube cometaria de Oort\index{Oort, nube de}. Nosotros defenderemos una tesis distinta, según la cual   la nube molecular en que se formó el disco protoplanetario ya contenía cometas capturados del medio interestelar. Y los planetesimales fueron formados por  los cometas más pequeños o minicometas  que se asentaron en el disco protoplanetario, y no al revés como lo sostiene la teoría clásica.
\begin{figure} 
\includegraphics[scale=0.65]{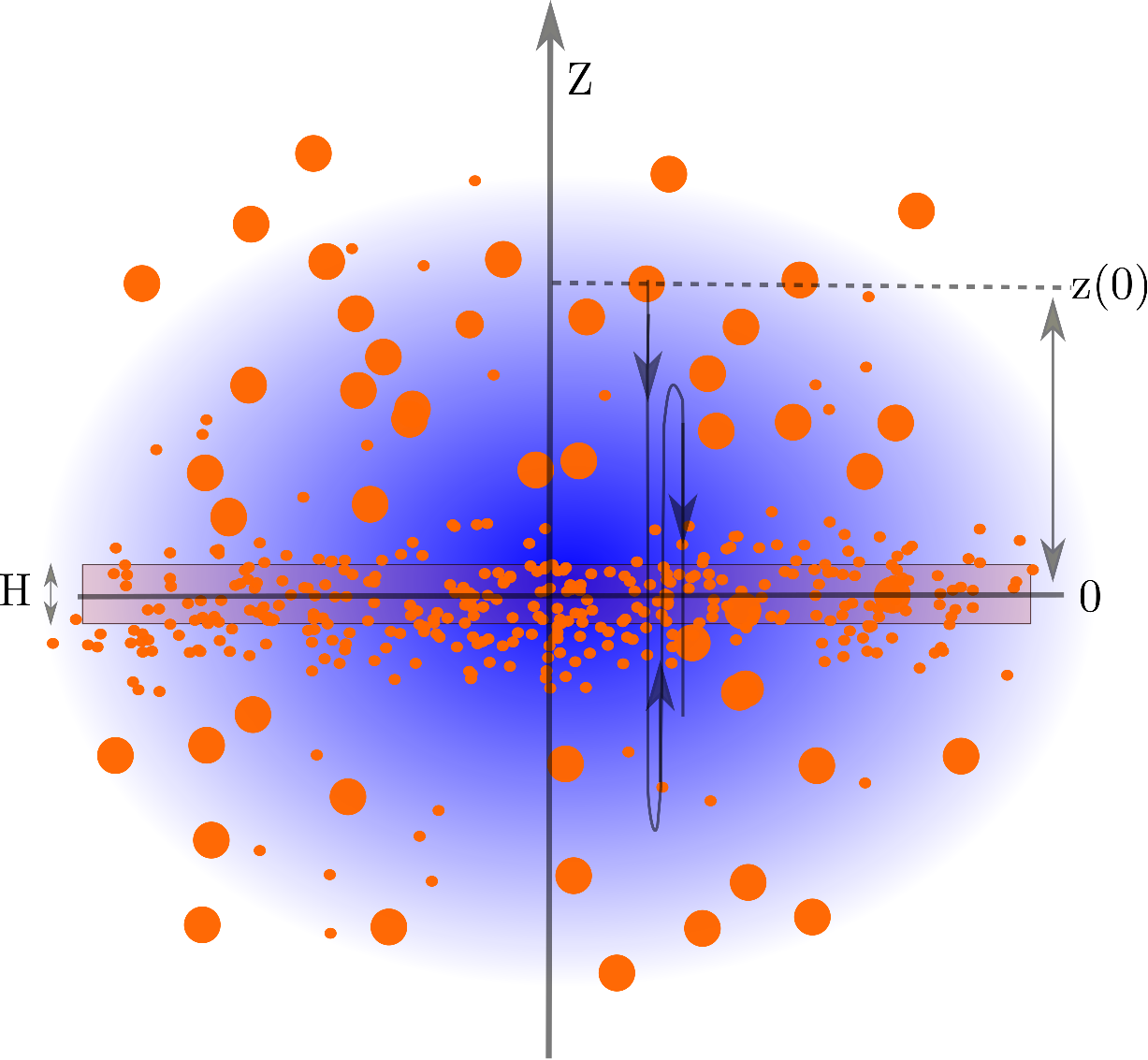} 
\caption{Modelo de un denso disco protoplanetario   inmerso en una nube interestelar de polvo, gas molecular y cometas. Los cometas (circulos naranjas) oscilan amortiguadamente en torno al disco protoplanetario y en consecuencia los cometas más  pequeños tienden a asentarse en el disco.}
\label{CometaryCloud}
\end{figure}

\begin{figure}   
\includegraphics[scale=0.99]{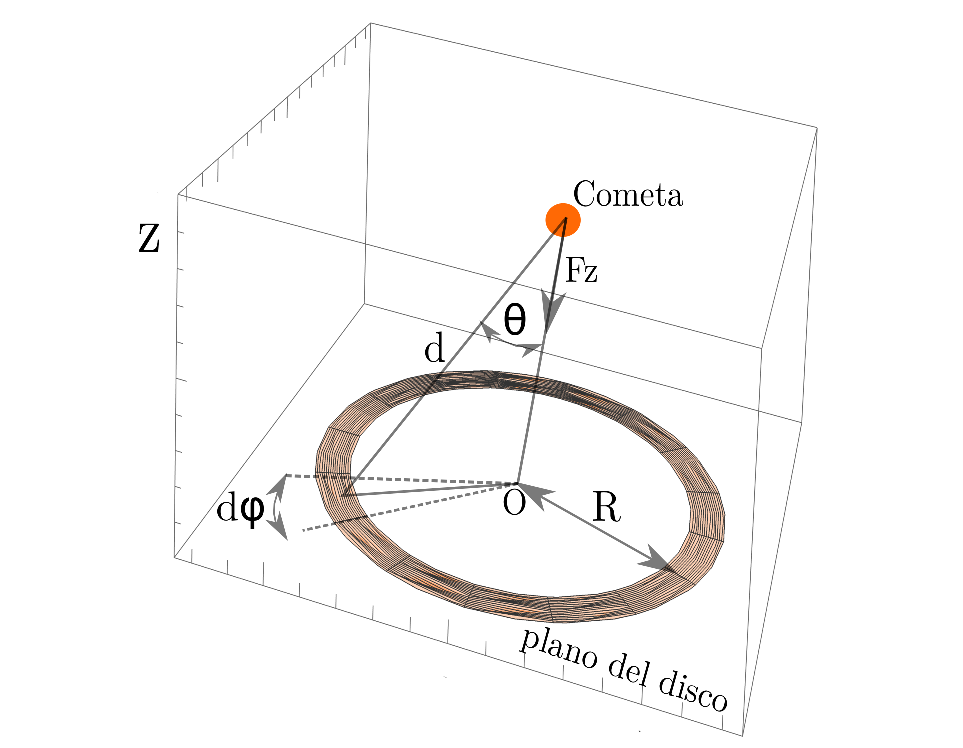} 
\caption{Fuerza gravitatoria F$_{Z}$ que ejerce un anillo del disco protoplanetario sobre un cometa que se encuentra  a una altura $Z$ sobre el centro del anillo.}
\label{GravedadAnillo}
\end{figure}

Por todo lo dicho, cabe la posibilidad de que  una plétora de minicometas y  cometas, como así también de meteoroides y asteroides, estén ligados gravitacionalmente a las nubes interestelares, hecho que nos conduce a replantearnos algunas ideas  sobre la formación de los sistemas planetarios. El modelo que proponemos es esquematizado en la Fig. \ref{CometaryCloud}, donde una extensa y densa capa de gas, formada dentro de un complejo molecular, atrae gravitacionalmente a los cometas de distintos tamaños que se encuentran en su entorno. A continuación, estudiamos los movimientos de los cometas en una dirección $z$, perpendicular a la capa o disco de gas. Para ello, dividiremos el disco de gas en un gran número de anillos cilíndricos concéntricos y consecutivos, numerados con el índice entero $i$. Los anillos tienen  el espesor $H$ del disco, un ancho $dR$ igual para todos y  cada uno con su propio radio medio $R_{i}$ (todos centrados en el punto $O$, ver Fig. \ref{GravedadAnillo}). Si el disco de gas tiene una densidad $\rho$ homogénea, la masa  contenida en una porción del anillo $i$ es $dm= \rho\, H\, d \varphi \, R_{i}\, dR $. Un cometa ubicado a una altura $z$ sobre el punto $O$ es atraído por $dm$  con una fuerza Newtoniana $d \vec{F}= G \frac{\cal M \it\, dm}{ \rm d \it ^{2}}$ , donde $G$ es la constante gravitatoria, $\cal M \it$  es la masa del cometa y $\rm d \it  \, (=\sqrt{R_{i}^{2}+z^{2}})$ es la distancia entre $dm$ y el cometa. La componente de  $d \vec{F}$ en la dirección de $z$ es $dFz=\mid  d \vec{F}\mid cos\, \theta$, donde $cos\,\theta=\frac{z}{\rm d \it }$. Las otras dos componentes ortogonales dependen de $\varphi$ y son opuestas a las que produce la porción del anillo con $\varphi+180^{\circ}$, por lo tanto se anulan. En consecuencia, la fuerza neta total que actúa sobre el cometa debido al anillo $i$ completo es $F_{z}=  G \cal M \it \,\frac{z}{\rm d \it_{i}^{3}} \sum  dm$.  El número de porciones $dm$ que tiene el anillo $i$ completo es $\frac{2 \pi}{d \varphi}$ y, por lo tanto, $\sum  dm= \frac{2 \pi}{d \varphi} dm= 2 \pi \rho\, H\, R_{i}\, dR$. Entonces, la fuerza que ejerce el anillo $i$ de gas, representado en la  Fig. \ref{GravedadAnillo},  sobre el cometa es 
\begin{equation}
F_{i}=  2 \pi G \cal M \it \rho H \, \frac{z\, R_{i}\, dR}{(R_{i}^{2}+z^{2})^{3/2}} .
\label{FuerzaZ}
\end{equation}

La fuerza total $F$ que ejerce el disco sobre el cometa  se puede obtener integrando  la ecuación (\ref{FuerzaZ}) con respecto a $R$. Sin embargo,  simplificaremos la ecuación para obtener una integral más simple. Dado que la distancia al cuadrado del cometa a las porciones del  anillo $i$ es $\rm d \it_{i}^{2}=(R_{i}^{2}+z^{2})$  y que a las del anillo $i+1$ es $\rm d \it_{i+1}^{2}=(\rm d \it _{i}+ d \rm d \it)^{2} =(R_{i}+ dR)^{2}+z^{2}$, resulta desarrollando los cuadrados:  $\rm d \it_{i}^{2}+ 2 \rm d \it_{i} d \rm d \it + d\rm d \it ^{2} = R_{i}^{2}+ 2 R_{i} dR+ dR^{2} +z^{2}$. Por lo tanto, $\rm d \it_{i}\, d \rm d\it =  R_{i} dR$, dado que $d \rm d \it ^{2}$ y $dR^{2}$ son valores muy pequeños, comparados con los otros miembros de la ecuación, y pueden despreciarse. Entonces, obtenemos de la última igualdad $d \rm d\it = \frac{R_{i} dR}{\rm d\it_{i}}$, con lo cual podemos escribir la ecuación (\ref{FuerzaZ}) del siguiente modo:
\begin{equation}
F_{i}=  2 \pi G \cal M \it \rho H \, z\, \frac{d \rm d\it}{\rm d\it_{i}^{2}},
\label{FuerzaZ2}
\end{equation} 
y la fuerza total $F$: 

\begin{equation}
F=\sum_{i=1}^{\infty}F_{i}= 2 \pi G \cal M \it \rho H \, z\, d \rm d\it \sum_{i=1}^{\infty} \frac{1}{\rm d\it_{i}^{2}}.
\label{FuerzaTotal1}
\end{equation}
Como la distancia mínima a los anillos es $z$, podemos escribir que  $\rm d\it_{i}= i \, d\rm d\it +z$. Fijando el  diferencial de  distancia $d \rm d\it$ en un valor pequeño, obtenemos un número $n=\frac{z}{d \rm d\it} \gg 1$. Por lo tanto, $\rm d\it_{i}= i \, d \rm d\it  + n\, d \rm d\it = d \rm d\it \, (i + n)$. Haciendo los reemplazos en (\ref{FuerzaTotal1}) encontramos

\begin{equation}
F= 2 \pi G \cal M \it \rho H \, \sum_{i=1}^{\infty} \frac{n}{(i+n)^{2}}=2 \pi G \cal M \it \rho H.
\label{FuerzaTotal2}
\end{equation} 
Note que $\sum_{i=1}^{\infty} \frac{n}{(i+n)^{2}}=1$, independientemente del valor de $n$.

También, la fuerza total $F$ puede obtenerse integrando la ecuación (\ref{FuerzaZ2}) entre la  distancia mínima $\rm d_{\it{min}}$ y la máxima $\rm d_{\it{máx}}$.
\begin{equation}
F=\sum F_{i}=  2 \pi G \cal M \it \rho H \, z\, \int_{\rm d\it_{min}}^{\rm d\it_{máx}} \frac{1}{\rm d\it^{2} } d \rm d\it.
\label{FuerzaZ3}
\end{equation}
La función primitiva de la integral de la expresión (\ref{FuerzaZ3}) es $P(\rm d\it)=-\frac{1}{\rm d\it}$  \footnote{La integral de una variable $x$ elevada a una potencia $p$ tiene como primitiva a $\frac{x^{p+1}}{p+1}$, en nuestro caso $p=-2$} y por lo tanto la integral es igual a $P(\rm d\it_{máx})-P(\rm d\it_{min})= -\frac{1}{\rm d\it_{máx}} +\frac{1}{\rm d\it_{min}}$. La distancia mínima del cometa es la altura $z$ sobre el plano del disco y, como consideramos un disco extenso,  la distancia máxima es muy grande o infinita. En consecuencia, $\int_{z}^{\infty} \frac{1}{\rm d\it^{2} } d\rm d\it =\frac{1}{z}$, con la cual reemplazada en (\ref{FuerzaZ3}) obtenemos el mismo resultado que (\ref{FuerzaTotal2}), y teniendo en cuenta que, de acuerdo con la segunda ley del movimiento de Newton, $F=\cal M \it g$ obtenemos que

\begin{equation}
 g  =2 \pi G \rho H.
\label{Gravedad} 
\end{equation} 
Cabe recordar  que (\ref{Gravedad}) es sólo una buena aproximación si la altura $z$ del cometa es mucho menor que el diámetro del disco. Por otra parte,  (\ref{Gravedad}) es sólo estrictamente válida para $\mid z \mid \geq \frac{H}{2}$, pues cuando el cometa pasa por el plano medio del disco $\mid g \mid =0$.
 
 Usaremos los siguientes valores para la densidad $\rho$ y el espesor $H$ del disco proto-planetario:  
 \begin{eqnarray}
 \rho & = & 1.67 \times 10^{-15} \,\, \rm g \,  cm ^{-3} \nonumber \\ 
 H & = & 0.01\,\, \rm pc, 
 \label{ParaDisco}
 \end{eqnarray} 
donde el valor de  $\rho$ adoptado corresponde a una densidad numérica $n=5 \times 10^{8}$ moléculas de hidrógeno  por cm$^{3}$. Con los parámetros de (\ref{ParaDisco}), se obtiene que la densidad superficial o columnar $\sum$ del disco es  $\sum=\rho H=52$ g cm$^{-2}$, la cual está dentro de los  valores estimados para los discos proto-planetarios (30-100 g cm$^{-2}$).
 
 De modo que el cometa cae hacia el disco de gas con una aceleración $g$ constante, en forma similar a la de caída libre de un objeto desde lo alto de una torre, verbigracia la emblemática torre inclinada de Pisa \footnote{Las primeras teorías sobre los mecanismos del movimiento se deben, por supuesto, a los antiguos griegos. El filósofo presocrático Parménides\index{Parménides}, el primer cosmólogo y epistemólogo, fue quien abordó el problema. Por un lado, la bien conocida paradoja  de Zenón\index{Zenón} y, por otro, el hecho de que el Universo esté lleno  de materia, parecen no dejar resquicios para el movimiento y llevaron a Parménides\index{Parménides} a negar la posibilidad del movimiento. La teoría atómica de Demócrito\index{Demócrito}, otro presocrático, nació como una refutación a la teoría de Parménides\index{Parménides}. Pues, la materia no es  un continuo sino que consiste en átomos inmersos en el vacío. La cinemática de la caída de los cuerpos en el aire se termina de comprender más de 2000 años después con las investigaciones de Galileo\index{Galileo}.}. Entonces, podemos aplicar las ecuaciones Galileanas de movimiento para estudiar los movimientos de los cometas en la dirección $z$ (perpendicular al plano medio del disco). Es claro que aquí la aceleración de la gravedad es dada por la fórmula (\ref{Gravedad}), y no por la aceleración sobre la superficie terrestre. Si, en el instante $t_{0}$, un cometa tiene la altura $z(t_{0})$, sobre el plano medio del disco, y la velocidad $v_{z}(t_{0})$, la posición y velocidad del cometa en un instante posterior $t$ es dado por las siguientes ecuaciones: 
 
\begin{equation}
z(t)= z(t_{0}) + v_{z}(t_{0})(t-t_{0}) \mp \frac{1}{2} g (t-t_{0})^{2},
\label{PositionZ}
\end{equation}

\begin{equation}
v_{z}(t)=v_{z}(t_{0})  \mp g (t-t_{0}).
\label{VelZ}
\end{equation}

Nuestro interés es estudiar solo la componente $z$ de movimiento, la cual en principio puede estudiarse independientemente de las otras componentes. El eje $z$ cruza perpendicularmente el disco y el punto de intersección  del eje $z$ con el plano medio del disco es adoptado como origen del eje coordenado $z$ (ver Figs. \ref{CometaryCloud} y \ref{GravedadAnillo}). Convenimos que la parte positiva del eje $z$ yace por encima del plano medio del disco, y la negativa por debajo. Cuando el cometa cae hacia el disco desde posiciones $z$ positivas, significa que $z$ disminuye con el tiempo y se hace cero cuando atraviesa el plano medio del disco. Dado que el valor de $v_{z}(t_{0})$ puede ser cualquiera y en particular cero, la fórmula (\ref{PositionZ}) indica que corresponde usar la opción con el signo menos $(-)$, en ambas fórmulas. Al contrario, debemos  usar la opción con el signo  más $(+)$ en las fórmulas (\ref{PositionZ}) y (\ref{VelZ}),   cuando el cometa se mueve debajo del disco  ($z<0$). 

Supongamos que un cometa se encuentra por primera vez en una posición $z(t_{0})>0$, donde la atracción del disco de gas predomina y se puede  despreciar la atracción gravitatoria del resto de la nube molecular. Por simplicidad,  supongamos  además que en dicha posición la velocidad del cometa $v_{z}(t_{0})=0$. Si contamos el tiempo a partir de dichas condiciones iniciales, $t_{0}=0$. Como ya sabemos que tratamos solo el movimiento en $z$, simplificaremos la notación quitando el sufijo $z$ de la velocidad. Si llamamos  $t_{1}$ al tiempo en el cual el cometa atraviesa por primera vez el disco, de acuerdo con la fórmula (\ref{PositionZ}),  $z(t_{1})= z(0) - \frac{1}{2} g\, t_{1}^{2}=0$. Entonces, 
podemos escribir:
\begin{equation}
t_{1}=\sqrt{2 \frac{z(0)}{g}}.
\label{Tiempo1}
\end{equation}
De acuerdo con la fórmula(\ref{VelZ}), $v(t_{1}) =-g\, t_{1}$ y por lo tanto
\begin{equation}
v(t_{1})= -\sqrt{2 \, z(0)\, g}.
\label{Velocidad1}
\end{equation}

Para obtener las fórmulas (\ref{Tiempo1}) y (\ref{Velocidad1}),  hemos supuesto que la fuerza de fricción que ejerce el  gas molecular sobre el cometa en su caída al disco es despreciable. Solo cuando $\mid z \mid \le H$, la fuerza de fricción debida a la gran densidad gaseosa del disco es muy importante. Como $z(0) \gg \frac{H}{2}$, podemos considerar que $t_{1}$ y  $v(t_{1})$ son el tiempo y la velocidad con que el cometa ingresa al disco. El cometa oscilará en torno al plano del disco y, por lo tanto, cruzará  repetidamente el disco, de arriba hacia abajo y de abajo hacia arriba (ver Fig. \ref{CometaryCloud}). Cada paso del cometa por el disco, lo indicaremos con el índice entero $j$. Como  $v(t_{1})$ es la velocidad con  que el cometa ingresa por primera vez al disco, simplificaremos su notación reemplazando $t_{1}$ por 1. Es decir, $v(t_{1})=v(1)$ y, en general,  $v(t_{j})=v(j)$.

 Al cruzar el disco, el cometa es desacelerado por la fuerza de fricción. Denotaremos con $\tilde{v}$ a la velocidad con que el cometa emerge del disco. Si $\Delta t_{1}$ es el tiempo que tarda el cometa en cruzar el disco, $v(t_{1}+ \Delta t_{1})=\tilde{v}(1)$ en el caso del primer cruce,  y en general $v(t_{j}+ \Delta t_{j})=\tilde{v}(j)$. 
 
La fuerza  $F_{r}$ de roce  que actúa sobre el cometa al cruzar el disco puede expresarse mediante la ecuación (1) de \cite{Olano5} y, de acuerdo con la segunda ley de Newton $F_{r}= \cal M \it \frac{dv}{dt}$, podemos escribir
 
 \begin{equation}
 \frac{dv}{dt} =-\kappa\, v^{2},
 \label{FriccionM}
 \end{equation}
donde $\kappa=
\epsilon \frac{\rho A}{\cal M \it}$, $A \,(=\pi a^{2})$ es el área del cometa que se proyecta en la dirección del movimiento y $\epsilon$ es el coeficiente de fricción. La solución de la ecuación diferencial (\ref{FriccionM}) es $v(t)=\frac{1}{\kappa\, t + \frac{1}{v(0)} }$ y, adaptada a nuestro caso $j=1$,
\begin{equation}
v(t)=\frac{1}{\kappa\, (t-t_{1}) + \frac{1}{v(1)} }, 
\label{Vcruce}
\end{equation}
para $t_{1}\leq t \leq t_{1}+\Delta t_{1}$. 

Como en el intervalo de tiempo $\Delta t_{1}$, el cometa recorre la distancia $H (>0)$, el espesor del disco, se debe cumplir que $H=\int_{t_{1}}^{t_{1}+\Delta t_{1}} \mid v(t) \mid\, dt= \int_{t_{1}}^{t_{1}+\Delta t_{1}} \frac{1}{\kappa\, (t-t_{1}) + \frac{1}{\mid v(1)\mid} }\, dt=\frac{ln(1 + \kappa\, \mid v(1)\mid \Delta t_{1})}{\kappa}$, de donde podemos despejar nuestra incógnita, $\Delta t_{1}$, y obtener que
\begin{equation}
\Delta t_{1}=\frac{e^{\kappa H}-1}{\kappa\, \mid v(1)\mid }.
\label{Deltat1}
\end{equation}
Otra de nuestras incógnitas es $\tilde{v}(1)=v(t_{1}+\Delta t_{1})$, la cual se obtiene haciendo $t=t_{1}+\Delta t_{1}$ en la fórmula (\ref{Vcruce}) y reemplazando $\Delta t_{1}$ por su expresión (\ref{Deltat1}). El resultado es 
\begin{equation}
\tilde{v}(1)=- \mid v(1) \mid \, e^{-\kappa\, H}.
\label{VPrima}
\end{equation}

El  cometa sale del disco con la velocidad  $\tilde{v}(1)$ hacia el semiespacio inferior; a ello se debe el signo negativo de  $\tilde{v}(1)$. Aquí la energía del cometa se conserva pues solo actúa la gravedad. El cometa llega hasta una altura máxima sobre el disco, donde su velocidad se invierte,  y cae hacia el disco. Un fenómeno similar ocurre cuando  arrojamos una piedra hacia el cielo: esta vuelve y llega al suelo con la misma velocidad con que la lanzamos pero moviéndose en sentido inverso. Siguiendo con nuestra notación, $v(2)$ es  la velocidad   con la cual el cometa ingresa al disco en el segundo cruce  ($j=2$) y, por lo dicho, cumple la igualdad $v(2)=-\tilde{v}(1)$.
Aplicando las fórmulas (\ref{Deltat1}) y (\ref{VPrima}) para este caso $j=2$, $\Delta t_{2}=\frac{e^{\kappa H}-1}{\kappa\, \mid v(2) \mid}$ y  $\tilde{v}(2)= e^{-\kappa\, H}\,\mid v(2) \mid$, pero en vista de que $v(2)=-\tilde{v}(1)= \mid v(1) \mid \, e^{-\kappa\, H}$ obtenemos

\begin{equation}
\Delta t_{2}=\frac{e^{\kappa H}-1}{\kappa\, \mid v(1) \mid} e^{\kappa H}
\label{Deltat2}
\end{equation}
y
\begin{equation}
\tilde{v}(2)=\mid v(1)\mid \, (e^{-\kappa\, H})^{2}.
\label{VPrima2}
\end{equation}
De (\ref{Deltat1}) y (\ref{Deltat2}), y de (\ref{VPrima}) y (\ref{VPrima2}), obtenemos la siguientes reglas generales:
\begin{equation}
\Delta t_{j}=\frac{e^{\kappa H}-1}{\kappa\, \mid v(1) \mid} (e^{\kappa H})^{j-1}
\label{Deltat3}
\end{equation}
y
\begin{equation}
\tilde{v}(j)=\mid v(1)\mid \,(-e^{-\kappa\, H})^{j} ,
\label{VPrima3}
\end{equation}
para $j=1,2,3,...$

Denominamos $\Delta t_{1,2}$ al intervalo de tiempo transcurrido entre la salida  del cometa del disco en el cruce $j=1$ y su posterior ingreso al mismo  en el cruce $j=2$. De la ecuación (\ref{VelZ}), obtenemos $v(2)= -\tilde{v}(1)=\tilde{v}(1) + g \Delta t_{1,2}$ y, por lo tanto, $\Delta t_{1,2}=-\frac{2\, \tilde{v}(1)}{g}$. Teniendo en cuenta (\ref{VPrima}),
\begin{equation}
\Delta t_{1,2}= 2 \frac{\mid v(1)\mid}{g} e^{-\kappa H}.
\label{Tmedio}
\end{equation}

De las ecuaciones (\ref{PositionZ}) y (\ref{VelZ}), se deduce que, en el instante $\frac{\Delta t_{1,2}}{2}$, el cometa alcanza el mayor alejamiento del disco,  que llamamos $\hat{z}_{1,2}$. Allí el cometa tiene velocidad cero y comienza a caer hacía el disco. Es decir,  $\hat{z}_{1,2}=z(\frac{\Delta t_{1,2}}{2})=\tilde{v}(1) \frac{\Delta t_{1,2}}{2} + \frac{1}{2} g (\frac{\Delta t_{1,2}}{2})^{2}$, donde hemos considerado que $z(0)=0$. Reemplazando las relaciones  (\ref{VPrima}) y  (\ref{Tmedio}) para $\tilde{v}(1)$ y $\Delta t_{1,2}$, obtenemos
\begin{equation}
 \hat{z}_{1,2}=-\frac{ (\mid v(1) \mid)^{2}}{2 g} (e^{-\kappa H})^{2}.
 \label{Amplitud1}
\end{equation}
 
Para la trayectoria del cometa entre $j=2$ y $j=3$, $\Delta t_{2,3}= \frac{2\, \tilde{v}(2)}{g}$ y, como  $\tilde{v}(2)$ es dada por (\ref{VPrima2}), 
\begin{equation}
\Delta t_{2,3}=2 \frac{\mid v(1)\mid}{g} (e^{-\kappa H})^{2} .
\label{Tmedio2}
\end{equation}
Procediendo   en forma similar a como se obtuvo $\hat{z}_{1,2}$, encontramos
 \begin{equation}
 \hat{z}_{2,3}=\frac{ (\mid v(1) \mid)^{2}}{2 g} (e^{-\kappa H})^{4}.
 \label{Amplitud2}
\end{equation}

Comparando  (\ref{Tmedio}) con (\ref{Tmedio2}), y  (\ref{Amplitud1}) con (\ref{Amplitud2}), obtenemos las siguientes reglas generales:

\begin{equation}
\Delta t_{j,j+1}=2 \frac{\mid v(1)\mid}{g} (e^{-\kappa H})^{j} 
\label{DeltaI}
\end{equation}
y
\begin{equation}
\hat{z}_{j,j+1}= (-1)^{j} \frac{(\mid v(1) \mid)^{2}}{2 g}(e^{-\kappa H})^{2 j}.
\end{equation}

\begin{figure} 
\includegraphics[scale=1.1]{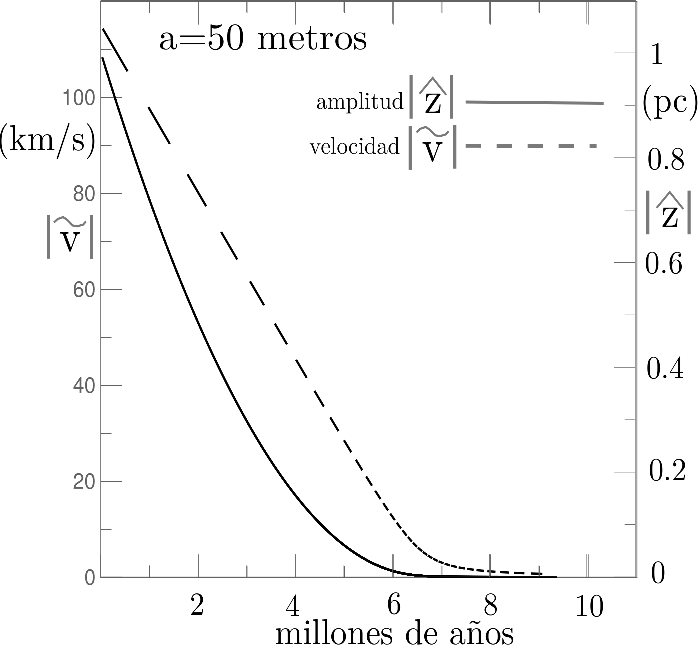} 
\caption{Amplitud $\hat{z}$ de las oscilaciones (ordenada derecha y línea llena) y velocidad $\tilde{v}$ al cruzar el disco protoplanetario (ordenada izquierda  y curva de línea a trazos) de  un cometa de radio $a$=50 metros con respecto al tiempo expresado en millones de años.}
\label{CaidaLibre1}
\end{figure}

Si $\tilde{t}_{1}$  es el tiempo transcurrido desde el momento en que el cometa comienza a caer sobre disco, lo penetra  por primera vez y emerge del mismo. $\tilde{t}_{1}= t_{1} + \Delta t_{1}$. Por lo tanto, la velocidad del cometa en el tiempo  $\tilde{t}_{1}$ es $\tilde{v}(1)$. Similarmente, en el tiempo $\tilde{t}_{2}= \tilde{t}_{1} + \Delta t_{1,2}+ \Delta t_{2}$, la velocidad del cometa es  $\tilde{v}(2)$. Generalizando, 
$\tilde{t}_{j+1}= \tilde{t}_{j} +\Delta t_{j,j+1}+ \Delta t_{j+1}$  y, con los reemplazos de (\ref{DeltaI}) y de (\ref{Deltat3}), obtenemos
\begin{equation}
\tilde{t}_{j+1}= \tilde{t}_{j} + 2 \frac{\mid v(1)\mid}{g} (e^{-\kappa H})^{j}+ \frac{e^{\kappa H}-1}{\kappa\, \mid v(1) \mid} (e^{\kappa H})^{j}.
 \end{equation}
Ello nos permite estudiar la evolución de $\tilde{v}(j)$, representando los pares ordenados  ($\tilde{t}_{j}, \tilde{v}(j))$ con $j=1,2,3,...$; tal como lo muestra la Fig. \ref{CaidaLibre1}. La curva representada en la Fig. \ref{CaidaLibre1} corresponde a un cometa  que se mueve en un disco gaseoso de las características  especificadas en (\ref{ParaDisco}) y que tiene un radio $a=50$ metros y una posición inicial $z(0)=1$ parsec, la cual determina $t_{1}$ y $v(t_{1})= v_{1}$ por medio de (\ref{Tiempo1}) y (\ref{Velocidad1}). 

 \begin{figure} 
\includegraphics[scale=0.9]{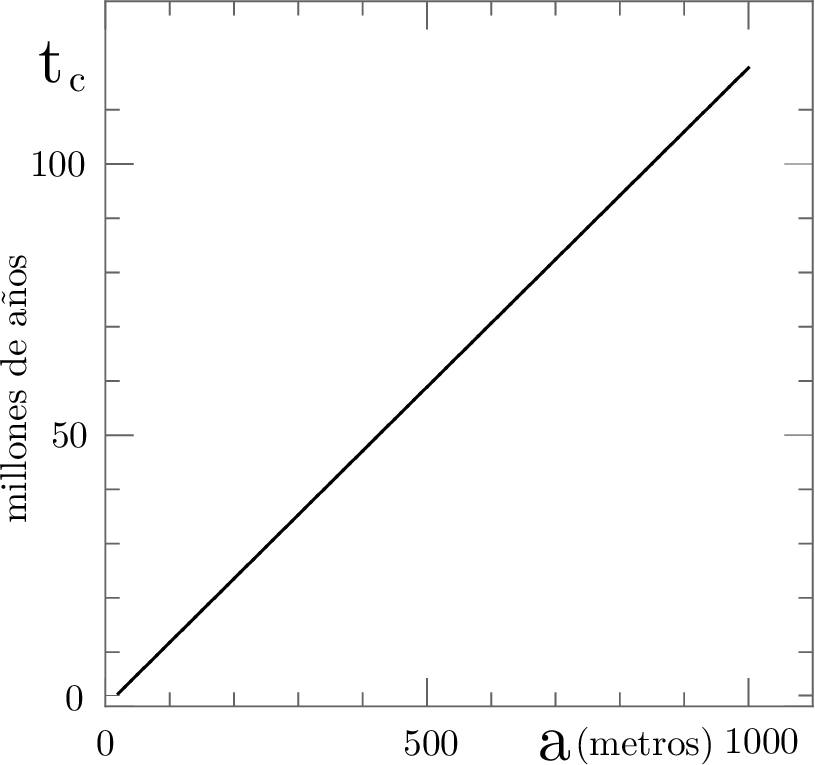} 
\caption{Tiempos $t_{c}$ de  decantación,  sobre un  disco protoplanetario, de cometas con diferentes radios $a$.}. 
\label{CaidaLibre2}
\end{figure}

Designamos $\hat{t}_{j}$ a los tiempos correspondientes a $\hat{z}_{j,j+1}$, de modo que $\hat{t}_{1}=\tilde{t}_{1}+ \frac{\Delta t_{1,2}}{2}$, $\hat{t}_{2}=\tilde{t}_{2}+ \frac{\Delta t_{2,3}}{2}$ y en general 
\begin{equation}
\hat{t}_{j}=\tilde{t}_{j}+ \frac{\Delta t_{j,j+1}}{2}.
\end{equation}
La representación de los pares ordenados  ($\hat{t}_{j}, \hat{z}_{j,j+1})$ con $j=1,2,3,...$ nos permite estudiar la evolución de la amplitud de oscilación del cometa en torno al plano medio del disco. A fin de comparar conjuntamente la variación de la amplitud  y de  la velocidad con el tiempo, del cometa de marras, $\hat{z}$ es también  representada en  la Fig. \ref{CaidaLibre1}. En dicha figura, observamos que  los valores absolutos de la velocidad  y de la amplitud de oscilación  decaen con el tiempo. En el ejemplo dado,  el minicometa de 50 metros de radio es frenado e incorporado al disco gaseoso  en 6-7 millones de años.
 
 Denotamos $t_{c}$ al tiempo transcurrido desde que comienza la oscilación amortiguada de un cometa hasta que su amplitud disminuye  al valor de $\frac{3}{2} H$, con lo cual el cometa pasa a formar parte del disco gaseoso.  En  la Fig. \ref{CaidaLibre2}, representamos $t_{c}$ en función del radio $a$ del cometa: relación  lineal que indica que los cometas pequeños son capturados por el disco más rápidamente que  los más grandes. La escala de tiempo para la formación de un sistema planetario sería del orden de 100-200 millones de años. Por lo tanto, según la Fig. \ref{CaidaLibre2}, cometas de hasta 1 km de radio pueden ser capturados por el disco proto-planetario y 
formar parte de los planetesimales que dan origen a los planetas. 

Los cometas mayores que se encuentran ligados gravitacionalmente al disco, y que  casi no son frenados por el disco, forman una nube cometaría esférica similar a la nube de Oort\index{Oort, nube de}. Nótese que en nuestro modelo, el disco protoplanetario es alimentado de planetesimales desde afuera. Para la formación de un sistema planetario, es además necesario que el disco protoplanetario rote. Ello es posible dado que el disco madre puede estar constituido de grandes celdas que se  formaron  al comprimir nubes en rotación, las cuales conservan sus momentos angulares originales. La rotación del disco protoplanetario arrastra a los cometas capturados y los hace participar de la rotación.

\section{Eras de hielo \label{ErasHielo}}

\begin{quote}
\small \it{El cielo, el aire y la roca fueron escrutados para hallar la solución de estos enigmas, pero a pesar de la amplitud de nuestros conocimientos, la causa de las congelaciones sigue siendo uno de los grandes misterios del globo...En algún lugar entre el laberinto de conjeturas astronómicas, meteorológicas y geológicas debe encontrarse la síntesis que ha de resolverlo todo.} \rm

John Hodgdon Bradley\index{Bradley, J.H.} (1898-1962) en el libro ``Autobiografía de la Tierra'' (Autobiography of Earth)
\end{quote}

Un capítulo importante de la historia de la Tierra lo constituye el estudio de los diversos episodios climáticos que se sucedieron a lo largo  de su historia. Las perturbaciones gravitatorias de los grandes planetas del Sistema Solar pueden variar la inclinación del eje de rotación de la Tierra y su órbita   en torno al Sol, fuente principal de calor. Por lo tanto,  la cantidad y distribución de la energía lumínica del Sol   que incide sobre la Tierra puede modificarse y así influir sobre el clima terrestre. Parte de la luz solar que incide sobre la Tierra calienta la atmósfera y los océanos, los cuales a su vez mediante vientos y corrientes marinas redistribuyen el calor sobre el planeta. Este sistema circulatorio de aire y agua sigue los circuitos que les permiten las masas continentales. Como las placas continentales tuvieron posiciones distintas a las actuales, por el fenómeno de la ``deriva continental'', el patrón de circulación también fue cambiando y con ello el clima. El sistema atmósfera-océano \footnote{Es un sistema fuertemente acoplado no lineal, susceptible de efectos ``mariposa'':  una variación pequeña de las condiciones iniciales puede provocar que el sistema salte de un estado a otro muy diferente. Esta es una de las  razones por la cual los pronósticos meteorológicos de largo alcance suelen ser inciertos.}, el cual  determina en última instancia las condiciones meteorológicas y el clima, depende de su entorno geológico y biológico, como así también de  influencias externas provenientes de sistemas astronómicos de mayor jerarquía: el Sistema Solar y la Galaxia. Los científicos aún tratan de reconstruir, en la forma más fidedigna posible,  el complejo pasado climático de la Tierra y de encontrar las causas probables de las eras de hielo. Una tarea difícil pero no imposible.

\subsection{¿Qué evita que la Tierra no esté eternamente congelada?: el efecto invernadero \label{Arrhenius83}}
Los primeros científicos que intentaron explicar las eras de hielo, como el gran matemático francés Jean-Baptiste Fourier\index{Fourier, J.B.}  (1768-1830) y  el gran físico y químico sueco Svante Arrhenius\index{Arrhenius, S.} (1859-1927) \footnote{Él recibió el Premio Nobel de Química en 1903.}, se encontraron con una paradoja: en vez de explicar por qué ocurren episódicamente las eras de hielo, deberíamos primero explicar por qué la Tierra no se encuentra permanentemente en una era de hielo. La explicación es que la Tierra está arropada por ``el efecto invernadero'' de nuestra atmósfera. Fourier\index{Fourier, J.B.}  fue el primero que explicó el efecto invernadero que produce la atmósfera terrestre. Fourier\index{Fourier, J.B.} se dio cuenta de que la superficie de la Tierra no está globalmente congelada  porque  la atmósfera se comporta como un aislante térmico  que retiene parte del calor generado por la luz solar que incide sobre la superficie del planeta \footnote{El calor interno de la Tierra influye muy poco en la temperatura superficial, la cual es regulada esencialmente por la luz solar.}. Svante Arrhenius\index{Arrhenius, S.}  dedujo la siguiente fórmula que estima la temperatura de la superficie de la Tierra $T_{s}$: 
 
 \begin{equation}
 T_{s}=255\,\, \sqrt[4]{\frac{1}{1-\epsilon/2}},
 \label{Arrhenius}
 \end{equation}
 donde $\epsilon$ es un número entre cero y uno ($0<\epsilon<1$), cuyo valor es especificado por la composición química de la atmósfera.  Sin la existencia de la atmósfera $\epsilon=0$, y por lo tanto  $T_{s}=255$ Kelvin ($=-18$ grados centígrados). Es decir, la superficie del planeta estaría totalmente congelada. Sin embargo,  la temperatura media actual de la Tierra es del orden de $288$ K ($=15$ grados centígrados). Si reemplazamos  $T_{s}=288$ en  la fórmula (\ref{Arrhenius}) y despejamos $\epsilon$, obtenemos   que $\epsilon=0.78$. Arrhenius\index{Arrhenius, S.} notó que el aumento de la concentración de dióxido de carbono ($CO_{2}$)  y de otros gases de efecto invernadero en la atmósfera  aumenta el valor de $\epsilon$,  con lo cual aumenta  $T_{s}$.  Arrhenius, \index{Arrhenius, S.} que  vivió en plena revolución industrial, advirtió  que  debido a la actividad industrial aumenta la contaminación atmosférica con gases de efecto invernadero, lo cual conduciría a un calentamiento global del planeta. Sin embargo, el  principal interés de Arrhenius\index{Arrhenius, S.} no era predecir  una catástrofe futura, sino demostrar que la causa de las glaciaciones del pasado podría ser el descenso del contenido de $CO_{2}$ de la atmósfera.
 
 Nuestro próximo paso es deducir la fórmula (\ref{Arrhenius}). Nuestra atmósfera  es  transparente (o altamente transparente) a la luz solar de ciertas longitudes de onda. Por una de esas   ``ventanas'' de transparencia pasa la radiación del espectro óptico,  al cual es sensible el ojo y de allí su nombre. La emisión máxima de energía radiante  del Sol se produce justamente  en el rango de longitudes de onda del espectro  óptico, centrado en la luz verde-amarilla \footnote{Interesantemente, el ojo humano tiene la mayor sensibilidad en la luz verde-amarilla. Evidentemente, esta es una adaptación evolutiva de nuestro ojo para captar la luz que con mayor intensidad ilumina a  los objetos. Intuyendo dicho fenómeno, el sabio alemán Johann Wolfgang von Goethe\index{von Goethe, J.W.} dijo que el ojo era  ``solar''.}. Si $S_{0}$ es  la energía solar por unidad de tiempo y de
superficie que llega al tope de la atmósfera de la Tierra. Entonces,  la potencia radiante  del Sol que  la Tierra intercepta es $\pi R_{T}^{2} S_{0}$, la cual dividida  por la superficie de la Tierra, da  la energía solar media, por unidad de tiempo y de superficie, que recibe la Tierra: $\frac{\pi R_{T}^{2} S_{0}}{4 \pi   R_{T}^{2}}=\frac{S_{0}}{4}$.

\begin{figure}
\includegraphics[scale=1.0]{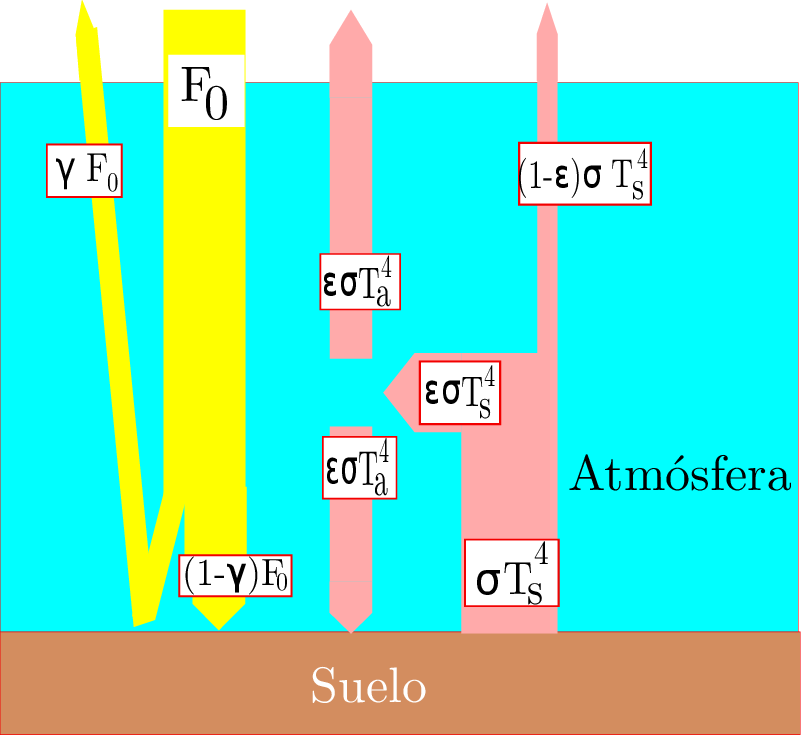} 
\caption{Modelo de Arrhenius para explicar el efecto invernadero.}
\label{ArrheniusModelo}
\end{figure}

La luz visible incidente, representada por $\frac{S_{0}}{4}$, traspasa la atmósfera, así como traspasa el vidrio de un invernadero. Una fracción de la luz incidente,  $\gamma\frac{S_{0}}{4}$,  es reflejada por la superficie de la Tierra y vuelve al espacio exterior porque la reflexión de la luz no varia sus  longitudes de onda y en esas longitudes de onda la atmósfera es transparente. El factor $\gamma$ ($0<\gamma<1$) se conoce como albedo del planeta y en este caso $\gamma\approx 0.3$.  La fracción restante de la luz incidente, $(1-\gamma) \frac{S_{0}}{4}$, es absorbida por la superficie de la Tierra, la cual  se calienta e emite radiación infrarroja. 
Si la temperatura de la superficie es $T_{s}$, la potencia de la radiación infrarroja que ésta emite es, según la ley de Stefan-Boltzmann\index{Stefan-Boltzmann, ley de}, $\sigma T_{s}^{4}$. 
A diferencia de la luz visible, la radiación infrarroja es en parte atrapada por la atmósfera (o en el caso de un invernadero,  por el vidrio que lo cubre). Por lo tanto, la potencia de la radiación infrarroja absorbida por la atmósfera puede expresarse como $\alpha \sigma T_{s}^{4}$, donde  $\alpha$  es  la absortividad de la atmósfera, y como consecuencia, la atmósfera  se calienta a una cierta temperatura $T_{a}$.

El espesor $H$ de la atmósfera es pequeño en comparación con el radio terrestre. De modo que podemos representar localmente a la atmósfera como una capa de gas con forma de  cilindro chato. La base del cilindro de área $A=\pi r^{2}$ apoya sobre la superficie de la Tierra. La potencia radiante que emite el gas atmosférico contenido dentro del cilindro es $\epsilon \sigma T_{a}^{4} S$, donde $\epsilon$ es la emisividad de la atmósfera y $S=2\pi r^{2} + 2 \pi r H$ es la superficie del cilindro. Como podemos considerar que $r \gg H$, $S=2\pi r^{2}=2 A$ y en consecuencia la potencia radiante total es igual a  $2 \epsilon \sigma T_{a}^{4} A$. Es decir, la mitad de la radiación que emite la atmósfera escapa al espacio vacío por el techo del cilindro y la otra mitad, $\epsilon \sigma T_{a}^{4} A$, retorna a la superficie de la Tierra, a través de la base del cilindro. Por balance energético, la potencia radiante que entra al cilindro y queda allí atrapada,  $\alpha \sigma T_{s}^{4} A$, debe ser igual a la potencia que irradia el gas atmosférico  contenido en el cilindro, $2 \epsilon \sigma T_{a}^{4} A$. Es decir,  $\alpha \sigma T_{s}^{4} A= 2 \epsilon \sigma T_{a}^{4} A$. La  ley de Kirchhoff\index{Kirchhoff, ley de} de la radiación térmica establece  que, en condiciones de equilibrio termodinámico, la emisividad  $\epsilon$ de un objeto es igual a su absortividad  $\alpha$. Por lo tanto,
 
\begin{equation}
 T_{s}^{4}= 2 \, T_{a}^{4}.
 \label{Kirchhoff}
\end{equation}

La potencia total (por unidad de superficie)   que absorbe la superficie de la Tierra  es igual a aquella de la luz óptica,  $(1-\gamma) \frac{S_{0}}{4}$, más aquella de la radiación infrarroja que la atmósfera retorna  a la superficie,  $\epsilon \sigma T_{a}^{4}$. Por otra parte, la superficie emite en el infrarrojo  con una potencia total (por unidad de superficie) igual a  $ \sigma T_{s}^{4}$. Entonces, el balance energético prescribe que la potencia total de entrada a la superficie debe ser igual a aquella de salida de la misma:
\begin{equation}
(1-\gamma) \frac{S_{0}}{4} + \epsilon \sigma T_{a}^{4}=\sigma T_{s}^{4}.
\label{BalanceE}
\end{equation}
Despejando $T_{a}$ de  la ecuación (\ref{Kirchhoff}) y  reemplazando su expresión  en la ecuación (\ref{BalanceE}), se  obtiene  una ecuación con una única incógnita: $T_{s}$, de donde resulta : 
\begin{equation}
T_{s}=\sqrt[4]{(1-\gamma) \frac{S_{0}}{4 \sigma}}\,\, \sqrt[4]{\frac{1}{1-\epsilon/2}}.
\label{Arrhenius2}
\end{equation}
Adoptando para $S_{0}$ el valor de la constante solar 1360  W m$^{-2}$, $\sqrt[4]{(1-\gamma) \frac{S_{0}}{4 \sigma}} =255$ K, con lo cual la fórmula (\ref{Arrhenius2}) coincide con la fórmula (\ref{Arrhenius}) obtenida por Arrhenius\index{Arrhenius, S.}.

\subsection{Las glaciaciones del Cuaternario}
El naturalista suizo-americano Louis Agassiz\index{Agassiz, L.} (1807-1873) fue quien profundizó la idea de  que Europa estuvo  cubierta por ``ríos de hielo'' (glaciares) \footnote{La parte inferior de una gruesa capa de hielo se derrite debido a la alta presión que soporta, con lo cual se  disminuye el roce con el suelo, permitiendo  un lento  deslizamiento  del glaciar.} en una antigua edad de hielo \footnote{Karl Friedrich
Schimper\index{Schimper, K.F.} (1803-1867), un eminente botánico y poeta alemán, acuñó la palabra
 ``Eiszeit'' (edad de hielo).}, con  lo cual se  explica  la presencia de enormes bloques de roca en las llanuras Europeas como trozos de roca  arrancados y arrastrados por el flujo de hielo desde  regiones montañosas alejadas \footnote{Esta idea, que  circulaba entre los aficionados al montañismo, fue  propuesta en el año 1821 en un artículo por Ignace Venetz\index{Venetz, I}, un ingeniero suizo.}. En su estadía en Paris,  Agassiz\index{Agassiz, L.} realizó estudios bajo la dirección  de Georges Cuvier\index{Cuvier, G.} (1769-1832), el  gran paleontólogo y zoólogo francés \footnote{Cuvier\index{Cuvier, G.} es considerado el padre del catastrofismo. Él mostró que las especies podían extinguirse y que  existían estratos geológicos  con fósiles de faunas extintas encima de las cuales aparecen estractos con especies nuevas. Ello  indicaría que la Tierra sufrió  una serie de grandes catástrofes. En el periodo entre dos  catástrofes consecutivas, según Cuvier\index{Cuvier, G.},  se crean nuevas especies de animales y plantas.}, y con ello Agassiz\index{Agassiz, L.} adquirió una visión catastrofista y creacionista de la historia de la Tierra.   

Los geólogos no solo corroboraron  la teoría glaciar de Agassiz\index{Agassiz, L.}, sino que  además descubrieron que hubo más de un suceso de glaciación. El estudio de sucesivas capas de sedimentos de glaciares separadas por suelo o turba demostró que la Tierra experimentó varias edades de hielo. Las primeras glaciaciones estudiadas fueron llamadas  Günz\index{Günz, glaciación de}, Mindel\index{Mindel, glaciación de}, Riss\index{Riss, glaciación de} y Würm\index{Würm, glaciación de} (la más reciente, que se extendió desde 110 mil años  a 10 mil años atrás): nombres de afluentes del Danubio\index{Danubio, río} en cuyos depósitos   se identificaron  dichas glaciaciones. Los geólogos y climatólogos  Albrecht Penck\index{Penck, A.} (1858-1945) y Eduard Brückner\index{Brückner, E.} (1862-1927), de origen alemán y austríaco respectivamente, fueron quienes establecieron esas cuatro glaciaciones clásicas, las cuales  se sucedieron durante el último millón de años. Sin embargo, los geólogos fueron encontrando  evidencias claras de que  muchas más glaciaciones habrían ocurrido. Las capas de hielo alcanzaron espesores de varios kilómetros y cubrieron hasta un tercio de la superficie de la Tierra (hoy los glaciares ocupan el 10 por ciento), mientras que  los niveles del mar descendieron hasta 150 metros.

Cada glaciación borra gran parte de las huellas  dejadas en las rocas por las glaciaciones anteriores. Entonces, por este método, resulta  muy difícil encontrar  pruebas de la existencia de glaciaciones muy antiguas. Afortunadamente, el extraordinario desarrollo de la física atómica permitió el análisis isotópico de muestras geológicas, mediante la espectroscopía de masas. Con esta  formidable herramienta, se pueden determinar en principio la edad absoluta de una muestra geológica y la característica medioambiental en la cual se formó. En la mitad de la década de los años cincuenta, el climatólogo ítalo-americano Cesare Emiliani\index{Emiliani, C.} (1922–1995) midió la relación isotópica  del oxígeno, $^{18}O$ : $
^{16}O$, en muestras de sedimentos marinos de micro-fósiles y descubrió una variación periódica de  dicha relación con la edad absoluta. Cuanto más fría está el agua de mar, el isótopo más pesado del oxígeno tiende a acumularse en las caparazones de los microorganismos marinos. Con lo cual, Emiliani\index{Emiliani, C.} demostró que hubo muchos más ciclos de glaciaciones que los cuatro clásicos conocidos antes del estudio de Emiliani\index{Emiliani, C.}. Durante el último millón de años, las edades  de hielo, alternadas con periodos interglaciares, se sucedieron con una periodicidad de $\approx 100$ mil años. 

Al periodo geológico que abarca los últimos 2.5 millones de años se lo denomina \it{ Cuaternario}\rm. El Cuaternario\index{Cuaternario} se divide a su vez en dos épocas: el Pleistoceno, casi sinónimo  de las glaciaciones  recurrentes \footnote{El Pleistoceno es también sinónimo de los bellos  paisajes de montaña. Por ejemplo, los enormes lagos de los  Andes Patagónicos  ocupan depresiones excavadas por los glaciares durante el Pleistoceno.},  y el Holoceno\index{Holoceno}, que se inició hace unos 12.000 años al comenzar el periodo interglaciar actual. Durante el Pleistoceno\index{Pleistoceno} apareció el hombre y por lo tanto se corresponde con el Paleolítico arqueológico. Durante las glaciaciones, Europa fue habitada por el Hombre de Neandertal, el cual estaba muy bien adaptado al clima frío. Un poco antes del final de la última glaciación (glaciación de Würm\index{Würm, glaciación de}), hace $\approx$ 30000 años, el Homo Sapiens ingresó a Europa desde Africa, o al menos una de las corrientes migratorias de éste, la del hombre de Cro-Magnon. Los Neanthertales\index{Neanthertales} y los  Homo Sapiens\index{Homo Sapiens}, ambas subespecies muy similares, probablemente convivieron y  se fundieron genéticamente. Por otro lado, durante la última glaciación,  el estrecho de Bering\index{Bering, estrecho} formaba un puente entre Asia y Norteamérica y se piensa que a través de dicho puente de hielo ingresó a América una de las corrientes migratorias del Hombre.

\subsection{Causas de las glaciaciones del Cuaternario: la teoría de Milankovitch y otras posibilidades \label{Cuaternario}}

El matemático francés Joseph Adhemar\index{Adhemar, J.} (1797–1862) y
el científico autodidacta escosés James Croll\index{Croll, J.} (1821-1890) fueron los primeros que entrevieron una relación de causa y efecto entre  cambios en la órbita terrestre en torno al Sol y las glaciaciones desencadenadas en el Cuaternario. Por otra parte, el climatólogo serbio Milutin Milankovitch\index{Milankovitch, M.} (1879-1958) fue quien primero realizó un modelo  paleoclimático detallado usando las variaciones de largo término de los parámetros astronómicos. Como consecuencia de los cambios en los parámetros orbitales,  la distribución geográfica de la radiación solar recibida en el pasado por la Tierra, o insolación para abreviar, habría variado cíclicamente de forma tal que provocó cambios climáticos drásticos como las glaciaciones.

Si bien la órbita de la Tierra  es principalmente determinada por la fuerza gravitatoria del Sol, los planetas restantes  juegan también un papel, alterando los elementos orbitales de la Tierra.
El cálculo de las perturbaciones planetarias sobre la Tierra, incluyendo las de la Luna, es extremadamente complejo. Exceptuando el problema de los dos cuerpos que tiene solución analítica, el problema de n-cuerpos solo puede resolverse por extensos desarrollos  numéricos: problema abordado por una rama de la Astronomía llamada  Mecánica Celeste. 

La Mecánica Celeste recibió un gran impulso durante los siglos  XVIII y XIX gracias a grandes matemáticos y astrónomos, entre los cuales se encuentran Joseph-Louis Lagrange\index{Lagrange, J.L.} (1736-1813) y Pierre-Simon Laplace\index{Laplace, P.S.} (1749–1827).  Uno de los momentos de mayor esplendor de la Mecánica Celeste ocurrió en el año 1846, cuando el matemático francés Urbain Le Verrier\index{Le Verrier, U.} (1811-1877) descubre el planeta Neptuno  por medio de cálculos matemáticos. En efecto, las posiciones orbitales  observadas del  planeta Urano no coincidían con las teóricas, lo cual llevó a Le Verrier\index{Le Verrier, U.} a proponer la existencia de un planeta, hasta entonces desconocido, que perturbaba la órbita de Urano. A partir del análisis de  las variaciones orbitales de Urano\index{Urano, planeta},   Le Verrier\index{Le Verrier, U.} dedujo la posición en el cielo del planeta perturbador (Neptuno, planeta) y, a menos de un grado de la posición predicha,  Neptuno\index{Neptuno, planeta} apareció en el campo de visión del telescopio. Obviamente, en ese entonces,  no se disponía de computadoras electrónicas. Esos complejos y extensos cálculos eran efectuados ``a mano'' mediante el uso de tablas de funciones trigonométricas y logarítmicas; una proeza de la inteligencia y tenacidad humanas.

Milankovitch\index{Milankovitch, M.} calculó la insolación  de la Tierra durante los pasados 600000 años, para lo cual utilizó las variaciones de los parámetros orbitales de la Tierra, calculados con el procedimiento de Le Verrier\index{Le Verrier, U.} \footnote{Milankovitch contó con la colaboración del  Prof. Vojislav Miskovitch (1892–1976)\index{Miskovitch, V.} para la realización de los cálculos astronómicos.}. Cada parámetro orbital contribuye a la variación de la insolación con una intensidad y periodicidad o frecuencia propia. Notablemente, la curva paleo-climática reconstruida a partir de los registros geológicos posee las mismas componentes periódicas que la curva de insolación calculada por Milankovitch\index{Milankovitch, M.}, a saber: 100000, 41000 y 26000 años. Ello significó un fuerte aval a la teoría astronómica. Sin embargo, la teoría presenta dificultades que los paleo-climatólogos aún tratan de resolver. A continuación, estudiaremos las influencias de los parámetros orbitales de la Tierra  en el comportamiento de la insolación.

\begin{figure} 
\includegraphics[scale=1.0]{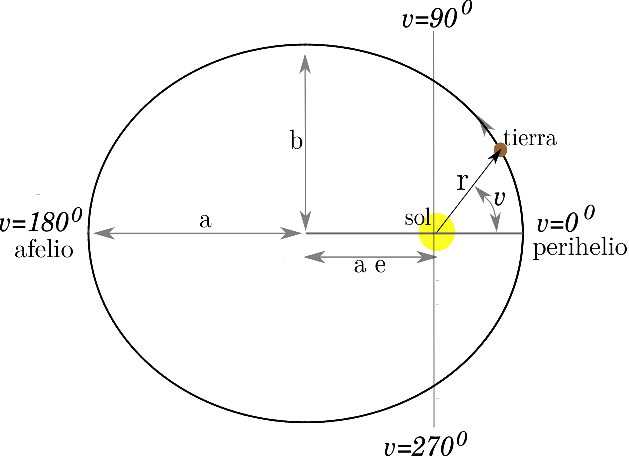} 
\caption{La órbita de la Tierra, y de los planetas en general, es una elipse con el Sol en uno de los focos. La figura indica los siguientes  elementos orbitales: el semieje mayor $a$ y el menor $b$ de la elipse, y la distancia entre el centro de la elipse y el Sol = $a\times e$, donde $e$ es la excentricidad de la órbita. Además, se indica la anomalía verdadera $v$, ángulo que varia con el tiempo y que permite ubicar  el radio vector $r$ y, por lo tanto,  determinar la posición de la Tierra en su órbita. La excentricidad $e$ de la órbita de la Tierra es mucho menor que la del dibujo.}
\label{Elipse}
\end{figure}

Sabemos por las leyes de Kepler\index{Kepler} que la Tierra, como  todos los demás planetas, describe una órbita elíptica con el Sol ubicado en uno de sus focos. En su recorrido anual alrededor Sol, la Tierra pasa por una distancia Tierra-Sol mínima llamada perihelio, y una distancia máxima llamada afelio.  En el perihelio,  la distancia Tierra-Sol es $a (1-e)$ y en el afelio $a (1+e)$, donde $a$ es el semieje mayor de la elipse y $e$ es su excentricidad. 
La excentricidad $e$ es un número, $0\le e <1$,  que mide el grado de  achatamiento de la elipse. Cuando $e=0$, la órbita planetaria es circular. Al estar la Tierra en el perihelio, el Sol se encuentra más cerca y por lo tanto la Tierra recibe una mayor potencia ``instantánea'' de la radiación solar. En el afelio, ocurre lo contrario. La  ecuación de una elipse en coordenadas polares es dada por

\begin{equation}
r=\frac{a\,(1-e^{2})}{(1 + e\, cos \, v)},
\label{Orbita}
\end{equation}
donde $r$ es el radio vector, que da la distancia entre el Sol y la Tierra, y $v$ es el ángulo, con vértice en el Sol, entre la recta que une al Sol con el perihelio  y el radio vector $r$. El ángulo $v$ es llamado anomalía verdadera (ver Fig. \ref{Elipse}).

La cantidad media de energía radiante solar que recibe el tope de la atmósfera terrestre por unidad de tiempo y superficie  perpendicular a la radiación incidente  es llamada irradiancia solar y la denotamos  $S_{m}$.   La irradiancia solar se obtiene mediante la expresión $S_{m} = \frac{L_{\odot}}{4 \pi r_{m}^{2}}$, donde $L_{\odot} (=3.85 10^{26}$ W) es la luminosidad del Sol y $r_{m}$ la distancia media Tierra-Sol. Aquí aceptamos que  $L_{\odot}$ permaneció constante durante los tiempos geológicos considerados.  Podemos definir  $r_{m}$ como el radio de un círculo, $\pi r_{m}^{2}$, que tiene el mismo área que la elipse que describe la órbita terrestre, $\pi a^{2} \sqrt{1-e^{2}}$. Por lo tanto, $r_{m}^{2}=a^{2} \sqrt{1-e^{2}}$.

Como la distancia Tierra-Sol varía de acuerdo con (\ref{Orbita}), donde $v$ es una función del tiempo,  la  irradiancia solar instantánea ($= \frac{L_{\odot}}{4 \pi r^{2}}$) cambia a lo largo de la órbita según  
 
 \begin{equation}
 S= S_{m} (\frac{r_{m}}{r})^{2}= S_{m} \frac{a^{2} \sqrt{1-e^{2}}}{{r}^{2}}.
 \label{Irradiancia} 
 \end{equation}
 Reemplazando $r$  por su  expresión (\ref{Orbita}) en (\ref{Irradiancia}), obtenemos
 \begin{equation}
  S=S_{m} \frac{\sqrt{1-e^{2}}}{(1-e^{2})^{2}} (1+ e\, cos \, v)^{2}.
  \label{Irradiancia2} 
 \end{equation}

\begin{figure} 
\includegraphics[scale=1.1]{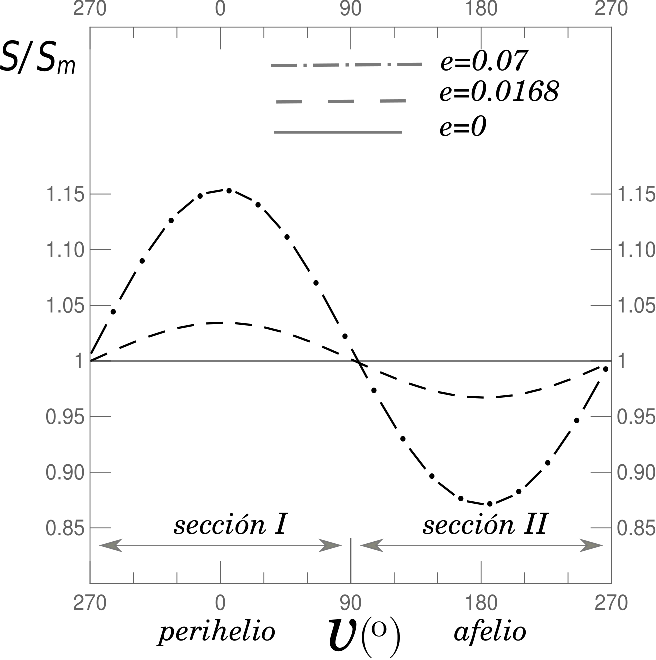} 
\caption{Irradiancia  solar (o insolación) $S$ relativa a la irradiancia solar anual media $S_{m}$ en función de  la anomalía verdadera $v$ de la órbita de la Tierra con las siguientes excentricidades de la órbita: $e=0.07$ (curva de puntos y rayas), $e=0.0168$ (curva de rayas) y $e=0$ (curva llena). Se indican las dos secciones en que se dividió la órbita de la Tierra: la sección I se centra en el perihelio, sobre el cual transita  la Tierra cuando $v=0^{\circ}$, mientras que la sección II se centra en el afelio ($v=180^{\circ}$).}

\label{MilankovitchAB2}
\end{figure}

 El valor del semieje mayor $a$ (=1 UA) de la órbita terrestre puede considerarse constante, con una muy buena aproximación, a través de los últimos millones de años. En cambio, su excentricidad oscila casi periódicamente entre $e=0$, órbita circular, y  $e=0.07$, levemente elíptica, con un período medio de  $\approx$ 100000 años. En el presente, $e=0.016773$, con lo cual la órbita actual de la Tierra es casi circular. La irradiancia solar actual, también llamada ``constante solar'',  es denotada $S_{0}$ (=1360 Watts m$^{-2})$ para diferenciarla de $S_{m}$  que puede tener otros valores  en tiempos geológicos. Usando la ecuación (\ref{Irradiancia2}), representamos  en la Fig. \ref{MilankovitchAB2} $\frac{S}{S_{m}}$ en función de $v$ para la excentricidad actual (=0.016773) y para la   mínima y  la máxima, 0 y 0.07, entre las cuales la excentricidad de la órbita terrestre puede oscilar. Lo interesante de la  Fig. \ref{MilankovitchAB2} es que, en $v=90^{\circ}$ y  $v=270^{\circ}$, $\frac{S}{S_{m}}\approx 1$, separando en dos  secciones  la órbita (ver también Fig. \ref{Elipse}). En la sección I, $\frac{S}{S_{m}}> 1$ y, en la sección II,  $\frac{S}{S_{m}}< 1$.
 
 La segunda ley de Kepler\index{Kepler, leyes} estipula que el radio vector $r$ de la órbita de un planeta barre áreas iguales en tiempos iguales, lo cual equivale a decir que  la variación del área $A$ barrida con respecto al tiempo $t$ es constante. En términos matemáticos,
\begin{equation}
\frac{dA}{dt}=\frac{1}{2} h,
\label{LeyKepler2}
\end{equation}
donde $h$ es una constante particular a la órbita planetaria considerada. De consideraciones dinámicas del movimiento orbital de la Tierra,  se demuestra que $h=\sqrt{G M_{\odot}\, a (1-e^{2})} \,\, $  \footnote{En la expresión exacta de $h$, a la masa del Sol se le suma la masa de la Tierra, pero como ésta es mucho menor que la masa del sol, podemos despreciarla en la suma.}, donde $G$ es la constante gravitatoria y  $M_{\odot}$ la masa del Sol. Cuando la Tierra completa un periodo $T$ de revolución, el radio vector $r$ barre el área total de la elipse: $A= \pi a^{2}\sqrt{1-e^{2}}$. La  integración trivial de (\ref{LeyKepler2}) arroja que $A=\frac{1}{2} h \int_{0}^{T} dt= \frac{1}{2} h T$ y,  reemplazando las expresiones para $A$ y $h$, obtenemos

\begin{equation}
\frac{T^{2}}{a^{3}}=\frac{4 \pi^{2}}{G M_{\odot}}.
\label{LeyKepler3}
\end{equation}
Esta ecuación es conocida como la tercera ley de Kepler  y muestra que $T$ solo depende del semieje mayor $a$, un cuasi-invariante orbital. En el caso de la Tierra, $T=365.25$ días, naturalmente. 

Si $ds$ es la longitud de arco de la órbita que recorre la Tierra en un intervalo muy pequeño de tiempo $dt$, $dA=\frac{1}{2} r ds$ y, de acuerdo con (\ref{LeyKepler2}), $\frac{dA}{dt}= \frac{1}{2} r \frac{ds}{dt}=\frac{1}{2} h$.  Ésta última ecuación puede escribirse como $\frac{1}{2} r V_{T}=\frac{1}{2}h$, dado que  $\frac{ds}{dt}=V_{T}$ es la velocidad tangencial, la componente de velocidad perpendicular al radio vector $r$. Por lo tanto, 
$V_{T}= \frac{h}{r}$  y, con el reemplazo de  $r$ por la expresión (\ref{Orbita}) y de $h$, obtenemos 
\begin{equation}
V_{T}= \frac{\sqrt{G M_{\odot}\, a (1-e^{2}) }}{a (1-e^{2})} (1+ e\, cos\, v).
\label{VT}
\end{equation}

 La fórmula (\ref{VT}) muestra que la velocidad orbital de la Tierra es máxima en el perihelio ($cos \, v=1$) y mínima en el afelio ($cos\, v=-1)$ \footnote{En el perihelio y afelio, la velocidad orbital total de la Tierra coincide con $V_{T}$,  puesto que cuando la Tierra transita por los extremos de la elipse,  su velocidad orbital radial es nula.}.
 El diferencial de ángulo  $dv$ que subtiende $ds$ es dado por $dv=\frac{ds}{r}$ y, teniendo en cuenta que $ds=V_{T} dt$, $dv= \frac{V_{T}}{r} dt$. A partir de esa última fórmula, podemos expresar $dt$ en función de $v$: $dt= \frac{r}{V_{T}} dv$ y reemplazando $r$ por su expresión (\ref{Elipse}) y $V_{T}$ por (\ref{VT}) obtenemos
\begin{equation}
dt=\frac{a^{2} (1-e^{2})^{2}}{\sqrt{G M_{\odot}\, a (1-e^{2})}} \frac{1}{(1+ e\, cos\, v)^{2}} dv.
\label{Exposicion}
\end{equation}

En el tiempo $dt$,  la Tierra se  expone a la irradiancia $S$ y acumula  por $m^{2}$ la energía  $dW= S\, dt$. Usando las expresiones (\ref{Irradiancia2}) y (\ref{Exposicion}) para $S$ y $dt$, respectivamente, resulta
\begin{equation}
dW= S_{m} \frac{a^{2}}{\sqrt{G M_{\odot}\, a}}\, dv.
\label{Wd}
\end{equation}
En la Fig. \ref{MilankovitchAB2}, dividimos la órbita en dos secciones. La sección I está centrada sobre el perihelio y se extiende entre $v=270^{\circ}$ y   $v=90^{\circ}$  y la sección II, centrada sobre el afelio, completa la órbita. La integración de $dW$ mediante la fórmula (\ref{Wd}) nos permite calcular la energía total $W_{I}$ acumulada por la Tierra en su trayectoria sobre la sección I, y la energía $W_{II}$ acumulada en la sección 
II. Dado que el ángulo que subtienden ambas secciones es el mismo,  $\Delta v=\pi$, ambas energías son iguales y dadas por
\begin{equation}
W_{I}=W_{II}=S_{m} \frac{a^{2}}{\sqrt{G M_{\odot}\, a}}\,\pi.
\label{SeccionesI-II}
\end{equation}
Este resultado es aparentemente contradictorio con el hecho de que la Tierra recibe mayor irradiancia $S$ en la sección I que en  la sección II (ver Fig.  \ref{MilankovitchAB2}). Sin embargo, debemos tener en cuenta que la Tierra tarda menos tiempo en recorrer la sección I que en recorrer la sección II. Entonces, en el caso de la sección I, la mayor irradiancia solar se compensa con un menor tiempo de exposición y, en el caso de la sección II,  ocurre lo  inverso.

A fin de probar la rigurosa compensación energética mencionada en el párrafo anterior, integraremos la ecuación (\ref{Exposicion}) para determinar los tiempos $T_{I}$ y $T_{II}$ que tarda la Tierra en transitar la sección orbital I y la II, respectivamente. La función que depende de la variable de integración $v$ es  $\frac{1}{(1+ e\, cos\, v)^{2}}$ y, dado que $e\, cos\, v$ es un número mucho menor que uno, puede aproximarse por un desarrollo en series de Taylor:  $\frac{1}{(1+ e\, cos\, v)^{2}} \approx 1- 2 e\, cos\, v$. Por lo tanto, con la integral $\int_{-\pi/2}^{\pi/2} (1- 2 e\, cos\, v) dv=(\pi- 4 e)$, multiplicada por el factor independiente de $v$ de la ecuación (\ref{Exposicion}), obtenemos que 
\begin{equation}
T_{I}=\frac{a^{2} (1-e^{2})^{2}}{\sqrt{G M_{\odot}\, a (1-e^{2})}} (\pi- 4 e).
\label{Exposicion2}
\end{equation}
Para calcular $T_{II}$, debemos utilizar la integral  $\int_{\pi/2}^{3\pi/2} (1- 2 e\, cos\, v) dv=(\pi+ 4 e)$, con lo cual obtenemos que
\begin{equation}
T_{II}=\frac{a^{2} (1-e^{2})^{2}}{\sqrt{G M_{\odot}\, a (1-e^{2})}} (\pi+ 4 e).
\label{Exposicion3}
\end{equation}

A partir de estas fórmulas y con la excentricidad actual de la Tierra ($e=0.01677$), $T_{I}$ y $T_{II}$ resultan, redondeando, iguales a 179 y 186 días, respectivamente. En el presente, y por pura coincidencia, la Tierra se encuentra al comienzo del año cerca del perihelio. Es decir, las estaciones de primavera y verano en el Sur, y las opuestas en el Norte, ocurren cuando la Tierra transita la sección orbital I, y por lo tanto la duración sumada de las dos estaciones es de  tan solo 179 días. La restante secuencia de estaciones ocurre durante el tránsito de la Tierra por la sección II,  cuya duración es de 186 días. Esta diferencia en la duración de las estaciones fue descubierta por el astrónomo griego Hiparco\index{Hiparco} \footnote{Los pueblos antiguos aprendieron a determinar el momento exacto del comienzo  de las estaciones observando las posiciones de salida y puesta del Sol en el horizonte. Si bien en la antigua Grecia se discutía el modelo  heliocéntrico de  Aristarco de Samos\index{Aristarco de Samos} (310-230 a. C.), en general se aceptaba el punto de vista  geocéntrico que permitía  ``salvar las apariencias''. En ese contexto, para  Hiparco\index{Hiparco}, la diferencia de duración de  las estaciones implicaba que la Tierra no estaba ubicada  en el centro de la órbita circular del Sol.} en el año 120 antes de Cristo. 

La energía solar $W_{T}$ acumulada por la Tierra al completar su órbita en torno al Sol se puede calcular haciendo  $dv= 2 \pi$ en la fórmula (\ref{Wd}), o teniendo en cuenta que $W_{T}=W_{I} + W_{II}$. La irradiancía promedio en un período $T$ de revolución es dada por $\overline{S_{T}}=\frac{W_{T}}{T}$. Utilizando  (\ref{Wd}) o (\ref{SeccionesI-II}) para obtener $W_{T}$, y dado que  de acuerdo con la ecuación (\ref{LeyKepler3}) $T=\frac{2 \pi}{\sqrt{G M_{\odot}}} a^{3/2}$,  la irradiancia media $\overline{S_{T}}$ resulta 
\begin{equation}
\overline{S_{T}}= S_{m},
\label{IrradianciaT}
\end{equation}
resultado que también se aprecia en la Fig. \ref{MilankovitchAB2}.
En forma similar, podemos calcular el valor medio de la irradiancia  en el intervalo de tiempo $T_{I}$ por medio de $\overline{S_{I}}=\frac{ W_{I}}{T_{I}}$ y en el  intervalo de tiempo $T_{II}$ por medio de $\overline{S_{II}}=\frac{ W_{II}}{T_{II}}$. Tal como lo hemos mencionado, la excentricidad de la órbita de la Tierra  oscila entre $e=0$ y $e=0.07$ con un periodo de aproximadamente 100000 años. En dicho rango de variación de $e$, el valor de $S_{m}$ es casi constante e igual a la irradiancia solar actual: $\overline{S_{T}}= S_{m}\approxeq S_{0}$.  En cambio,  $\overline{S_{I}}$ siempre supera a $\overline{S_{T}}$ por una diferencia $\Delta \overline{S_{I}}=\overline{S_{I}}-\overline{S_{T}}>0$. Esta diferencia crece en forma lineal con $e$, desde $\Delta \overline{S_{I}}=0$ cuando $e=0$ a $\Delta \overline{S_{I}}\approx 0.1 S_{0}$ cuando $e=0.07$. En el caso de la sección II, 
$\Delta \overline{S_{II}}=\overline{S_{II}}-\overline{S_{T}}<0$  y, al igual que en sección I, alcanza  la diferencia máxima  en $e=0.07$: $\Delta \overline{S_{II}}\approx - 0.1 S_{0}$.

En resumen, las estaciones que transcurren mientras la Tierra transita la sección I de su órbita reciben un poco más de energía solar y,  por lo tanto, son un poco más cálidas o templadas, que las restantes estaciones, las cuales  transcurren en la sección orbital II.
En el presente, $e=0.01677$ y, por lo tanto,  $\Delta \overline{S_{I}}\approx - 0.02 S_{0}$ y $\Delta \overline{S_{II}}\approx - 0.02 S_{0}$. Tan exigua variación de la irradiancia en el transcurso del año muestra que la excentricidad de la órbita terrestre no determina las estaciones. 
Por otra parte, el hecho intrigante es que la periodicidad de las glaciaciones durante los últimos 800 mil años es coincidente con la periodicidad astronómica de la excentricidad. Sin embargo, a partir de los 800000 años hasta al menos dos millones de años atrás, las glaciaciones se sucedieron cada aproximadamente 41000 años. Al problema de explicar  el ciclo de los 100000 años de las glaciaciones, sin aparentemente una conexión directa con el ciclo astronómico de la excentricidad,  y el cambio en la periodicidad de las glaciaciones más antiguas que 800000 años, se lo conoce como ``el problema  de los 100000 años''.

\begin{figure} 
\includegraphics[scale=0.85]{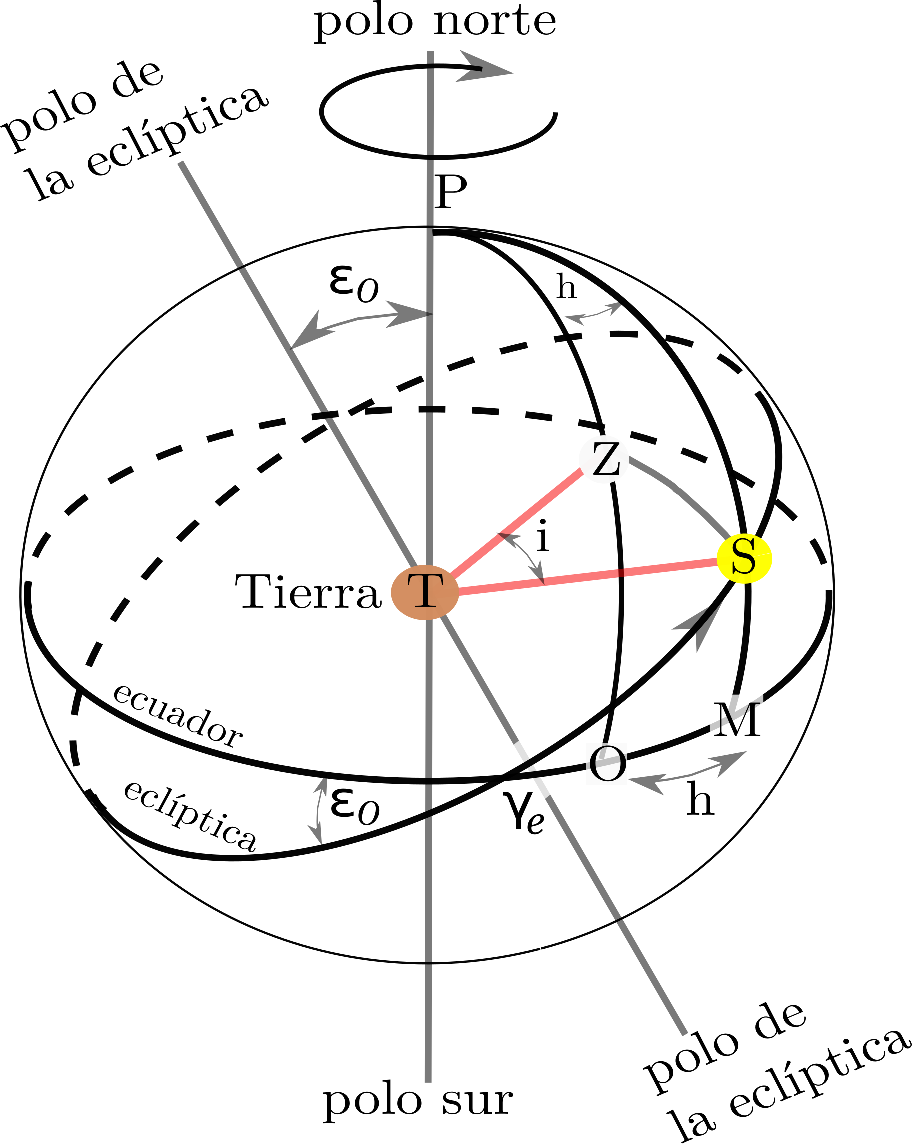} 
\caption{Movimientos aparentes del Sol desde una perspectiva geocéntrica. La rotación del eje polar indicada se refiere a la rotación aparente de la esfera celeste, la cual naturalmente es inversa a la rotación de la Tierra.}
\label{EsferaCeleste}
\end{figure}

El parámetro orbital $\epsilon_{O}$, llamado oblicuidad de la eclíptica\index{eclíptica}, determina las estaciones del año y, además, juega un papel central en la teoría de Milankovitch, como veremos a continuación. 
El Sol presenta dos movimientos aparentes superpuestos: el movimiento diario de salida por el este y puesta por el oeste,  ``reflejo'' de la rotación de la Tierra sobre su eje, y el movimiento de aproximadamente $1^{\circ}$ por día hacia el este con respecto a las estrellas fijas del fondo, reflejo del movimiento de traslación de la Tierra en torno al Sol. Al desplazarse  la Tierra en torno al Sol, la línea de la visual al Sol cambia de dirección  y, en consecuencia, el fondo de estrellas sobre el cual el Sol se proyecta cambia también, dando la sensación de que el Sol se mueve con respecto a las estrellas fijas \footnote{La luz solar dispersada por la atmósfera impide  ver las estrellas durante el día, excepto si ocurre un eclipse total de Sol. Sin embargo, observando la región del cielo que rodea al Sol, un poco antes del amanecer o un poco después de la puesta del Sol,  se puede reconocer el grupo de estrellas o constelación sobre  la cual el Sol está transitando en ese momento. Las diferentes constelaciones que atraviesa el Sol en el año forman una banda centrada en la trayectoria aparente del Sol llamada \it{zodíaco}\index{zodíaco}\rm.}. A dicha trayectoria aparente del Sol se la llama \it{eclíptica}\index{eclíptica} \rm \footnote{El nombre proviene del hecho de que en ese plano,  donde el Sol,  la Tierra  y  la Luna  pueden alinearse,  se producen los eclipses.}. En otras palabras, la eclíptica es el círculo máximo que forma la intersección del plano orbital de la Tierra con la esfera celeste (ver Fig.\ref{EsferaCeleste}).

Un sistema de coordenadas esféricas con origen en la Tierra es muy conveniente para determinar  posiciones y movimientos de los astros. En la Fig. \ref{EsferaCeleste}, representamos la eclíptica y el ecuador celeste, proyección del ecuador terrestre sobre la esfera celeste, y sus respectivos ejes polares. La oblicuidad $\epsilon_{O}$ de la eclíptica es el ángulo entre el plano de la eclíptica y el plano ecuatorial  o, lo que es lo mismo,  
el ángulo entre el polo de la eclíptica y el eje de rotación de la Tierra. Actualmente, 
$\epsilon_{O}=23^{\circ}.5$, y oscila entre $22^{\circ}.1$ y  $24^{\circ}.5$, con un período de aproximadamente 41000 años. Entonces, la variación de $\epsilon_{O}$ explicaría muy bien el ciclo de 41000 años de las glaciaciones que se sucedieron entre 800000 y 2 millones de años atrás. 

El punto $\gamma_{e}$ que yace en la intersección de la eclíptica y el ecuador celeste se llama \it{punto vernal} \rm  y es empleado como origen de las coordenadas celestes (ver Fig.\ref{EsferaCeleste}). En dicha figura, se indican, con la letra $S$,   la posición del Sol en un cierto momento  y, con una flecha, la dirección de su movimiento aparente en la eclíptica. El ángulo $\gamma_{e} S$ se llama  longitud del Sol y se designa $\lambda_{\odot}$. 
El ángulo $M S$, donde $M$ yace en la intersección del ecuador con  el meridiano $P M$ que pasa por el Sol, se llama declinación del Sol y se designa $\delta_{\odot}$. Cuando el Sol se encuentra en el punto $\gamma_{e}$ ($\delta_{\odot}=\lambda_{\odot}=0$),   la línea que une $\gamma_{e}$ y el centro de la Tierra es perpendicular al eje polar. Por lo tanto, los rayos del Sol inciden  perpendicularmente al mediodía sobre la superficie terrestre de las regiones ecuatoriales,  la duración del día 
es igual a la duración de la noche  en todas las latitudes \footnote{Si fuese $\epsilon_{O}=0$, el día sería igual a la noche en toda posición geográfica y en todo tiempo del año. Además, no habría  casi estaciones.} y el Sol sale exactamente por el este. Por ello, al punto vernal se lo llama también punto equinoccial,  de primavera para el norte y de otoño para el sur. Cuando el Sol se encuentra en el otro equinoccio ($\delta_{\odot}=0$, $\lambda_{\odot}=180^{\circ})$, comienzan el otoño en el norte y la primavera en el sur.

El máximo apartamiento del Sol sobre el ecuador ocurre cuando $\delta_{\odot}=\epsilon_{O}$, allí los rayos solares apuntan hacia el hemisferio norte y, al mediodía,  son perpendiculares a la superficie terrestre en el paralelo del Trópico de Cáncer, cuya latitud es naturalmente igual a $\epsilon_{O}$. A ese momento del año, se lo denomina solsticio, de verano en el norte y de invierno en el sur. Similarmente, cuando  $\delta_{\odot}=-\epsilon_{O}$,  el Sol yace al mediodía en el zenit del  Trópico de Capricornio, y comienza el verano en el sur y el invierno en el norte. El primero que midió el valor de $\epsilon_{O}$, con notable precisión,  fue  Eratóstenes\index{Eratóstenes} (276-194 a. C.), un destacado astrónomo, geógrafo y geodesta griego principalmente conocido por ser quien primero calculó la circunferencia y diámetro de la Tierra. Eratóstenes\index{Eratóstenes} sabía de la gran distancia a la que se encuentra el Sol, de hecho hizo precisas estimaciones de la misma, y por lo tanto sabía que los rayos de luz solar que llegan a la Tierra son paralelos entre sí  \footnote{Siendo director de la biblioteca de Alejandría, Eratóstenes\index{Eratóstenes}  se enteró de que  el agua en el fondo de un pozo muy profundo de  Syena\index{Syena} (Egipto\index{Egipto}) reflejaba, al mediodía en el solsticio de verano, la imagen del Sol. En otras palabras, los rayos solares incidían perpendicularmente al suelo local. Mientras tanto,  el faro de Alejandría\index{Alejandría}, que se encontraba más al norte y aproximadamente sobre  el mismo meridiano  (distaban en $\approx3^{\circ}$), proyectaba una sombra. Hecho que solo podía explicarse por la curvatura de la superficie terrestre,  dado que los rayos solares son paralelos entre sí.  A partir de la longitud de la sombra y la altura del faro, se calcula el ángulo $\alpha$ entre los rayos solares y el faro. Por simples relaciones geométricas, el ángulo $\alpha$ es igual al ángulo que subtiende, desde el centro de la Tierra,  el arco $AS$ de circunferencia que une Alejandría con Syena\index{Syena}, una longitud que se conocía. Por lo tanto, si C es a longitud de la circunferencia de la Tierra, $C= \frac{2 \pi}{\alpha} AS $ y el diámetro $D=\frac{C}{\pi}$.}. Este concepto es importante para explicar las diferencias en la irradiancia, o insolación instantánea, proyectadas sobre la superficie terrestre que hace que en un hemisferio sea verano y en el otro invierno. Aquí, consideraremos los términos irradiancia e  insolación instantánea   como sinónimos.

Si un haz cilíndrico de luz solar, con un radio $r$, incide perpendicularmente sobre el suelo, el sector del suelo iluminado es un círculo de radio $r$. Por lo tanto, la potencia energética que recibe el suelo es $ W_{0}= \pi r^{2} S$, donde $S$ la irradiancia instantánea que llega a la parte superior de la atmósfera. En cambio, si  dicho haz  de luz incidente está inclinado en un  ángulo $i$ con respecto a la vertical o perpendicular al suelo, el sector del suelo iluminado es una elipse, de semieje menor $b=r$ y semieje mayor $a=\frac{r}{cos \,i}$ . En este caso, la potencia que recibe el suelo es también $W_{0}$, pero repartida en una superficie mayor. En efecto, el área  de la elipse es  $A_{e}=\pi a\, b= \pi \frac{ r^{2}}{cos\, i}$ y, por lo tanto, la irradiancia sobre una posición en la superficie terrestre, o irradiancia local,  es $S_{l}= \frac{ W_{0}}{A_{e}}$, con lo cual
\begin{equation}
S_{l}= S\, cos \, i, 
\label{inclinacion}
\end{equation}
La insolación local $S_{l}$  depende de la posición geográfica. En cambio, la irradiancia instantánea $S$ al tope de la atmósfera,  dada por la ecuación (\ref{Irradiancia2}),  es la misma para ambos hemisferios.

En el verano, la insolación es obviamente mayor que en el invierno y ello, tal como lo expresa (\ref{inclinacion}),  es en parte consecuencia de que los rayos solares inciden sobre el correspondiente  hemisferio con ángulos $i$ relativamente menores que en el invierno. Ello se manifiesta también en que, al mediodía, las sombras son más cortas en el verano  que en el invierno. Por otra parte, en el verano, los días  son  más largos que las noches, mientras que   en el invierno sucede lo inverso. Por lo tanto, la cantidad de energía solar que recibe el suelo en el verano es mayor  que la que recibe en el invierno.
 
La línea $TZ$ representa la vertical de un sitio sobre la superficie terrestre  y su intersección con la esfera celeste es llamado zenit del lugar, representado con la letra $Z$ en la Fig. \ref{EsferaCeleste}. La línea $TS$, la cual une el centro de Tierra con el Sol,  es la dirección de incidencia de los rayos solares. Por lo tanto, el ángulo entre las rectas $TZ$ y $TS$ (dibujadas en rojo la figura) es $i$, el ángulo de incidencia de los rayos solares sobre el lugar. A fin de obtener el valor de $i$ en función de la posición geográfica y  los  momentos  del día y del año, debemos resolver el triángulo esférico $PZS$ de la Fig. \ref{EsferaCeleste}. Aplicando el teorema del coseno de la trigonometría esférica, tenemos que $ cos\, (ZS)=cos\, (PZ)\, cos\,(PS)\,+ \,sen\, (PZ)\, sen\,(PS)\, cos\, (h)$, donde el ángulo $h$ con vértice en $P$ esta relacionado con la hora del día. Idealizando a la superficie de la tierra como perfectamente esférica, el ángulo que subtiende el arco $ZO$ desde el centro de la Tierra es la latitud geográfica $b$ del lugar considerado y, por lo tanto, $ZO=b$  y  $PZ=90^{\circ}- b$, dado que $PO=90^{\circ}$. Similarmente, $ZS=i$ y $PS=90^{\circ}-\delta_{\odot}$, con lo cual fórmula derivada del teorema del coseno  se convierte en 
\begin{equation}
cos\, i=sen \,b\, sen\,\delta_{\odot} \,+ \, cos\,b \, cos\,\delta_{\odot}\, cos\, h.
\label{Tcos}
\end{equation}
La resolución del triángulo esférico $\gamma_{e} S M$ nos permite determinar  $\delta_{\odot}$ en función de $\epsilon_{O}$ y $\lambda_{\odot}$. En efecto, dado que conocemos dos arcos, $\gamma_{e} S$ y $MS$, y sus respectivos ángulos opuestos, $\epsilon_{O}$ con vértice en $\gamma_{e}$ y el ángulo recto con vértice en $M$, podemos aplicar el teorema del seno: $\frac{sen (SM)}{sen\, \epsilon_{O}} = \frac{sen (\gamma_{e} S)}{sen\,90^{\circ}}$, y dado que 
$SM=\delta_{\odot}$ y $\gamma_{e} S= \lambda_{\odot}$ obtenemos
\begin{equation}
sen\, \delta_{\odot}=sen\, \epsilon_{O}\, sen\, \lambda_{\odot}.
\label{Tsen}
\end{equation}
De la identidad trigonométrica $(sen\,  \delta_{\odot})^{2} + (cos\,  \delta_{\odot})^{2}=1 $ se obtiene que $cos \,  \delta_{\odot}= \sqrt{1-(sen\,  \delta_{\odot})^{2}}$ y usando la relación (\ref{Tsen}) resulta que
\begin{equation}
cos \, \delta_{\odot}= \sqrt{1-(sen\, \epsilon_{O}\, sen\, \lambda_{\odot})^{2}}.
\label{Dsol}
\end{equation}

La variable $h$ que aparece en la fórmula (\ref{Tcos}) es el ángulo horario del Sol, ángulo entre  el meridiano que pasa sobre el zenit, meridiano local,  y  el mediano que  pasa  sobre el Sol, y se mide sobre el ecuador celeste desde el meridiano local hacia el oeste. Cuando el Sol se encuentra sobre el meridiano local, $h=0$ y es mediodía.  Cuando el Sol se encuentra al este del meridiano local, consideraremos que $h<0$. Generalmente, $h$ se expresa en horas y,  para aplicarlo en la fórmula (\ref{Tcos}), debemos tener en cuenta que 1 hora equivale   a $15^{\circ}$. La hora solar local, que denominaremos  $H$, es dada por $H (horas)=12 + h$.
\begin{figure} 
\includegraphics[scale=1.2]{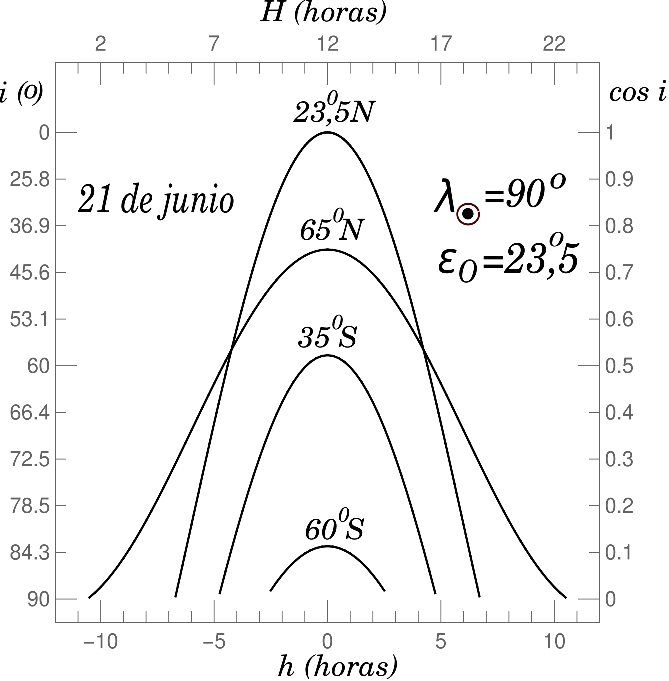} 
\caption{Ángulos de incidencia  de los rayos solares con respecto a la vertical del lugar a lo largo del día de  21 de junio ($\lambda_{\odot}=90^{\circ}$),  para distintas latitudes geográficas, anotadas sobre cada curva $(23^{\circ}.5, \,\, 65^{\circ},-35^{\circ}$ y $-60^{\circ})$. El ángulo de incidencia $i$ expresado en grados se representa sobre el eje  vertical izquierdo y el $cos\,i$ sobre el eje  vertical derecho. En el eje horizontal inferior, se representa el ángulo horario en horas y, en el el eje horizontal superior, se representa la hora solar local.}
\label{Oblicuidad}
\end{figure}

Cuando $cos\, i=0$, significa que $i=90^{\circ}$ y que el Sol está saliendo o poniéndose en el horizonte. Por lo tanto, con la ecuación (\ref{Tcos}), podemos obtener el ángulo horario $h_{0}$ de salida o puesta del Sol:

\begin{equation}
cos\,  h_{0}= -tan\, b \,\, tan\, \delta_{\odot}=-tan\, b \, \frac{sen\, \epsilon_{O}\, sen\, \lambda_{\odot}}{\sqrt{1-(sen\, \epsilon_{O}\, sen\, \lambda_{\odot})^{2}}}.
\label{AngH0}
\end{equation}
Si el Sol sale y se pone en la latitud geográfica  y el día del año considerados,  $cos \, h_{0}$ debe ser un número entre -1 y 1. Si  $cos \, h_{0}>1$, el Sol no sale y si  $cos \, h_{0}<-1$, el Sol no se pone. 

Reemplazando las expresiones (\ref{Tsen}) y (\ref{Dsol}) en la fórmula (\ref{Tcos}), obtenemos
\begin{equation}
cos\, i= sen \,b\, sen\, \epsilon_{O}\, sen\, \lambda_{\odot} \,+ \, cos\,b \, 
\sqrt{1-(sen\, \epsilon_{O}\, sen\, \lambda_{\odot})^{2}} \, cos\, h.
\label{Tcos2}
\end{equation}
Haciendo uso de (\ref{Tcos2}), representamos, en la Fig. \ref{Oblicuidad},  $i$ y $cos\,  i$ en función del ángulo horario para diferentes latitudes durante el solsticio de verano del hemisferio norte (y de invierno en el hemisferio sur). El 21 de junio, día del solsticio,  $\delta_{\odot}=\epsilon= 23^{\circ}.5$ y, de acuerdo con (\ref{Tsen}), $\lambda_{\odot}=90^{\circ}$. Por lo tanto,   en ese día al mediodía $(h=0)$, los rayos solares  inciden  perpendicularmente $(cos \,i =1)$  sobre  las regiones ubicadas a la latitud norte $b= 23^{\circ}.5$ (ver Fig. \ref{Oblicuidad}). Entonces, el 21 de junio, cada latitud del  norte tiene su día  más largo y, en consecuencia, su noche más corta del año, mientras que en el sur ocurre lo inverso. Los extremos del  rango de $h$ que abarca la curva de  la latitud indicada en la  Fig. \ref{Oblicuidad}  corresponden a $h_{0}$ y $-h_{0}$ y, por lo tanto, la duración del día en esa  latitud es $2 h_{0}$. Es decir, para una curva dada, $h_{0}$ o $-h_{0}$ es el $h$ en el cual $i=90^{\circ}$ o $cos \, i=0$. Por ejemplo, de la curva de la Fig. \ref{Oblicuidad} correspondiente a la latitud $b=65^{\circ} N$ y al día del solsticio, obtenemos que $2\, h_{0}=21$ horas;  es decir,  los lugares a lo largo del paralelo  tienen 21 horas con Sol y apenas de 3 horas de noche.

La idea principal de la teoría de Milankovtich\index{Milankovitch, M.} es que los veranos frescos en las altas latitudes del norte pueden no derretir toda la nieve acumulada en el casquete polar ártico  durante el invierno y,  como consecuencia, la superficie helada puede crecer desatando una era glacial. El hecho de que el casquete polar ártico  esté rodeado de grandes superficies continentales favorece la expansión del hielo polar. Milankovitch\index{Milankovitch, M.} prestó atención fundamentalmente al paralelo de latitud $b=65^{\circ} N (norte)$ y calculó explícitamente   los efectos de la variación temporal de todos los parámetros astronómicos sobre la insolación de verano en dicha latitud, a fin de encontrar coincidencias entre las eras de hielo y  los períodos de mínima insolación.

La insolación media diaria en un cierto tiempo y lugar geográfico puede calcularse  empleando  la fórmula (\ref{inclinacion}) mediante la siguiente operación:
\begin{equation}
\overline{S_{l}}=S \frac{\int_{-h_{0}}^{\,\,\, h_{0}} cos \, i\,\, dh}{2 \pi},
\label{Insolacion}
\end{equation}
donde la irradiancia $S$ al tope de la atmósfera  es considerada constante durante el día considerado y
la división por $2 \pi$ corresponde a las 24 horas horas del día expresado en radianes, pues 1 hora $= 15^{\circ}$ y 24 horas $=24 \times 15=360^{\circ}=2 \pi$. Teniendo en cuenta la relación (\ref{Tcos}), $\int_{-h_{0}}^{\,\,\, h_{0}} cos \, i=
sen \,b\, sen\, \epsilon_{O}\, sen\, \lambda_{\odot} \, \int_{-h_{0}}^{\,\,\, h_{0}} dh + \, cos\,b \, \sqrt{1-(sen\, \epsilon_{O}\, sen\, \lambda_{\odot})^{2}} \int_{-h_{0}}^{\,\,\, h_{0}} \, cos\, h\,\, dh$. Dado que  $\int_{-h_{0}}^{\,\,\, h_{0}} dh=2\, h_{0}$ y 
 $\int_{-h_{0}}^{\,\,\, h_{0}} \, cos\, h\,\, dh=2 \,sen\, h_{0}$, la expresión (\ref{Insolacion}) se reduce a 
 
\begin{equation}
\overline{S_{l}}=S  \frac{2 \, h_{0} \, \, sen \,b\, sen\, \epsilon_{O} \,  sen\, \lambda_{\odot}  \,  +  \,2 \, sen\, h_{0} \,\, cos\, b \sqrt{1-(sen\, \epsilon_{O}\, sen\, \lambda_{\odot})^{2}} }{2 \pi}, 
\label{Insolacion2}
\end{equation}
 donde $h_{0}$ se obtiene a través de la fórmula (\ref{AngH0}).

 Ahora nuestro propósito es estudiar, en una forma un tanto idealizada,  la evolución de la insolación, de verano,  de la latitud $b=65^{\circ} N$ a lo largo del último millón de años, a fin de ilustrar los principales conceptos de la teoría de  Milankovitch\index{Milankovitch, M.}. Para ello, utilizaremos la fórmula (\ref{Insolacion2}) y fijaremos  $\lambda_{\odot}=120^{\circ}$, que corresponde a un día representativo del verano del norte. 
Dado que $S$ depende de  la conjugación de dos de los variables orbitales seculares, la excentricidad $e$ y la precesión del equinoccio, que estudiaremos debajo, primero consideramos  la relación $\frac{\overline{S_{l}}}{S}$ obtenida de (\ref{Insolacion2}). Ello nos permite estudiar, en forma separada, el comportamiento de la insolación que depende del parámetro orbital restante, la oblicuidad $\epsilon_{O}$, que  es la variable temporal  que falta determinar en la relación ($\frac{\overline{S_{l}}}{S}$). La variación de la oblicuidad con respecto al tiempo $t$, $\epsilon_{O} (t)$, como así también  la de los restantes parámetros orbitales de la Tierra, se obtiene por integraciones numéricas de las perturbaciones que ejercen los principales planetas. La solución numérica para  $\epsilon_{O} (t)$ se la aproxima analíticamente mediante una expansión en series trigonométricas (senos y cosenos). Los extremos superiores de la oscilación de  $\epsilon_{O} (t)$ alcanzan como máximo el valor de $24^{\circ}.5$ y los extremos inferiores de la oscilación alcanzan como mínimo el valor de $22^{\circ}.1$. Es decir, la amplitud de la oscilación no es exactamente constante y tampoco lo es la frecuencia de oscilación, por ello decimos que la variación de  $\epsilon_{O} (t)$ es cuasi-periódica con un periodo de $\approx$ 41000 años. Sin embargo, para nuestro propósito, es suficiente considerar que $\epsilon_{O} (t)$ oscila armónicamente en torno a su valor medio, $\overline{\epsilon_{O}}$, con una amplitud $A_{\epsilon}=\frac{24^{\circ}.5-22^{\circ}.1}{2}=1^{\circ}.2$ y un periodo $T=41000$ años. En consecuencia,

\begin{equation}
\epsilon_{O}(t)= \overline{\epsilon_{O}} - A_{\epsilon} sin ( \frac{2 \pi }{T} t + F_{\epsilon}),
\label{EpsilonO}
\end{equation}
donde $\overline{\epsilon_{O}}=\frac{24^{\circ}.5+22^{\circ}.1}{2}=23^{\circ}.3$ y  la  fase de la onda $F_{\epsilon}=-0.167\, radianes$. Como  referimos a tiempos pasados,  $t$ es negativo. El signo menos que antecede a $A_{\epsilon}$ en (\ref{EpsilonO}) se debe a que $\epsilon_{O}$ está actualmente disminuyendo ($\frac{d\epsilon_{O}(t) }{dt}< 0$ en $t=0$). El valor de la fase $F_{\epsilon}$ se obtuvo con la condición de que $\epsilon_{O}(0)=23^{\circ}.5$.

\begin{figure} 
\includegraphics[scale=0.7]{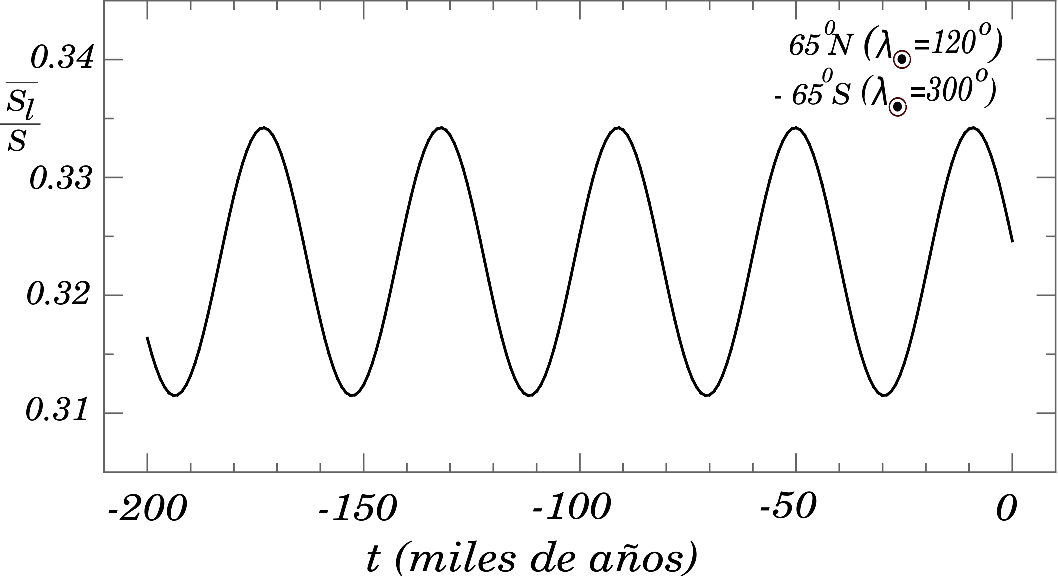} 
\caption{Relación $\frac{\overline{S}_{l}}{S}$ en los últimos  200000 años para dos posiciones geográficas opuestas, $65^{\circ}$ y $-65^{\circ}$, en sus respectivos días de verano, $\lambda_{\odot}=120^{\circ}$ y $\lambda_{\odot}=300^{\circ}$. A diferencia de $\overline{S}_{l}$, la insolación media diaria $S$ sobre un punto de la superficie de la Tierra, o sea  la insolación inmediatamente fuera de la atmósfera, no depende de la latitud geográfica ni de la oblicuidad de la eclíptica. Por lo tanto, las dos  curvas aquí representadas, las cuales resultan coincidentes, reflejan el efecto de la variación de la oblicuidad de la eclíptica.}
\label{OblicuidadPlot}
\end{figure}

Reemplazando en la ecuación (\ref{Insolacion2})  $\epsilon_{O}$ por  $\epsilon_{O}(t)$, de la ecuación (\ref{EpsilonO}), podemos expresar $\frac{\overline{S_{l}}}{S}$  como una función del tiempo. La Fig. \ref{OblicuidadPlot} muestra el resultado de nuestra simulación de $\frac{\overline{S_{l}}}{S}(t)$ para la latitud $b=65^{\circ} N$ en el día de verano determinado por $\lambda_{\odot}=120^{\circ}$. Aquí  solo representamos los últimos 200 mil años, pues el mismo patrón de oscilaciones se repite a lo largo del cuaternario, es decir un ciclo de 41000 años. Si representamos la misma relación $\frac{\overline{S_{l}}}{S}(t)$ pero para la posición geográfica opuesta, latitud del sur  $b=-65^{\circ} S$,  en el día de verano determinado por $\lambda_{\odot}=300^{\circ}$, que es equivalente al del norte, obtenemos la misma curva. Milankovitch\index{Milankovitch, M.} predijo  un ciclo  predominante de glaciaciones  de 41000 años, el cual está de acuerdo con los datos geológicos correspondientes al período entre 0.8 y 2 millones de años atrás. Sin embargo, en los últimos 800 mil años, el período predominante de las glaciaciones es de 100 mil años (``el problema de los 100 mil años''). No obstante, ello no invalida la teoría de Milankovitch\index{Milankovitch, M.}, pues probablemente el ciclo de 41 mil años se encuentre enmascarado  
por otros ciclos que no dependen directamente de los parámetros astronómicos, sino de factores internos de la Tierra. El análisis espectral de los datos geológicos \footnote{Si la historia climática de la Tierra fue  producto de la superposición de varios factores cíclicos, las frecuencias de dichos ciclos no se ponen de manifiesto de una simple inspección  de la secuencia temporal de los eventos geológicos  que caracterizan a las glaciaciones. Una herramienta matemática conocida como transformadas de Fourier\index{Fourier, J.B.}, derivadas de las famosas series de Fourier, permite rescatar las frecuencias predominantes que configuran los datos paleoclimáticos. } relativos a las glaciaciones recurrentes muestra la existencia del ciclo de  41000 años, y de otros posiblemente también relacionados con la variación de los parámetros orbitales de la Tierra.

\begin{figure} 
\includegraphics[scale=0.7]{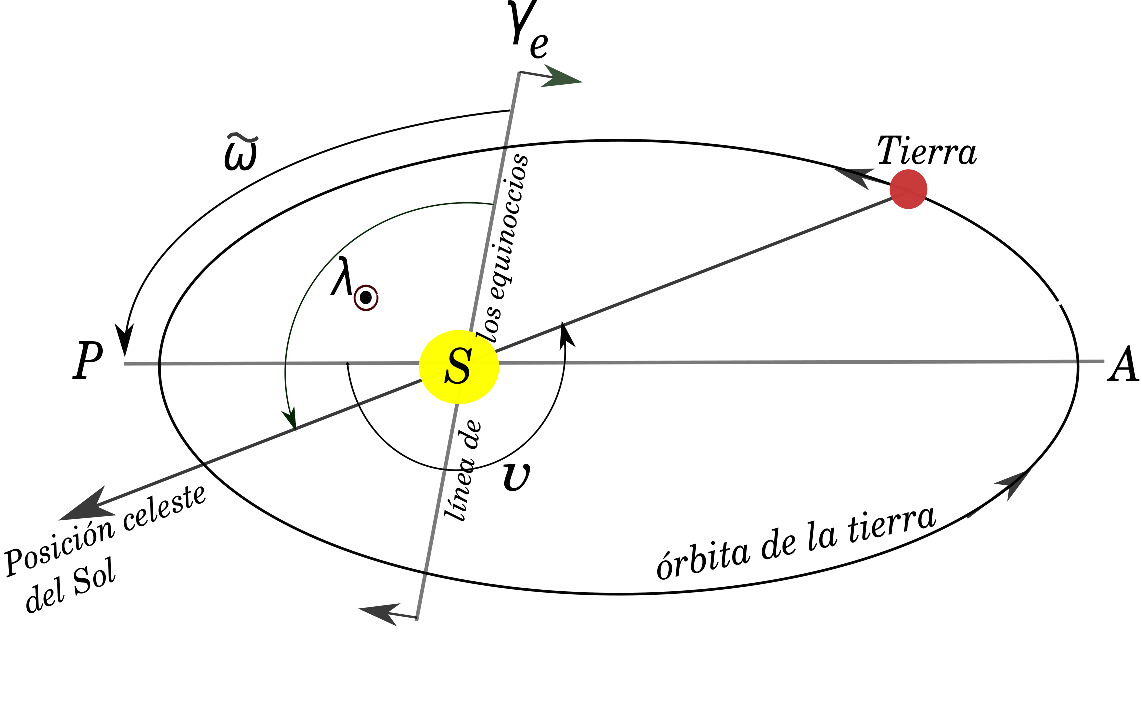} 
\caption{Vista esquemática de la órbita de la Tierra desde el polo norte de la eclíptica. Con fines de visualización, se exagera la excentricidad de la órbita. La letra $P$ indica el perihelio y la letra $A$ el afelio. Este esquema muestra la relación de la anomalía verdadera $v$ con  las longitudes del perihelio, $\tilde{\omega}$, y del Sol, $\lambda_{\odot}$.}
\label{LongPerihelio}
\end{figure}

El equinoccio vernal  $\gamma_{e}$, que  se utiliza como  el origen de referencia de las posiciones celestes (ver Fig. \ref{EsferaCeleste}), se desplaza hacia el oeste y, por lo tanto, las posiciones de los objetos celestes  cambian. Hiparco\index{Hiparco} notó  que  las posiciones de ciertas estrellas brillantes habían variado, comparadas con viejas posiciones, pero que las posiciones relativas de  esas estrellas fijas permanecieron  invariables. Ello condujo a Hiparco\index{Hiparco} al descubrimiento de que el origen del sistema de referencia, el  punto imaginario $\gamma_{e}$,  se desplaza un poco más de un $1^{\circ}$ por siglo \footnote{Cuando en la civilización babilónica se dividió  el ``camino del Sol'' en las 12 constelaciones  del zodiaco, el punto $\gamma_{e}$ se encontraba en Taurus, el Toro. En la época de Hiparco, $\gamma_{e}$ se ubicaba en la constelación de Aries y en la actualidad se encuentra en Piscis.}. Ello implica que las constelaciones se desplazan lentamente con respecto a las estaciones, fenómeno que se conoce como \it{precesión de los equinoccios}\rm.

La causa física de la precesión de la Tierra se  debe a la atracción gravitatoria conjunta de la Luna y del Sol sobre el ensanchamiento ecuatorial de la Tierra, provocado por su rotación. En consecuencia,  las fuerzas gravitatorias que actúan sobre  el abultamiento de la Tierra ejercen un momento de cupla, similar al que se realiza al girar una canilla, que intenta acercar el eje de rotación de la Tierra al eje polar de la eclíptica. La acción de este momento es siempre perpendicular al plano que forman el eje polar, o  de rotación, de la Tierra y el eje polar de la eclíptica. El momento angular de la Tierra es, como todo momento, una magnitud vectorial cuya dirección casi coincide con el eje de rotación  de la Tierra y su magnitud  depende  de la velocidad angular y del momento de inercia de la Tierra. Bajo la acción de la cupla, se modifican la dirección del vector del momento angular y,  por lo tanto,  la dirección del eje de rotación terrestre. Sin embargo,  se conserva constante  la magnitud del vector del momento angular. En consecuencia,   no se modifica la duración del día. El resultado es que  la dirección del eje de rotación terrestre, o eje polar, gira en torno al eje polar de la eclíptica,  describiendo un cono con un ángulo de apertura constante, en principio, e igual a $\epsilon_{O}$. El eje polar completa un giro en torno al polo de la eclíptica cada 26000 años, como así también   la línea de los equinoccios (intersección del plano de la eclíptica con el plano del ecuador celeste). Con un giróscopo  o trompo en rotación, se visualiza vívidamente el fenómeno de la precesión, razón por la cual los libros de física refieren a dichos artefactos para explicar el fenómeno. En este sentido, la Tierra es un giróscopo gigante.

El esquema representado en la  Fig. \ref{LongPerihelio} muestra una vista   de la órbita de la Tierra desde el polo norte de la eclíptica  que permite establecer las relaciones entre  el punto equinoccial $\gamma_{e}$, la longitud del Sol $\lambda_{\odot}$, la longitud del perihelio $\tilde{\omega}$, donde los ángulos de ambas longitudes se miden  desde el punto $\gamma_{e}$ sobre la eclíptica, y la anomalía verdadera $v$ de la Tierra. Debido a la precesión de los equinoccios, el punto   $\gamma_{e}$ se desplaza sobre la órbita de la Tierra hacia el oeste y, en consecuencia,  la línea equinoccial gira con  eje en Sol en sentido horario. El movimiento de precesión de   $\gamma_{e}$  hacia el oeste aumenta la longitud del perihelio $\tilde{\omega}$ con el tiempo $t$. Entonces, podemos escribir 
\begin{equation}
\tilde{\omega}(t)=\tilde{\omega}_{0}+ \frac{360^{\circ}}{T_{p}} t,
\label{LongitudP}
\end{equation}
donde $T_{p}=26000$ años es el período de la precesión de los equinoccios y $\tilde{\omega}_{0}=102^{\circ}.972$ es la longitud actual de $\tilde{\omega}$. Para obtener la fórmula (\ref{LongitudP}), hemos supuesto que el perihelio permanece inmóvil. Si bien no es así, el avance o precesión del perihelio de la Tierra tiene un periodo de 34 millones de años y  puede aquí despreciarse. El avance del perihelio de los planetas es un efecto relativista pequeño, que es un poco mayor  en  la órbita de Mercurio, debido a su mayor cercanía a la masa solar. Los astrónomos de finales del siglo XIX  sabían que la órbita de Mercurio tenía un desplazamiento anómalo de 43’’ por siglo y no pudieron explicarlo en términos de perturbaciones de un planeta mayor. Le Verrier\index{Le Verrier, U.}  fue el primero en notar la pequeña precesión de la órbita de Mercurio. El primer \it{test} \rm de la teoría general de la relatividad \footnote{En la teoría de la relatividad general, a diferencia de la mecánica clásica, la fuerza de gravedad no existe, solo el efecto gravitacional causado por la métrica (o curvatura)  del espacio-tiempo. Este giro Copernicano producido por Einstein\index{Einstein, A.} condujo a descubrimientos de fenómenos  sorprendentes que revolucionaron la física moderna.} lo realizó el propio Einstein\index{Einstein, A.}, quien mostró que su teoría sí podía explicar el desplazamiento del perihelio de Mercurio.

En la Fig. \ref{LongPerihelio}, se ve que se cumple la siguiente relación:
\begin{equation}
v=\lambda_{\odot}-\tilde{\omega}(t) +180^{\circ},
\label{AnomaliaV}
\end{equation}
donde $\tilde{\omega}(t)$ es dada por  la fórmula (\ref{LongitudP}). La anomalía verdadera $v$ expresada por (\ref{AnomaliaV}) proporciona  la posición orbital de la Tierra en función del tiempo $t$.   La longitud $\lambda_{\odot}$ determina la estación y día del año. Por ejemplo, cuando  $\lambda_{\odot}=90^{\circ}$, el día 21 de junio, comienza el verano en el norte y el invierno en el sur (ver, Fig. \ref{Oblicuidad} ). Si fijamos $\lambda_{\odot}$, por ejemplo en $90^{\circ}$ en la fórmula  (\ref{AnomaliaV}), obtenemos que  la posición orbital de la Tierra en el solsticio  no es siempre la misma. Siguiendo con el ejemplo,  actualmente $v=167^{\circ}$ en el solsticio de verano del norte y, por lo tanto, la Tierra se encuentra en la sección II de la órbita (ver la Fig. \ref{MilankovitchAB2}). Si no fuera por la precesión de los equinoccios, el hemisferio norte tendría  eternamente veranos frescos y, de acuerdo con la teoría de Milankovitch\index{Milankovitch, M.}, ello tendería a  extender  el casquete polar ártico. Sin embargo, dentro de 13000 años, los veranos del norte coincidirán con la Tierra ubicada en la sección I. Por ello, para calcular la insolación de un determinado lugar a lo largo del tiempo, debemos combinar los efectos de la precesión de los equinoccios con las variaciones de la excentricidad de la órbita de la Tierra.

A fin de calcular la irradiancia al tope de la atmósfera en función del tiempo, teniendo en cuenta la precesión de los equinoccios y la variación de la excentricidad de la órbita de la Tierra en tiempos geológicos, empleamos la fórmula (\ref{Irradiancia2}). Si reemplazamos en esa fórmula la anomalía verdadera por la relación (\ref{AnomaliaV}), obtenemos

\begin{equation}
  S(t)=S_{m} \frac{\sqrt{1-e(t)^{2}}}{(1-e(t)^{2})^{2}} (1- e(t)\, cos ( \lambda_{\odot}-\tilde{\omega}(t)))^{2},
  \label{Irradiancia3} 
 \end{equation}
donde denotamos la excentricidad $e$ como  una función del tiempo. Simulamos la función $e(t)$ mediante la siguiente expresión:
\begin{equation}
e(t)=\overline{e}-A_{e} sen (\frac{2 \pi}{T_{e}} t+ F_{e}),
\label{FunExc}
\end{equation}
donde $T_{e}=100000$ años es el período de oscilación de $e$, $\overline{e}=0.035$  el valor medio, $A_{e}=0.035$ la amplitud de la oscilación, y  $F_{e}=0.547$ la fase de la oscilación. Note que  la función (\ref{FunExc}) para $t=0$ da $e=0.0168$, que es la excentricidad actual, y que $e$ fluctúa entre 0 y 0.07, de acuerdo con el extremo mínimo y máximo que proveen  los cálculos de la mecánica celeste. 

\begin{figure} 
\includegraphics[scale=0.7]{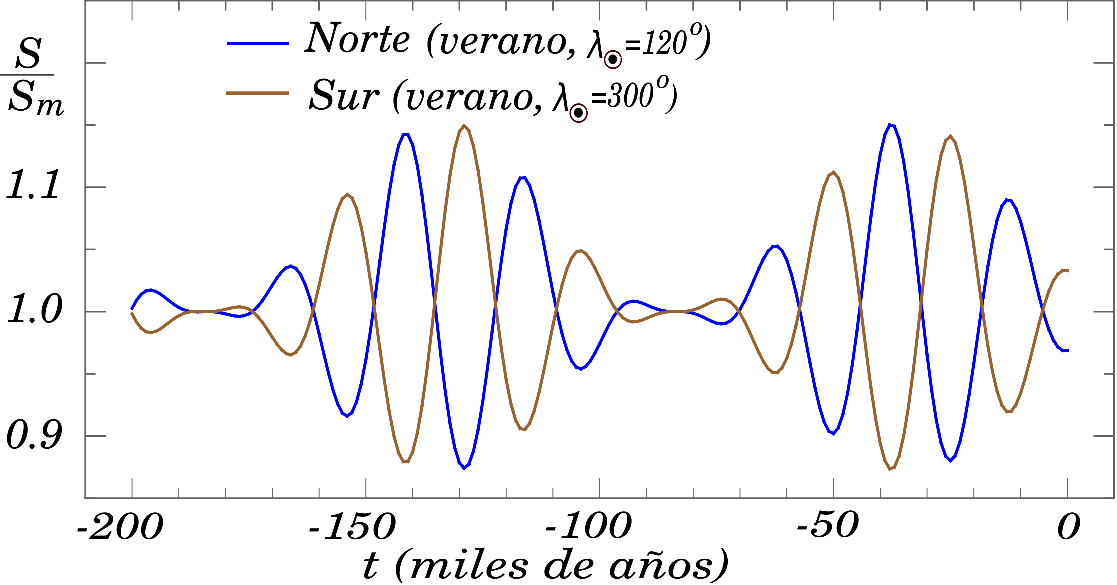} 
\caption{Irradiancia $S$, al tope de la atmósfera, relativa a la irradiancia anual media $S_{m}$ durante los últimos 200000 mil años.  La curva azul corresponde  al hemisferio norte en un día particular de verano, especificado por $\lambda_{\odot}=120^{\circ}$, y la curva marrón corresponde  al hemisferio sur en un día  de verano, equivalente al del norte y especificado por $\lambda_{\odot}=300^{\circ}$.}
\label{contrafase}
\end{figure}

Reemplazando  $e(t)$ y $\tilde{\omega}(t)$ en la ecuación (\ref{Irradiancia3}) por sus expresiones dadas  en (\ref{FunExc}) y (\ref{LongitudP}) respectivamente, podemos representar $\frac{S}{S_{m}}$ en función de tiempos geológicos para un cierto día del verano, tanto del hemisferio Norte como del Sur. Nosotros hemos adoptado como un día representativo del verano del Norte al que corresponde a $\lambda_{\odot}=120^{\circ}$, y  para el verano del Sur $\lambda_{\odot}=300^{\circ}$. Con dichas condiciones, se representa  $\frac{S}{S_{m}} (t)$ en la Fig. \ref{contrafase}, la cual muestra que la amplitud de las oscilaciones debidas a la precesión es  modulada por la amplitud de las oscilaciones de la excentricidad de la órbita de la Tierra. La Fig. \ref{contrafase} también muestra que las variaciones de la irradiación en el verano del Norte están a contrafase  de  las correspondientes del Sur. El mismo patrón de oscilaciones de $\frac{S}{S_{m}} (t)$ se extiende más allá del tiempo  representado en la figura de marras. 

\begin{figure} 
\includegraphics[scale=0.6]{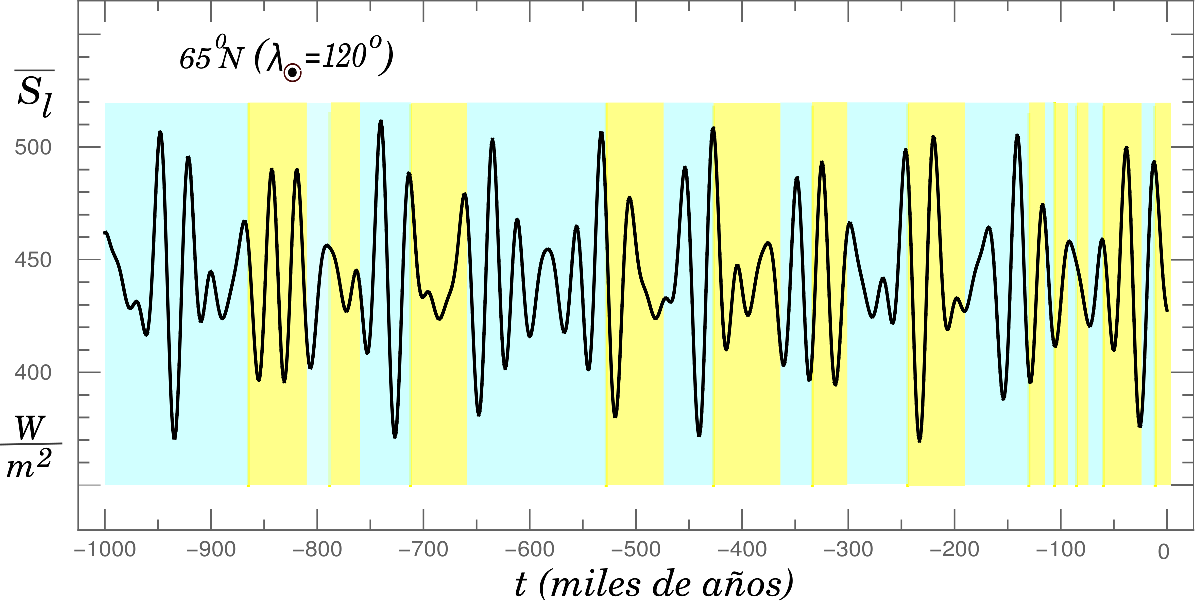} 
\caption{Insolación media $S_{l}$, expresada en $Watts/m^ {2}$, de la posición geográfica $65^{\circ}$ norte  durante el día de verano correspondiente a $\lambda_{\odot}=120^{\circ}$ para un millón de años en el pasado. Los sombreados celestes y amarillos indican los períodos fríos y templados de la Tierra, respectivamente, también  llamados ``estadios isotópicos marinos''  (en inglés, se abrevia con la sigla MIS por marine isotope stages), puestos de manifiesto por la relación isotópica  del oxígeno, $^{18}O$ : $
^{16}O$, de muestras del fondo marino que permite deducir las temperaturas del pasado.}
 \label{ModeloConceptual}
\end{figure}

Ahora estamos en condiciones de determinar la evolución completa de la insolación de un cierto lugar geográfico y, siguiendo el criterio de Milankovitch\index{Milankovitch, M.}, calculamos $\overline{S_{l}}$ en un día de verano para la latitud norte de $65^{\circ}$.  Para ello, empleamos la ecuación (\ref{Insolacion2}) donde reemplazamos $S$ por  la expresión (\ref{Irradiancia3}) de $S(t)$. Además, seguimos adoptando $\lambda_{\odot}=120^{\circ}$ que corresponde un día representativo del verano del norte, y $S_{m}=S_{0}$. Recordemos que la expresión para $\tilde{\omega}(t) $ es dada por (\ref{LongitudP}). El resultado de nuestra simulación se presenta en la Fig. \ref{ModeloConceptual}, y es ilustrativo compararla con la Fig. 5 de la referencia \cite{Laskar}, donde las variaciones de los parámetros orbitales 
son aproximadas por expansiones en series.

Llama la atención, tanto en nuestra simulación como en los cálculos más exactos, que los máximos y mínimos de la insolación $\overline{S_{l}}$ son pulsos muy angostos debidos al efecto conjunto de la precesión de los equinoccios y la variación de la excentricidad orbital y no hay una correlación evidente entre los mínimos con los períodos glaciales  y los máximos con los períodos interglaciales. De hecho, se alternan máximos y mínimos tanto dentro de las  eras glaciales como de las interglaciales. Sin embargo, debemos tener en cuenta que la insolación es un estímulo externo y nos falta determinar la reacción del sistema climático. El sistema climático es uno muy complejo, integrado por cuatro subsistemas acoplados: la atmósfera, la biosfera, los océanos y las masas continentales, y su reacción no es directa ni inmediata. Los climatólogos aún no comprenden  como tan sutiles variaciones de la insolación pudieron desencadenar tan drásticos cambios climáticos.

Se han tomado muestras de hielo en estractos profundos de la capa de hielo Antártica, correspondientes a nevadas del Pleistoceno\index{Pleistoceno} que abarcan los últimos 800000 años. En dichas muestras de hielo, se midió la relación isotópica  del oxígeno, $^{18}O$ : $
^{16}O$, lo cual  permitió determinar las temperaturas ambientales de la Tierra (ver Sección 10.2) a lo largo del período en que ocurrieron las glaciaciones clásicas. También se midió el contenido de $CO_{2}$ de pequeñas burbujas de gas atmosférico atrapadas en las sucesivas capas de hielo que se forman al compactarse la nieve de la superficie. Notablemente, se encontró que las temperaturas y el $CO_{2}$ atmosférico descendieron en consonancia con los períodos glaciales y aumentaron en los períodos interglaciales. Ello indica que el efecto invernadero inverso jugó un papel importante  en la gestación de las eras de hielo del cuaternario. Quizá, algunos pulsos de los mínimos  de insolación fueron amplificados por el efecto de realimentación positiva del aumento del albedo y por el efecto invernadero inverso y, a partir de allí, el sistema climático tomó vida propia, generando una edad de hielo.

La profesora Caroline Lear  y otros científicos  de la Universidad de  Cardiff  \cite{Lear} descubrieron, por medio de un estudio de  la composición química de  micro-fósiles del lecho marino, que la cantidad de  $CO_{2}$ de las profundidades oceánicas aumentaba durante las eras de hielo  de los últimos 800000 años. Ello sugiere que a través del proceso de fotosíntesis una rica flora marina extrajo  $CO_{2}$ de la atmósfera y lo acumuló en las profundidades oceánicas. La disminución de un gas de efecto invernadero como el  $CO_{2}$ provocó un descenso de la temperatura, es decir un efecto invernadero inverso. La profesora Carolina Lear  y su equipo explican  que se produjo una suerte de respiración de los océanos, inhalación y exhalación de $CO_{2}$ con un ciclo de 100000 años, que  podría ser   en parte la solución al ``problema de los 100000 años''. El gran descenso del nivel del mar durante una era de hielo expuso a la atmósfera sedimentos marinos en descomposición que desprendieron gases de efecto invernadero, aumentando la temperatura y dando fin a la edad de hielo y comienzo a un período interglacial.

 El registro de 3 millones de años de la temperatura sobre la superficie terrestre, revelado por  la  relación isotópica  del oxígeno, $^{18}O$ : $^{16}O$, indica, como hemos dicho, variaciones que se repiten con una periodicidad de $\approx 100000$ años durante los últimos 800000 años. Sin embargo, después de una transición, el ciclo de la variación de temperatura mutó a una periodicidad de 41000 años que se mantuvo desde 1 a 3 millones de años atrás (ver Fig. 1 de la referencia \cite{Raymo}). El ciclo de 41000 años coincide con el ciclo de Milankovitch de la variación de la oblicuidad de la eclíptica, pero se pierden los restantes ciclos de Milankovitch. Podemos concebir que a partir de  800000 mil años atrás hacia el presente  hubo cambios en el sistema climático donde el efecto invernadero inverso ocultó los ciclos de Milankovitch, pero entre 1 y 3 millones de años atrás  deberían aparecer los pulsos de la intensidad de la insolación debidos a la precesión de los equinoccios y a la  variación de la excentricidad orbital. Una explicación es que  dicho  ciclo de Milankovitch\index{Milankovitch, M.} no está ausente, solo  que no se refleja en las temperaturas registradas por medio de la  relación isotópica  del oxígeno, $^{18}O$ : $^{16}O$: relación   que representa  la temperatura global de la superficie terrestre. La Fig. \ref{contrafase} muestra que los pulsos de insolación de verano que recibe el norte están en contrafase con aquellos que recibe el sur, con una diferencia en el  tiempo insignificante de solo 6 meses.  Es decir, ambos efectos se cancelan y no influyen sobre la temperatura global. En cambio, la Fig. \ref{OblicuidadPlot} muestra que la periodicidad de la insolación en el verano del norte debida a la  la variación de la oblicuidad de la eclíptica está en fase con la correspondiente al sur y, por lo tanto, ambas se potencian. Ello explicaría por qué  predomina la periodicidad de 41000 años en los registros de temperatura \cite{Raymo}.

Las condiciones climáticas globales durante el Cuaternario\index{Cuaternario} eran propicias para que las sutiles variaciones de la insolación debidas a los ciclos de Milankovitch\index{Milankovitch, M.}  surtieran efecto en  las recurrentes glaciaciones del Cuaternario. En cambio, durante el Período Terciario, las temperaturas medias de la Tierra fueron excepcionalmente altas y, por lo tanto, los ciclos de Milankovitch\index{Milankovitch, M.} no influían sobre el clima. Esta fase climática templada comenzó en el  Período Cretácico\index{Cretácico} y alcanzó la máxima temperatura promedio,  $\approx 29^{\circ}$, entre las épocas del Paleoceno\index{Paleoceno} y el Eoceno\index{Eoceno}, hace $\approx 57.8$ millones de años. Los casquetes polares se derritieron y el nivel de los océanos se elevó notablemente. Desde entonces, la temperatura descendió gradualmente y, a partir de $\approx 36.6$ millones de años atrás, el continente  Antártico comenzó a cubrirse de hielo. Al comienzo del Eoceno\index{Eoceno},  la disposición de los continentes fue muy similar a la actual, sólo que el océano Atlántico no era tan ancho y la India recién comenzaba a colapsar contra el continente Asiático. Recién alrededor de 5 millones de años atrás,  regiones del norte de   América y del norte de  Europa comenzaron a congelarse.

El máximo de temperatura del Paleoceno-Eoceno\index{Eoceno}\index{Paleoceno} fue consecuencia  de un  fuerte efecto invernadero, producido por una atmósfera  que entonces contenía  una muy alta concentración de gases de efecto invernadero, varias veces  mayor que los valores modernos del dióxido de carbono $(CO_{2})$. Ello ocurrió en el contexto de una crisis biológica y geológica originada 10 millones de años antes por la extinciones masivas  de las especies en el límite K-T (Cretácico-Terciario) \footnote{Está bien documentado que esa catástrofe medio ambiental fue provocada por la colisión de un gran asteroide o cometa  contra la Tierra. Pruebas de ello son la  capa de iridio que cubre el estracto geológico del límite K-T y la estructura de impacto  de Chicxulub (México)\index{Chicxulub, cráter}.}. El principal reservorio del  $CO_{2}$ y del metano $(CH_{4})$, otro gas de efecto invernadero aún más poderoso que el  $CO_{2}$, no es  la atmósfera sino el lecho de los océanos. Burbujas de gas metano producidas por la descomposición de microbios marinos que yacen en los sedimentos del fondo oceánico  afloran continuamente a la superficie y alimentan a la atmósfera. En la atmósfera, el metano se oxida rápidamente y se convierte en  $CO_{2}$. Por otra parte, los organismos fotosintéticos   extraen  $CO_{2}$ de la atmósfera. En condiciones de equilibrio, el $CO_{2}$  
ganado debe ser igual al perdido. Sin embargo, en el  Paleoceno-Eoceno\index{Eoceno}\index{Paleoceno}, debieron ocurrir eventos que abruptamente liberaron en la atmósfera enormes cantidades de gases de efecto invernadero. Una posibilidad es que grandes bolsas de gas metano emergieron desde el fondo marino removidas por el retemblor de  movimientos sísmicos.

La extinción masiva K-T, en la cual  los dinosaurios desaparecieron entre muchas otras  especies, marca el inicio de una nueva era geológica, la Era Cenozoica (``vida nueva''), dividida en dos períodos: Terciario\index{Terciario} y Cuaternario\index{Cuaternario}. En el Cenozoico\index{Cenozoico}, la vida desarrolló una sorprendente variedad de plantas y animales, entre ellos los mamíferos, y  todos los rasgos de la Tierra actual tomaron forma. Sin embargo, en dicho período, la evolución de la vida no estuvo exenta de dificultades. En efecto, en el Terciario, tanto en las épocas del Eoceno\index{Eoceno} como del Mioceno\index{Mioceno}, hubo eventos menores de  extinción de especies que muestran evidencias de estar asociados a impactos cósmicos \cite{Glass} \cite{Olano10} \cite{Olano1} \cite{Olano2b}.

\subsection{Las grandes glaciaciones antiguas: la primera gran glaciación y el período Criogénico\label{criogenico1}}

Con la Tierra recién formada y en un medio templado \footnote{Durante la era  Arqueozoica\index{Arqueozoica}, la luminosidad del Sol fue el 75-82 por ciento de su valor actual, según ciertos modelos teóricos. Ello implica una Tierra congelada. Sin embargo, se sabe que en ese entonces el agua se encontraba en estado líquido. A esta contradicción se la conoce como la paradoja del Sol pálido. La solución más aceptada es que el calentamiento por efecto invernadero contrarrestó la menor luminosidad solar.}, arropado por una atmósfera rica en gases de efecto invernadero, las formas de vida unicelulares emergieron rápidamente. La vida no sólo se adapta al medio, sino que también lo modifica. El  intercambio gaseoso de esos seres vivos con la atmósfera primitiva modificó de tal manera la composición química atmosférica que se originaron cambios climáticos y una serie de  glaciaciones globales muy severas durante el Precámbrico\index{Precámbrico} (4500 a 550 millones de años). Las glaciaciones del Cuaternario parecen benignas comparadas con aquellas del Precámbrico\index{Precámbrico}, las cuales estuvieron a punto de interrumpir el curvo evolutivo de la incipiente vida. Depósitos glaciares característicos,  encontrados en rocas antiguas de todos los continente, indican que la Tierra sufrió tempranamente glaciaciones de gran extensión  y duración.
 
 Como sabemos, la vida se inició en los océanos primitivos y los primeros seres fotosintéticos  aeróbicos fueron las cianobacterías\index{cianobacterías}, las cuales  expuestas a la luz solar combinaban moléculas de agua  con moléculas de dióxido de carbono ($CO_{2} $) atmosférico y como consecuencia liberaban moléculas de oxígeno ($O_{2})$  a la atmósfera \footnote{Hay reacciones químicas de origen no biológico, como la fotólisis del agua, que también pueden liberar oxígeno. Sin embargo, se piensa que la principal fuente de oxigeno atmosférico fue la fotosintética.}. El $CO_{2}$ extraído de la atmósfera, no volvía a ella, sino que era  retenido en sedimentos marinos. Se argumenta que al reducirse  el contenido de $CO_{2}$ en la atmósfera, un gas de efecto invernadero, la superficie de la Tierra se fue enfriando hasta el punto de desencadenar un periodo glacial. A medida que se extiende la superficie cubierta de hielo, aumenta el albedo de la Tierra $\gamma$ (ver la fórmula \ref{Arrhenius2} en la Sección \ref{Arrhenius83}) y en consecuencia aumenta aún más el enfriamiento. Es decir, se produce un efecto de  realimentación positiva.

El aumento del oxígeno molecular en la atmósfera condujo a lo que se conoce como  ``La Catástrofe del Oxígeno'' o `` La Gran Oxidación''. Los organismos anaerobios, para los cuales el oxígeno es veneno, debieron refugiarse en   nichos reducidos, libres del aire atmosférico. Con ello disminuyó dramáticamente la población  de dichos  organismos y la provisión del  metano ($CH_{4}$) atmosférico, un gas de efecto invernadero aún más poderoso que el $CO_{2}$, debida en su mayor parte a los organismos  anaerobios. Por otra parte, el metano junto al oxígeno molecular y la radiación ultravioleta  se oxida rápidamente, formando dióxido de carbono ($CO_{2}$) \footnote{La oxidación del metano recién disminuyó cuando la abundancia del oxígeno fue suficiente para formar una capa de ozono ($O_{3}$) en la atmósfera y así bloquear el paso de la radiación ultra violeta.}. La conversión del $CH_{4}$ al $CO_{2}$ en la atmósfera redujo aún más el efecto invernadero y por lo tanto la temperatura global. Ello desencadenó la primera glaciación, llamada Huroniana\index{Huroniana, glaciación}, en la era Paleoproterozoico\index{Paleoproterozoico} entre 2400-2100 millones de años atrás.

Otro factor que puede iniciar y determinar el desarrollo de una era glacial es la distribución de las masas continentales. Alfred Wegener\index{Wegener, A.} y otros pioneros de la teoría de la deriva continental  notaron  que los continentes no están quietos y que  hace $\approx 250$ Ma  todos ellos estaban unidos en un supercontinente, llamado  Pangea\index{Pangea, supercontinente}. Sin embargo, Pangea\index{Pangea, supercontinente} no fue  el supercontinente original, sino que éste resultó de la fusión de fragmentos (continentes) de un supercontinente más antiguo que se había disgregado. Hay evidencias de que, a lo largo de toda la historia de la Tierra,   se formaron  varios supercontinentes sucesivos, donde cada uno se construyó con los ``escombros'' del anterior. El geólogo canadiense John Tuzo Wilson\index{Wilson, J.T.} (1908-1993) propuso  que la ruptura  y reunificación de las placas continentales  es un proceso cíclico, el cual en su honor se llama  Ciclo super-continental de Wilson o simplemente Ciclo de Wilson. En coincidencia con la Glaciación Huroniana\index{Huroniana, glaciación},  existió un supercontinente,  Kenorlandia\index{Kenorlandia, supercontinente}, el cual para algunos científicos fue el primer supercontinente  genuino.

Los casquetes polares de hielo no siempre han existido. Si bien las regiones polares son las que menos energía solar reciben en forma directa, ello  por sí solo  no es suficiente para la formación de capas ``eternas''  de hielo.  El continente Antártico  es uno de los desprendimientos de la Pangea\index{Pangea, supercontinente} y su presente posicionamiento sobre el Polo Sur impide que las corrientes marinas cálidas de las zonas tropicales ingresen a la zona polar y la calienten. Las nevadas se depositan sobre un continente muy frío, formando  sucesivas capas de hielo ``eterno''. El casquete Ártico no está montado sobre un continente, pero está rodeado por masas continentales cercanas que obstaculizan  el ingreso de aguas cálidas. 

Como hemos dicho, la Glaciación Huroniana\index{Huroniana, glaciación}  es  la más antigua  que se tenga conocimiento,  también  la de más larga duración, $\approx 300$ Ma, y se piensa que casi  convirtió  a la Tierra en una bola de nieve. Estás características de la era glacial  Huroniana\index{Huroniana, glaciación} pueden explicarse por la combinación de la disminución del efecto invernadero, como lo hemos descripto arriba, y de la existencia del supercontinente Kenorlandia\index{Kenorlandia, supercontinente}.

Probablemente, parte de Kenorlandia\index{Kenorlandia, supercontinente} cubrió uno de los polos y desde el casquete polar el hielo se extendió a todo el supercontinente. A medida que el hielo se extiende, el albedo del planeta aumenta y por lo tanto la temperatura media de la superficie de la Tierra disminuye aún más (ver la fórmula \ref{Arrhenius2} en la Sección \ref{Arrhenius83}). El vapor de agua atmosférico extraído del océano se precipitó como nieve sobre el supercontinente, formando con el tiempo una gruesa y pesada capa de hielo sobre él.

Por suerte para la supervivencia de la vida, la era de hielo  Huroniana\index{Huroniana, glaciación} tuvo su fin. Ahora nos preguntamos : ¿cuáles fueron las causas que llevaron al fin de la gran glaciación? Una posibilidad es que el gran peso del hielo hundió al supercontinente,  agrietándolo y generando calor en su base.  La gran superficie del supercontinente retuvo también en su base el calor que emana del interior de la Tierra. El calor acumulado debajo del supercontinente formó bolsones de magma que, como  en una olla a presión, se escurrieron a través de las grietas de  la superficie,  empujando  hacia arriba y hacia los costados. Ello desencadenó  la separación de los fragmentos  del supercontinente y  un vulcanismo masivo. Los volcanes eyectan a la atmósfera polvo y gases de efecto invernadero, los cuales al comienzo no influyen mayormente sobre el calentamiento de la superficie de la Tierra, pues tienen efectos contrarios y se neutralizan mutuamente. Terminado el paroxismo volcánico, el polvo volcánico se decantó, transparentando  la atmósfera a la luz solar. De modo que  los gases de efecto  invernadero aportados por las violentas erupciones volcánicas restablecieron  el equilibrio con  temperaturas  que descongelaron al planeta. 

 En el Precámbrico, 1350 Ma después de la glaciación Huroniana y 200 Ma antes de la aparición  de los metazoos macroscópicos, se sucedieron otras glaciaciones extremas. En efecto, al final del Precámbrico en el período Criogénico, entre 750 a 550 millones de años atrás, la Tierra sufrió una serie de glaciaciones intermitentes que la envolvieron de hielo. Durante el Precámbrico Medio se formó el supercontinente llamado  Rodinia, cuya fragmentación  tuvo lugar en coincidencia con las glaciaciones del  Criogénico. Como consecuencia de la actividad tectónica, probablemente  sobrevino una intensa actividad volcánica 
El vulcanismo masivo puede por sí mismo provocar una era de hielo y, si la glaciación fue iniciada por el efecto invernadero inverso, puede al comienzo agravarla, pero a la postre pone fin a la era glacial y a la amenaza sobre la vida  (ver Sección \ref{Huroniana}). Por la ausencia de fósiles de microorganismos sin esqueletos, es difícil evaluar el daño efectuado a la población de bacterias. De todos modos, las bacterias sobrevivieron y probablemente sobrevivirán hasta el día final de la Tierra, que ocurrirá dentro de 5000 millones de años cuando el Sol convertido en una estrella gigante roja engulla a la Tierra.
 
 Otro escenario posible, aunque no excluyente del descripto arriba, es que la Tierra sufrió intensos bombardeos de cometas interestelares durante el período Criogénico. En efecto, en coincidencia con el período Criogénico,  el Sol habría sido capturado por una nube gigante interestelar, progenitora del Sistema Local de gas y estrellas dentro del cual se encuentra actualmente el Sol (ver Sección \ref{TioVivo} y referencias\cite{Olano1}, \cite{Olano2b},   \cite{Olano4}, \cite{Olano7} ,\cite{Olano8}, \cite{Olano9}). En ese contexto, es posible el encuentro penetrante del Sol con  densas nubes moleculares que tendría las siguientes consecuencias: 1) Si un  denso núcleo de una nube molecular envuelve al Sol y a la Tierra,  oscurecería  la luz solar por un período  largo de tiempo (ver Sección \ref{Penetrante1}); 2) Si el Sol penetra lentamente una  densa nube molecular con un alto contenido de cometas interestelares, el Sol puede atrapar gran cantidad de cometas interestelares con potenciales consecuencias catastróficas para la Tierra (ver Secciones \ref{Modelos} y \ref{Reposición}). El choque de un gran cometa sobre la Tierra puede impregnar  la atmósfera de una espesa capa de polvo y sumir a la Tierra  en una larga noche cerrada y extremadamente fría, como así también puede desencadenar  actividad volcánica. 
  
  Otra posible consecuencia de las grandes glaciaciones es que el enfriamiento y la fragmentación de la corteza terrestre pueden  alterar las corrientes convectivas  del interior de la Tierra e inducir la inversión magnética de los polos. En el proceso de inversión magnética,  la intensidad del campo magnético disminuye y en consecuencia se debilita el blindaje contra los rayos cósmicos solares y galácticos (ver Sección 5 de \citep{Olano5}). Por lo tanto, en los períodos glaciales prolongados, el flujo de rayos cósmicos que alcanza la atmósfera y la superficie terrestre  aumentaría.

\subsection{Efectos térmicos del polvo volcánico o cósmico en la atmósfera\label{Modelos}}
El vulcanismo masivo o el choque de un gran asteroide o cometa contra la Tierra pueden elevar a la estratosfera una gran cantidad de  polvo sub-milimétrico, tanto terrestre como  del propio cuerpo cósmico, y formar una espesa envoltura de polvo en torno  a la atmósfera. Un fenómeno similar puede también provocarlo la inyección directa de polvo  por el encuentro de la Tierra con  una densa nube de polvo interestelar o la coma de un gran cometa. El bloqueo de la luz solar por la capa o envoltura de polvo produce un brusco descenso de la temperatura de la superficie terrestre. El enfriamiento de la Tierra puede prolongarse lo suficiente como para originar al menos una pequeña edad de hielo o ``un invierno nuclear'' \footnote{las explosiones atómicas de una conflagración nuclear mundial pueden inyectar, en la alta atmósfera, polvo y vapor de agua y producir un descenso global de las temperaturas.}. A fin de estimar el descenso de las temperaturas, planteamos aquí tres modelos semi-cualitativos, similares al de Arrhenius que tiene la virtud de captar la esencia del fenómeno (ver Sección \ref{Arrhenius83}), en los cuales incorporamos la presencia de granos de polvo en la atmósfera. Estos modelos están basados en consideraciones de balance radiativo y no incluyen los efectos del transporte convectivo del calor.

{\bf Modelo I}. Con $F_{0}$, representamos la energía radiante solar media por unidad de tiempo y de superficie que incide sobre la superficie exterior de la capa de polvo que envuelve la baja atmósfera. Con 
 $\gamma_{1}$, representamos el albedo de dicha capa de polvo que se expresa como la fracción de $F_{0}$ que es reflejada hacia el espacio exterior, esencialmente por una relativamente delgada capa de la superficie exterior de la capa de polvo.

Por lo tanto, la parte  reflejada es  $\gamma_{1} F_{0}$ y suponemos que  el resto $(1-\gamma_{1}) F_{0}$ es totalmente absorbido en la capa de polvo. 
Es decir, suponemos que la profundidad óptica $\tau_{1}$ de la capa de polvo es muy grande ($\tau_{1}\gg 1$)  y por lo tanto la parte de $(1-\gamma_{1})F_{0}$ que consigue atravesar la capa de polvo es $(1-\gamma_{1}) F_{0} \, e^{-\tau_{1}}=0$ (ver por ejemplo la ecuación 52 de \citep{Olano5}). En otras palabras, la luz solar visible que incide sobre el hemisferio iluminado de la Tierra no consigue atravesar  completamente la capa de polvo, y en consecuencia toda la superficie de la Tierra queda en absoluta oscuridad. La energía radiante solar absorbida en la capa por sus granos de polvo es irradiada en longitudes de ondas infrarrojas, invisibles al ojo humano.

La profundidad óptica de una nube de polvo resulta de la siguiente relación:
\begin{equation}
\tau=\pi\, a^{2}\, Q_{ex}\, {\cal N}_{p},
\label{ProfundidadOpt}
\end{equation}
donde $a$ es el radio de las partículas de polvo, $Q_{ex}$ el coeficiente de extinción y ${\cal N}_{p} (=\rho_{p} L)$ la densidad columnar, la cual es dada por la multiplicación de la densidad numérica $\rho_{p}$ de los granos de polvo y del espesor $L$ de la capa de polvo (ver Sección 7.1 de la primera parte de esta monografía \citep{Olano5}). En particular, a la profundidad óptica de la capa de polvo del modelo I, la llamamos $\tau_{1}$. Si $\tau_{1}=20$, el brillo $B_{\odot} $ del Sol es atenuado por el factor $e^{-\tau_{1}}=e^{-20}=2\times  10^{-9}$. En otras palabras, el brillo del Sol que veríamos desde la superficie de la Tierra sería igual a $2 \times  10^{-9} B_{\odot}$, lo cual significa que la capa de polvo  casi tapa  totalmente al Sol y que toda la superficie terrestre queda bajo  una noche cerrada. El coeficiente de extinción $Q_{ex}$ depende de la variable $x=\frac{2 \pi a}{\lambda}$, donde $\lambda$ es la longitud de onda,  y del índice de refracción $m$ del material de las partículas de polvo. Si la capa de polvo está compuesta de partículas de hielo sucio, propio del material cometario, $m=1.33-0.09i$. Suponiendo que las  partículas  de hielo sucio son esferas de radio  $a \sim 0.1 \mu m$ y son iluminadas por longitudes de ondas ópticas $\lambda=\lambda_{opt}\sim 0.6 \mu m $,  $x\approx 1$ y $Q_{ex} (x=1)\approx 0.5$ \cite{Hulst}. Despejando  ${\cal N}_{p}$  en (\ref{ProfundidadOpt}) y reemplazando  los valores de $a=0.1 \mu m$ y $Q_{ex}= 0.5$ y $\tau=\tau_{1}=20$, obtenemos el valor de  ${\cal N}_{p}$. Multiplicando la densidad columnar por la superficie de la Tierra $4 \pi R_{T}^{2}$ obtenemos el número total $N_{t}$  de partículas que contiene la  envoltura de polvo. Si adoptamos $\rho_{iced} \approx 2.35 \, \frac{gr}{cm^{3}}$ como la densidad del hielo sucio,  la masa de cada partícula es $\rho_{iced} \, \frac{4}{3} \pi a^{3}$, la cual  multiplicada  por $N_{t}$ da la masa total $M_{e}$ de la envoltura de polvo:
\begin{equation}
M_{e}=\frac{16\pi}{3}  \frac{\tau_{1}}{Q_{ex}} \rho_{iced} \, a R_{T}^{2}\approx 6 \times 10^{9}\, \rm toneladas.
\label{MasaCapa}
\end{equation}
$M_{e}$ es mucho menor que la masa de un cometa típico y por lo tanto el encuentro de la Tierra con un cometa o parte de él puede crear una envoltura de polvo que oscurezca totalmente al Sol. En efecto, la mayor parte de las masas de los cometas que ingresan al interior del Sistema Solar tienen masas en el rango de $10^{11}-10^{14}$ toneladas \cite{Bailey}.

Ahora nuestro propósito es comparar la profundidad óptica de la capa de polvo en longitudes de ondas visibles $\lambda_{opt}$, profundidad óptica que hemos llamado $\tau_{opt}= \tau_{1}$, con aquella  en ondas infrarrojas $\lambda_{infr}$, que denominaremos $\tau_{infr}$. Las longitudes de ondas largas son menos absorbidas y difractadas por el polvo que las ondas más cortas, por lo tanto $\tau_{infr} \ll \tau_{opt}$. Como en general $x=\frac{2 \pi a}{\lambda_{infr}} \ll 1$, podemos calcular los coeficientes de dispersión $Q_{s}$ y de absorción $Q_{a}$ para $\lambda_{infr}$  mediante las siguientes fórmulas \cite{Hulst}:

\begin{eqnarray}
Q_{s} & = & \frac{8}{3}\, x^{4} \,\Bigg|  \frac{m^{2}-1}{m^{2}+2} \Bigg| ^{2}, \label{Qs} \\
Q_{a} & = &-4\, x\,  Im\Bigg(  \frac{m^{2}-1}{m^{2}+2} \Bigg).
\label{Qa} 
\end{eqnarray}
El índice de refracción $m$ es un número complejo  y para las partículas de hielo sucio $m=1.33-0.09i$, con lo cual $\Big|  \frac{m^{2}-1}{m^{2}+2} \Big| ^{2}=0.04$ y, tal  como lo requiere la fórmula (\ref{Qa}), la parte imaginaria de $\frac{m^{2}-1}{m^{2}+2}=-0.05$.
Las expresiones (\ref{Qs}) y (\ref{Qa}) muestran que $Q_{s}\propto x^{4}$, donde $x \ll 1$, mientras $Q_{a} \propto x$. En consecuencia, $Q_{s} \ll Q_{a}$ y el coeficiente de extinción total es $Q_{ex}=Q_{s}+Q_{a}\approx Q_{a}$. Entonces, mediante  la fórmula (\ref{Qa}) y sus valores numéricos arriba calculados,  $Q_{ex}$ para $\lambda_{infr}$ puede escribirse como sigue;

\begin{equation}
Q_{ex}(\lambda_{infr})= 0.2\,  \frac{2 \pi a}{\lambda_{infr}}.
\label{QeInfr}
\end{equation}

Teniendo en cuenta la definición (\ref{ProfundidadOpt}), el cociente entre la profundidad óptica de la capa de polvo en  longitudes de ondas infrarrojas  y aquella en  longitudes de ondas visibles   es 
\begin{equation}
\frac{\tau_{infr}}{\tau_{opt}}=\frac{Q_{ex}(\lambda_{infr})}{Q_{ex} (\lambda_{opt})}
\label{CocienteTau}
\end{equation}
Hemos visto que $Q_{ex} (\lambda_{opt})=0.5$ con $a=0.1 \mu m$ y 
$\lambda_{opt} \sim 0.6 \times 10^{-3}\,\mu m$ y que $Q_{ex}(\lambda_{infr})$ es dada por la fórmula (\ref{QeInfr}). Haciendo esos reemplazos en (\ref{CocienteTau}) y teniendo en cuenta que hemos definido $\tau_{opt}=\tau_{1}$, obtenemos
\begin{equation}
\tau_{infr}= 0.25 \frac{\tau_{1}}{\lambda_{infr}},
\label{TauInfr}
\end{equation}
donde $\lambda_{infr}$ es expresada en $\mu m$.

\begin{figure}
\includegraphics[scale=0.65]{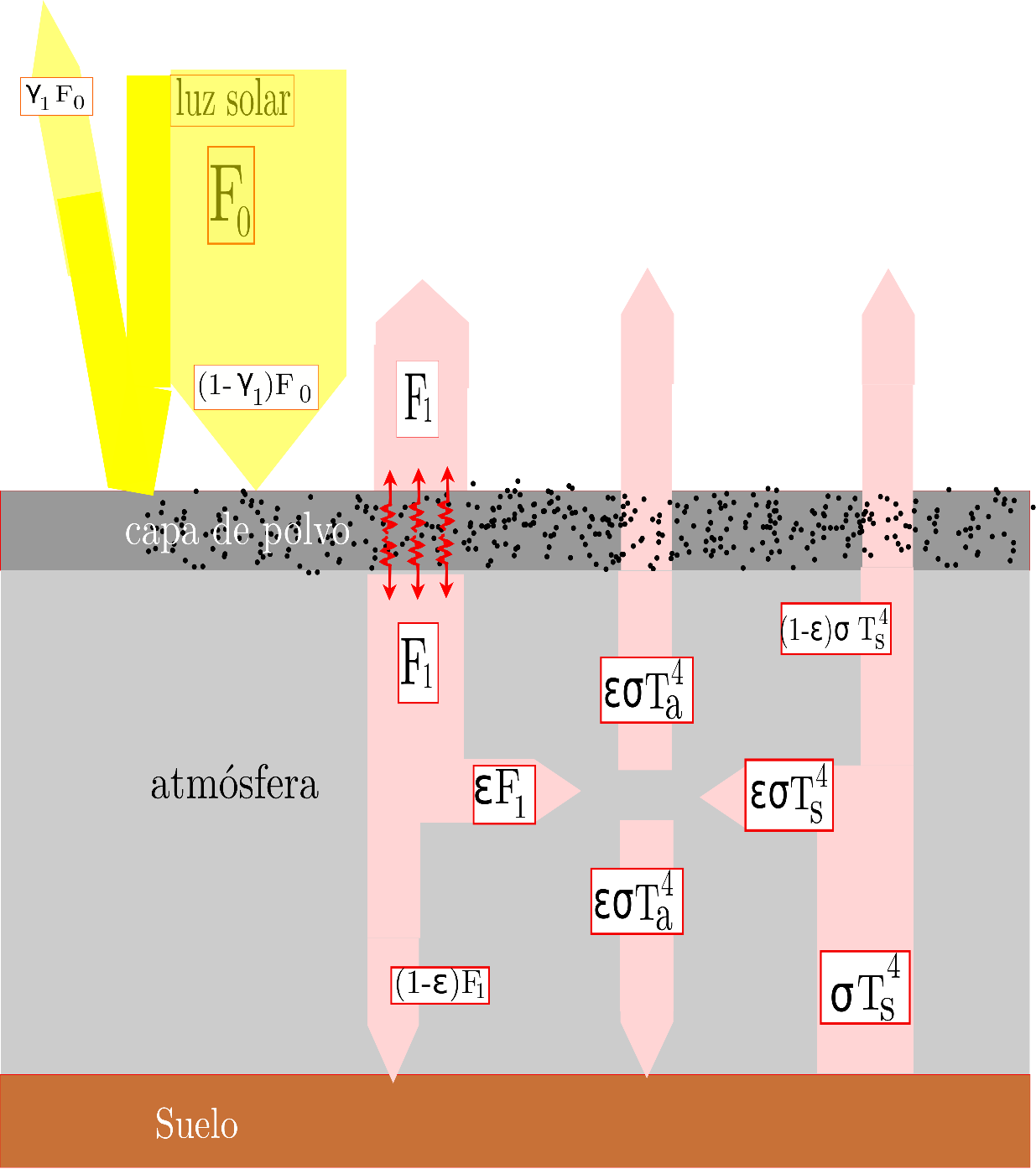} 
\caption{Diagrama de flujo de la radiación infrarroja de la envoltura de polvo y de la atmósfera y superficie de la Tierra. Aquí consideramos que el polvo es transparente a la radiación infrarroja ($\beta=0$).}.
\label{FlujoRadiacion}
\end{figure}

Si el flujo de energía infrarroja emitida por la capa de polvo hacía  el interior de la atmósfera y la superficie terrestre  es $F_{1}$,  el flujo de energía que se emite hacía el espacio exterior es también igual a $F_{1}$. 
Una parte del $F_{1}$ que ingresa a la atmósfera, $\alpha \, F_{1}$,  es absorbido por ella. Dado que el coeficiente de absorción $\alpha$ y el de emisión de un material $\epsilon$ son iguales según la ley de Kirchhoff\index{Kirchhoff, ley de} de radiación, la parte absorbida por la atmósfera puede expresarse  $\epsilon F_{1}$. Suponemos que la parte que no es absorbida por la atmósfera $(1-\epsilon)\, F_{1}$ llega al suelo y es absorbido totalmente por este (ver en la Fig. \ref{FlujoRadiacion} la línea de flujo izquierda). Esto supone que el albedo $\gamma_{i}$ del suelo  para la radiación infrarroja es cero, lo cual no es rigurosamente cierto, sin embargo luego demostraremos que el resultado no es muy sensible al valor de  $\gamma_{i}$ y este puede despreciarse.

La línea de flujo derecha de la Fig. \ref{FlujoRadiacion} representa el flujo de la radiación que emite la superficie de la Tierra hacia la atmósfera y el espacio exterior. La potencia de dicha radiación es $\sigma T_{s}^{4}$, según la ley de Stefan-Boltzmann\index{Stefan-Boltzmann, ley de}, donde $T_{s}$ es la temperatura del suelo. Por lo tanto, la parte del flujo que absorbe la atmósfera es $\epsilon\, \sigma\, T_{s}^{4}$ y la parte que se dirige hacia arriba e ingresa a la capa de polvo es $(1-\epsilon)\, \sigma\, T_{s}^{4}$. El flujo total de energía radiante absorbido por la atmósfera, representado esquemáticamente en el centro de la Fig. \ref{FlujoRadiacion} por la confluencia 
de la  parte correspondiente del flujo entrante y aquel del saliente, es re-irradiado por la atmósfera  con una temperatura $T_{a}$ que debe satisfacer la siguiente igualdad
derivada de  la ley de Stefan-Boltzmann\index{Stefan-Boltzmann, ley de}:
\begin{equation}
2 \epsilon \sigma\, T_{a}^{4}= \epsilon F_{1}+\epsilon\, \sigma\, T_{s}^{4}.
\label{EmisionAtmosfera}
\end{equation}

El factor 2 del primer término de  la ecuación (\ref{EmisionAtmosfera})  se  debe  al hecho que representamos a nuestra atmósfera como una  delgada capa con dos superficies: por cada una de ellas se irradia el flujo $\epsilon \sigma\, T_{a}^{4}$, al cual  debemos multiplicarlo por 2 para obtener  el flujo total irradiado. A través de la parte inferior de la atmósfera, se retorna al suelo el flujo $\epsilon \sigma\, T_{a}^{4}$  (ver Fig. \ref{FlujoRadiacion}). Como hemos dicho, el suelo emite el flujo $\sigma\, T_{s}^{4}$, el cual debe ser igual a la suma de los flujos que el suelo absorbe:

\begin{equation}
\sigma\, T_{s}^{4}= (1-\epsilon)\, F_{1} + \epsilon \sigma\, T_{a}^{4}.
\label{EmisionSuelo}
\end{equation}

La energía que recibe la capa de polvo, $2 F_{1}$, (y que re-emite)  proviene de la luz solar absorbida $(1-\gamma_{1}) F_{0}$ y de parte de la radiación  infrarroja que retorna la atmósfera $\beta \epsilon \sigma T_{a}^{4}$ y la Tierra $\beta (1-\epsilon) \sigma T_{s}^{4}$. El factor $\beta$ tiene en cuenta la fracción de la radiación infrarroja que es absorbida por la capa de polvo y es dado por
\begin{equation}
\beta= (1-e^{-\tau_{infr}}),
\label{Beta}
\end{equation}
 donde  $\tau_{infr}$ se obtiene de (\ref{TauInfr}).
 Por lo tanto,
 
\begin{equation}
 2 F_{1}= (1-\gamma_{1}) F_{0} + \beta \epsilon \sigma T_{a}^{4} + \beta (1-\epsilon) \sigma T_{s}^{4},
\label{F1equation}
\end{equation}
donde $F_{0}= \frac{S_{0}}{4}$, y $S_{0}$ es la constante solar (ver Sección 6.1)
 
Reemplazando la expresión para $F_{1}$, a partir de (\ref{F1equation}), en las ecuaciones (\ref{EmisionAtmosfera}) y (\ref{EmisionSuelo}),  podemos despejar nuestras incógnitas, $T_{s}$ y $T_{a}$, con lo cual obtenemos

\begin{equation}
T_{s}=T_{a}=  \sqrt[4]{\frac{1}{2}(1-\gamma_{1}) \frac{ S_{0} }{4 \sigma}} \,\, \sqrt[4]{ \frac{1}{ (1-\frac{\beta}{2})}}.
\label{InvernaderoInverso}
\end{equation}

Los términos que contenían  $\epsilon$ se cancelaron y por lo tanto $T_{s}$ es  independiente del contenido  atmosférico de  gases de efecto invernadero. Si adoptamos $\gamma_{1}=0.3$, como el albedo actual de la Tierra, $\sqrt[4]{\frac{1}{2}(1-\gamma_{1}) \frac{ S_{0} }{4 \sigma}}=214$  y la fórmula (\ref{InvernaderoInverso}) resulta 
\begin{equation}
T_{s}=T_{a}= 214 \,\, \sqrt[4]{ \frac{1}{ (1-\frac{\beta}{2})}}.
\label{InvernaderoInverso2}
\end{equation} 
Note la similitud de (\ref{InvernaderoInverso}) y (\ref{InvernaderoInverso2}) con las fórmulas de  Arrhenius\index{Arrhenius, S.} (\ref{Arrhenius}) y  (\ref{Arrhenius2}). Aquí $\beta$ juega un papel similar al de  $\epsilon$ en las fórmulas de Arrhenius\index{Arrhenius, S.}. Sin embargo, la temperatura mínima, que resulta de  (\ref{InvernaderoInverso2}) con $\beta=0$, para la temperatura de la superficie de la Tierra es de solo 214 Kelvin (=-59 grados centígrados), mucho menor que la temperatura de 255 K que es la que correspondería a la Tierra sin  atmósfera en la fórmula de Arrhenius. Ello es consecuencia del factor $\frac{1}{2}$ que aparece en el radicando de la primera raíz cuarta de (\ref{InvernaderoInverso}), que la  diferencia de la fórmula de Arrhenius\index{Arrhenius, S.}. Dicho factor proviene del hecho de que la mitad de la energía radiante solar que ingresa a la capa de polvo es re-irradiada en el infrarrojo por el polvo hacia el espacio exterior. Cuando $\beta=0$,  la capa de polvo es transparente  a la radiación infrarroja que emite la atmósfera y el suelo de la Tierra y  el efecto invernadero que puede ejercer en este caso ($\beta=0$) la capa de polvo es nulo.

En la  obtención de la solución (\ref{InvernaderoInverso}), $\beta$  es considerado implícitamente como una constante, escondiendo debajo de la alfombra el hecho de que en realidad $\beta$ depende de $T_{s}$  o de  $T_{a}$, dado que aquí ambas son iguales. En efecto, $\beta$ es una función de $\tau_{infr}$ (ver \ref{Beta}) que depende de $\tau_{1}$ y de $\lambda_{infr}$ (ver \ref{TauInfr}). Consideramos a $\tau_{1}$ como una variable independiente, es decir adoptaremos ciertos valores para ella, pero $\lambda_{infr}$ depende de $T_{s}$. Aplicando la ley de Wien\index{Wien, ley de}, el máximo de la radiación que emite la atmósfera y  la superficie de la Tierra  yace en torno a 
\begin{equation}
\lambda_{infr} =\frac{2897.6}{T_{s}}\, \mu m K .
\label{Wien}
\end{equation}

Entonces, la ecuación (\ref{InvernaderoInverso2}) tiene dos incógnitas, $\beta$ y  $ T_{s}$, las cuales pueden resolverse por un método iterativo. Como $0\leq \beta \leq 1$, supondremos inicialmente que $\beta(1)=0$, con lo cual  mediante (\ref{InvernaderoInverso2}) obtenemos  $T_{s}(1)=214$ K como primera aproximación. Reemplazando  $T_{s}(1)$ en (\ref{Wien}) obtenemos $\lambda_{infr}$, con lo cual y un dado $\tau_{1}$ calculamos $\tau_{infr}$ mediante (\ref{TauInfr}). Reemplazando el valor de $\tau_{infr}$ en (\ref{Beta}) obtenemos la segunda aproximación de $\beta$,  $\beta(2)$, la cual reemplazada en (\ref{InvernaderoInverso2}) nos permite calcular la segunda aproximación de $T_{s}$,  $T_{s}(2)$. Repitiendo el procedimiento $n$ veces, encontramos que $T_{s}(n)$ y $\beta(n)$ convergen a la solución (ver la Tabla \ref{SolucionesIterativas}).

\begin{table*}
\centering
 \begin{minipage}{140mm}
  \caption{Soluciones de la ecuación (\ref{InvernaderoInverso2}) por el método iterativo}
\centering
\begin{tabular}{cccccc} \hline \hline
$\tau_{1}$      & $T_{s} (K)$     & $\beta$  &  $\lambda_{infr} (\mu m)$  & $\tau_{infr}$ \\ \hline\hline
5     & 216.5    & 0.09          &  13.38   &   0.09   \\ \hline
10   & 218.0    & 0.17          &  13.23    & 0.19    \\ \hline
15   & 221.3 & 0.25          & 13.09      & 0.29  \\ \hline
20    & 223.6     & 0.32          & 12.96   & 0.38   \\  \hline \hline
\end{tabular}
\label{SolucionesIterativas}
\end{minipage}
\end{table*}

En condiciones de equilibrio,  la energía solar  que entra al sistema de la capa de polvo, atmósfera y suelo, $(1-\gamma_{1}) \frac{S_{0}}{4}$, debe ser igual a la energía total que sale de la capa de polvo y escapa al espacio exterior (ver Fig. \ref{FlujoRadiacion}):
\begin{equation}
(1-\beta)(1-\epsilon) \sigma T_{s}^{4} + (1-\beta) \epsilon \sigma T_{a}^{4} + F_{1}.
\label{BalanceEnergetico}
\end{equation}
 Reemplazando  en la expresión (\ref{BalanceEnergetico}), $T_{s}= T_{a}$ por la solución (\ref{InvernaderoInverso}) y $F_{1}$ derivado de (\ref{F1equation}), demostramos que (\ref{BalanceEnergetico}) es igual $(1-\gamma_{1}) \frac{S_{0}}{4}$.  Por lo tanto, el balance energético se satisface.

Lo interesante de la fórmula (\ref{InvernaderoInverso}) es que ésta no depende de $\epsilon$ ni de $\gamma_{i}$. La no dependencia de  $\gamma_{i}$ es obvia puesto que la hemos despreciado  por hipótesis. Sin embargo, la no dependencia de  $\epsilon$ es una consecuencia. Dado que el flujo infrarrojo que incide sobre el suelo es $(1-\epsilon) F_{1}$, el flujo reflejado es $\gamma_{i} (1-\epsilon) F_{1}$, y el flujo que absorbe la Tierra  $(1-\gamma_{i}) (1-\epsilon) F_{1}$. Una parte del flujo reflejado, $\epsilon\gamma_{i} (1-\epsilon) F_{1}$,  es absorbido por la atmósfera y  una parte del  resto, $\beta \gamma_{i} (1-\epsilon)^{2} F_{1}$, es absorbida por  la capa de polvo y la otra parte escapa al espacio exterior. Para ver si  es legitimo despreciar  $\gamma_{i}$, escribiremos en forma completa las ecuaciones equivalentes a (\ref{EmisionAtmosfera}) y (\ref{EmisionSuelo}): 

\begin{eqnarray}
2 \epsilon \sigma\, T_{a}^{4} & = & \epsilon F_{1}+   \epsilon \gamma_{i} (1-\epsilon) F_{1} +  
\epsilon \sigma\, T_{s}^{4},  \nonumber\\
\sigma\, T_{s}^{4} & = & (1-\gamma_{i}) (1-\epsilon)\, F_{1} + \epsilon \sigma\, T_{a}^{4},
\label{EmisionSueloAtmosfera}
\end{eqnarray}
donde ahora, $F_{1}$ se deriva  de una ecuación equivalente a (\ref{F1equation}), a saber:
\begin{equation}
 2 F_{1}= (1-\gamma_{1}) F_{0} + \beta \epsilon \sigma T_{a}^{4} + \beta (1-\epsilon) \sigma T_{s}^{4} + \gamma_{i} \beta (1-\epsilon)^{2} F_{1}.
\label{F1equation2}
\end{equation}
Note que (\ref{F1equation2}) solo difiere de (\ref{F1equation}) en el último término.
Reemplazando en  (\ref{EmisionSueloAtmosfera}) la  expresión de $F_{1}$, despejada de (\ref{F1equation2}), obtenemos un sistema de ecuaciones que depende solo de $T_{s}$ y $T_{a}$, cuya solución es
\begin{eqnarray}
T_{a} & = &  \sqrt[4]{\frac{1}{2}(1-\gamma_{1}) \frac{ S_{0} }{4 \sigma}} \,\, \sqrt[4]{ \frac{1}{ (1-\frac{\beta}{2})}}, \nonumber\\
T_{s} & = & f \sqrt[4]{\frac{1}{2}(1-\gamma_{1}) \frac{ S_{0} }{4 \sigma}} \,\, \sqrt[4]{ \frac{1}{ (1-\frac{\beta}{2})}},
\label{InvernaderoInverso3}
\end{eqnarray}
donde $f=\sqrt[4]{1-\gamma_{i}(1-\epsilon)}$. El factor $f$  varía entre $0.95$ y 1 para $ 0\leq \gamma_{i} \leq 0.3$ y  $0.4 \leq \epsilon \leq 1$. Por lo tanto, $f\approx 1$ para un rango amplio de variación de $\gamma_{i}$ y $\epsilon$ que incluye estados de glaciación global y en consecuencia la fórmula (\ref{InvernaderoInverso}) y el esquema de la Fig. \ref{FlujoRadiacion} son  una buena aproximación.

Hasta aquí hemos considerado que $\tau_{1} \gg 1$, ahora estudiaremos el caso más general que incluye profundidades ópticas delgadas $\tau_{1} \ll 1$. El albedo de la capa de polvo, $\gamma_{1}$,  se debe a la luz solar que dispersan las partículas de polvo de una relativamente delgada capa, sobre cuya superficie externa inciden los rayos solares. Si $\tau_{s}$ es la profundidad óptica  con la cual la luz solar es dispersada por el polvo de dicha delgada capa, la parte del flujo que no fue dispersado es $F_{0} e^{-\tau_{s}}$ y por lo tanto la parte dispersada es $(1- e^{-\tau_{s}}) F_{0}$. Como la capa donde se produce esta dispersión de la luz visible es considera delgada, una mitad de la luz dispersada escapa al espacio. Con lo cual el albedo es dado por
\begin{equation}
\gamma_{1}= \frac{1}{2} (1- e^{-\tau_{s}}).
\label{Albedo}
\end{equation}

Si $\gamma_{1}=0.3$ es  el máximo valor que puede tomar el albedo, entonces de la ecuación (\ref{Albedo}) derivamos que $\tau_{s}\leq 1$. Ello significa que la luz dispersada que escapa al espacio exterior proviene de profundidades  $\tau_{s}$ menores  a uno. Dado que $\frac{\tau_{s}}{\tau_{1}}=\frac{Q_{s}}{Q_{ex}}$, donde   $Q_{s}\approx 0.1$ y    $Q_{ex}\approx 0.5$. Por lo tanto, $\tau_{s}=0.2 \tau_{1}$, y mediante la ecuación (\ref{Albedo}), podemos expresar el albedo en función de $\tau_{1}$ del
siguiente modo:
\begin{eqnarray}
\gamma_{1} & = &  \frac{1}{2} (1- e^{-0.2\, \tau_{1}}) \,\,\,\, \rm  para \, \tau_{1} < 5, \nonumber\\
\gamma_{1} & = & 0.3  \, \,\,  \,\, \,\,\,\, \,\, \,\,  \,\, \,\, \,\,\,\, \,\, \,\,  \,\, \,\, \,\,\rm \,  para   \,\, \tau_{1} \geq 5.
\label{gamma1}
\end{eqnarray}
     
 En el caso general el sistema de ecuaciones que debemos resolver es:
 \begin{eqnarray}
 \sigma T_{s}^{4} & = & (1-\epsilon) F_{1} + \epsilon \sigma T_{a}^{4}+ (1-\gamma)(1-\gamma_{1}) F_{0} e^{-\tau_{1}}),  \nonumber\\
 2 \epsilon \sigma T_{a}^{4} & = & \epsilon F_{1} + \epsilon \sigma T_{s}^{4}, 
 \label{SistemaG}
 \end{eqnarray}
 donde ahora $F_{1}$ se obtiene de 
 \begin{eqnarray}
 2 F_{1} & = & (1-\gamma_{1}) F_{0} (1-e^{-\tau_{1}}) + \beta \epsilon \sigma T_{a}^{4} + \beta (1-\epsilon) \sigma T_{s}^{4} + \nonumber\\
         &  & \gamma (1-\gamma_{1}) F_{0} e^{-\tau_{1}} (1-e^{-\tau_{1}}).
\label{F1G}
\end{eqnarray}
 La solución del sistema (\ref{SistemaG}) para $T_{s}$ es entonces
 
 \begin{equation}
 T_{s} =\sqrt[4]{\frac{(1-\gamma_{1}) F_{0}}{4 \sigma}} \sqrt[4]{\frac{2(1-\epsilon/2) (1-\gamma e^{-2 \tau_{1}}) + (1-\gamma) (2 + (1-\beta)\epsilon) e^{-\tau_{1}} }{(1-\beta/2)(1-\epsilon/2)}}.
 \label{SolG}
 \end{equation}
 
En una atmósfera sin polvo,  $\tau_{1}=0$ y por lo tanto  $\gamma_{1}=0$ y $\beta=0$, con lo cual  la fórmula (\ref{SolG})   coincide con la fórmula de  Arrhenius (\ref{Arrhenius2}). En cambio si  $\tau_{1}  \gg 1$, (\ref{SolG}) se convierte en (\ref{InvernaderoInverso}). 
 \begin{figure}
\includegraphics[scale=0.65]{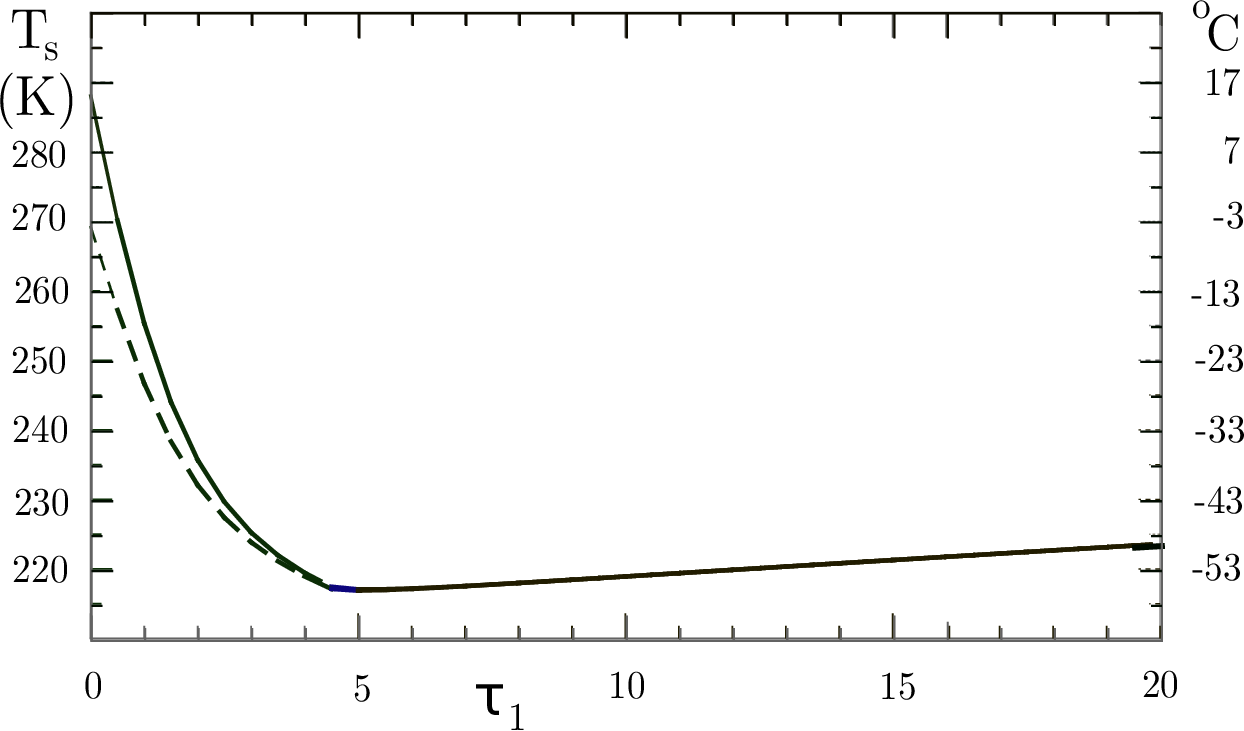} 
\caption{Modelo I: se representa la temperatura de la superficie terrestre ($T_{s}$)  en función de $\tau_{1}$ con $\epsilon=0.78 $ (línea llena) y con $\epsilon=0.4$ (línea a trazos). En la ordenada izquierda, $T_{s}$ es expresada en grados Kelvin y en la derecha en grados centígrados.} 
\label{Modelo1}
\end{figure}
 En la Fig. \ref{Modelo1}, representamos $T_{s}$ en función de $\tau_{1}$, usando la expresión (\ref{SolG}) con $\gamma_{1}$ dada por (\ref{gamma1}). Ambas  curvas de la  Fig. (\ref{Modelo1}) se calcularon con $\gamma=0.3$, pero con diferentes valores para $\epsilon$,  una con $\epsilon=0.78$ y la otra con $\epsilon=0.4$. Sólo en el rango $\tau_{1}<5$, las curvas son muy sensibles a los valores adoptados para $\gamma$ y  $\epsilon$, en cambio  ambas curvas coinciden en el resto ($\tau_{1}\geq 5$). Dicha figura muestra que  la superficie de la Tierra alcanza una temperatura mínima de  217 grados Kelvin o -56 grados centígrados, en $\tau_{1} \approx 5$.   Entre $\tau_{1}=5$ y $\tau_{1}=20$, $T_{s}$ aumenta levemente pero siempre permanece debajo de  -48 grados centígrados. Vemos que la fórmula  (\ref{InvernaderoInverso2}) es una buena aproximación en el rango de $\tau_{1}$ de 5 a 20 y que a los efectos prácticos es suficiente considerar $\beta=0$. Según la fórmula (\ref{MasaCapa}), una envoltura de polvo con  $\tau_{1}=5$ tiene una masa de $1.5 \times 10^{9}$ toneladas. Por lo tanto, una capa de polvo que rodee toda la atmósfera con una masa de  $1.5 \times 10^{9}$ a $ 6 \times 10^{9}$ toneladas puede causar un enfriamiento de la superficie total del  planeta a temperaturas polares e iniciar así una edad de hielo.  Si el contenido de agua de toda la atmósfera, que  es del orden de  13600 km$^{3}$, se precipitara como nieve,  su volumen aumentaría en un factor  $\approx$5, teniendo en cuenta la diferencia de densidad entre el agua y la nieve. Luego, el volumen total de nieve sería de 70000 km$^{3}$, el cual repartido sobre la toda superficie terrestre arroja una capa de nieve de 50 centímetros de espesor. Además, los lagos y ríos de todo el mundo se congelarían en muy poco tiempo.

{\bf Modelo II}. A diferencia del modelo anterior, aquí suponemos que los granos de polvo están distribuidos en toda la atmósfera por debajo de la capa de polvo del modelo I . Con el modelo II, contemplamos una etapa posterior a la del escenario del modelo I. Los granos de polvo suspendidos al principio en la alta atmósfera (mesósfera y estratósfera) van decantando lentamente hacia los estratos atmosféricos  inferiores. Supondremos que la capa de polvo esparcida en toda la atmósfera inferior tiene una profundidad óptica $\tau_{2}$ y un albedo  $\gamma_{2}$. Por lo tanto, $\gamma_{2} F_{0}$ es la parte del flujo de luz solar reflejado hacia el espacio exterior y  $(1-\gamma_{2})  F_{0}$ es la parte que ingresa a la atmósfera. 
Sólo  llega al suelo el flujo $(1-\gamma_{2}) F_{0} e^{-\tau_{2}}$ pues el resto, $(1-\gamma_{2}) F_{0} (1- e^{-\tau_{2}})$, fue  absorbido por el polvo  atmosférico. Debido al albedo  $\gamma$ de la Tierra,   la superficie terrestre refleja  el flujo $\gamma F_{0}(1-\gamma_{2}) e^{-\tau_{2}}$ y el resto del flujo incidente,   $(1-\gamma) (1-\gamma_{2})F_{0} e^{-\tau_{2}}$, es absorbido por el suelo. Sólo una parte del flujo reflejado, $\gamma (1-\gamma_{2}) F_{0} e^{-2\tau_{2}}$, retorna al espacio exterior y el resto, $\gamma (1-\gamma_{2}) F_{0} e^{-\tau_{2}} (1-e^{-\tau_{2}})$, es absorbido  por el polvo atmosférico  en el camino  de vuelta.

La absorción  total del flujo de luz visible por el polvo atmosférico es   $(1-\gamma_{2}) F_{0} (1- e^{-\tau_{2}})+ \gamma (1-\gamma_{2}) F_{0} e^{-\tau_{2}} (1-e^{-\tau_{2}})= (1-\gamma_{2}) F_{0} (1-e^{-\tau_{2}}) (1+  \gamma e^{-\tau_{2}})$, el cual se reirradia  en ondas infrarrojas. Una parte de  dicha radiación infrarroja, $\epsilon (1-\gamma_{2}) F_{0} (1-e^{-\tau_{2}}) (1+  \gamma e^{-\tau_{2}})$, es absorbida por la atmósfera y suponemos que la mitad del resto, $\frac{1}{2}(1-\epsilon) (1-\gamma_{2}) F_{0} (1-e^{-\tau_{2}}) (1+  \gamma e^{-\tau_{2}})$, escapa al espacio exterior mientras que la otra mitad es absorbida por el suelo.

El flujo que emite, en ondas infrarrojas, la superficie de la Tierra a la temperatura $T_{s}$ es  $\sigma T_{s}^{4}$,  del cual la fracción $\epsilon\sigma T_{s}^{4}$ es absorbida por la atmósfera y el resto, $(1-\epsilon) \sigma T_{s}^{4}$,  escapa al espacio exterior. Como, el flujo infrarrojo que recibe  la atmósfera debe ser igual  al flujo infrarrojo que ésta emite:
\begin{equation}
\epsilon\sigma T_{s}^{4}+ \epsilon (1-\gamma_{2}) F_{0} (1-e^{-\tau_{2}}) (1+  \gamma e^{-\tau_{2}})=2 \epsilon \sigma T_{a}^{4},
\label{Modelo2A}
\end{equation}
donde $T_{a}$ es la temperatura de la atmósfera.
El flujo en ondas ópticas e infrarrojas que recibe  la superficie de la Tierra debe ser igual al flujo en ondas infrarrojas que la superficie emite, es decir

\begin{equation}
(1-\gamma)(1-\gamma_{2}) F_{0} e^{-\tau_{2}} + \frac{1}{2}(1-\epsilon) (1-\gamma_{2}) F_{0} (1-e^{-\tau_{2}}) (1+  \gamma e^{-\tau_{2}})+ \epsilon \sigma T_{a}^{4}= \sigma T_{s}^{4}.
\label{Modelo2B}
\end{equation}
 Resolviendo las ecuaciones (\ref{Modelo2A}) y (\ref{Modelo2B}) para $T_{s}$ y teniendo en cuenta que $F_{0}=\frac{S_{0}}{4}$,
 
\begin{equation}
 T_{s}= \sqrt[4]{\frac{(1-\gamma) (1-\gamma_{2}) S_{0} e^{-\tau_{2}}}{4 \sigma (1-\frac{\epsilon}{2})}+\frac{1}{2}  \frac{(1-\gamma_{2}) S_{0} (1-e^{-\tau_{2}})(1+\gamma e^{-\tau_{2}})}{4 \sigma(1-\frac{\epsilon}{2}) } }.
 \label{Tsup}
\end{equation} 
\begin{figure}
\includegraphics[scale=0.7]{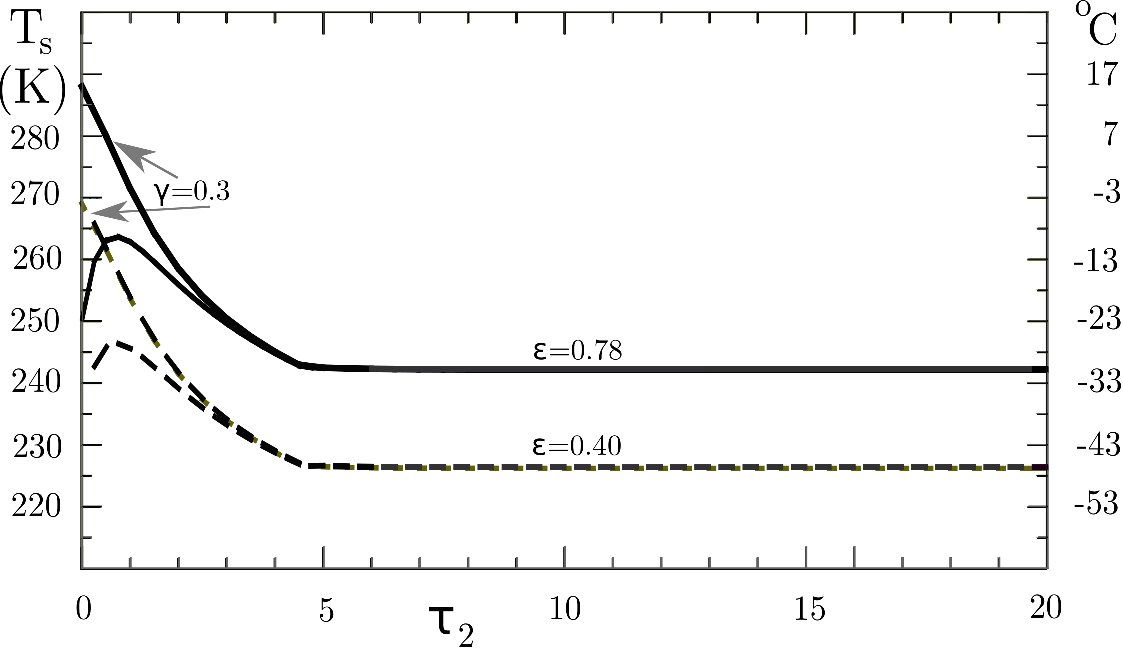} 
\caption{Modelo II: se representa $T_{s}$ en función de $\tau_{2}$ para diferentes valores de $\epsilon$ y $\gamma$, los cuales son  indicados sobre las curvas. Las curvas de línea llena  se calcularon con el mismo $\epsilon(=0.78)$, pero una con $\gamma=0.3$ y la otra con  $\gamma=0.6$. Ambas curvas solo difieren para  $\tau_{2}<5$.  Las curvas  a trazos   se calcularon con el mismo $\epsilon(=0.4)$, pero una con $\gamma=0.3$ y la otra con  $\gamma=0.6$. Ambas curvas solo difieren para  $\tau_{2}<5$.  En la ordenada izquierda, $T_{s}$ es expresada en grados Kelvin y en la derecha en grados centígrados.}
\label{Modelo2}
\end{figure}
Note que cuando $\tau_{2}=0$, es decir la atmósfera no contiene polvo, la fórmula (\ref{Tsup}) se convierte en la de Arrhenius (\ref{Arrhenius2}). Si $\tau_{2} \gg 0$, la fórmula (\ref{Tsup}) se convierte en 
\begin{equation}
 T_{s}= \sqrt[4]{\frac{1}{2}  (1-\gamma_{2}) \frac{S_{0}}{4 \sigma}} \sqrt[4]{\frac{1}{(1-\frac{\epsilon}{2}) } },
 \label{Tsup2}
\end{equation} 
con lo cual sólo depende de $\epsilon$ y no de $\gamma$. Note la semejanza de  (\ref{Tsup2}) con (\ref{InvernaderoInverso}). En la Fig. \ref{Modelo2},  representamos  $T_{s}$ de la fórmula (\ref{Tsup}) en función de $\tau_{2}$  para dos diferentes valores de $\epsilon$ (0.4 y 0.78) y de $\gamma$ (0.3 y 0.6). Para el cálculo de $\gamma_{2}$, utilizamos las relaciones (\ref{gamma1}) donde reemplazamos $\tau_{1}$ por $\tau_{2}$.   La espesa envoltura de polvo suspendida en la alta atmósfera (modelo I) enfría la superficie de la Tierra a la temperatura media de $220\,\rm K \, (-53\,\, ^0 C) $ (ver Fig. \ref{Modelo1}). Cuando dicha envoltura de polvo decanta completamente hacía la atmósfera inferior, $ T_{s}=242\,\rm K (-31\,\, ^0 C)$, según la curva de la   Fig. \ref{Modelo2} correspondiente a $\epsilon=0.78$, valor del coeficiente de efecto invernadero actual de la  Tierra. Para el caso de $\epsilon=0.4$,  $ T_{s}=226\,\rm  K \, (-47\,\, ^0 C)$. Es decir, aunque la temperatura media de la superficie terrestre se eleva, aún se mantienen temperaturas extremadamente bajas.

{\bf Modelo III}. Este es un modelo de dos capas de polvo, uno en el alta  atmósfera (capa I) y el otra debajo (capa II), es decir es una combinación del modelo I y II. Supondremos de forma más  general que las profundidades ópticas de la capa superior $\tau_{1}$ e inferior  $\tau_{2}$ pueden tener cualquier valor. Siguiendo similares  argumentos a los  empleados en el modelo I y II, obtenemos,

\begin{eqnarray}
T_{s} & = & \Big[\frac{1}{2}\,\, \frac{S_{0}}{4 \, \sigma (1-\frac{\epsilon}{2})} \Big((1-\gamma_{3}) (1-\frac{\epsilon}{2}) + (1-\gamma) (1-\gamma_{3}) e^{-(\tau1+\tau2)}  -  \nonumber\\ 
      &   & \gamma  (1-\gamma_{3}) (1-\frac{\epsilon}{2}) e^{-2 (\tau1+\tau2)} + (1-\gamma_{3}) \frac{\epsilon}{2} e^{-(\tau1)} - \nonumber\\
      &   &  \gamma (1-\gamma_{3}) \frac{\epsilon}{2} e^{-(\tau1+ 2 \tau2)} \Big)\Big]^{1/4}, 
\label{ModeloIII}
\end{eqnarray}
donde  $\gamma_{3}$ es el albedo combinado de las dos capas de polvo, el cual se obtiene  utilizando  las relaciones (\ref{gamma1}) con  $\tau_{1}$ reemplazado por $\tau_{1}+\tau_{2}$.

Con las condiciones del Modelo I, $\tau_{1} \gg 0$ y  $\tau_{2}=0$, la ecuación  (\ref{ModeloIII}) se convierte  en la fórmula (\ref{InvernaderoInverso}) del Modelo I y con las condiciones del Modelo II,  $\tau_{1}=0$, la ecuación  (\ref{ModeloIII}) se convierte  en la fórmula (\ref{Tsup}) del Modelo II. Para una atmósfera sin polvo, $\tau_{1}=0$ y  $\tau_{2}=0$, la fórmula (\ref{ModeloIII}) coincide con aquella de  Arrhenius (\ref{Arrhenius2}).

 \subsection{Congelamiento de los mares y océanos \label{Oceano}}
Mediante el planteo de  condiciones de equilibrio  energético radiativo, hemos obtenido en la sección anterior que la temperatura de la superficie terrestre desciende a temperaturas muy bajas cuando una envoltura de polvo cubre la atmósfera. En el suelo sólido, el balance energético  se establece en unos pocos días, pero no ocurre lo mismo en los océanos donde la inercia térmica de la gran masa de  agua de los  océanos es muy grande. Por lo tanto, las fórmulas obtenidas en la sección anterior son sólo aplicables  cuando el balance energético se establece en toda la superficie de la Tierra.

Debido al alto calor especifico del agua salada ($c_{m}=$ 4200 J/(kg K)), los océanos son potentes acumuladores del calor solar, y en consecuencia cuesta mucho congelarlos. 
A profundidades de 1000 metros o mayores, el agua de los océanos se mantiene  a la temperatura de 4ºC, independientemente  de la posición geográfica y de  la temperatura del agua superficial.  Ello se explica porque el agua tiene su máxima densidad  a la temperatura de 4ºC  y porque el  agua más densa es la que se hunde.

Ahora procedemos a estimar el tiempo que demanda  el congelamiento de la superficie de los océanos, bajo las condiciones de extremo frío y oscuridad que produce la envoltura de polvo que cubre la atmósfera.  Para ello, adoptamos como ejemplo un modelo simple  del perfil térmico \it{actual} \rm de los océanos, a saber;  a) entre  la superficie  (profundidad $h_{0}=0$)   y la profundidad $h_{M}=1000$ metros, la temperatura media $T_{0}$ del agua es igual 10ºC, y b) para profundidades mayores a 1000 metros, la temperatura del agua  se mantiene a valores cercanos a 4ºC. En otras palabras, los océanos pueden  representarse  por dos capas diferenciadas de agua: una capa superior que tiene   aguas templadas y un espesor $E=h_{M}-h_{0}=1000$ metros y una  capa inferior de agua con temperaturas casi constantes  de  4ºC que se extiende en profundidad desde  $h_{M}$ hasta el fondo oceánico.

Si dividimos la capa superior de agua en $N$ sub-capas iguales, $\Delta E=\frac{E}{N}$ es el espesor de cada sub-capa. Identificaremos a cada subcapa con el índice $i=1,2,3,...N$, donde $i=1$ corresponde a la que inicialmente contiene la superficie del océano y $i=N$ a la que yace a la profundidad $h_{M}$. La subcapa $i=1$ expone su superficie externa a la atmósfera y pierde calor por radiación. Nuestro propósito es calcular cuánto  tiempo tarda  el agua de la subcapa  $i=1$,  que inicialmente está a la temperatura  $T_{0}=283$ K (10ºC),  en enfriarse a  la temperatura $T=277$ K  (4ºC). Para ello usamos la siguiente fórmula derivada de  (\ref{tiempoE}), que nos da ese tiempo de enfriamiento:

\begin{equation}
t_{1}=-\frac{\rho_{m}\,  \Delta E\, c_{m}}{\sigma} \left(\frac{1}{3 T_{0}^{3}}-\frac{1}{3 T^{3}}\right).
 \label{tiempoE1}
  \end{equation}
  
 Transcurrido el tiempo  $t_{1}$, el agua de la subcapa $i=1$  alcanza la temperatura de 4ºC y la máxima densidad, mayor que la densidad de las subcapas que subyacen, pues tienen temperaturas mayores. En consecuencia, la subcapa $i=1$ se hunde hasta la profundidad  $h_{M}$, y cada una  las subcapas restantes sube un escalón de altura $\Delta E$ al ser atravesada por la subcapa $i=1$ en su descenso. La subcapa $i=1$ está estructurada en celdas que permiten la interpenetración con las otras subcapas. Ahora, el lugar original de la subcapa $i=1$ es ocupado por la  subcapa $i=2$, con lo    cual expone su superficie superior a la atmósfera y pierde calor por radiación.  Transcurrido el tiempo  $t_{2} (=t_{1})$, la subcapa $i=2$ aumenta su densidad y también decanta hacia el fondo hasta ubicarse sobre la subcapa $i=1$, y por  lo tanto la subcapa inmediatamente subyacente, $i=3$, emerge a la superficie. Naturalmente, las subcapas subsiguientes $i=4,5,6,..N$ también suben un nuevo escalón $\Delta E$. El proceso se repite hasta que la subcapa $i=N$ llega a la superficie del océano. Note que ahora el ordenamiento de las subcapas es inverso al original. Al comienzo  el índice $i$ ordena las subcapas de menor a mayor  profundidad. Cuando la subcapa $i=N$ alcanza la temperatura 4ºC, no se puede hundir porque todas las subcapas tienen la temperatura de 4ºC y por lo tanto la misma densidad. El tiempo que demandó llegar a ese estado es la suma de los tiempos de enfriamiento de cada subcapa:
 \begin{equation}
  t_{e}=\sum_{i=1}^{N} t_{i}=N t_{1}.
  \label{TiempoE2}
 \end{equation}
 Reemplazando $t_{1}$ dado por (\ref{tiempoE1}) en  (\ref{TiempoE2}), y teniendo en cuenta que $N \Delta E=E$, obtenemos que 
 \begin{equation}
 t_{e}=-\frac{\rho_{m}\,  E\, c_{m}}{\sigma} \left(\frac{1}{3 T_{0}^{3}}-\frac{1}{3 T^{3}}\right).
 \label{TiempoE3}
 \end{equation}
 Recordando que $E=1000$ metros, $T_{0}=283$ K (10ºC) y $T=277$ K  (4ºC) y adoptando $\rho_{m}= 1026$ kg/m$^{3} (=1.026$ gr/cm$^{3})$ y $c_{m}=$ 4200 J/(kg K) para la densidad y el calor especifico del agua salada respectivamente, la fórmula (\ref{TiempoE3}) nos proporciona el siguiente valor para $t_{e}$:
 \begin{equation}
 t_{e}= 2.35 \,\,\rm años.
 \label{Tenf}
 \end{equation}
 Este valor para  $t_{e}$ es naturalmente una aproximación dado que, si bien el mecanismo de  pérdida de calor por emisión radiativa juega un papel muy  importante, no es el único mecanismo que influye. Por ejemplo, las fuertes nevadas y el azote de vientos fríos e  intensos sobre la superficie de los océanos pueden también contribuir grandemente al enfriamiento de las aguas. Por ello, el valor de $t_{e}$ dado por (\ref{Tenf}) debería ser tomado como un límite superior.
 
 Cuando la subcapa $i=N$ se enfría en la superficie oceánica a  la temperatura de  4ºC, en la cual alcanza su máxima densidad,  y aun así no puede hundirse,  por lo tanto el enfriamiento por radiación  de la subcapa $i=N$ continua y provoca  el congelamiento de  dicha subcapa. En consecuencia, la subcapa  $i=N$ se convierte en una capa de hielo que flota en la superficie del océano debido a que la densidad del hielo es menor que la del agua. Utilizando la ecuación (\ref{tiempoE1}), estimaremos el tiempo $t_{N}^{\star}$ que tarda la capa de hielo en alcanzar la temperatura de equilibrio $T=214$ K (-59 ºC) que predice el modelo I. En este caso, $T_{0}=277$ K (4ºC) y si adoptamos $\Delta E=1$ metros, obtenemos que 
 \begin{equation}
 t_{N}^{\star}= 16 \,\, \rm dias.
 \end{equation} 
 Entonces, el tiempo total que demanda el congelamiento superficial de los océanos es
 $t_{e} +  t_{N}^{\star}                                                                                                      \approx 2.4$ años. Es decir, la Tierra envuelta durante al menos 2 años y medio por una  espesa de capa de polvo se convierte en una bola de hielo.

\subsection{Eras de hielo causadas por el efecto invernadero inverso, grandes colisiones cósmicas y erupciones volcánicas masivas: modelos \label{Huroniana}}

Dejado atrás el trauma de la gran colisión que dio lugar a la formación de la Luna (ver Sección 3) y finalizado  el proceso de acreción de planetesimales, la Tierra entró en un período de relativa calma que permitió la formación del sistema  atmósfera-océano. Se cree que ello ocurrió entre 4100 y 3800  millones de años atrás, la última parte del período geológico llamado informalmente Hádico\index{Hádico, período geológico},  el cual se corresponde con la primera división y la época más antigua  del Precámbrico.  En ese entonces, la Tierra había completado  su formación como planeta y adquirido una masa lo suficientemente  grande como para que su fuerza de gravedad pueda retener los gases que se liberaban en su superficie. En consecuencia, los gases vertidos por los antiguos volcanes, esencialmente $CO_{2}$, $CH_{4}$ y $H_{2}O$, originaron en parte la atmósfera primigenia. En ese entonces, nuestra Luna y probablemente también la Tierra sufrieron  impactos de grandes cuerpos cósmicos, fenómeno que suele llamarse  ``el último bombardeo intenso''. Por lo tanto, los gases que se desprendieron del material fundido por las colisiones  pasaron también a formar parte del gas atmosférico. También, se piensa que una parte del agua terrestre fue aportado por los cometas que chocaron contra la Tierra. Cuando las condiciones  atmosféricas fueron las adecuadas para la condensación del  vapor de agua, se produjeron intensas lluvias que llenaron  las grandes  depresiones de la corteza terrestre, con lo cual se formaron los primeros océanos y mares. Al final del Hádico\index{Hádico, período geológico}, los océanos estaban estables.

 Los primeros signos de vida aparecieron en el período Arcaico, que abarca
 aproximadamente desde 3800 a 2500 millones de años atrás y que por lo tanto sigue al período Hádico. La atmósfera y los océanos primitivos jugaron un papel fundamental en el origen y evolución de la vida. La reconstrucción del paleoclima mediante modelos geoquímicos revela que, durante el período Arcaico, la temperatura media de la superficie de la Tierra fue  de $\approx 303 $ K (30 ºC) (ver Fig. 15 de \cite{Jaffres}). Las aguas poco profundas de los mares y océanos a temperaturas tan templadas  fueron un caldo de cultivo para la proliferación de la vida primitiva. Esas temperaturas relativamente altas se explican por un intenso efecto invernadero debido a que la atmósfera primitiva estaba saturada de gases de efecto invernadero. Es decir, en el modelo de  Arrehenius, esa condición atmosférica la podemos representar con $\epsilon=1$. Reemplazando  $T_{s}=303$ K y  $\epsilon=1$  en la fórmula de Arrehenius (\ref{Arrhenius2}), podemos despejar nuestra incógnita $\gamma$: el valor del albedo de la Tierra en el período Arcaico\index{Arcaico, período geológico}. De este modo, obtenemos que $\gamma=0.3$, el cual coincide con el albedo de la Tierra actual.
 
\begin{figure}
\includegraphics[scale=0.47]{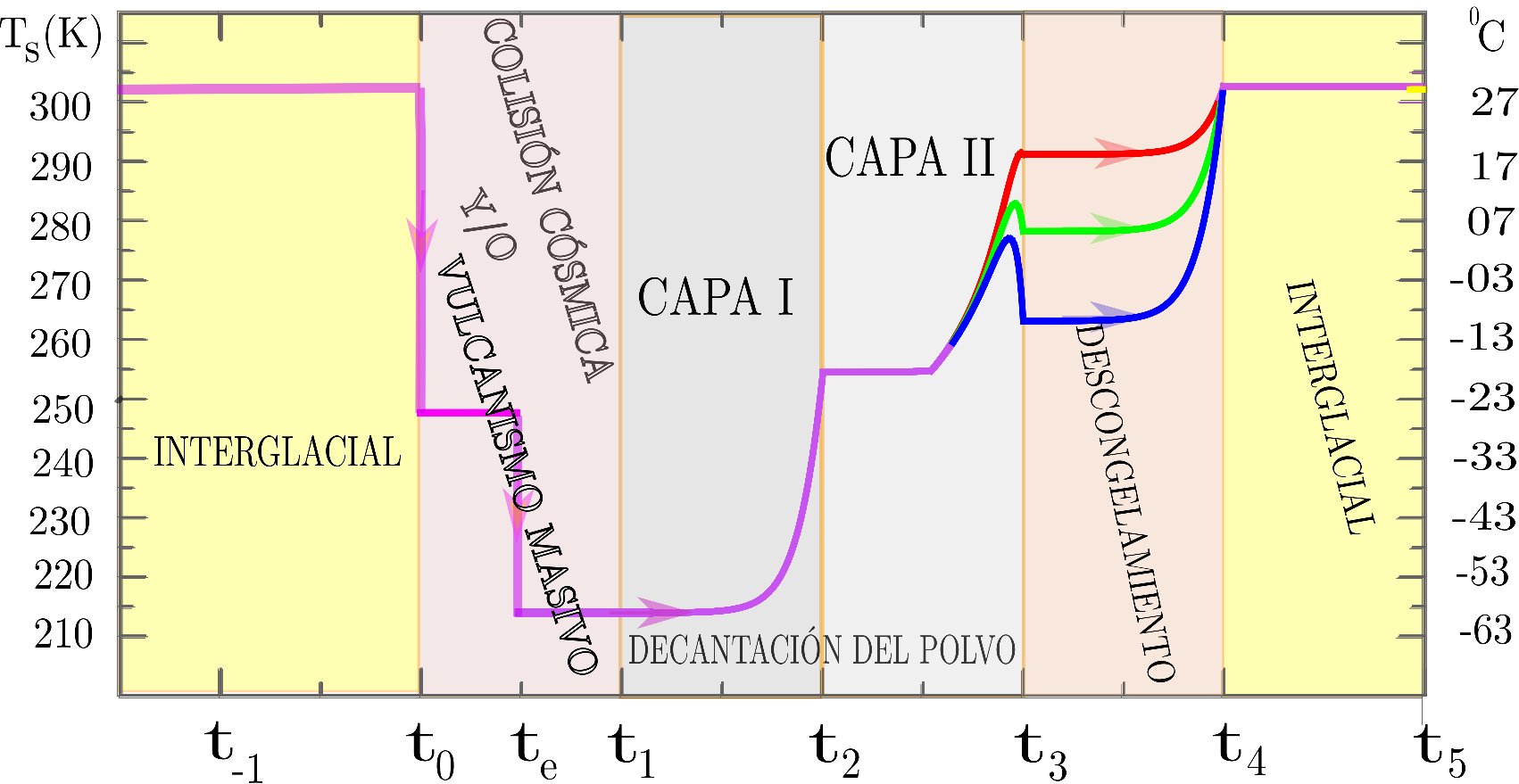} 
\caption{Caso A de glaciación: Evolución de la temperatura $T_{s}$ de la superficie terrestre en las distintas etapas del Caso A.  Las áreas coloreadas indican las distintas etapas.  Las temperaturas se expresan en grados Kelvin (ordenada izquierda) y en grados centígrados (ordenada derecha).}
\label{CasoAA} 
\end{figure}

Ahora nuestro propósito  es aplicar los modelos que hemos desarrollado en la Sección \ref{Modelos}, en diferentes escenarios  dados por los casos que denominamos A, B y  C. Los Casos A y B se desarrollan en el contexto de una  alta concentración de gases de efecto invernadero ($\epsilon=1$), tal como la atmósfera primitiva  que hemos descripto en el parágrafo anterior. En cambio, en el Caso C, la condición atmosférica es más parecida a la actual.
\\
\\
{\bf Caso A: efectos térmicos de grandes colisiones cósmicas y del vulcanismo masivo,  en una Tierra templada.
\label{SecCasoA}}

En el caída de un cometa de gran tamaño sobre la Tierra, la relativamente tenue nube de  polvo que rodea a la parte densa del cometa es frenada en la parte alta de la atmósfera terrestre. En cambio, el denso núcleo del   cometa puede  penetrar la atmósfera y chocar contra  la superficie terrestre, generando una fuerte onda de choque que esparce, en la atmósfera, partículas de polvo del cuerpo desintegrado del cometa y del suelo impactado. Un efecto similar puede ser causado por el choque de un gran asteroide. Naturalmente, la consecuencia es que se forma, en un corto período de tiempo,  una envoltura de polvo que rodea a la Tierra e impregna a la atmósfera.

La corteza terrestre  a modo de un  caparazón rígido puede  resquebrajarse  en fragmentos o placas bajo el golpe violento de un cometa o asteroide. En efecto, hay indicios de la existencia de  placas tectónicas en la juventud  de la Tierra. Las grietas o fallas de la corteza permiten que el magma del manto emerja  hacia la superficie y se produzcan erupciones volcánicas  a todo lo largo de las grietas. Así es como la colisión de un gran cuerpo cósmico contra la Tierra podría  inducir un episodio de vulcanismo masivo. Por otra parte, las placas tectónicas se apoyan sobre un substrato relativamente  plástico y por lo tanto pueden deslizarse sobre la superficie de la Tierra. Los movimientos de deriva de esas placas tectónicas pueden a su vez desencadenar la actividad volcánica. Las  cenizas arrojadas profusamente a la atmósfera por la violenta acción colectiva de los volcanes forman una capa de polvo que puede quedar suspendida en la atmósfera por un tiempo prolongado. En el caso del vulcanismo inducido por un impacto cósmico, a la capa de polvo originada previamente por el impacto  cósmico  se le agrega la contribución volcánica.

El objetivo de esta sección es estudiar las consecuencias climáticas de 
una densa capa de polvo creada en torno a la Tierra por alguno de los sucesos arriba descriptos, o por una combinación encadenada de ellos,
en una Tierra templada ($T_{s}=303$ K) por un efecto invernadero intenso $(\epsilon=1)$. Supongamos que la colisión cósmica y/o el polvo eyectado por los volcanes  impregna densamente,  en poco tiempo,  la atmósfera superior como inferior y absorbe la luz solar visible, oscureciendo por completo a la Tierra por un periodo prolongado. Si el vulcanismo fue inducido por el choque de un gran cometa o asteroide, una parte del polvo que impregna la atmósfera se debe al material  del cuerpo cósmico desprendido  en su ingreso  a la atmósfera y al material  eyectado en el choque contra el suelo. Entonces,  en esas condiciones, si la superficie de Tierra fuera toda sólida, en unos pocos días,  está se enfriaría hasta alcanzar una temperatura de equilibrio  (ver Sección anterior) y  partir de allí podríamos aplicar las fórmulas (\ref{ModeloIII}),  (\ref{Tsup2}) o  (\ref{InvernaderoInverso}) para calcular $T_{s}$. Sin embargo, en el pasado como en el presente, la superficie cubierta por los océanos profundos fue probablemente  importante y por lo tanto se tarda un tiempo $t_{e}$ para que toda la superficie terrestre alcance temperaturas de equilibrio (ver Sección anterior).

Supongamos que la mitad de la superficie terrestre es ocupada por el suelo sólido y las aguas de baja profundidad, con lo cual la emisión de esta parte de la superficie es   $\frac{1}{2} \sigma T_{s}^{4}$. Si la otra mitad de la superficie terrestre es ocupada por los océanos profundos, su emisión es $\frac{1}{2} \sigma T_{oc}^{4}$, donde 
$T_{oc}$ es la temperatura superficial de los océanos. Suponemos además que, debido a los vientos, la energía total emitida por ambas partes de la superficie se mezcla en toda  la atmósfera. Dado que  en este caso $\tau_{1} \gg 1$ (la profundidad óptica de la capa superior de polvo), podemos emplear las fórmulas   (\ref{EmisionAtmosfera}) y (\ref{EmisionSuelo}) para obtener $T_{s}$, pero debemos reemplazar el término  $\epsilon\sigma T_{s}^{4}$  de la fórmula (\ref{EmisionAtmosfera}) por $\frac{1}{2} \epsilon\sigma (T_{s}^{4} +T_{oc}^{4})$. Como durante el tiempo de enfriamiento de los océanos, su temperatura media superficial $T_{oc}$ se mantiene paradójicamente constante (ver Sección anterior). Por lo tanto, también  la temperatura $T_{s}$ del suelo sólido permanece constante  pero a valores muy por debajo de 0 ºC. Si suponemos que antes de desatarse el vulcanismo masivo $T_{oc}=T_{s}=T_{0}=303$ K, durante  el proceso de enfriamiento de los océanos $T_{oc}\approx \frac{T_{0} + T}{2}=290$ K, donde  T=277 K (4 ºC) (ver Sección anterior). Entonces para este caso, $t_{e} \approx 9$ años. Empleando las fórmulas (\ref{EmisionAtmosfera}) y (\ref{EmisionSuelo}) con las modificaciones indicadas arriba, obtenemos que  $T_{s}=248$ K  durante los primeros 9 años desde el inicio del vulcanismo masivo. 

Apenas, se traspasa el tiempo $t_{e}$ de enfriamiento,  ambas temperaturas, $T_{oc}$ y $T_{s}$,  decaen abruptamente a un valor común mucho más bajo: 214 K, tal como lo predice el modelo I en condiciones de equilibrio térmico. De modo que ya no es necesario distinguirlas y las llamamos simplemente $T_{s}$. La Fig. \ref{CasoAA} describe la secuencia evolutiva del Caso A, dada por la variación de $T_{s}$ en función de una variable a-dimensional $t$ (o número puro) que muestra  el orden en que se suceden  las diferentes etapas del proceso. La duración temporal de cada etapa puede ser diferente y no es aquí definida. En el intervalo $(t_{0},t_{1})$ los volcanes eyectan polvo y gases de efecto invernadero a la atmósfera (ver  Fig. \ref{CasoAA}). La  cantidad de gases de efecto invernadero en la atmósfera aumenta, pero  el valor del coeficiente $\epsilon$ no varía porque ya se encontraba en el máximo ($\epsilon=1$) en  $t_{0}$. Es decir, la atmósfera está saturada de gases de efecto invernadero. Entre $(t_{0},t_{e})$, $T_{s}=248$ K y $T_{oc}=290$ K. En $t_{e}$, los océanos se congelan y las temperaturas $T_{s}$ y $T_{oc}$ descienden abruptamente a $T_{s}=T_{oc}=214$ K, de modo que la Tierra se convierte en una bola de nieve.

Suponemos que  al finalizar vulcanismo y en consecuencia la inyección de polvo en la atmósfera, indicado por $t_{1}$ en la Fig. \ref{CasoAA}, las profundidades ópticas de la capa superior (I) e inferior (II) de polvo son    $\tau_{1}=10$ y  $\tau_{2}=10$, respectivamente. Con esas profundidades ópticas, que no son extremadamente altas, la atenuación del brillo solar es $B_{\odot}\, e^{-(\tau_{1}+\tau_{2})}=2 \times 10^{-9} B_{\odot}$, es decir el Sol está totalmente oscurecido y la Tierra inmersa en una noche cerrada. De $t_{1}$ en adelante, empleamos la fórmula (\ref{ModeloIII}) del modelo III a fin de estimar la evolución de $T_{s}$. Suponemos que, desde $t_{1}$ a $t_{2}$, el polvo de la capa  superior se precipita hacia  la capa inferior a un ritmo constante  y que queda totalmente vacía en $t_{2}$. Por lo tanto,  $\tau_{1}$ decae desde 10 a cero. Suponemos  también que $\tau_{2}=10$ en el periodo $(t_{1},t_{2})$. Es decir, $\tau_{2}$ permanece constante, lo cual significa que la cantidad de polvo de  la capa inferior que se deposita en el suelo es igual a la cantidad de polvo que le ingresa desde la capa superior.

Al acercarse a $t_{2}$,  $\tau_{1}$ tiende a cero y $T_{s}$ se eleva abruptamente a  $254\, K$ (-19  ºC) de acuerdo con la fórmula (\ref{Tsup2})  del Modelo II. En el intervalo entre $t_{2}$ y $t_{3}$, $\tau_{1}=0$ y $\tau_{2}$ decrece linealmente desde el valor de 10 en $t_{2}$ a cero en $t_{3}$. En la Fig.  \ref{CasoAA}, representamos tres diferentes curvas para  $T_{s}$, todas fueron calculadas con $\epsilon=1.0$ pero con diferentes albedos;  $\gamma=0.60$ (curva azul) ,$\gamma=0.50$ (curva verde) y $\gamma=0.40$ (curva roja). En el intervalo $(t_{1},t_{3})$, la temperatura es insensible al valor  del albedo $\gamma$ de la Tierra, excepto cuando  $\tau_{2}<3$ (ver Fig. \ref{CasoAA}) cerca de $t_{3}$.  Es decir, las tres curvas de la Fig. \ref{CasoAA} coinciden en la mayor parte del intervalo $(t_{1}, t_{3})$. Ello se explica porque  si la luz visible es casi completamente absorbida  antes de llegar al suelo, la luz reflejada por el suelo es casi nula independiente del valor de  $\gamma$. 

El proceso de decantación del polvo finaliza en $t_{3}$,  y de allí en adelante la atmósfera está libre de polvo ($\tau_{1}=\tau_{2}=0$). En estas condiciones, las fórmulas (\ref{ModeloIII}) y (\ref{Arrhenius2}) (de  Arrhenius) coinciden; y $T_{s}$ depende fuertemente de $\gamma$.  En efecto, vemos que las tres curvas de la Fig. \ref{CasoAA} se separan en el intervalo $(t_{3}, t_{4})$, debido solo a los distintos valores de $\gamma$ ya que el valor de $\epsilon$ es el mismo  en los  tres casos ($\epsilon=1)$. Los tres albedos elegidos representan a la Tierra en estado de bola de nieve con distintos grados en que  las cenizas volcánicas  ensuciaron el hielo.

En el inicio del intervalo $(t_{3}, t_{4})$, las temperaturas medias del suelo $T_{s}$ son aproximadamente -5 ºC, 7 ºC y 20 ºC para $\gamma=0.6$,  $\gamma=0.5$ y $\gamma=0.4$, respectivamente. Todas son temperaturas de descongelamiento. Si bien en el caso de $\gamma=0.6$, la temperatura media   es de -5 grados centígrados,  en las regiones tropicales las temperaturas están por encima de cero grado. A medida que la Tierra se descongela  el albedo disminuye, con lo cual el proceso se acelera pues  la temperatura aumenta de acuerdo con  (\ref{Arrhenius2}). Simulamos el proceso de descongelamiento mediante la siguiente fórmula para el albedo de la Tierra;

\begin{equation}
\gamma=\gamma_{0}- (\gamma_{0}-\gamma_{f}) e^{const. \times \, (t-t_{4})} \,\, \rm para \,\,\, t_{3}< t < t_{4},
\label{derretimiento}
\end{equation}
donde $\gamma_{0}=\gamma$ en  $t_{3}$  y $\gamma_{f}=0.3$, tal como el albedo  actual de la Tierra.

Por último, a partir de  $t_{4}$ se inicia un periodo que podríamos  llamarlo \it{interglaciar}\rm, en el cual la atmósfera y la superficie terrestre  retoman las condiciones previas al comienzo de la era glacial.
\\

{\bf Caso B: efectos del vulcanismo masivo, inducido por causas endógenas
o por colisiones cósmicas,  en una Tierra congelada por el efecto invernadero inverso.}

Se piensa que la Tierra a lo largo de su historia poseyó  periodos con alta concentración de gases de efecto invernadero que mantuvieron, por un largo tiempo, climas templados y propicios para la expansión de vida. Entonces, la exuberante vida fotosintética fue incorporando  lentamente  el dióxido de carbono de la atmósfera, el cual finalmente quedó capturado en  los restos sedimentarios de sus cuerpos. Como consecuencia de la disminución de los gases de efecto invernadero, la temperatura de la superficie de la Tierra  disminuye; fenómeno que suele denominarse efecto invernadero inverso. A medida que la Tierra se enfría por el efecto invernadero inverso, crece la superficie cubierta de nieve y hielo y con ello aumenta el albedo y el enfriamiento de la Tierra se potencia. La Tierra entra en una era de hielo que tiende a eternizarse.

\begin{figure}
\includegraphics[scale=0.47]{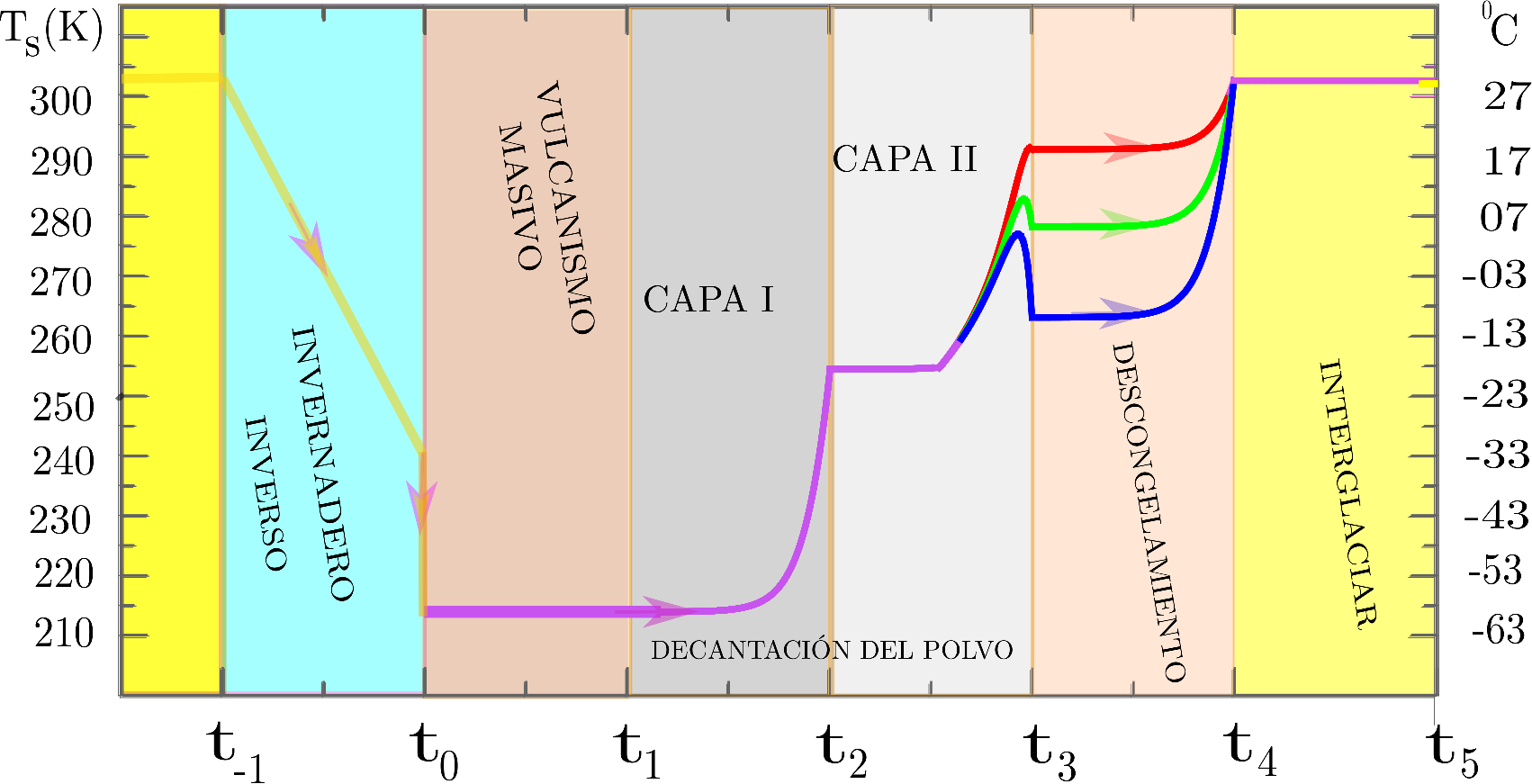} 
\caption{Caso B de glaciación: Evolución de la temperatura $T_{s}$ de la superficie terrestre en las distintas etapas del Caso B.  Las temperaturas se expresan en grados Kelvin (ordenada izquierda) y en grados centígrados (ordenada derecha).}
\label{CasoB}
\end{figure}

Un interrogante que planteamos es qué fenómenos pueden poner fin a una Tierra convertida en una bola de nieve por el efecto invernadero inverso y restablecer condiciones climáticas que permitan la continuidad de la vida.  A fin de dar una respuesta, imaginamos que en el contexto de una era de hielo originada  por el efecto invernadero inverso ocurren episodios de vulcanismo masivo, desencadenados  por movimientos tectónicos o por traumas de impactos cósmicos. A este escenario lo llamaremos Caso B. La Fig. \ref{CasoB} describe la secuencia evolutiva del Caso B, dada por la variación de $T_{s}$ en función de una variable a-dimensional $t$ (o número puro) que muestra  el orden en que se suceden  las diferentes etapas del proceso. La duración temporal de cada etapa puede ser diferente y no es aquí definida.

La primera etapa comienza en  $t_{-1}$, donde debido al alto contenido de gases de efecto invernadero $\epsilon=1.0$. Suponemos además  que la Tierra tenía  en ese entonces un albedo  como el actual, $\gamma=0.3$, con lo cual según la fórmula de Arrehenius (\ref{Arrhenius2}) $T_{s}=303$ K (+30ºC) en  $t_{-1}$. Por el efecto invernadero inverso suponemos que en  $t_{0}$, $\epsilon=0.6$ y $\gamma=0.6$ (el albedo del hielo es del orden de un 60 por ciento), y por lo tanto según la fórmula de Arrehenius (\ref{Arrhenius2}) $T_{s}=242$ K (-31  ºC). Hemos representado esquemáticamente el descenso de temperatura por una linea recta, cuando en realidad estos procesos  no son en general lineales.

Imaginemos que en $t_{0}$ se desata un episodio de vulcanismo masivo que inyecta durante un tiempo prolongado polvo (cenizas) y gases de efecto invernadero en la atmósfera, tanto inferior como superior. Si el vulcanismo fue inducido por el choque de un gran cometa o asteroide, una parte del polvo que impregna la atmósfera se debe al material  del cuerpo cósmico desprendido  en su ingreso  a la atmósfera y al material  eyectado en el choque contra el suelo. La espesa envoltura de polvo hace descender aún más la temperatura de la superficie terrestre a valores del orden de 214 K, tal como se infiere de la fórmula  (\ref{InvernaderoInverso2}) del modelo I. Cabe mencionar que esa temperatura de equilibrio se alcanza rápidamente dado el estado de congelamiento global del suelo, previo al suceso del vulcanismo (ver sección anterior).

En el intervalo $(t_{0},t_{1})$ en el que se desarrolla el vulcanismo (ver  Fig. \ref{CasoB}), aumenta la cantidad de gases de efecto invernadero en la atmósfera y por lo tanto el valor del coeficiente $\epsilon$ que lo cuantifica. Sin embargo, a causa de la formación de una espesa capa de polvo en la atmósfera superior, la temperatura $T_{s}$ es insensible al valor de   $\epsilon$ y de $\gamma$ y se mantiene constante en 214 K tal como lo predice el modelo I. Vamos a suponer que entre  $t_{0}$ y $t_{1}$, la atmósfera recupera la cantidad de gases de efecto invernadero que tenia en $t_{-1}$, es decir antes de comenzar el efecto invernadero inverso. Por lo tanto,  adoptaremos $\epsilon=1.0$ para todas la etapas que siguen a partir de $t_{1}$. Además suponemos que al terminar el vulcanismo y en consecuencia la inyección de polvo en la atmósfera, las profundidades de la capa superior (I) e inferior (II) de polvo son    $\tau_{1}=10$ y  $\tau_{2}=10$, respectivamente. De modo que, a partir de  $t_{1}$, las condiciones son las mismas que las del Caso A y en consecuencia la evolución de $T_{s}$ tal como se aprecia en la  Fig. \ref{CasoB}.

Por último, a partir de  $t_{4}$ se inicia un periodo que podríamos  llamarlo \it{interglaciar}\rm, en el cual la atmósfera y la superficie terrestre  retoman las condiciones previas al comienzo de la era glacial por el efecto invernadero inverso. La vida que logró sobrevivir tiene ahora condiciones propicias para prosperar  nuevamente, pero  puede lamentablemente iniciarse el ciclo representado en la Fig. \ref{CasoB} y así  una nueva era de hielo. Estas  largas eras de hielo iniciadas por el efecto invernadero inverso y separadas por benignos periodos interglaciares se habrían sucedido en la historia temprana de la Tierra.

 A diferencia del Caso A en que el vulcanismo masivo ocurre en una Tierra con temperaturas templadas, el vulcanismo que describimos en esta sección  ocurre en una Tierra totalmente congelada. Aquí el vulcanismo no inició la era de hielo pero si fue el principio del fin de una era de hielo (la inyección de gases de efecto invernado hizo posible el fin); mientras que en el Caso A, el vulcanismo causa la edad de hielo.
Esto nos muestra el doble papel que puede jugar el vulcanismo: como causa del origen o del  fin  de una era de hielo.
\\ 
 
{\bf Caso C: congelamiento de la Tierra por el choque de un cometa o de un asteroide, sin alcanzar el congelamiento completo de los mares y océanos.}

\begin{figure}
\includegraphics[scale=0.7]{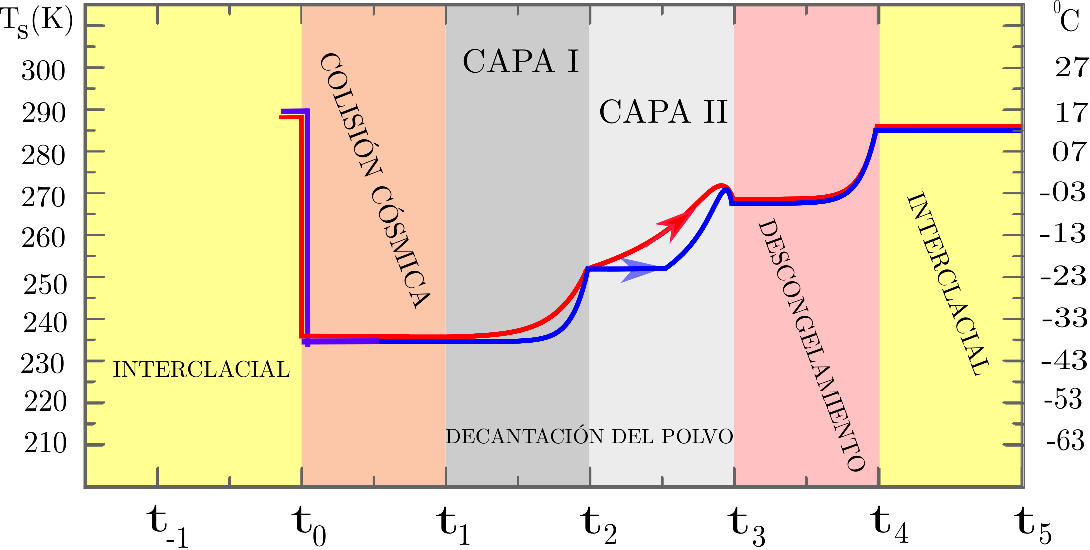} 
\caption{Caso C de glaciación: Evolución de la temperatura $T_{s}$ de la superficie terrestre en las distintas etapas del Caso C. Las áreas coloreadas indican las distintas etapas.  Las temperaturas se expresan en grados Kelvin (ordenada izquierda) y en grados centígrados (ordenada derecha).}
\label{NoEquilibrio}
\end{figure}
Aquí suponemos que el impacto de un gran cuerpo cósmico forma, en unas pocas semanas,  una espesa capa de polvo 
en e inmediatamente  sobre toda la atmósfera  terrestre y suponemos además que las condiciones atmosféricas son similares a las actuales, con $\epsilon=0.78$. También consideramos que  el tiempo de decantación de la capa de polvo\footnote{Los granos pequeños de polvo  tardan  de 1 a  2 años en llegar a la superficie terrestre\cite{Yabushita}.} es menor o igual al tiempo de enfriamiento de los océanos, el cual en este caso $t_{e}\approx 2.5$ años (ver Sección \ref{Oceano}). Es decir $t_{3}-t_{0}\le t_{e}$. La temperatura media superficial de los océanos $T_{oc}$ permanece constante durante el tiempo de enfriamiento y, en las condiciones actuales de la Tierra, $T_{oc}=\frac{T_{0}+T}{2}=\frac{283+277}{2}=280$ K, donde hemos considerado que la mitad de la superficie terrestre es ocupada por los mares y océanos. Por lo tanto, solo varía la temperatura $T_{s}$ de la superficie terrestre sólida. En la Sección \ref{SecCasoA}, explicamos como incorporar $T_{oc}$ en las ecuaciones de los modelos desarrollados en la Sección \ref{Modelos}. La evolución  $T_{s}$ en cada etapa se calculó  empleando  el mismo procedimiento   que utilizamos  en los Casos A y B. 

En la Fig. \ref{NoEquilibrio}, representamos $T_{s}$ para las distintas etapas y para distintas profundidades ópticas iniciales de las dos capas de polvo. La curva roja de  la Fig. \ref{NoEquilibrio} se obtuvo con $\tau_{1}=5 $ y $\tau_{2}=5$ para las profundidades ópticas iniciales de la capa I y la capa II, respectivamente, 
y la curva azul de  la Fig. \ref{NoEquilibrio} se obtuvo con  $\tau_{1}=\tau_{2}=10$. Ambas curvas se calcularon con $\gamma=0.5$ para los estados de congelamiento de la superficie sólida de la Tierra y,  para la etapa de descongelamiento,  $\gamma_{0}=0.5$ y $\gamma_{f}=0.3$ (ver fórmula \ref{derretimiento}). La Fig. \ref{NoEquilibrio} muestra que, aun con  las temperaturas  morigeradas por la inercia térmica de los océanos, la superficie sólida de la Tierra alcanza una temperatura media mínima  del orden de $-37^{\circ}$ C. 

A diferencia de los Casos A y B, donde  el tiempo de deposición del polvo, $(t_{1}-t_{0})$, es largo o al menos del orden $t_{e}$, en el Caso C suponemos  lo contrario. Es decir, en el Caso C representamos una única colisión donde el polvo se dispersa en la atmósfera rápidamente, en unas pocas semanas. En cambio, en el caso de choques múltiples con réplicas volcánicas en una atmósfera como la actual ($\epsilon=0.78$) y la duración de la etapa inicial $(t_{0}, t_{1})$ mayor que  $t_{e}$, la evolución de $T_{s}$ es similar a los de los Casos A y B. En ese caso, la temperatura media mínima de \it{toda} \rm la superficie terrestre es igual a 214 K (-59$^{\circ}$C), si $\tau_{1}\geq 5$ (ver Fig. \ref{Modelo1}). Como en esta etapa predomina el Modelo I, $T_{s}$ no depende de $\epsilon$ (ver ecuación \ref{InvernaderoInverso2}). En las otras etapas, solo difieren de los Casos A y B en que las temperaturas disminuyen en $\approx 10-15$ K.

\section{El paso del Sol a través de densas nubes interestelares \label{Penetrante1}}

\begin{quote}
\small \it{I am a part of all that I have met.} \rm

Alfred Tennyson\index{Tennyson, A.}
\end{quote}

El gran avance de la astronomía galáctica en las primeras décadas del siglo XX permitió delinear la estructura y cinemática de la Vía Láctea, como así también determinar nuestra posición dentro de la Galaxia. El Sol se encuentra  en la periferia de la Galaxia  a $\approx$ 8 kpc del centro Galáctico. El Sol  tiene una velocidad peculiar con respecto a las estrellas y nubes interestelares vecinas de $\approx$ 20  km s$^{-1}$  y, en conjunto con las estrellas y la materia interestelar del entorno solar,  rota alrededor del centro galáctico con una  velocidad de $\approx$220  km s$^{-1}$. Es decir, somos viajeros temerarios del espacio y, en consecuencia, es probable que la Tierra haya soportado vicisitudes a lo largo de su  viaje galáctico. Las llamadas nubes moleculares gigantes (NMGs) \footnote{La sigla en inglés es GMC por giant molecular cloud.}, extensas y densas nubes interestelares de polvo y moléculas (ver Sección \ref{MedioInterestelar}),  son una de las componentes de la materia interestelar que entraña peligros para la Tierra cuando el Sistema Solar se acerca o atraviesa una NMG. 

\subsection{Efectos no-gravitacionales \label{no-gravitacionales}}
Los posibles efectos no-gravitacionales de los encuentros del Sistema Solar con tales nubes interestelares han llamado la atención de los astrónomos como posible causa de las eras de hielo. Una reseña de las principales ideas sobre el tema hasta el año 1981, con sus referencias bibliográficas,  se encuentra en un  artículo del astrónomo británico Sir Willian Hunter McCrea (1904-1999)\index{McCrea, W.H.} \footnote{McCrea visitó la Argentina\index{Argentina} en diciembre del año 1983 \cite{McCrea2}, y los jóvenes astrónomos de entonces tuvimos el honor de dialogar con él.}\index{McCrea, W.H.}, que incluye además sus  propias investigaciones \citep{McCrea1}. 

Al internarse el Sistema Solar en una NMG, se producen  dos efectos no-gravitacionales que pueden influir sobre el clima terrestre: 1) La deposición del gas y polvo interestelar en la alta atmósfera terrestre oscurece la luz solar; 2) El flujo de la materia de la nube interestelar que intercepta el Sol disipa energía en el choque y  así aumentaría la luminosidad del Sol y, por lo tanto, la constante solar.  
En principio, ambos efectos se contraponen. Ahora procedemos a analizar el primer efecto. Consideremos una NMG con densidad numérica $n$ de moléculas de hidrógeno $(H_{2})$ a fin de calcular la densidad numérica $\rho_{polvoN}$ de granos de polvo de la NMG. La masa total de polvo de una nube interestelar es solo el 1 por ciento de la masa total del gas  de dicha  nube (ver Sección \ref{MedioInterestelar}). Por lo tanto,
$M_{polvo}=\frac{1}{100} M_{gas}=\frac{1}{100} m_{H_{2}}\, n\, V$, donde $m_{H_{2}}$ es la masa de la molécula de $H_{2}$ y  $V$ es el volumen de la nube. La masa $m_{grano}$ de un grano de polvo de radio $a$ y densidad  $\rho_{iced}$ (ver Sección \ref{Modelos}) es dada por $m_{grano}=\rho_{iced} \frac{4}{3} \pi a^{3}$ y,  entonces, la densidad numérica de granos de polvo es $\rho_{polvoN}= \frac{M_{polvo}}{ V\, m_{grano}}=\frac{1}{100} \frac{m_{H_{2}}}{m_{grano}} n$. Si adoptamos $\rho_{iced}=2.35 \frac{gr}{cm^{3}}$ y $a=0.1 \mu m$,

\begin{equation}
\rho_{polvoN}= 3.39\times 10^{-12}\, n\,\,\, (\frac{granos}{cm^{3}}).
\label{polvoN}
\end{equation}

 Si el Sol atraviesa la NMG con una velocidad media $v_{\odot}$ relativa al gas y polvo de la NMG, el flujo del gas y polvo ingresa al Sistema Solar con la dirección del vector velocidad  $\vec{v_{\odot}}$  e incide sobre la Tierra con la dirección del vector velocidad  $\vec{v_{E}}= \vec{v_{\odot}}- \vec{v_{\oplus}}$, donde  $\vec{v_{\oplus}}$ es el vector  velocidad de traslación de la Tierra en torno al Sol. El módulo de $\vec{v_{E}}$ es dado por $v_{E}=\sqrt{(v_{\odot}- v_{\oplus} cos \, \lambda)^{2}+(v_{\oplus} sin\, \lambda)^{2} }$, donde  $\lambda$ es el ángulo entre $\vec{v_{\odot}}$ y  $\vec{v_{\oplus}}$. La Tierra completa su órbita en un año y por lo tanto, en ese tiempo,  $\lambda$ varía entre 0 y $2 \pi$. Si tomamos como representativa de $v_{E}$ su media cuadrática,
$v_{E}= \frac{1}{2 \pi} \int_{0}^{2 \pi} ((v_{\odot}- v_{\oplus} cos \, \lambda)^{2}+(v_{\oplus} sin\, \lambda)^{2})\, d\lambda$, con lo cual

\begin{equation}
v_{E}=\sqrt{v_{\odot}^{2}+v_{\oplus}^{2}}.
\label{vE}
\end{equation}

   La Tierra barre la siguiente cantidad de granos de polvo en el intervalo de tiempo $\Delta t$: $ \rho_{polvoN} A_{\oplus}\, v_{E}\, \Delta t$,  
 donde $A_{\oplus}=\pi R_{\oplus}^{2}$ es el área que la Tierra de radio $R_{\oplus}$ expone ante el flujo de partículas de la NMG. Como la Tierra rota sobre su eje,  toda la superficie terrestre queda expuesta al flujo del polvo interestelar. Por lo tanto,  el flujo ${\cal F}_{p}$ o cantidad promedio de granos de polvo, por  unidad de área y de  tiempo,  que incide sobre la superficie de la Tierra $(S_{\oplus}=4 \pi R_{\oplus}^{2})$  es $\frac{\rho_{polvoN} A_{\oplus}\, v_{E}}{S_{\oplus}}$, de lo cual resulta 
 \begin{equation}
 {\cal F}_{p}=\frac{1}{4}\rho_{polvoN}\, v_{E}.
 \label{FlujoPolvo}
 \end{equation} 
 Del mismo modo, podemos calcular el flujo de moléculas de hidrógeno ($H_{2}$) que choca en la alta atmósfera:
 
 \begin{equation}
 {\cal F}_{H_{2}}=\frac{1}{4} n \, v_{E}.
 \label{FlujoH2}
 \end{equation}
 
 El flujo de polvo interestelar es casi frenado en la alta atmósfera y desciende a través de la atmósfera muy lentamente hacía la superficie terrestre. Suponiendo que $v_{D}$ es la velocidad de decantación del polvo, el flujo del polvo interestelar  en la atmósfera es $\rho_{polvoA}\, v_{D}$, donde $\rho_{polvoA}$ es la densidad numérica de los granos de  polvo en la atmósfera. Como el flujo  del polvo interestelar que incide sobre la Tierra, dado por la ecuación (\ref{FlujoPolvo}), tiene que ser igual al flujo de decantación del polvo en la atmósfera,  $\frac{1}{4}\rho_{polvoN}\, v_{E}=\rho_{polvoA}\, v_{D}$. Entonces,
 
 \begin{equation}
 \rho_{polvoA}= \frac{1}{4}  \frac{v_{E}}{v_{D}}\, \rho_{polvoN}.
 \label{polvoA}
 \end{equation}
Como $v_{E} \gg v_{D}$, la ecuación (\ref{polvoA}) indica que $\rho_{polvoA} \gg \rho_{polvoN}$. Reemplazando $\rho_{polvoN}$ por su expresión (\ref{polvoN}) en la ecuación (\ref{polvoA}), obtenemos

\begin{equation}
\rho_{polvoA}= 8.48 \times 10^{-13}\,\,\, \frac{v_{E}}{v_{D}}\,n \,\,\,\,\,(\frac{granos}{cm^{3}}).
\end{equation}
A fin de calcular la profundidad óptica $\tau_{A}$ de la capa de polvo interestelar que decanta lentamente  en la atmósfera, aplicamos la fórmula (\ref{ProfundidadOpt}). Aquí, la densidad columnar del polvo  en la atmósfera es ${\cal N}_{p}= \rho_{polvoA}\,\, H$, donde $H$ es el espesor de la atmósfera, y por lo tanto $\tau_{A}= \pi\, a^{2}\, Q_{ex} \,\, \rho_{polvoA}\,\, H $. Considerando que los granos de polvo comienzan a  frenarse  al entrar en la parte superior de la mesósfera, cuya altura sobre el suelo es $\approx$ 80 km, adoptamos $H=80$ km. Además, teniendo en cuenta que $a=0.1\mu$m y $Q_{ex}=0.5$, obtenemos

\begin{equation}
\tau_{A}= 1.07 \times 10^{-15} \frac{v_{E}}{v_{D}}\,n.
\label{TauA}
\end{equation}

No solo el polvo depositado en la atmósfera oscurece la luz del Sol, sino también el polvo que yace entre el Sol y la Tierra, ya que todo el Sistema Solar se encuentra inmerso en la NMG. La densidad columnar del polvo que yace entre el Sol y la Tierra es ${\cal N}_{p}= \rho_{polvoN}\,\, d_{TS}$, donde $d_{TS}=151 \times 10^{6}$ km es la distancia Tierra-Sol. Por lo tanto, la correspondiente profundidad óptica $\tau_{TS}$ es $\pi\, a^{2}\, Q_{ex}  \rho_{polvoN}\,\, d_{TS}$
y, con los valores adoptados para $a$ y $Q_{ex}$ y con $\rho_{polvoN}$ dado por la fórmula (\ref{polvoN}), obtenemos

\begin{equation}
\tau_{TS}=8 \times 10^{-9} n. 
\label{TauTS}
\end{equation}

Para evaluar la profundidad óptica total $\tau (=\tau_{A}+\tau_{TS})$,  debemos conocer los valores de $v_{E}$, $v_{D}$ y $n$. Si adoptamos  la velocidad peculiar del Sol de 20 km s$^{-1}$ como la velocidad relativa $v_{\odot}$ del Sol con las NMG, y teniendo en cuenta que $v_{\oplus}\approx 30$  km s$^{-1}$, obtenemos a partir de la fórmula (\ref{vE}) que $v_{E}=36$ km s$^{-1}$.
Se estima que el tiempo $t_{s}$ de suspensión de los granos de polvo en la atmósfera, es decir el tiempo que le demanda a un grano de polvo descender desde la alta atmósfera hasta el suelo, es de unos pocos años \cite{Yabushita}. Nosotros adoptamos $t_{s}=$ 2 años y, por lo tanto, $v_{D}=\frac{H}{t_{s}}=1.3\times 10^{-6}$ km s$^{-1}$. Para la densidad de la NMG,  adoptamos $n=300$ moléculas de $H_{2}$ por cm$^{3}$ (\footnote{Las nubes interestelares de  tales densidades pueden acercarse a la Tierra sin ser frenadas significativamente por el viento solar.}), las densidades típicas de las NMG se encuentran  entre 100 y 300 moléculas por cm$^{3}$. Utilizando dichos números y las fórmulas
(\ref{TauA}) y (\ref{TauTS}), encontramos que $\tau_{A}=9.0 \times 10^{-6}$ y $\tau_{TS}=2.4 \times 10^{-6}$ y que por lo tanto $\tau=1.1 \times 10^{-5}$ y  $ e^{-\tau}$, el factor de atenuación en la luz solar visible, es prácticamente 1. En consecuencia, la luz solar que incide sobre la superficie terrestre  casi no es atenuada por el paso del Sol a través de una NMG.

Ahora analizamos el segundo efecto no-gravitatorio, originalmente propuesto por los astrónomos ingleses Fred  Hoyle\index{Hoyle, F.} y Raymond Lyttleton\index{Lyttleton, R.A.}, que puede aumentar la luminosidad solar.  El flujo de moléculas de $H_{2}$ que choca contra la superficie del Sol es dado por la fórmula (\ref{FlujoH2}), donde debemos reemplazar $v_{E}$ por $v_{\odot}$, y la energía cinética de cada molécula de $H_{2}$ que se deposita sobre el Sol es $\frac{1}{2} m_{H_{2}} v_{\odot}^{2}$. Por lo tanto, la energía cinética  por unidad de tiempo y de área que se disipa en la superficie del Sol y se transforma en energía radiante es $(\frac{1}{2} m_{H_{2}} v_{\odot}^{2})( \frac{1}{4} n v_{\odot})$, la cual multiplicada por la superficie del Sol, $S_{\odot}$, nos da que la variación de la luminosidad solar es .
 \begin{equation}
 \Delta L_{\odot}=\frac{1}{8} n\, m_{H_{2}}\, v_{\odot}^{3}\, S_{\odot}.
 \label{VariacionLuminosidad}
 \end{equation}
Con $n=300$ $H_{2}$ por cm$^{3}$ y $v_{\odot}=20$ km s$^{-1}$, valores arriba adoptados,   obtenemos de la fórmula (\ref{VariacionLuminosidad}) que $\Delta L_{\odot}=6\times 10^{17}$ Watts y, por lo tanto, $\Delta L_{\odot}$ debida a este proceso de acreción es insignificante comparada con la luminosidad total del Sol $(L_{\odot}=3.85 10^{26}$ Watts).

Si bien los dos efectos no gravitacionales arriba analizados no afectan el balance energético y la temperatura de la Tierra, las moléculas y partículas de polvo interestelares que ingresan a la atmósfera pueden  producir  efectos  químicos en el gas atmosférico. Por ello, a continuación, analizamos este aspecto. La masa total de polvo  ${\cal M}_{p}$ y de gas molecular ${\cal M}_{H_{2}}$  que la superficie de la Tierra, $S_{\oplus}$, recoge de la NMG en un intervalo de tiempo $\Delta t$ son 

\begin{eqnarray}
{\cal M}_{p} & = & {\cal F}_{p}\, S_{\oplus}\, \Delta t\,\, m_{polvo}, \nonumber \\ 
{\cal M}_{H_{2}} & = &  {\cal F}_{H_{2}}\, S_{\oplus}\, \Delta t\,\,  m_{H_{2}}, 
\label{MasaPolvoGas}
\end{eqnarray}
donde ${\cal F}_{p}$ es dado por (\ref{FlujoPolvo}) y ${\cal F}_{H_{2}}$  por (\ref{FlujoH2}). Si tomamos $\Delta t=$ 1 año e introducimos en (\ref{MasaPolvoGas}) el valor  de la masa $m_{polvo}$ del grano de polvo, que hemos adoptamos arriba, y el de la  masa $m_{H_{2}}$ de la molécula de $H_{2}$, 
obtenemos que ${\cal M}_{p}=1462$  y ${\cal M}_{H_{2}}=145505$ toneladas anuales.
 Esto muestra que ${\cal M}_{p}$ colectada de la NMG es relativamente pequeña, comparada con  la cantidad de polvo interplanetario que recibe la Tierra anualmente. Al chocar  con alta velocidad en la  atmósfera, las moléculas de $H_{2}$ de la NMG se excitan y reaccionan con el oxígeno atmosférico formando moléculas de agua $( H_{2} + \frac{1}{2} O_{2}=H_{2} O)$. Si dicha reacción química se produce eficientemente, la masa  ${\cal M}_{H_{2}O}$ de agua que se produce anualmente es
 \begin{equation}
  {\cal M}_{H_{2}O}=\frac{{\cal M}_{H_{2}}}{m_{H_{2}}} m_{H_{2}O},
  \label{MasaAgua}
 \end{equation}
 donde $m_{H_{2}O}$ es la masa de la molécula de agua $(=2.98\times 10^{-23} $ gramos). De la relación (\ref{MasaAgua}), obtenemos que ${\cal M}_{H_{2}O}$=13123 toneladas por año, con lo cual la producción anual de agua debido al encuentro con una NMG es pequeña  con respecto a  la masa total de la hidrósfera $(\approx1.4\times 10^{18}$ toneladas) y no alteraría el clima terrestre. La producción de agua se hace a expensas del oxígeno terrestre  y la cantidad ${\cal M}_{O}$ de  oxígeno libre extraído de la atmósfera anualmente se puede estimar mediante la siguiente expresión:
\begin{equation}
{\cal M}_{O}=\frac{{\cal M}_{H_{2}}}{m_{H_{2}}}\,\,  m_{O},
\label{MasaO}
\end{equation}
donde $m_{O}$ es la masa del átomo de oxígeno ($=2.656 \times 10^{-23}$ gramos). 
Por lo tanto, a partir de  la fórmula (\ref{MasaO}), obtenemos que ${\cal M}_{O}=11696$ toneladas por año, un valor insignificante frente a la masa total de la atmósfera terrestre, $\approx5.1 \times 10^{15}$ toneladas. En condiciones normales, durante el año, la biosfera podría reponer esa disminución del oxigeno en la atmósfera.  Debemos recordar, sin embargo, que estos procesos transcurren durante todo el tiempo que tarda el Sol en cruzar la NMG, que en un encuentro central es del orden de $4.9 \times 10^{6}$ años. En efecto,  los tamaños típicos $L$ de las NMG son  de 100 parsec y, por lo tanto, el tiempo de cruce  es $t_{cruce}=\frac{L}{v_{\odot}} \approx 4.9 \times 10^{6}$ años.

Las NMG son lugares donde se forman las estrellas y, en consecuencia,  contienen subestructuras gaseosas de muy altas densidades, denominadas también núcleos moleculares densos. Estas densas concentraciones de moléculas  tienen tamaños de 0.1 a 1 parsec y densidades típicas entre  $10^{4}$ y $10^{6}$ $H_{2}$ cm$^{-3}$ y aún mayores. Si, al atravesar una NMG, el Sol  se encuentra con un núcleo de  densidad $n \geq 3 \times 10^{7}$ $H_{2}$ cm$^{-3}$ se alteraría significativamente el clima global de la Tierra. Repitiendo los cálculos con las fórmulas (\ref{TauA}) y (\ref{TauTS}) para $\tau_{A}$ y $\tau_{TS}$, pero  ahora con $n \geq 3 \times 10^{7}$ $H_{2}$ cm$^{-3}$, obtenemos que $\tau =\tau_{A}+\tau_{TS} \geq 1.12$. Asimilando $\tau$ a $\tau_{1}$ o $\tau_{2}$ de las Figs. \ref{Modelo1} y \ref{Modelo2}, notamos que en esas condiciones se inicia una edad de hielo.  Si el diámetro del núcleo es $D_{n}=0.1$ pc, el tiempo de cruce del Sol es como máximo $t_{cruce}=\frac{D_{n}}{v_{\odot}}$ y, por lo tanto, $t_{cruce} \approx$ 4900 años, tiempo que puede considerarse como una estimación de  la duración de la edad de hielo. Por otro lado,  el aumento temporario  de  la luminosidad solar o constante solar, debido a  la acreción de gas del núcleo denso  por parte del Sol, es despreciable aun para   densidades $n$ del  núcleo molecular del orden de $10^{9}$ moléculas por cm$^{-3}$, similar a las densidades de los discos protoplanetarios. En efecto, de acuerdo con la fórmula (\ref{VariacionLuminosidad}), $\Delta L_{\odot}\sim 10^{-7}  L_{\odot}$.

\subsection{Efectos gravitacionales: fuerzas de marea \label{gravitacionales1}}

El encuentro del Sistema Solar con una nube interestelar masiva causa fuerzas de marea sobre la nube de Oort\index{Oort, nube de} que perturban las órbitas de los cometas. El propósito de esta sección   es presentar   las herramientas matemáticas básicas para el estudio de  las perturbaciones  gravitatorias sobre las órbitas de los cometas de la nube de Oort\index{Oort, nube de}, a causa de  un  encuentro  del Sistema Solar con una nube molecular gigante (NMG).

\begin{figure}
\includegraphics[scale=0.7]{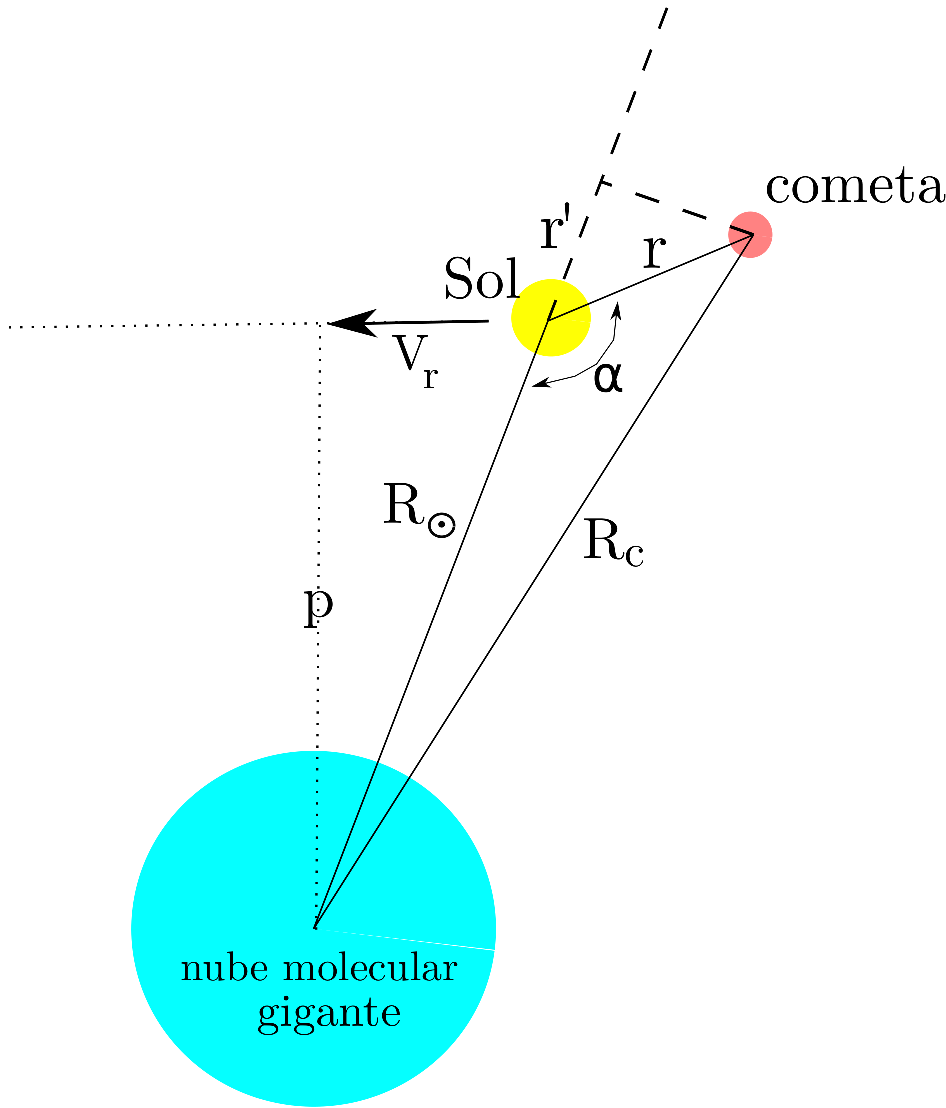} 
\caption{Esquema del encuentro del Sol (círculo amarillo) con una NMG (círculo celeste), donde se indican los principales parámetros y variables del encuentro que permiten calcular la órbita perturbada de un cometa de la nube de Oort (círculo rojo). }
\label{Marea} 
\end{figure}
Supongamos la situación mostrada en la Fig. \ref{Marea} en la cual el Sol se mueve en una trayectoria lineal con una velocidad  $V_{r}$ relativa a una NMG, y  la  distancia  más corta al centro de la NMG que el Sol alcanza en su trayectoria es $\cal P$. La distancia $\cal P$ es llamada parámetro de impacto. Suponemos además que el radio de la NMG es menor o igual a $\cal P$. Designamos con $R_{\odot}$ y $R_{c}$ las distancias al Sol y a un cometa, respectivamente, con respecto  al  centro de la NMG. Con $r$, denotamos la distancia del cometa con respecto al Sol. Las distancias $R_{\odot}$,  $R_{c}$ y $r$ formar los lados de un triángulo plano. Si llamamos $\alpha$  al ángulo interno con vértice en la posición del Sol,  el teorema del coseno  establece la siguiente relación: $R_{c}^{2}=R_{\odot}^{2}+ r^{2} - 2 R_{\odot} r \, cos \alpha
$. Dividiendo ambos miembros de dicha relación por $R_{\odot}^{2}$ y, teniendo en cuenta que $\frac{r}{R_{\odot}} \ll 1$, el término $ (\frac{r}{R_{\odot}})^{2} $ puede despreciarse y por lo tanto  $\frac{R_{c}}{R_{\odot}}=\sqrt{1-2 \frac{r}{R_{\odot}}cos \alpha} \approx 1- \frac{r}{R_{\odot}}cos \alpha$. Es decir, $R_{c}= R_{\odot}-r\, cos \alpha$ y si escribimos $r^{'}=-r\, cos \alpha$ (ver Fig. \ref{Marea}), obtenemos

\begin{equation}
R_{c}= R_{\odot}+r^{'}.
\label{RcMarea}
\end{equation}

Como consecuencia de la fuerza gravitatoria que ejerce la NMG sobre el Sol, la aceleración $a_{\odot}$ del Sol relativa a la NMG  es $a_{\odot}=-G \frac{M_{NMG}}{R_{\odot}^{2}}$, donde $M_{NMG}$ es la masa de la NMG. Por otra parte, la aceleración $a_{c}$ del  cometa relativa a la NMG es $a_{c}=-G \frac{M_{NMG}}{R_{c}^{2}}$, y teniendo en cuenta la fórmula (\ref{RcMarea}) $a_{c}=-G \frac{M_{NMG}}{(R_{\odot}+r^{'})^{2}}=-G \frac{M_{NMG}}{ R_{\odot}^{2}} \frac{1}{(1+\frac{r^{'}}{R_{\odot}})^{2}}$. Dado que   $\frac{r^{'}}{R_{\odot}} \ll 1$,  el factor $\frac{1}{(1+\frac{r^{'}}{R_{\odot}})^{2}} \approx (1-\frac{2 r^{'}}{R_{\odot}})$ y por lo tanto 
$a_{c}=-G \frac{M_{NMG}}{R_{\odot}^{2}} (1-\frac{2 r^{'}}{R_{\odot}})= -G \frac{M_{NMG}}{R_{\odot}^{2}} + 2 G \frac{M_{NMG}}{R_{\odot}^{3}} r^{'}$. Es decir,

\begin{equation}
a_{c}- a_{\odot} = 2 G \frac{M_{NMG}}{R_{\odot}^{3}} r^{'}.
\label{aSolac}
\end{equation}

En el caso representado en la Fig. \ref{Marea}, $ \alpha >  90^{\circ}$ y por lo tanto $r^{'}$ es positivo, con lo cual la fórmula (\ref{aSolac}) nos indica que el cometa tiende a ser alejado del Sol. En otras palabras, el cometa tiende a caer hacia la NMG más lentamente que el Sol. Cuando el cometa se encuentra  más cercano a la NMG  que el Sol,  $\alpha <  90^{\circ}$ y por lo tanto $r^{'}$ es negativo. En este caso  la aceleración de caída del cometa es mayor que la del Sol, y también el cometa tiende a ser  alejado del Sol. Este es el efecto de marea que estira la nube de Oort\index{Oort, nube de}   en la dirección que une al Sol con la NMG. Como dicha dirección rota a medida que el Sol se desplaza en su trayectoria, también lo hace la marea afectando   al conjunto de los cometas de  la nube de Oort\index{Oort, nube de} .  

Si multiplicamos la aceleración del cometa con respecto al Sol, $a= a_{c}-a_{\odot}$, por un intervalo de tiempo pequeño $\Delta t$, obtenemos la variación o perturbación de la velocidad del cometa $\Delta V_{c}=2 G \frac{M_{NMG}}{R_{\odot}^{3}} r^{'} \Delta t$. Dado que   $\Delta t=\frac{\Delta s}{V_{r}}$, donde $\Delta s$ es un pequeño tramo de la trayectoria del Sol,  podemos escribir $\Delta V_{c} = 2 G \frac{M_{NMG}}{R_{\odot}^{3}} r^{'} \frac{\Delta s}{V_{r}}$. La mayor perturbación a las órbitas de los cometas se produce cuando la distancia entre el Sol y la NMG es mínima, o sea cuando $R_{\odot}= \cal P$,  en ese momento 

\begin{equation}
\Delta V_{c} = 2 G \frac{M_{NMG}}{{\cal P}^{3}} r^{'} \frac{\Delta s}{V_{r}}.
\label{ImpulsiveV}
\end{equation}

La fórmula  (\ref{ImpulsiveV}) expresa la  velocidad impulsiva impartida a un cometa y es muy útil para evaluar la influencia de los principales parámetros de un encuentro. Por ejemplo, cuanto menor es  la velocidad relativa $V_{r}$ del encuentro, mayores son las perturbaciones orbitales de los cometas. Sin embargo, la aproximación impulsiva aplicada arriba no es apropiada cuando la duración del encuentro es larga en comparación con el periodo orbital del cometa. Debemos tener en cuenta  que mientras actúan las fuerzas de marea el cometa órbita en torno al Sol, y por lo tanto debemos considerar también la atracción que ejerce el Sol sobre el cometa.

\begin{figure}
\includegraphics[scale=0.7]{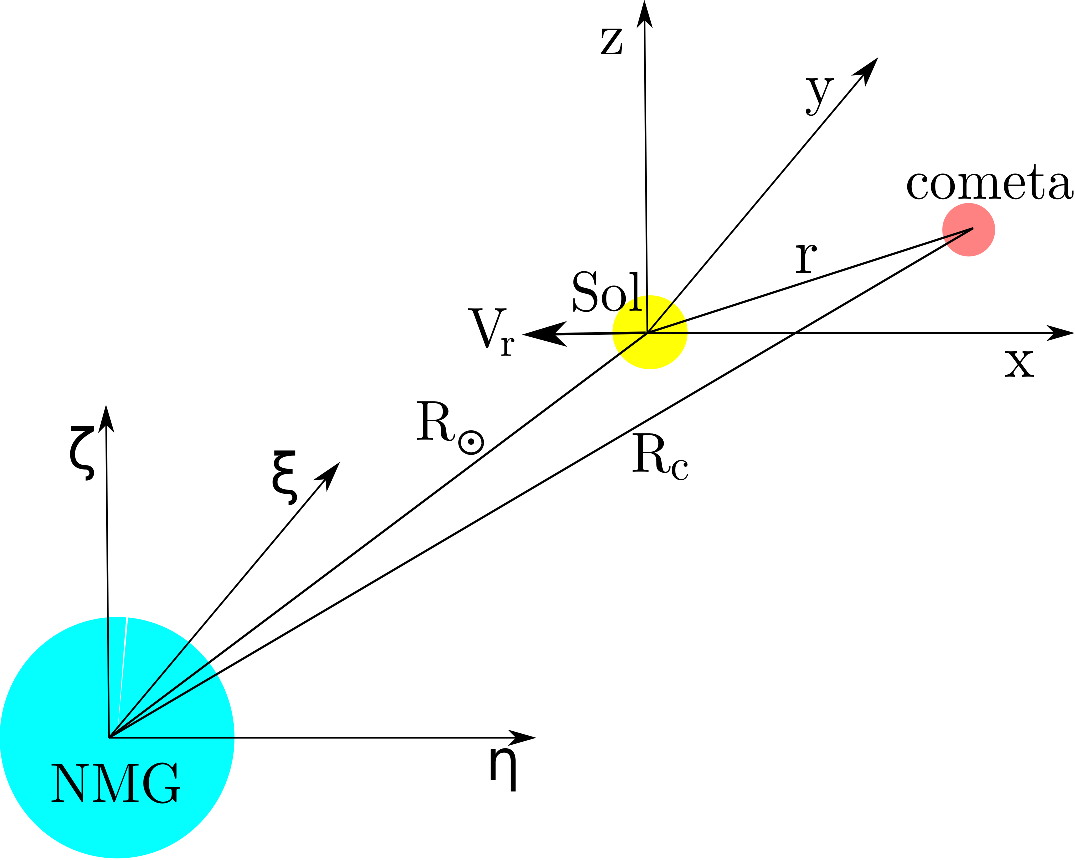} 
\caption{Representación de los dos sistemas Cartesianos utilizados para referir los movimientos de los cometas de la nube de Oort\index{Oort, nube de}: El sistema $(\eta, \xi, \zeta)$ con origen en el centro de la NMG y el sistema $(x,y,z)$ solidario con el Sol. Los detalles restantes son similares a aquellos de la Fig. \ref{Marea}.}
\label{Marea2} 
\end{figure}

En este contexto, la órbita de un cometa, perturbada por el encuentro con una NMG, puede  calcularse  planteando las ecuaciones de movimiento  del cometa en un sistema  inercial de referencia Cartesiano $(\eta, \xi, \zeta)$ con  origen en   el centro de la NMG (ver Fig. \ref{Marea2}). La aceleración del cometa debida a la fuerza que ejerce la NMG es un vector de magnitud $  G \frac{M_{NMG}}{R_{c}^{2}}$ y dirección dada por el vector unidad $-\frac{\vec{R_{c}}}{R_{c}}$ y similarmente la aceleración debida al Sol es  $ - G \frac{M_{\odot}}{r^{2}} \frac{\vec{r}}{r}$. Por lo tanto, la aceleración total del cometa es 

\begin{equation}
\vec{a_{c}}=- G \frac{M_{NMG}}{R_{c}^{2}} \frac{\vec{R_{c}}}{R_{c}} - G \frac{M_{\odot}}{r^{2}} \frac{\vec{r}}{r}.
\label{EcMovVec}
\end{equation}
Si $\eta, \xi, \zeta$  denotan las coordenadas Cartesianas del cometa y  sus derivadas  con respecto al tiempo se expresan con la notación de Newton\index{Newton, notación de} \footnote{Por ejemplo, en la notación de Newton,    la derivada primera y la derivada segunda de $\eta$ se expresan $\dot{\eta}$ y  $\ddot{\eta}$, respectivamente, las cuales  equivalen a  $\frac{d\eta}{dt}$ y $\frac{d}{dt} (\frac{d\eta}{dt})$ en la notación de  Leibniz\index{Leibniz, notación de}.}, $\vec{a_{c}}=(\ddot{\eta},\ddot{\xi},\ddot{\zeta})$, $\vec{R_{c}}=(\eta, \xi, \zeta)$, $\vec{r}=(\eta-\eta_{\odot}, \xi-\xi_{\odot}, \zeta-\zeta_{\odot})$, $R_{c}=\sqrt{\eta^{2}+ \xi^{2}+ \zeta^{2}}$, $r=\sqrt{(\eta-\eta_{\odot})^{2}+ (\xi-\xi_{\odot})^{2}+ (\zeta - \zeta_{\odot})^{2}}$, donde  $\eta_{\odot}$, $\xi_{\odot}$, $\zeta_{\odot}$ son las coordenadas Cartesianas del Sol. 

En el encuentro del Sol o Sistema Solar con una NMG, el Sol describe una órbita hiperbólica con foco en el centro de la NMG. Consideramos que el plano definido por los ejes coordenados $\eta$ y $ \zeta$ coincide con el plano orbital del Sol en torno a  la NMG. Por lo tanto,  $\zeta_{\odot}=0$, $\vec{r}=(\eta-\eta_{\odot}, \xi-\xi_{\odot}, \zeta)$ y $r=\sqrt{(\eta-\eta_{\odot})^{2}+ (\xi-\xi_{\odot})^{2}+ \zeta^{2}}$. Haciendo los reemplazos correspondientes en la ecuación (\ref{EcMovVec}), las ecuaciones de movimiento del cometa considerado resultan
\begin{eqnarray}
\ddot{\eta}+ G \frac{M_{NMG}}{R_{c}^{3}} \eta + G \frac{M_{\odot}}{r^{3}} (\eta-\eta_{\odot}) & = & 0, \nonumber \\
\ddot{\xi}+ G \frac{M_{NMG}}{R_{c}^{3}} \xi + G \frac{M_{\odot}}{r^{3}} (\xi-\xi_{\odot})  & = & 0, \nonumber \\
\ddot{\zeta}+ G \frac{M_{NMG}}{R_{c}^{3}} \zeta + G \frac{M_{\odot}}{r^{3}} \zeta & = & 0.
\label{EcsMov}
\end{eqnarray}

 La integración numérica del  sistema de ecuaciones (\ref{EcsMov}) provee una solución auto-consistente que incorpora automáticamente los efectos de marea. Note que para resolver las ecuaciones (\ref{EcsMov}),  debemos primero determinar las  posiciones $\eta_{\odot}$, $\xi_{\odot}$ del Sol en función del tiempo. Para ello, se deben resolver  las siguientes ecuaciones de movimiento
 \begin{eqnarray}
 \ddot{\eta}_{\odot}+ G \frac{M_{NMG}}{R_{\odot}^{3}} \eta_{\odot}  & = & 0, \nonumber \\
  \ddot{\xi}_{\odot}+ G \frac{M_{NMG}}{R_{\odot}^{3}} \xi_{\odot}  & = & 0, 
  \label{OrbSol}
\end{eqnarray}
 donde $R_{\odot}=\sqrt{\eta_{\odot}^{2}+ \xi_{\odot}^{2}}$. El sistema de ecuaciones (\ref{OrbSol}) para la atracción gravitatoria entre dos cuerpos, el Sol y la NMG, tiene una solución analítica tal como lo hemos dicho más arriba. Sin embargo, para nuestros propósitos  es suficiente resolver las ecuaciones (\ref{OrbSol}) numéricamente.
 
 En lo que sigue, supondremos, como representativos del encuentro del Sol con una NMG, los siguientes parámetros: $M_{NMG}=10^{4} M_{\odot}$ para la masa de la NMG y $\cal P$= 20 pc para el parámetro de impacto (ver Fig. \ref{Marea}). Para la posición  inicial  ($t=0$) del Sol adoptamos  $\eta_{\odot}(0)= 3\, \cal P$ y $\xi_{\odot}(0)= \cal P$, y para la velocidad inicial $\dot{\eta}_{\odot}(0)= -20$ km s$^{-1}$  y $\dot{\xi}_{\odot}(0)=0$ km s$^{-1}$. Ello implica que la dirección de  la velocidad $V_{r} (t=0) $ del Sol relativa a la NMG es  paralela al eje coordenado $\eta$ y que $V_{r}(0)=\dot{\eta}_{\odot}(0)= -20$ km s$^{-1}$.
 
 \begin{figure}
\includegraphics[scale=0.60]{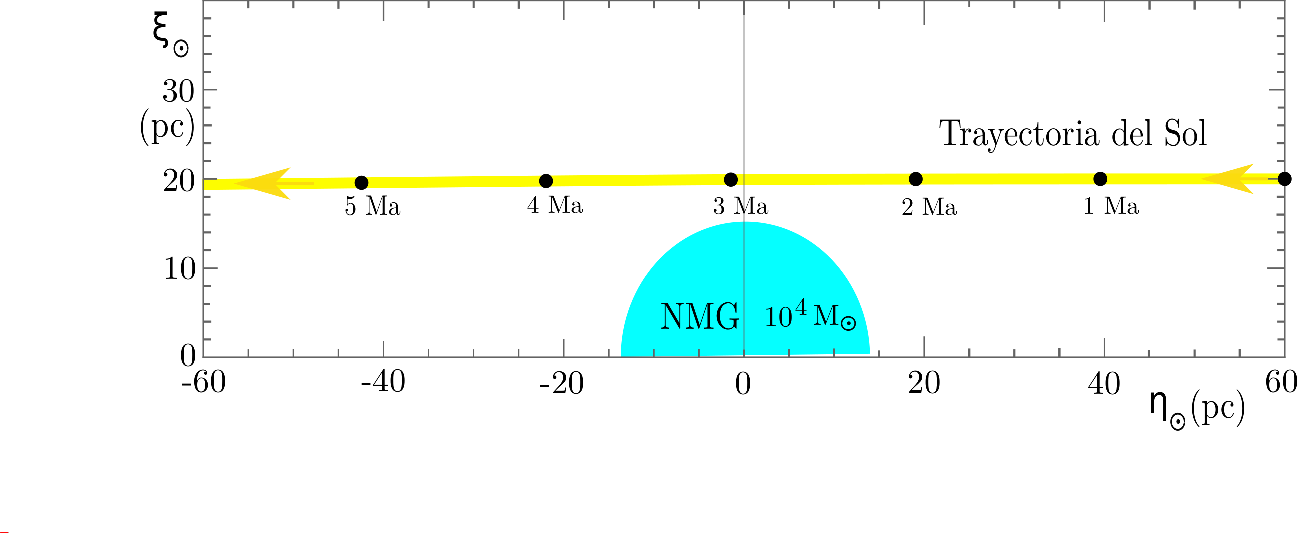} 
\caption{Órbita del Sol bajo el campo gravitatorio de una NMG de $10^{4} M_{\odot}$ con la cual el Sol se encuentra con un parámetro de impacto ${\cal P} =20$ pc y una velocidad relativa $V_{r}=20$ km s$^{-1}$. La órbita esta referida a un sistema de coordenadas $(\eta, \xi)$ con origen en el centro de la NMG. Los puntos negros muestran la posición del Sol cada millón de años.}
\label{SolNMG} 
\end{figure}
 
 La órbita del Sol obtenida con las ecuaciones (\ref{OrbSol}) y los parámetros y condiciones iniciales adoptados es representada en la Fig. \ref{SolNMG}, la cual muestra que la trayectoria del Sol apenas se desvía de una línea recta. También, la velocidad $V_{r}$ se mantiene casi constante. Por lo tanto,  la posición $(\eta_{\odot} ,\xi_{\odot} ,\zeta_{\odot})$ del Sol en función del tiempo $t$ es muy bien aproximada  por las siguientes fórmulas:
 
 \begin{eqnarray}
 \eta_{\odot} & = & 3\, {\cal P} + V_{r}\ t, \nonumber \\
 \xi_{\odot} & = & \cal P, \nonumber \\
 \zeta_{\odot} & = & 0.
 \label{PosSol}
 \end{eqnarray}
 
 Las posiciones y velocidades de los cometas  son normalmente referidas al Sol. Por ello es necesario definir un sistema de referencia centrado en el Sol, que se mueva solidario con el Sol y que nos permita relacionarlo con el sistema coordenado $(\eta, \xi, \zeta)$ centrado en la NMG. El sistema Cartesiano $(x,y,z)$ representado en la Fig. \ref{Marea2} cumple con dicho propósito. El eje $x$ y el eje $\eta$ son paralelos y tienen el mismo sentido y lo mismo ocurre con  los ejes  $y$ y $z$ y los respectivos ejes $\xi$ y $\zeta$. Además,  los ejes $x$ e $y$ yacen en el plano definido por los ejes $\eta$ y $\xi$, y la dirección del eje $x$ coincide con la trayectoria del Sol (ver Fig. \ref{SolNMG}). 
Por lo tanto, las fórmulas  de transformación de un sistema de referencia al otro son
\begin{eqnarray}
 \eta  & = & x+ \eta_{\odot}, \nonumber \\
 \xi   & = & y+ \xi_{\odot},  \nonumber \\
 \zeta & = & z. 
 \label{TransCoor}
 \end{eqnarray}
 
 Reemplazando en las fórmulas (\ref{TransCoor}) las expresiones para $\eta_{\odot}$ y $\xi_{\odot}$ dadas por las fórmulas  (\ref{PosSol}), obtenemos

\begin{eqnarray}
\eta  & = & x+ 3\, {\cal P} + V_{r} t, \nonumber \\
\xi   & = & y+ \cal P,  \nonumber \\
\zeta & = & z. 
\label{TransCoor2}
\end{eqnarray} 
Las relaciones entre las velocidades de los dos sistemas coordenados se obtienen derivando con respecto al tiempo ambos miembros de las ecuaciones (\ref{TransCoor2}):
\begin{eqnarray}
\dot{\eta}  & = & \dot{x} + V_{r}, \nonumber \\
\dot{\xi}   & = & \dot{y},  \nonumber \\
\dot{\zeta}  & = & \dot{z}. 
\label{TransCoor3}
\end{eqnarray} 

 Un modo de determinar la órbita de un cometa es integrar numéricamente las ecuaciones (\ref{EcsMov}), para lo cual necesitamos conocer la posición  $(\eta(0),\xi(0), \zeta (0))$ y velocidad  $(\dot{\eta}(0), \dot{\xi}(0), \dot{\zeta}(0))$ iniciales, es decir en $t=0$. Si la posición y la velocidad iniciales del cometa  están dadas en el sistema $x$, $y$ y $z$, utilizamos las relaciones (\ref{TransCoor2}) y (\ref{TransCoor3}) para convertirlas:
, $(\eta(0),\xi(0), \zeta (0))=(x(0)+ 3 {\cal P},  y(0)+ {\cal P}, z(0))$ y $(\dot{\eta}(0), \dot{\xi}(0), \dot{\zeta}(0))=(\dot{x}(0) + V_{r}, \dot{y}(0), \dot{z}(0))$. Por otra parte, debemos reemplazar 
en  las ecuaciones (\ref{EcsMov}) las expresiones (\ref{PosSol}) para  $\eta_{\odot}$, $\xi_{\odot}$ y $\zeta_{\odot}$.

 Otro modo de proceder es convertir  las ecuaciones diferenciales (\ref{EcsMov}) 
en ecuaciones diferenciales de $x$, $y$ y $z$, utilizando las relaciones 
 (\ref{TransCoor}) y (\ref{TransCoor2}). En efecto, diferenciando con respecto al tiempo ambos miembros de las ecuaciones (\ref{TransCoor3}), obtenemos que $\ddot{\eta}=\ddot{x}$, $\ddot{\xi}=\ddot{y}$ y $\ddot{\zeta}=\ddot{z}$. Si además tenemos en cuenta las ecuaciones (\ref{TransCoor2}) y que  $ \eta - \eta_{\odot} = x$,  $\xi-\xi_{\odot} =  y$,  $\zeta  = z$,  las ecuaciones diferenciales (\ref{EcsMov})  se convierten en el siguiente sistema de ecuaciones:

 \begin{eqnarray}
\ddot{x}+ G \frac{M_{NMG}}{R_{c}^{3}} (x+ 3 {\cal P} + V_{r} t) + G \frac{M_{\odot}}{r^{3}} x & = & 0, \nonumber \\
\ddot{y}+ G \frac{M_{NMG}}{R_{c}^{3}} (y + {\cal P})  + G \frac{M_{\odot}}{r^{3}} y  & = & 0, \nonumber \\
\ddot{z}+ G \frac{M_{NMG}}{R_{c}^{3}} z + G \frac{M_{\odot}}{r^{3}} z & = & 0,
\label{EcsMov2}
\end{eqnarray}
 donde  $R_{c}=\sqrt{(x+ 3 {\cal P} + V_{r} t)^{2}+ (y + {\cal P})^{2}+ z^{2}}$ y $r=\sqrt{x^{2}+ y^{2}+ z^{2}}$.
 
 \begin{figure}
\includegraphics[scale=0.80]{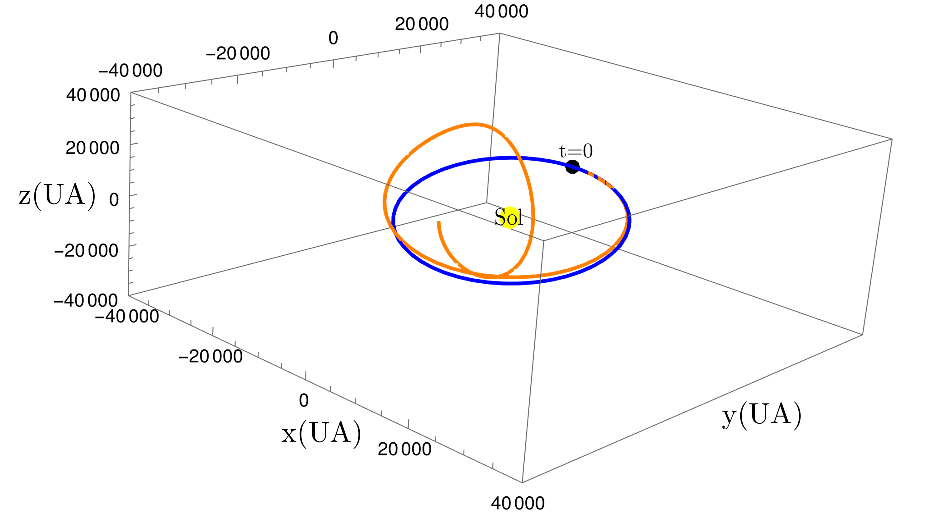} 
\caption{Comparación de la órbita de un cometa de la nube de Oort\index{Oort, nube de}. libre de perturbaciones gravitatorias externas (línea azul), con la órbita del mismo cometa pero perturbada por el encuentro del Sistema Solar con una  NMG (línea roja). El punto negro indica la posición inicial del cometa, con una distancia al Sol $r=20000$ UA. La órbita trazada en azul (no perturbada)  es circular con el semieje $a=r=20000$ UA. La masa de la NMG, el parámetro de impacto y la velocidad relativa aquí adoptados son $10^{4} M_{\odot}$, 20 pc y 20 km s$^{-1}$, respectivamente.}
\label{ComSolGMC} 
\end{figure}

\begin{figure}
\includegraphics[scale=0.60]{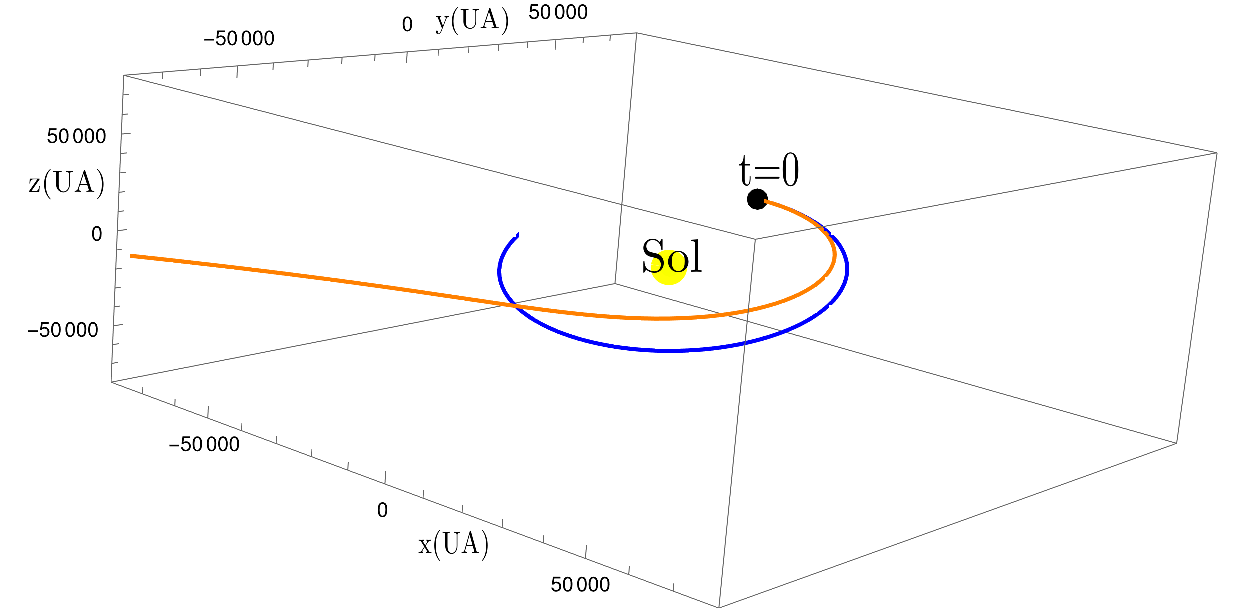} 
\caption{Igual a la Fig. \ref{ComSolGMC} pero con $r=40000$ UA para la distancia inicial entre el cometa y el Sol. La órbita trazada en azul (no perturbada)  es circular con el semieje $a=r=40000$ UA.}
\label{ComSolGMC2} 
\end{figure}

Como ejemplo, mostraremos los efectos del encuentro del Sol con la NMG sobre un cometa en órbita circular previo al encuentro. En esa situación, la velocidad circular del cometa es $V_{c}=\sqrt{\frac{G M_{\odot}}{r}}$, la cual  se puede derivar de la fórmula  (\ref{Vtotal0}) (ver Sección siguiente) haciendo $a=r$.  Por simplicidad, suponemos que en el tiempo $t=0$ la posición del cometa  se encuentra  en el plano formado por los ejes coordenados $y$ y $z$ y el radio vector  $\vec{r}$ del cometa forma un ángulo de $45^{\circ}$ con el eje $y$. Por lo tanto, la posición inicial del cometa es $(x(0),y(0), z(0))= (0, r \, cos\, 45^{\circ},  r\, sen \, 45^{\circ})$. Dado que el vector velocidad $\vec{V}_{c}$ debe ser perpendicular al radio vector $\vec{r}=(0, r\, cos\, 45^{\circ},  r\, sen \, 45^{\circ})$, $\vec{V}_{c}=(V_{c},0,0)$ y por lo tanto la velocidad inicial del cometa es $(\dot{x}(0),\dot{y}(0),\dot{z}(0))=(V_{c},0,0)$. Especificando el valor de  $r$ inicial en las condiciones iniciales dadas, podemos aplicar el sistema de ecuaciones (\ref{EcsMov2}) para determinar la órbita del cometa durante el encuentro. En la Fig. \ref{ComSolGMC}, se representa la órbita de un cometa con $r=20000$ UA en el inicio del encuentro, junto a la órbita del cometa si éste no hubiera sido perturbado (línea azul). En forma similar, se  representan  en la Fig.  \ref{ComSolGMC2} las órbitas de un cometa con $r=40000$ UA. En ambas figuras, se aprecia que las perturbaciones orbitales causadas por el encuentro son importantes.

\subsection{Efectos gravitacionales: lluvias de cometas sobre el interior del Sistema Solar \label{ gravitacionales2}}

\begin{figure}
\includegraphics[scale=0.5]{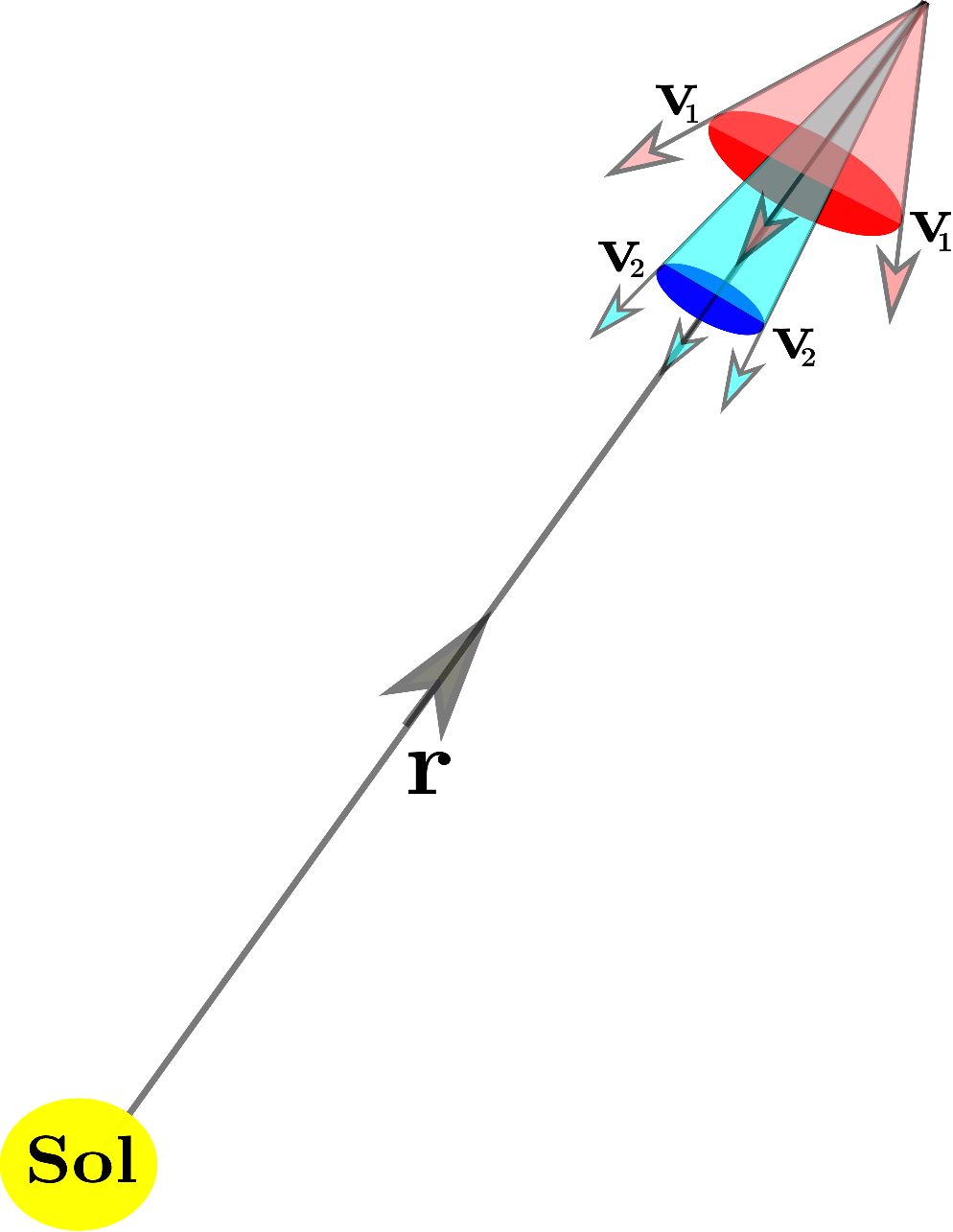} 
\caption{Esquema de dos conos de pérdida de cometas que se encuentran en la posición dada por el radio vector $\vec{r}$ con origen en el Sol. Los cometas con  velocidades $\vec{V_{1}}$ que yacen dentro del cono rojo y aquellos con  velocidades $\vec{V_{2}}$ que yacen dentro del cono azul tienen órbitas que cruzan el Sistema Solar interno y pueden por lo tanto colisionar con la Tierra. La figura muestra que el ángulo de apertura del cono de pérdida en una dada distancia $\mid \vec{r} \mid$ depende de la velocidad  $\vec{V}$ de los cometas, a mayor velocidad menor es el ángulo de apertura de los conos.}
\label{ConoDisparo} 
\end{figure}

En esta sección estudiamos la posibilidad de que los encuentros del Sistema Solar con nubes moleculares gigantes (NMG) perturben gravitacionalmente  la nube de cometas de Oort\index{Oort, nube de}, e induzcan  una lluvia de cometas sobre el Sistema Solar y, en particular, sobre la Tierra. Para ello, antes estudiamos el concepto de ``conos de pérdida''. Dentro de una cierta relativamente pequeña porción de la nube de Oort\index{Oort, nube de}, cuya distancia al Sol es especificada por el vector posición o radio vector $\vec{r}$, los cometas con órbitas que cruzan el Sistema Solar tienen  vectores velocidad $\vec{v}$  que generan conos con un eje común coincidente con la dirección de $\vec{r}$ (ver Fig. \ref{ConoDisparo}). Sin embargo, eso implica que las órbitas de los cometas con velocidades dentro  dichos conos de velocidades pueden  ser alteradas al interactuar gravitacionalmente con los grandes planetas, principalmente con Júpiter y Saturno. En consecuencia,  luego de varias revoluciones orbitales de dichos cometas, los conos de velocidades se vaciarían y el flujo de cometas sobre el Sistema Solar disminuiría significativamente. El paso de estrellas  y de NMGs por las cercanías del Sol agita la  nube de Oort\index{Oort, nube de} y rellenaría los conos de pérdidas de la nube de cometas, con lo cual el flujo de cometas sobre el Sistema Solar aumentaría abruptamente.

\begin{figure}
\includegraphics[scale=0.65]{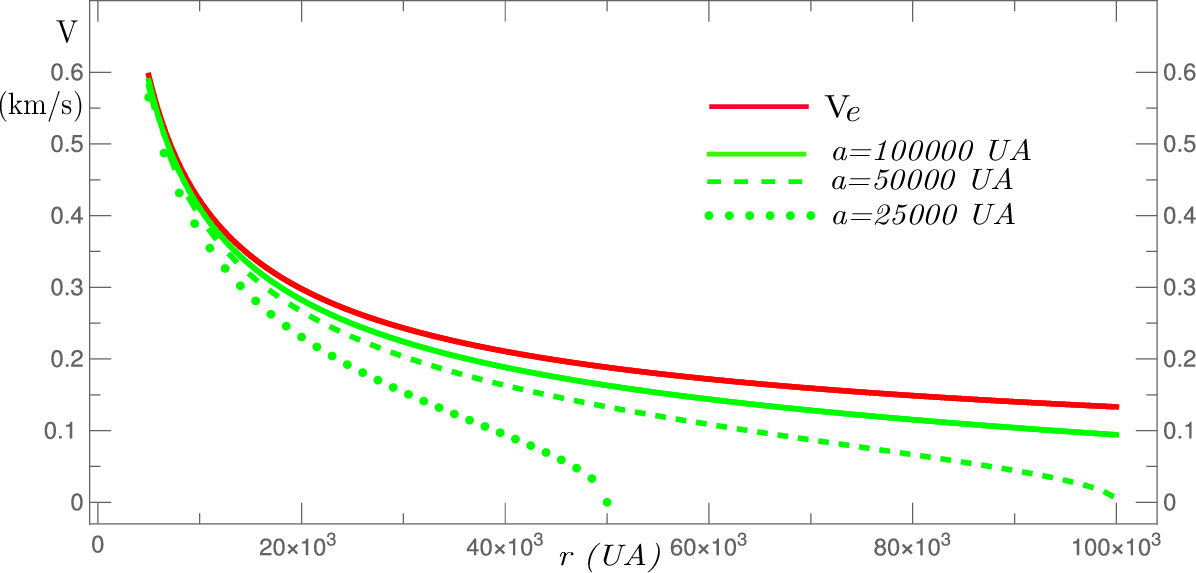} 
\caption{La curva roja muestra la velocidad de escape $V_{e}$ de los cometas de la nube de Oort\index{Oort, nube de} en función de las distancias $r$ de los cometas al Sol. Las demás curvas (en verde) muestran la velocidad $V$ en función de $r$ para diferentes valores de los semiejes $a$ de sus órbitas. La línea  llena, de a trazos y de puntos corresponden a $a=100000$, $a= 50000 $ y $a= 25000$ UA, respectivamente.}
\label{VescapeFig} 
\end{figure}

\begin{figure}
\includegraphics[scale=0.7]{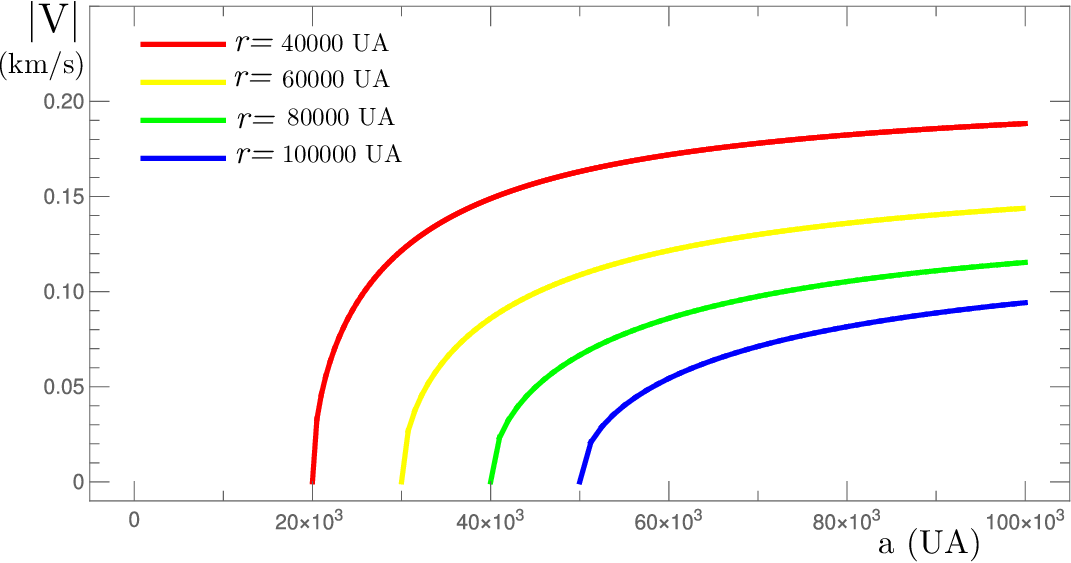} 
\caption{Curvas que representan $V$ en función de $a$ para diferentes valores de $r$. La curva roja, amarilla, verde y azul corresponden a los siguientes valores de $r$: 40000, 60000, 80000 y 100000 UA, respectivamente.}
\label{Vescape2bFig} 
\end{figure}

La órbita elíptica de un cometa en torno al Sol satisface la siguiente fórmula que relaciona la velocidad total $V (=\mid \vec{v} \mid)$ con el módulo del radio vector  $(r =\mid \vec{r} \mid)$ del cometa:
\begin{equation}
  V^{2}= \mu \left( \frac{2}{r}-\frac{1}{a} \right),
  \label{Vtotal0}
 \end{equation}
 donde $\mu= G M_{\odot}$, $G$ es la constante gravitatoria, $M_{\odot}$ la masa del Sol y $a$ el semieje mayor de la elipse. La fórmula (\ref{Vtotal0}) se puede aplicar al caso de una órbita circular de radio $r_{c}$, donde $a=r_{c}$, y por lo tanto $V_{c}=\sqrt{\frac{G M_{\odot}}{r_{c}}}$.  Con $a \rightarrow \infty$, la fórmula $(\ref{Vtotal0})$  nos proporciona la velocidad de escape $V_{e}$ de un cometa en la posición $r$, a saber
 
\begin{equation}
 V_{e}=\sqrt{\frac{2 G M_{\odot}}{r}}.
 \label{Vescape}
\end{equation}
Nótese  que  $V_{e}=\sqrt{2}\,\, V_{c}$. En la Fig. \ref{VescapeFig}, representamos   $V_{e}$ y $V$ en función de $r$, empleando las fórmulas $(\ref{Vtotal0})$ y $(\ref{Vescape})$. Como   $V=f(r,a)$, si fijamos un valor para $a$, podemos expresar $V$ en función de sólo $r$, es decir  $V=f(r)$. Las curvas $V=f(r)$ representadas  corresponden a cometas con $a=25000,  50000$ y $100000$ unidades astronómicas. La Fig. \ref{VescapeFig} muestra que los cometas con valores grandes de $a$ que se acercan mucho al Sol   tienen  velocidades que tienden a la velocidad de escape, o en otras palabras, tienen órbitas casi parabólicas con excentricidades $e\rightarrow 1$. Nótese además que los semiejes $a$ de cometas con órbitas elípticas que pasan por  una determinada posición $r$ cumplen la relación $a>\frac{r}{2}$.  Dicha relación se aprecia  en  la Fig.\ref{Vescape2bFig}, donde se representa $V$ en función de $a$ para   valores fijos de $r$.

 Sabemos que si se conocen $\vec{r}$ y $\vec{v}$ de un cometa en un instante de  tiempo, se pueden determinar los parámetros orbitales, $a$ y $e$, y el plano orbital del cometa. Si despejamos $a$ de la fórmula $(\ref{Vtotal0})$, obtenemos que 
 \begin{equation}
a = \left( \frac{2}{r}-\frac{V^{2}}{\mu} \right)^{-1}, 
\label{semiejeA}
\end{equation}
fórmula con la cual podemos determinar $a$ con solo  conocer $V$ y naturalmente $r$. En cambio, para determinar $e$, debemos conocer además la velocidad tangencial $V_{T}$, es decir la componente de la velocidad  $\vec{v}$ perpendicular a $\vec{r}$. A partir de la segunda ley de Kepler (ver Sección \ref{Cuaternario} y fórmula $\ref{LeyKepler2}$),  se deduce que 
\begin{equation}
V_{T}=\frac{h}{r},
\label{VelTang}
\end{equation}
donde $h$ es una constante particular para  cada órbita. Por otra parte, se demuestra que $h^{2}=\mu \, a (1-e^{2})$, de la cual se puede despejar  $e$ obteniéndose
\begin{equation}
e  = \sqrt{1-\frac{h^{2}}{\mu a}}. 
\label{Excentricidad}
\end{equation}
Para nuestros propósitos, es conveniente expresar $V_{T}$ en función del ángulo  $\alpha$ entre $\vec{r}$ y $\vec{v}$, dado que  $V_{T}=V sin\, \alpha$, con lo cual obtenemos de la fórmula $(\ref{VelTang})$ que $h=r V sin\, \alpha$. Con esta expresión  para $h$ reemplazada en ($\ref{Excentricidad}$), obtenemos que
\begin{equation}
e  = \sqrt{1-\frac{(r V sin\, \alpha)^{2}}{\mu a}}. 
\label{Excentricidad2}
\end{equation}

Introduciendo el valor de  $r$, $V$ y $\alpha$ de un cometa en las fórmulas $(\ref{semiejeA})$ y $(\ref{Excentricidad2})$, determinamos sus parámetros orbitales $a$ y $e$ y, por lo tanto, la órbita del cometa. Sin embargo, queda indeterminado el plano orbital del cometa. El vector de la posición $\vec{r}$ y de la velocidad $\vec{v}$ se encuentran sobre el plano orbital y por lo tanto lo definen. Para ello, expresamos las componentes de dichos vectores en un sistema de coordenadas cartesianas X-Y-Z centrado en el Sol. Con respecto a dicho sistema X-Y-Z, referimos también las coordenadas esféricas $l$ y $b$ (longitud y latitud),  que especifican  la posición celeste del cometa.
\begin{figure}
\includegraphics[scale=0.7]{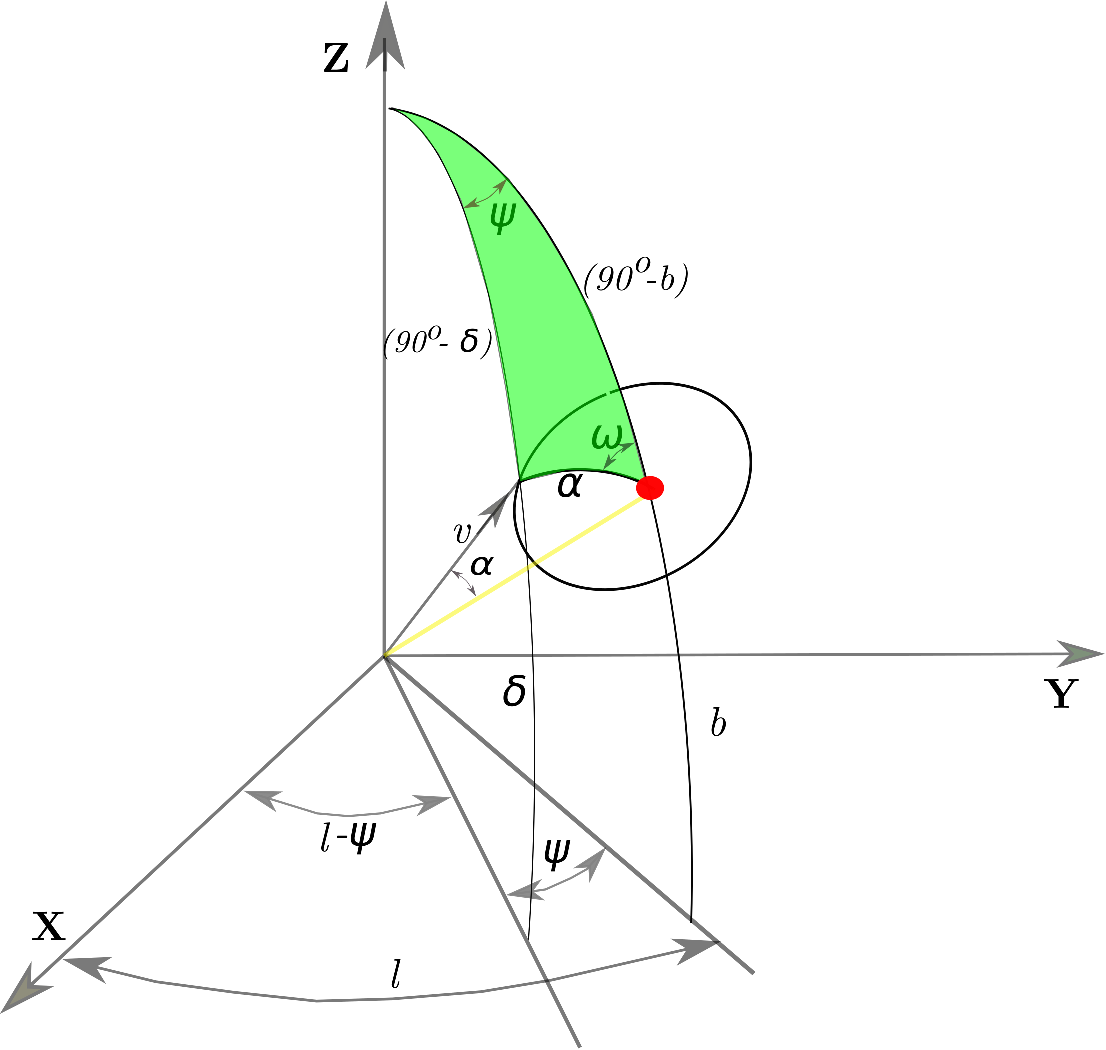} 
\caption{Relación entre las coordenadas cartesianas X-Y-Z con centro en el Sol  y las esféricas de un cometa cuyo radio vector $r$ apunta en la dirección indicada por la recta amarilla y el punto rojo. La solución del triángulo esférico sombreado en verde permite expresar las componentes cartesianas de la velocidad $\vec{v}$ del cometa en función de las coordenadas celestes $(l,b)$, y de los  ángulos $\alpha$ y $\omega$ del cometa.}
\label{ConoVelocidades} 
\end{figure}

Entonces, las componentes de $\vec{r}$ pueden escribirse del siguiente modo

\begin{eqnarray}
X & = & r\, cos \, b \,\,cos \, l, \nonumber \\
Y & = & r\, cos \, b \,\, sen  \, l, \nonumber \\
Z & = & r\, sen \, b.
\label{VecR}
\end{eqnarray}

Para escribir las componentes de $\vec{v}$ en forma conveniente debemos resolver el triángulo esférico de la Fig. \ref{ConoVelocidades} sombreado en verde, cuyos lados son $(90^{\circ}-b)$, $(90^{\circ}-\delta)$ y $\alpha$ y sus ángulos $\omega$ y $\psi$. El ángulo $\omega$ representa el ángulo entre el plano orbital, el cual contiene el lado $\alpha$, y el plano meridiano que pasa por  la posición ($l,b$). Adoptamos como variables independientes $l$, $b$, $\alpha$ y $\omega$ y expresamos las variables restantes, $\delta$ y $\psi$, en función de las variables independientes.  Aplicando el teorema del coseno al triángulo  esférico de la Fig.  \ref{ConoVelocidades}, sombreado en verde, obtenemos 
\begin{eqnarray}
sin\, \delta & = & cos \,\alpha \,\, sin\, b + sin\, \alpha \,\,cos\, b \,\,cos\, \omega, \nonumber \\
cos\, \delta & = & \sqrt{1-(sin \, \delta)^{2}},
\label{SenCosDelta}
\end{eqnarray}
donde la segunda fórmula se deriva de la identidad trigonométrica fundamental.
Aplicando el teorema del seno al  triángulo  esférico de marras, tenemos 

\begin{eqnarray}
sin \,\psi & = &\frac{sin \,\alpha \,\,sin \,\omega}{cos\, \delta}, \nonumber \\
cos\, \psi & = & \sqrt{1-(sin \, \psi)^{2}},
\label{Cospsi}
\end{eqnarray}
donde ${cos\, \delta}$ se calcula con las fórmulas (\ref{SenCosDelta}) y $cos\, \psi$ de la relación de identidad trigonométrica.

Ahora estamos en condiciones de obtener las componentes de $\vec{v}$. Para ello, proyectamos $\vec{v}$ sobre los ejes coordenados X, Y y Z, teniendo en cuenta que $(l-\psi)$ y $\delta$ son la longitud y latitud respectivamente de la dirección de $\vec{v}$, con lo cual $v_{X}= V\,\, cos\, \delta \,\, cos\, (l-\psi)$, $v_{Y}= V \,\, cos\, \delta\,\, sen\, (l-\psi)$ y $v_{Z}=V\,\, sin \, \delta$. Recordemos que $V=\mid \vec{v} \mid$.
Utilizando las identidades trigonométricas $cos\, (l-\psi)=cos \,l\,\, cos \, \psi + sin\, l \,\, sin \psi$ y $sin\, (l-\psi)=sin \,l\,\, cos \, \psi - cos\, l \,\, sin \psi$, podemos escribir

\begin{eqnarray}
v_{X} & = & V\,\, cos\, \delta\,\, (cos \,l\,\, cos \, \psi + sin\, l \,\, sin \psi), \nonumber \\
v_{Y} & = & V\,\, cos\, \delta\,\,(sin \,l\,\, cos \, \psi - cos\, l \,\, sin \psi), \nonumber \\
v_{Z} & = & V\,\, sin \, \delta.
\label{VecV}
\end{eqnarray}

\begin{figure}
\includegraphics[scale=1.0]{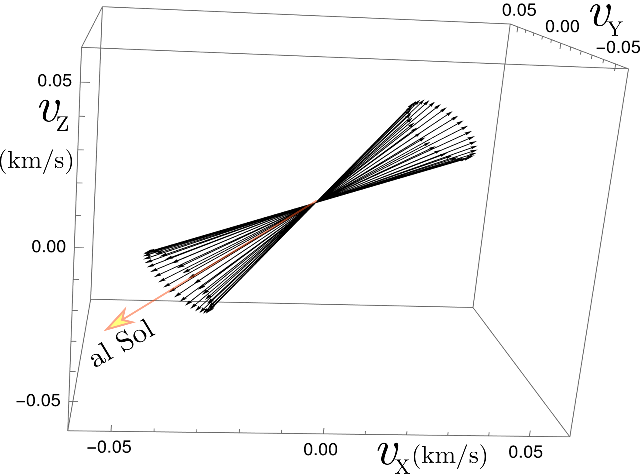} 
\caption{Vectores de velocidad $\vec{v}$ de cometas que comparten el mismo radio vector $\vec{r}$, el mismo módulo de velocidad  $V=\mid \vec{v} \mid (=0.05$ km s$^{-1})$ y el mismo ángulo de apertura $\alpha\, (=15^{\circ})$, entre $\vec{v}$ y $\vec{r}$.}
\label{ConoEjemplo} 
\end{figure}

Mediante las ecuaciones (\ref{VecV}) y auxiliares (\ref{SenCosDelta} y \ref{Cospsi}) calculamos distintos vectores   $\vec{v}$ de  un conjunto de cometas con el mismo vector posición $\vec{r}$. Por lo tanto, se fijan una  dirección $(l,b)$  y un modulo $r$  para   $\vec{r}$. Supongamos, como ejemplo, que todos los vectores   $\vec{v}$ tienen el mismo módulo o velocidad total $V=0.05$ km s$^ {-1}$ (o $V=-0.05$ km s$^ {-1}$) y forman el mismo ángulo $\alpha=15^{\circ}$ con $\vec{r}$. Suponemos además que el ángulo $\omega$  varía entre $0^{\circ}$ y $360^{\circ}$ (ver Fig. \ref{ConoVelocidades}). Los resultados de este ejemplo se muestran en la Fig. \ref{ConoEjemplo}, donde todos los vectores $\vec{v}$ tienen distintas direcciones y, sin embargo,  los elementos orbitales $a$ y $e$ de los correspondientes cometas son iguales. En efecto, de acuerdo con las fórmulas (\ref{semiejeA}) y (\ref{Excentricidad2}), $a$ y $e$ dependen solo de los valores de $r$, $V^{2}$ y $\alpha$, que se mantuvieron fijos. En cambio, los planos orbitales  son distintos, pues ellos son determinados por  $\omega$, que variamos entre $0^{\circ}$ y $360^{\circ}$. Un cometa puede orbitar en el sentido de las agujas de reloj o en sentido contrario, hecho que puede contemplarse  considerando que el signo de  $V$  puede ser tanto negativo como positivo. Por lo tanto, los vectores $\vec{v}$ generan dos superficies cónicas o conos de revolución que tienen el mismo eje en la dirección de  $\vec{r} $ y  un mismo  vértice, pero un cono  apunta en dirección al Sol (generado con $V=-0.05$ km s$^ {-1}$) y el otro apunta en la dirección opuesta (ver la Fig. \ref{ConoEjemplo}).
 
 La distancia $r_{p}$ entre el Sol y  el perihelio de la órbita elíptica de un cometa se relaciona con sus elementos orbitales $a$ y $e$  del siguiente modo:  
 \begin{equation}
 r_{p}=a\, (1-e),
 \label{Rperihelio}
\end{equation}  
y la distancia $r_{a}$ del afelio es dada por
  \begin{equation}
 r_{a}=a\, (1+e).
 \label{Rafelio}
\end{equation}  

\begin{figure}
\includegraphics[scale=0.9]{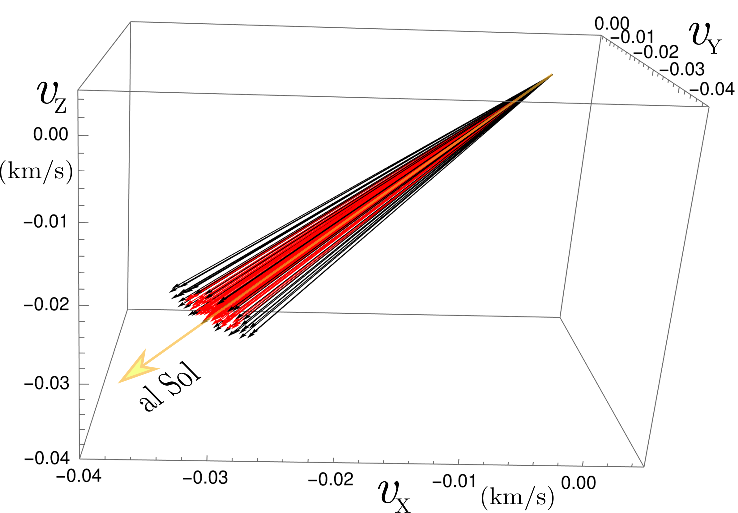} 
\caption{Lo mismo que la Fig. \ref{ConoEjemplo} pero con varios valores para $\alpha$. Aquí omitimos por claridad los conos que apuntan en dirección opuesta a la del Sol. Los vectores en rojo corresponden a cometas que se encuentran dentro del cono de pérdida y que por lo tanto tienen $\alpha< \alpha_{a}$, donde $\alpha_{a}$ es el ángulo de apertura que define el cono de pérdida.}
\label{ConoEjemplo2} 
\end{figure}

\begin{figure}
\includegraphics[scale=0.75]{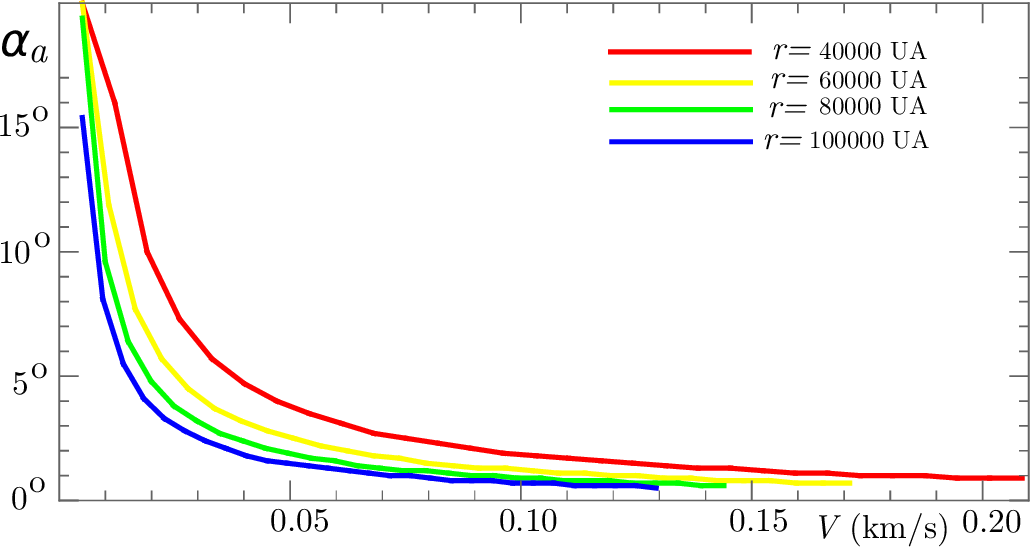} 
\caption{Ángulo de apertura $\alpha_{a}$ de los conos de pérdida en función de la velocidad $V$. Cada curva identificada por un color corresponde a una diferente distancia $r$ de los cometas.}
\label{AngApertura} 
\end{figure}
Las velocidades del cono representado en la Fig. \ref{ConoEjemplo} tienen   órbitas con los  mismos valores de $a$ y $e$  y, por lo tanto, con las mismas distancias $r_{p}$. Para que  un cono de velocidades defina o sea  parte de un cono de pérdida  exigimos que $r_{p} \leq 10$ UA, región en la cual  las órbitas de los cometas  pueden ser perturbadas por Júpiter y  Saturno. Si, en  el caso representado en  la Fig. \ref{ConoEjemplo}, variamos solo $\alpha$, obtendremos conos de la misma longitud $V$, y por lo tanto órbitas de igual $a$,  pero diferentes ángulos $\alpha$ de apertura, y por lo tanto órbitas  con diferentes $e$ (ver la fórmula \ref{Excentricidad2}). Llamaremos ángulo de apertura  $\alpha_{a}$ del cono de pérdida,  al ángulo $\alpha$ que implica un valor de $e$  con el cual  $r_{p}=10$ UA. Todos los conos interiores tienen  $\alpha<\alpha_{a}$ y $r_{p}< 10$ UA y son representados con vectores en rojo en la Fig. \ref{ConoEjemplo2}. En dicha figura, omitimos por claridad los conos de velocidades que apuntan en la dirección opuesta al Sol, donde los cometas se  acercan al Sol luego de dar la vuelta en el afelio.

Si el segundo término del radicando de la fórmula  (\ref{Excentricidad2}) es mucho menor que 1 , $e \approx  1- \frac{1}{2}\frac{(r V sin\, \alpha)^{2}}{\mu a}$.  Si escribimos $\epsilon=  \frac{1}{2}\frac{(r V sin\, \alpha)^{2}}{\mu a}$, el valor máximo de $\epsilon$,  $\epsilon_{\alpha_{a}}$, 	se  produce cuando $\alpha=\alpha_{a}$ y, en consecuencia, $r_{p}=10$ UA. Por lo tanto, $1-\epsilon_{\alpha_{a}} \leq e \leq 1$. En el ejemplo representado en las Figs. \ref{ConoEjemplo} y \ref{ConoEjemplo2}, $V=0.05$ km s$^{-1}$ y $r=40000$ UA, implicando que $a=21195$ UA, $\alpha_{a}=3.8^{\circ}$ y $\epsilon_{\alpha_{a}}=4.7 \times 10^{-4}$. Igual resultado para $e$, se obtiene despejando $e$ de la fórmula (\ref{Rperihelio}): $e=1- \frac{r_{p}}{a}$. En consecuencia, $\epsilon=\frac{r_{p}}{a}$ y, con  $r_{p}=10$ UA y  $a=21195$ UA, obtenemos que  $\epsilon=\epsilon_{\alpha_{a}}=4.7 \times 10^{-4}$. Es decir, las órbitas de esos cometas, antes de ser dispersados en el interior del Sistema Solar, son  cuasi parabólicas. 

La Fig. \ref{AngApertura} muestra la variación de $\alpha_{a}$ en función de $V$ y $r$. Fijado $r$, el ángulo de apertura  $\alpha_{a}$ aumenta al disminuir $V$, efecto que  también muestra esquemáticamente  la Fig. \ref{ConoDisparo}. Las curvas representadas en la Fig. \ref{AngApertura} para $\alpha_{a}$ se ajustan muy bien a la siguiente expresión 
\begin{equation}
\alpha_{a} (r,V) = c_{1} \frac{ e^{-c_{2} \mid V \mid}}{\mid V \mid^{c_{3}}},
\label{alfacp}
\end{equation}
donde $c_{1}$, $c_{2}$ y $c_{3}$ dependen solo de $r$ (ver Cuadro \ref{tablaAlfa}). Como $\alpha_{a} \leq 90^{\circ}$,  $\mid V \mid$  debe ser mayor que un cierto valor mínimo $\mid V_{min} \mid $, el cual dependiendo del valor de $r$ varia entre 0.001 y 0.01 km/s.

\begin{table*}
\centering
 \begin{minipage}{140mm}
  \caption{Valores de los parámetros $c_{1}$, $c_{2}$ y  $c_{3}$, para distintas distancias  $r$, que permiten  evaluar el ángulo $\alpha_{a}$ en grados de acuerdo con la fórmula (\ref{alfacp}), expresando $V$  en km/s. }
\centering
\begin{tabular}{ccccc} \hline \hline \\
$r(UA)$ & $c_{1} $ & $c_{2}$ & $c_{3}$   \\ \hline\hline
10000   & 0.5633  &-0.6054  & 1.0962          \\ \hline
20000   & 0.3627  & 0.0192  & 1.0162          \\ \hline
40000   & 0.1899  & 0.2618  & 1.0018        \\ \hline
60000   & 0.1296  & 0.5471  & 0.9966          \\ \hline
80000   & 0.0973  & 0.6232  & 0.9978        \\ \hline
100000  & 0.0768  & 0.6974  & 1.0000        \\  \hline \hline
\end{tabular}
\label{tablaAlfa}
\end{minipage}
\end{table*}

El ángulo de apertura  $\alpha_{a}$ de un cono de pérdida de velocidades $V$ nos permite calcular el ángulo sólido, $\Omega_{cp}$,  que subtiende el cono de pérdida. Por definición  $\Omega_{cp}=\frac{S}{V^{2}}$, donde $S$ es la superficie de la sección esférica definida por la intersección del cono con  la  esfera de radio $V$ centrada en el vértice del cono. Dado que $dS= V^{2} sen\,\alpha\, d\alpha\, d\omega$, $S= V^{2} \int_{0}^{\alpha_{a}} sen\,\alpha \, d\alpha\, \int_{0}^{2\pi} \,d\omega$ y, por lo tanto, 

\begin{equation}
\Omega_{cp}= 2 \pi (1-cos\, \alpha_{a}).
\label{omegacp}
\end{equation}
Si se supone que la nube de Oort\index{Oort, nube de}  está esféricamente distribuida  en torno al Sol y que ella no se encuentra  externamente perturbada, los conjuntos de conos de pérdida de diferentes partes del cielo son iguales y permanecen vacíos de cometas con las velocidades que abarcan los conos. En otras palabras, los conos de pérdida, que identificamos con los ángulos sólidos $\Omega_{cp}$, no dependen de la posición celeste $(l,b)$ pero si de la velocidad $V$ y del módulo del radio vector $\vec{r}$ ($i.e.\,\, \Omega_{cp} (r,V)$). 

A pesar de que la fórmula (\ref{omegacp}) es  sólo  estrictamente aplicable si la nube de Oort se encuentra en un estado estacionario, la aplicaremos en una forma simplista para estimar a grandes  rasgos la intensidad de la lluvia de cometas  que  se precipitaría sobre el interior del Sistema Solar    a causa de  la perturbación  gravitatoria ejercida sobre la nube de Oort durante  el encuentro cercano de  una estrella o de una nube interestelar con el Sistema Solar. Luego abordaremos el problema con mayor profundidad teniendo en cuenta la interacción dinámica entre el Sol, la nube de Oort y  una NMG.

Supongamos que al final del encuentro con  una estrella o una NMG, las variaciones de los parámetros  orbitales de la nube de Oort\index{Oort, nube de} son tales que los conos de pérdida, antes vacíos, se  llenaron de cometas. Como ahora se trata de un problema de solo dos cuerpos, el Sol y el cometa considerado, podemos aplicar la ecuación (\ref{omegacp}) para obtener la fracción  de cometas que forman parte de un cono de pérdida. Si $n_{c} (r)$ es el número total  de cometas de un elemento  volumen  ubicado en la posición $r$, el número de cometas cuyas velocidades yacen entre la esfera exterior de radio $V+dV$ y la esfera interior de radio $V$ es 
\begin{equation}
dn_{c} (r)= f(V)\, dV,
\label{dnc}
\end{equation}
donde $f(V)$ es la función distribución de velocidades, considerada isotrópica. Como el ángulo sólido de la esfera de radio $V$ es igual a $4 \pi$,  y el ángulo sólido del cono de pérdida a la velocidad $V$ es $\Omega_{cp}(r,V)$, la porción $dN_{cp}(r)$ de cometas  que se encuentra dentro de $\Omega_{cp}(r,V)$ es $dN_{cp}(r)= \frac{\Omega_{cp}(r,V)}{4 \pi} dn_{c} (r)$. Reemplazando en esa última fórmula $dn_{c} (r)$ por la ecuación ($\ref{dnc}$), obtenemos
\begin{equation}
dN_{cp}(r, V)= \frac{\Omega_{cp}(r,V)}{4 \pi} f(V) dV.
\label{dNcp}
\end{equation} 

La distribución $f(V)$ se puede en principio inferir  de la distribución $g(a)$ de los semiejes mayores $a$ de las órbitas de los cometas de la Nube de Oort, a través de la relación entre $V$ y $a$ dada por la fórmula (\ref{Vtotal0}). Si bien las distribuciones  de los parámetros orbitales de los cometas son pobremente conocidos, se estima que $g(a) \propto a^{-\gamma}$ con un valor de $\gamma$ entre 2 y 1. A semejanza de la fórmula (\ref{dnc}), podemos escribir

 \begin{equation}
 dn_{c} (r)= g(a)\, da.
 \label{dnca}
 \end{equation}

Utilizando la ecuación (\ref{semiejeA}), podemos expresar $g(a)$ y $da$ en función de $V$. En efecto, $g(a) \propto a^{-\gamma}=  \left( \frac{2}{r}-\frac{V^{2}}{\mu} \right)^{\gamma}$ y diferenciando la ecuación (\ref{semiejeA}) con respecto a $V$ obtenemos $da= \frac{2 V}{\mu \left( \frac{2}{r}-\frac{V^{2}}{\mu}  \right)^{2}}\,\, dV$, con lo cual la fórmula (\ref{dnca}) resulta

\begin{equation}
 dn_{c} (r)= C \left( \frac{2}{r}-\frac{V^{2}}{\mu} \right)^{\gamma} \, \frac{2 V}{\mu \left( \frac{2}{r}-\frac{V^{2}}{\mu} \right)^{2}}\, \, dV,
 \label{dncV}
 \end{equation}
 donde $C$ es una constante dentro del elemento de volumen considerado.
 Comparando las fórmulas (\ref{dnc}) y (\ref{dncV}), encontramos que
 
 \begin{equation}
 f(V)= C \left( \frac{2}{r}-\frac{V^{2}}{\mu} \right)^{\gamma} \, \frac{2 V}{\mu \left( \frac{2}{r}-\frac{V^{2}}{\mu} \right)^{2}}.
 \label{fv}
 \end{equation}
 La condición de  que  la integración de la ecuación (\ref{dnc}) con la expresión (\ref{fv}) para  $f(V)$ debe ser igual a $n_{c} (r)$ nos permite obtener el valor de $C$ : $C= \frac{n_{c} (r)}{ \int_{V_{min}}^{V_{max}} \left( \frac{2}{r}-\frac{V^{2}}{\mu} \right)^{\gamma} \, \frac{2 V}{\mu \left( \frac{2}{r}-\frac{V^{2}}{\mu} \right)^{2}}\,\, dV}$, con lo cual $f(V)$ resulta
 
 \begin{equation}
  f(V)= n_{c} (r) \frac{ \left( \frac{2}{r}-\frac{V^{2}}{\mu} \right)^{\gamma} \, \frac{2 V}{\mu \left( \frac{2}{r}-\frac{V^{2}}{\mu} \right)^{2}}}{\int_{V_{min}}^{V_{max}} \left( \frac{2}{r}-\frac{V^{2}}{\mu} \right)^{\gamma} \, \frac{2 V}{\mu \left( \frac{2}{r}-\frac{V^{2}}{\mu} \right)^{2}}\,\, dV}.
 \label{fvT}
 \end{equation}
 
Integrando la ecuación (\ref{dNcp}) con el reemplazo de la ecuación (\ref{fvT}) obtenemos

\begin{equation}
N_{cp}(r)=\frac{ n_{c} (r)}{4 \pi} \frac{\int_{V_{min}}^{V_{max}} \left( \frac{2}{r}-\frac{V^{2}}{\mu} \right)^{\gamma} \, \frac{2 V}{\mu \left( \frac{2}{r}-\frac{V^{2}}{\mu} \right)^{2}} \Omega_{cp}(r,V) dV}{\int_{V_{min}}^{V_{max}} \left( \frac{2}{r}-\frac{V^{2}}{\mu} \right)^{\gamma} \, \frac{2 V}{\mu \left( \frac{2}{r}-\frac{V^{2}}{\mu} \right)^{2}}\,\, dV},
\label{Ncpr}
\end{equation}
donde $V_{min}=0$ y $V_{max}$ es función de $r$ y se puede  inferir de las curvas de la Fig. \ref{Vescape2bFig}.

Obsérvese que  la expresión anterior contiene el valor medio de los conos de pérdida en la posición $r$. En efecto, 

\begin{equation}
\overline{\Omega_{cp}(r)}=\frac{\int_{V_{min}}^{V_{max}} \left( \frac{2}{r}-\frac{V^{2}}{\mu} \right)^{\gamma} \, \frac{2 V}{\mu \left( \frac{2}{r}-\frac{V^{2}}{\mu} \right)^{2}} \Omega_{cp}(r,V) dV}{\int_{V_{min}}^{V_{max}} \left( \frac{2}{r}-\frac{V^{2}}{\mu} \right)^{\gamma} \, \frac{2 V}{\mu \left( \frac{2}{r}-\frac{V^{2}}{\mu} \right)^{2}}\,\, dV}.
\label{ConoPerdidaM}
\end{equation}

Por lo tanto, la expresión (\ref{Ncpr}) puede escribirse del siguiente modo:

\begin{equation}
N_{cp}(r)=n_{c} (r) \frac{\overline{\Omega_{cp}(r)}}{4 \pi}.
\label{NroCP}
\end{equation}

 En las últimas décadas, ha ganado fuerza la idea de que la nube de Oort posee un oculto y masivo núcleo interior. La existencia de ese hipotético reservorio de cometas permitiría resolver problemas relacionados con la formación del disco protoplanetario, con el flujo de cometas que ingresan hacia al interior del Sistema  Solar convirtiéndose  en cometas de corto periodo y con la resistencia a la desintegración de la nube de Oort\index{Oort, nube de}. Aquí veremos  que el propuesto núcleo interno de cometas o también llamado nube de Hills\index{Hills, nube de} \footnote{El astrónomo norteamericano Jack Gilbert  Hills\index{Hills, J.G.} fue uno de los primeros en proponer la existencia del núcleo interno de la nube de Oort\index{Oort, nube de}.} juega  también un papel importante en la formación de  lluvias de cometas inducidas por los encuentros con estrellas o con nubes interestelares. En  este contexto, asumimos que dentro de la nube de Hills\index{Hills, nube de}, $r \leq 40000$ UA, el exponente de la función distribución $g(a)$ es $\gamma=2$,  y que en la parte externa de la nube de Oort, $r> 40000$ UA,   el exponente de $g(a)$ es $\gamma=1$ . Los valores de  $\gamma$ adoptados están dentro del rango  predicho por ciertos modelos (ver  \citep{Bailey2} y las referencias que allí se encuentran).

\begin{figure}
\includegraphics[scale=0.75]{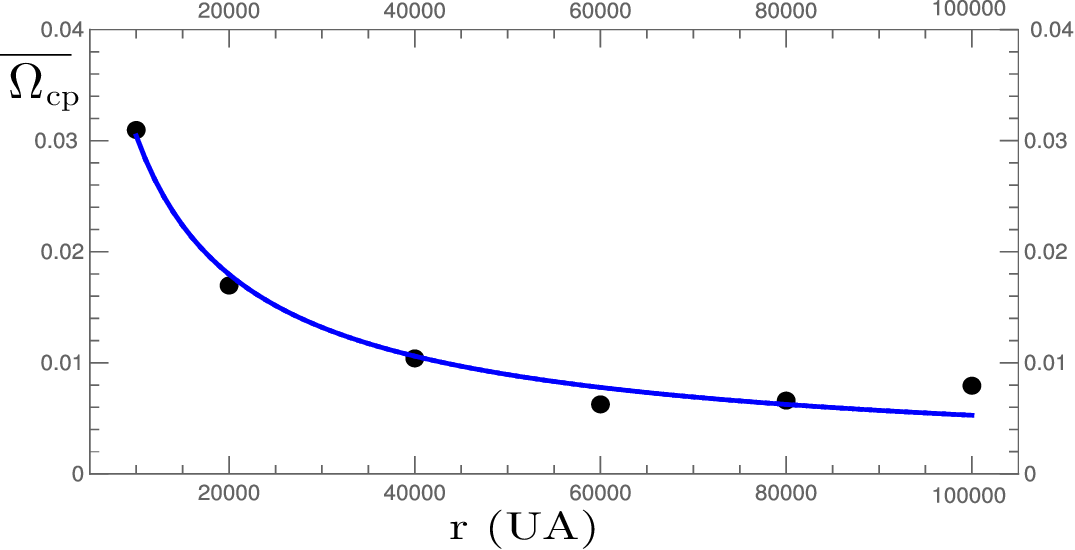} 
\caption{ $\overline{\Omega_{cp}}$ en función de $r$. Los puntos representan los valores de $(r, \overline{\Omega_{cp}})$ tomados de la columna 1 y 3 del cuadro \ref{tablaOmega}. La línea azul representa la función ajustada.}
\label{lluviaDibu1}
\end{figure}

 Los valores medios de $\Omega_{cp}$ para distintas posiciones $r$ de la Nube de Oort, calculados con la fórmula (\ref{ConoPerdidaM}) usando la ecuación (\ref{omegacp}) para $\Omega_{cp}(r,V)$ con $\alpha_{a}$ dada por la ecuación (\ref{alfacp}), son anotados en la tercera columna del Cuadro \ref{tablaOmega}. Con los valores $r$ de la columna (1) y  sus correspondientes $\overline{\Omega_{cp}}$ de la columna (3) del   Cuadro \ref{tablaOmega}, obtenemos los puntos $(r, \overline{\Omega_{cp}})$  representamos  en  la Fig. \ref{lluviaDibu1}, junto a la siguiente  función   que ajusta los puntos  razonablemente bien:
 \begin{equation}
 \overline{\Omega_{cp}(r)}  = 33.6074\,\,  r^{-0.7607},
 \label{ajusteOmegaCP}
\end{equation}
 donde $r$ se expresa en UA y $\overline{\Omega_{cp}}$ en estereorradianes.
 
 Se piensa que  la distribución de los cometas  es,  a grandes rasgos,  esférica con un  núcleo denso que rodea al Sol, la nube de Hills\index{Hills, nube de}, y con la densidad numérica de cometas que decrece radialmente hacia afuera de la nube de Oort.
 Se propone que la densidad numérica de cometas $n_{c}(r)$ se ajusta a una ley de potencia del tipo $r^{-n}$, donde $n>4$ \citep{Bailey2}. Nosotros adoptaremos conservadoramente  $n=4$, con lo cual $n_{c} (r)= r_{0}\, r^{-4}$  . La constante $r_{0}$ se obtiene de la condición $ N_{t}= \int_{r_{min}}^{r_{max}} n_{c} (r) 4 \pi r^{2} dr$, donde $N_{t}$ es el número total de cometas que contiene la nube de Oort\index{Oort, nube de}. Por lo tanto, 
 
 \begin{equation}
 n_{c} (r)= \frac{N_{t}}{4 \pi (\frac{1}{r_{min}}- \frac{1}{r_{max}})}\,\, r^{-4}.
 \label{Densir}
 \end{equation}
 
 En cada elemento de volumen, tenemos dos conos de pérdida de cometas, uno que apunta hacia el Sol y el otro en sentido contrario, y por lo tanto el número de cometas contenidos en dichos conos es $2\, N_{cp} (r)$ (ver ecuación \ref{NroCP}). En consecuencia,  el número de cometas contenidos en los conos de pérdida de un cascarón  de radio $r$ y de espesor $dr$ es $ 2\, N_{cp}(r)\, 4 \pi r^{2} \,dr$,  porción del total de  cometas  $ N_{ll}$ que integrará la ``lluvia''   de cometas sobre el interior del Sistema Solar. Reemplazando  $N_{cp}(r)$ por   la ecuación (\ref{NroCP}), obtenemos
 
 \begin{equation}
 dN_{ll}(r)= 2\, n_{c} (r)\,\, \overline{\Omega_{cp}(r)}\,\, r^{2} dr.
 \label{LluviaC}
\end{equation} 

 Reemplazando $n_{c} (r)$ y $\overline{\Omega_{cp}(r)}$ por la ecuaciones (\ref{Densir}) y 
(\ref{ajusteOmegaCP}), respectivamente, en la fórmula (\ref{LluviaC}) e integrando la expresión resultante, obtenemos
\begin{equation}
N_{ll}= \frac{33.6074\,\, N_{t}}{2 \pi (\frac{1}{r_{min}}- \frac{1}{r_{max}})} \int_{r_{min}}^{r_{max}}  r^{-2.7607} dr.
\label{Nlluvia}
\end{equation} 

Si adoptamos para el número total de  cometas, el radio mínimo y el radio máximo de la nube de Oort\index{Oort, nube de},  $ N_{t}=10^{14}$ cometas, $r_{min}=10000$ UA y $r_{max}=150000$ UA respectivamente, la fórmula (\ref{Nlluvia}) proporciona $N_{ll}=2.9 \times 10^{11}$ cometas, número   que refiere al total de cometas que confluyen hacia interior del Sistema Solar con perihelios $r_{p}<10$ UA. Si a pesar de las perturbaciones gravitatorias de los grandes planetas, cada cometa logra en promedio hacer  $n_{rr}$ revoluciones  en torno al Sol con $r_{p}<10$ UA, entonces el número de cometas que ingresan a la esfera de radio $r_{p}=10$ UA centrada en el Sol es $n_{rr}\, N_{ll}$. Si la distribución de las posiciones de los perihelios, transitadas a lo largo del tiempo, es uniforme dentro de la mencionada esfera, la densidad  de dichas posiciones es $ \frac{n_{rr}\, N_{ll}}{\frac{4}{3}\pi (10 UA)^{3}}$. La cantidad $N_{ll\oplus}$  de cometas con $r_{p}\leq 1$ UA, y que por lo tanto ingresan a la esfera de una UA de radio que circunscribe la órbita de la Tierra en torno al Sol,  es  $ \frac{n_{rr}\, N_{ll}}{\frac{4}{3}\pi (10 UA)^{3}} \frac{4}{3}\pi (1 UA)^{3}$. Es decir,

\begin{equation}
N_{ll\oplus}= \frac{n_{rr}\, N_{ll}}{10^{3}}.
\label{CometasTierra1}
\end{equation}
 
 En otras palabras, al terminar el proceso de vaciamiento de los conos de pérdida, el número total de cometas que incidieron  desde afuera en todas direcciones sobre la imaginaria superficie de la esfera de 1 UA de radio  es igual a $N_{ll\oplus}$. En general, cada cometa que ingresó al interior de dicha esfera egresó también de la misma. Si pensamos a cada cometa como un proyectil, el número de impactos que recibió la superficie que circunscribe la órbita terrestre es $2 \,\, N_{ll\oplus}$, es decir la superficie fue golpeada desde afuera y desde adentro. En consecuencia,  la Tierra  misma puede haber sido blanco de algunos de  dichos impactos. Podemos definir la probabilidad $p_{i}$ de que un cometa impacte sobre la Tierra como  la fracción del área que expone la Tierra como parte del  área de la esfera sometida al bombardeo:
 \begin{equation}
 p_{i}=\frac{\pi R_{T}^2}{4 \pi (1 UA)^2}  \,\,, 
 \label{Pi}
 \end{equation}
 donde $ R_{T}$ es el radio de la Tierra. Por lo tanto, el número probable de impactos ${\cal N}_{i}$ sobre la Tierra es ${\cal N}_{i}= 2 \,N_{ll\oplus} \,\, p_{i}$ y haciendo los reemplazos correspondientes
 
 \begin{equation}
 {\cal N}_{i}= \frac{R_{T}^2}{2 (1 UA)^2} \frac{n_{rr}\, N_{ll}}{ 10^{3}}.
 \label{NumImp1}
 \end{equation}
 
 Si adoptamos $n_{rr}=2$, y recordando que $N_{ll}=2.9 \times 10^{11}$ cometas, la fórmula (\ref{NumImp1}) da ${\cal N}_{i}=0.47$, es decir ``medio impacto'', resultado que debe interpretarse en el marco de un proceso donde ocurren dos posibles resultados:  impactos de igual probabilidad $p_{i}$ y no impactos con probabilidad $q_{i}=1-p_{i}$. En este caso, podemos aplicar la fórmula de Bernoulli\index{Bernoulli, fórmula de}, $P_{n}(k)=\frac{n!}{k! (n-k)!} p_{i}^k q_{i}^{n-k}$, para calcular la probabilidad de que ocurran $k$ impactos cometarios  sobre la Tierra  cuando el número total de cometas que potencialmente pueden chocar contra la Tierra   es 
 \begin{equation}
 n=2 \,\,N_{ll\oplus}.
 \label{tentativas}
 \end{equation}
 Dado que aquí $p_{i}=4.02
 \times 10^{-10}$ y $n=1.17 \times 10^{9}$, obtenemos  $P_{n}(0)=0.626$, $P_{n}(1)=0.293$, $P_{n}(2)=0.069$ y $P_{n}(3)=0.011$. Es decir, la probabilidad de que no ocurra ningún impacto es del $62.6 \%$ y la probabilidad de que ocurran uno o más impactos, $P_{n}(k\geq 1)= P_{n}(1)+P_{n}(2)+P_{n}(3)+...$, es del $37.5 \%$.  Note también que el valor medio de los impactos posibles es $1 P_{n}(1)+2 P_{n}(2)+3 P_{n}(3)={\cal N}_{i}=0.47$,  es decir a ``medio impacto''. En conclusión, bajo esa lluvia de cometas, la probabilidad de un impacto es significativa.
 
 El análisis  arriba desarrollado es en principio solo apropiado para describir el remanente de la tormenta de cometas. A continuación analizamos   la parte de  la tormenta que se desencadena  mientras tiene efecto  la perturbación gravitatoria externa sobre la nube de Oort\index{Oort, nube de}. Para ello, simularemos el encuentro del Sistema Solar con una típica NMG, utilizando el sistema de  ecuaciones (\ref{EcsMov2}) (o el sistema de  ecuaciones \ref{EcsMov}) desarrollada  en la Sección \ref{gravitacionales1} y las condiciones  del encuentro allí dadas, a saber: $M_{NMG}=10^{4} M_{\odot}$, ${\cal P}=20$ pc y $V_{r}=20$ km s$^{-1}$. 
 
 Suponemos que al comenzar el encuentro los conos de pérdida están vacíos de cometas. Sin embargo,  cometas con velocidades que yacen fuera de los conos de pérdida,  varían sus parámetros órbitales durante el transcurso del encuentro y parte de ellos pueden formar el equivalente a conos de pérdida y encaminar sus órbitas hacia el interior del Sistema Solar. 
 Por lo tanto, podemos aplicar los conceptos arriba desarrollados al caso de la nube de Oort en pleno estado de perturbación por el encuentro  con una NMG. En efecto, podemos considerar el  sistema coordenado X,Y,Z, definido en esta sección, coincidente con el sistema x,y,z definido en la Sección \ref{gravitacionales1}, y utilizar las ecuaciones  (\ref{VecR}) y (\ref{VecV}) para establecer las posiciones y velocidades iniciales de los cometas. Es decir, al comenzar el encuentro, $t=0$, $(x(0),y(0), z(0))=(X(0),Y(0),Z(0))$ y $(\dot{x}(0),\dot{y}(0), \dot{z}(0))=(v_{X}(0), v_{Y}(0),v_{Z}(0))$.

Si conocemos las posiciones y velocidades de los cometas de un elemento de volumen, identificado por su posición celeste $(l,b)$ \footnote{Las coordenadas esféricas $(l,b)$ se refieren al sistema particular definido en la sección (\ref{gravitacionales1}) y no se relacionan con la longitud y latitud de otros sistemas de coordenadas.} y su distancia $r$ al Sol, podemos determinar mediante el sistema de  ecuaciones (\ref{EcsMov2}) (o el sistema de  ecuaciones \ref{EcsMov}) las órbitas de esos cometas y así determinar si algunos de los cometas atraviesan el interior del Sistema Solar. Las posiciones iniciales de los cometas dentro del elemento de volumen  difieren relativamente muy poco  y por lo tanto sus componentes cartesianas se pueden considerar coincidentes e iguales a las dadas las fórmulas(\ref{VecR}). Las componentes de velocidades iniciales de los cometas pueden  obtenerse de la fórmulas (\ref{VecV}), variando $V$, $\alpha$ y $\omega$ de modo de cubrir todas las velocidades y direcciones posibles.

\begin{figure}
\includegraphics[scale=1.1]{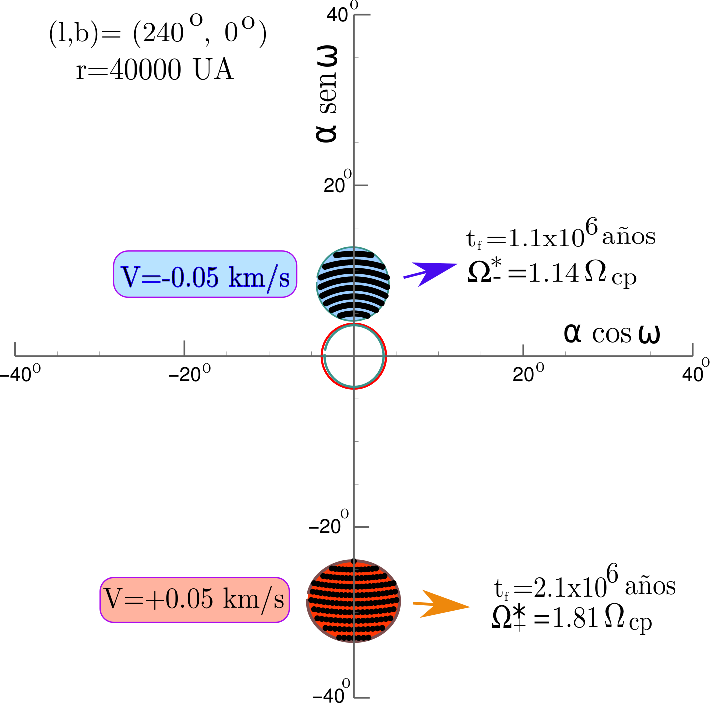} 
\caption{Conos de pérdida inducidos en la dirección $(l,b)=(240^{\circ}, 0^{\circ})$ y distancia $r=40000$ UA correspondientes a la velocidad $V=-0.05$ km s$^{-1}$ (sombreados en azul) y $V=+0.05$ km s$^{-1}$ (sombreados en rojo). Los dos círculos vacíos, uno de contorno azul y el otro rojo, ambos con centros coincidentes  en el origen del gráfico y el mismo radio indican los conos de pérdida $\Omega_{cp}$, correspondientes a la distancia $r=40000$ UA y la velocidad $V=-0.05$ km s$^{-1}$ (contorno azul)  y $V=+0.05$ km s$^{-1}$ (contorno rojo). Los tiempos $t_{f}$ indicados corresponden a los tiempos transcurridos desde el inicio del encuentro hasta los respectivos ingresos de los cometas al interior del Sistema Solar.} 
\label{LLuviaDibu2A}
\end{figure}

Un ejemplo del procedimiento es dado en la Fig. \ref{LLuviaDibu2A}, donde fijamos $r=40000$ UA, $(l,b)=(240^{\circ}, 0^{\circ})$ y $\mid V\mid=0.05$ km s$^{-1}$, y  variamos $\alpha$ entre  $\alpha_{a}$ y $90^{\circ}$ con un cierto paso y  similarmente para $\omega$ donde $0^{\circ}<\omega<360^{\circ}$. Es decir, obtenemos una red de puntos o posiciones $(\alpha,\omega)$  en el inicio del encuentro $(t=0)$ y calculamos para cada una de dichas  posiciones, teniendo en cuenta los parámetros fijados,  la órbita del cometa correspondiente.  Las posiciones $(\alpha,\omega)$ iniciales  que determinan órbitas de   cometas que  se acercan al Sol con $r_{ph}<10$ UA son indicadas por puntos en la  Fig. \ref{LLuviaDibu2A}. En dicha figura, el área que contiene un conjunto de puntos cercanos define un ángulo sólido o cono, que llamamos cono de pérdida  inducido y denotamos con un asterisco como superíndice, $\Omega^{\star}$. Suponemos que  los cometas contenidos en un cono $\Omega^{\star}$ se descargan completamente al ingresar por primera vez al interior del Sistema Solar.

\begin{figure}
\includegraphics[scale=0.9]{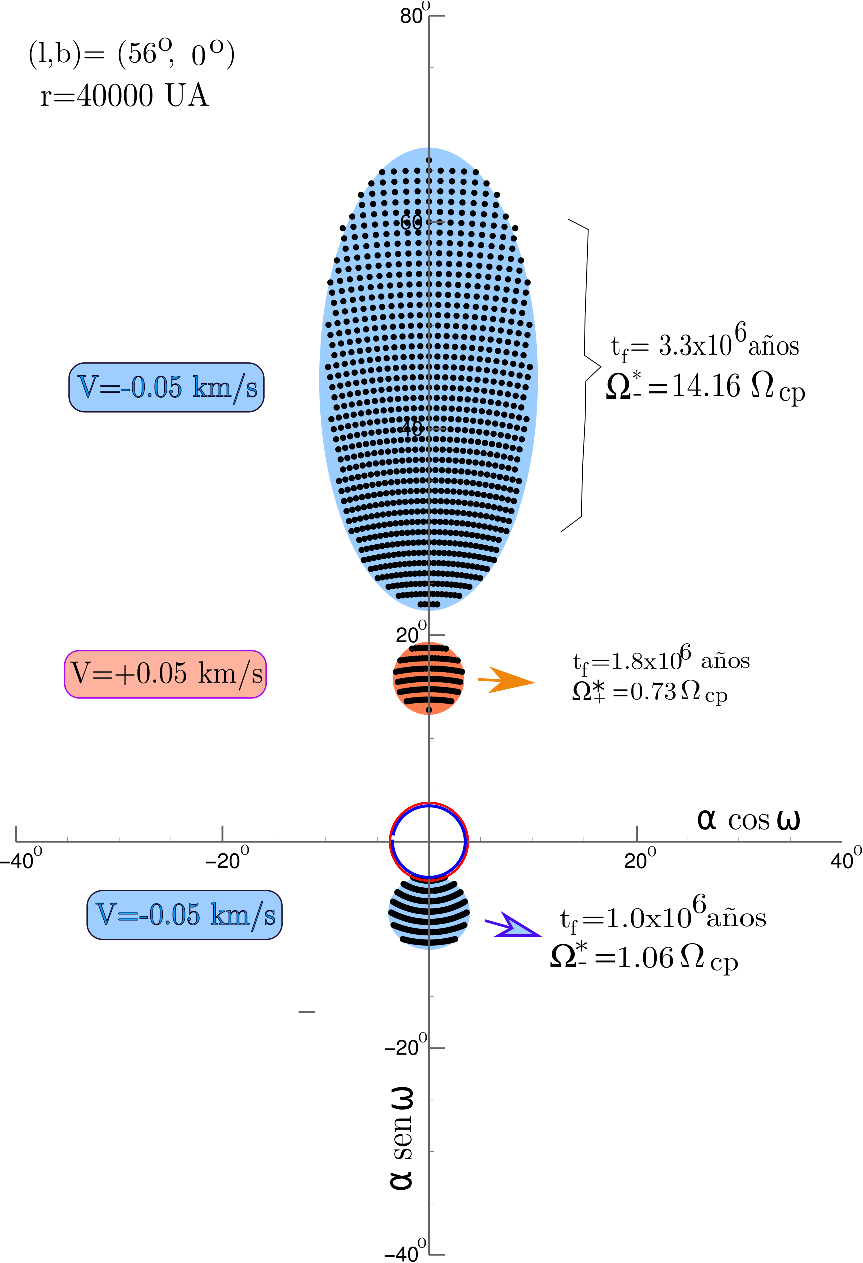} 
\caption{Lo mismo que la Fig. \ref{LLuviaDibu2A} pero para la dirección $(l,b)=(56^{\circ}, 0^{\circ})$.}
\label{LluviaDibu2B}
\end{figure}

A fin de calcular el flujo de cometas a través  del Sistema Solar, originado por los conos de pérdida inducidos, comenzamos por aplicar  la fórmula (\ref{ConoPerdidaM}), donde debemos reemplazar $\Omega_{cp}$ por  $\Omega^{\star}$. Para ello, debemos notar que, a diferencia de $\Omega_{cp}$, 
$\Omega^{\star}$ depende además de la dirección $(l,b)$. Es decir $\Omega^{\star} (r,V,l,b)$, tal como lo muestran las Figuras \ref{LLuviaDibu2A} y \ref{LluviaDibu2B}. Si fijamos las variables independientes  $r$ y  $V$ en ciertos valores, podemos calcular $\Omega^{\star}$ para una red de puntos $(l,b)$ que cubran el cielo con un espaciamiento en $b$ de $\Delta b\, (=15^{\circ})$ y en $l$ de $\frac{\Delta b}{cos b}$. y por lo tanto podemos calcular el ángulo de apertura promedio $ \overline{\Omega^{\star} (r,V)}$. El valor medio se calcula solo sobre  uno de los  hemisferios, $b>0$ o $b<0$, pues la distribución de $\Omega^{\star}$ es simétrica con respecto al plano $b=0$. Reemplazando $\overline{\Omega^{\star} (r,V)}$ por $\Omega_{cp} (r,V)$ en la fórmula (\ref{ConoPerdidaM}), obtenemos $\overline{\Omega^{\star} (r)}$ en lugar de  $\overline{\Omega_{cp} (r)}$. Como los conos de pérdida inducidos tienen en general distintos ángulos de apertura según apunten hacia el Sol (con velocidad V negativa) o en sentido contrario, los diferenciamos denotándolos  $\overline{\Omega^{\star}_{-} (r)}$ y $\overline{\Omega^{\star}_{+} (r)}$ a aquellos con $V<0$ y $V>0$ respectivamente. En la Tabla \ref{tablaOmega}, se dan los resultados de 
$\overline{\Omega^{\star}_{-}}$,  $\overline{\Omega^{\star}_{+}}$ 
 y la suma de ambos $\overline{\Omega^{\star}_{t}}\,  (= \overline{\Omega^{\star}_{-}}+ \overline{\Omega^{\star}_{+}}$) para diferentes valores de $r$. Note que el número de conos $\Omega^{\star}$ y sus amplitudes  por posición $(l,b)$ pueden ser mayor que las que corresponden a  $\Omega_{cp}$, tal como lo muestran las Figuras  \ref{LLuviaDibu2A} y \ref{LluviaDibu2B} y la última columna de la Tabla \ref{tablaOmega}.

 \begin{table*}
\centering
 \begin{minipage}{140mm}
  \caption{Conos de pérdida promedios con la nube de Oort no perturbada externamente (columna 3) y conos de pérdidas promedios, inducidos por el encuentro con una NMG (columnas 4, 5 y 6).  En la columna 2, se dan los exponentes $\gamma$ de la ley de potencia que utilizamos para la función distribución $g(a)$. Los valores de $\overline{\Omega_{cp}}$, $\overline{\Omega_{-}^{\star}}$, $\overline{\Omega_{+}^{\star}}$ y $\overline{\Omega_{t}^{\star}}$ se expresan en estereorradianes.}
\centering
\begin{tabular}{ccccccc} \hline \hline \\
$r(UA)$ & $\gamma$  & $\overline{\Omega_{cp}}$ & $\overline{\Omega_{-}^{\star}}$ & $\overline{\Omega_{+}^{\star}}$  &  $\overline{\Omega_{t}^{\star}}$ & $\frac{\overline{\Omega_{t}^{\star}}}{2 \overline{\Omega_{cp}}} $  \\ \hline\hline
10000 & 2 & 0.0310  & 0.0461  & 0.0467  & 0.0928   &  1.50\\ \hline
20000 & 2 & 0.0170  & 0.0397  & 0.0429  & 0.0822  &  2.42 \\ \hline
40000 & 2 & 0.0104  & 0.0141  & 0.0102  & 0.0243   &   1.17  \\ \hline
60000 & 1 & 0.0063  & 0.0037  & 0.0011  & 0.0048   &    0.38 \\ \hline
80000 & 1 & 0.0066  & 0.0025  & 0.0003  & 0.0028   &     0.21 \\ \hline
100000 & 1 & 0.0079 & 0.0007  & 0.0001  & 0.0008   &    0.05  \\  \hline \hline
\end{tabular}
\label{tablaOmega}
\end{minipage}
\end{table*} 
 
 Por semejanza con la fórmula (\ref{NroCP}), el número de cometas por unidad de volumen $N^{\star}(r)$ que se encuentran a la distancia $r$ y que integran los conos de pérdida inducidos es dado por
 
 \begin{equation}
N^{\star}(r)=n_{c} (r) \frac{\overline{\Omega_{t}^{\star}(r)}}{4 \pi}.
\label{NroCPI}
\end{equation}
 En consecuencia, la cantidad de cometas que ingresan al interior del Sistema Solar procedentes de un cascarón de radio $r$ y espesor $dr$ es $N^{\star}(r)\, 4 \pi r^{2} \,dr$, y reemplazando $N^{\star}(r)$ por su expresión (\ref{NroCPI}) y adaptando la expresión (\ref{LluviaC}) a este  caso de conos de pérdida inducidas podemos escribir: 
$dN_{ll}^{\star}= n_{c} (r)\,\, \overline{\Omega_{t}^{\star}(r)}\,\, r^{2} dr$. Por lo tanto,
\begin{equation}
N_{ll}^{\star}=  \int_{r_{min}}^{r_{max}} n_{c} (r)\,\, \overline{\Omega_{t}^{\star}(r)}\,\, r^{2}  dr.
\label{NlluviaCPI}
\end{equation} 

La integración que indica la fórmula (\ref{NlluviaCPI}) la realizamos con la función $ \overline{\Omega_{t}^{\star}(r)}$  obtenida con los datos de la sexta columna de la Tabla \ref{tablaOmega} y con la función $n_{c} (r)$ dada por la fórmula (\ref{Densir}), con lo cual obtenemos $N_{ll}^{\star}= 5.4 \times 10^{11}$ cometas. Por similitud con la fórmula (\ref{CometasTierra1}), el número $N^{\star}_{ll\oplus}$ de cometas contenidos en los conos de pérdida inducidos que atravesaron la esfera de 1 UA de radio  que contiene a la Tierra  puede escribirse:

\begin{equation}
N^{\star}_{ll\oplus}= \frac{N^{\star}_{ll}}{10^{3}}.
\label{CometasTierra2}
\end{equation}
Note que aquí $n_{rr}=1$, pues hemos supuesto  que cada cometa que ingresó al interior del Sistema Solar ($r<10$ UA) durante la interacción fuerte con la NMG, lo hizo solo una vez. Dado que $N_{ll}^{\star}= 5.4 \times 10^{11}$, $N^{\star}_{ll\oplus}=5.4 \times 10^{8}$ cometas. En forma similar al caso expresado por la fórmula (\ref{tentativas}), el número de ``tentativas'' de impactar sobre el blanco (la Tierra) es dado por $n=2 \,\,N^{\star}_{ll\oplus}=1.08 \times 10^{9}$ y la probabilidad $p_{i}$ de cada impacto es dada por la fórmula (\ref{Pi}). Aplicando nuevamente la fórmula de Bernoulli\index{Bernoulli, fórmula de}, obtenemos en este caso que   $P_{n}(0)=0.65$, $P_{n}(1)=0.28$, $P_{n}(2)=0.06$ y $P_{n}(3)=0.009$. Es decir, la probabilidad de que no ocurra ningún impacto es del $65 \%$ y la probabilidad de que ocurran uno o más impactos, $P_{n}(k\geq 1)= P_{n}(1)+P_{n}(2)+P_{n}(3)+...$, es del $35.2 \%$.  

En resumen, hemos considerado que el encuentro con una NMG origina a grandes rasgos dos períodos de lluvias de cometas. En el primer periodo, se establece lo que podríamos  llamar el frente de la tormenta originado por la formación de conos de pérdida inducidos. Al terminar la perturbación gravitatoria sobre la nube de Oort\index{Oort, nube de}, suponemos que los conos de pérdida que se forman en condiciones estacionarias, libre de toda perturbación, están llenos de cometas e inician otro período de lluvias de cometas, que podríamos llamar la cola de la tormenta. La probabilidad de que ocurran uno o más impactos mientras transcurre el frente de la tormenta y aquella mientras transcurre la cola de la tormenta son muy similares,  $35.2 \%$ y  $37.5 \%$ respectivamente. Si integramos las lluvias de cometas de ambos períodos, $n= 2 \,\,N_{ll\oplus} + 2 \,\,N^{\star}_{ll\oplus}=2.25 \times 10^{9}$                                 
 cometas y la probabilidad de una o más impactos cometarios sobre la Tierra es del $59.3 \%$.

La probabilidad de colisión de un gran cometa con la Tierra puede ser aun mucho mayor si tenemos en cuenta que el Sistema Solar puede capturar, principalmente por la influencia gravitatoria de Júpiter,  una fracción de los cometas de la lluvia convirtiéndolos en cometas de corto período. Los cometas que pasan en forma rasante a través del cinturón de Kuiper\index{Kuiper, cinturón de} pueden perder energía por la fricción física y dinámica de los cuerpos del 
cinturón de Kuiper\index{Kuiper, cinturón de} \footnote{Parte del Sistema Solar externo que forma un anillo  de $10^{8}$ a $10^{10}$ pequeños cuerpos, centrado sobre el mismo plano que los planetas pero mucho  más allá de la órbita de Plutón.}    y quedar ligados al Sistema Solar como cometas de corto período. Por lo tanto, la población de cometas de corto período y el número de cometas que se acercan al Sol periódicamente, con posibilidad de chocar contra la Tierra,  pueden aumentar dramáticamente.

Los resultados de los cálculos arriba realizados dependen del modelo de la nube de Oort\index{Oort, nube de} adoptado, y en consecuencia son necesariamente aproximados,  sin embargo tienen la virtud de mostrar que el encuentro del Sistema Solar con una nube molecular gigante puede desencadenar  una  intensa lluvia de cometas  sobre los planetas y en particular sobre la Tierra, con una alta probabilidad de al menos un impacto cometario.

\section{Pérdidas y reposiciones de los cometas del Sistema Solar: ¿la nube de Oort rejuvenecida durante el período Criogénico?\label{Reposición}}

\begin{quote}
\small \it{
Our solar system moves
through space with the velocity of 20 km/sec. In 15,000 years
it covers the distance of one lightyear. It comes from the
region of Orion where accumulations of meteoric matter are
suspected. There is nothing illogical therefore in assuming that
comets were acquired by our solar system in its passage through
a meteoric cloud about one million years ago. Indeed, such was
the theory of the origin of comets put forward by F. Nölke \index{Nölke, F.}  in
1909}. \rm 

Nicholas  Theodore Bobrovnikoff\index{Bobrovnikoff, N.T.}, en el artículo ``The cosmological significance of comets (año 1929)'' 
\end{quote}

En la teoría estándar, los cometas de la nube de Oort\index{Oort, nube de} se formaron  en el disco proto-planetario del Sol. Por nuestra parte, argumentamos que la formación de los cometas de la nube de Oort\index{Oort, nube de} precedieron a la formación del Sol y del disco proto-planetario. Las nubes moleculares son sitios donde se forman las estrellas y donde embriones cometarios interestelares pueden ser capturados y crecer. Al formarse  un denso disco autogravitante  de  gas y polvo dentro de una densa nube  molecular, aquellos  cometas de la nube molecular capturados en la esfera de influencia gravitatoria del disco habrían originado una nube de cometas  como la de Oort (ver Sección \ref{DiscoProtoplanetario}). Por ende, en ambas teorías, los cometas de la nube de Oort han acompañado al Sistema Solar durante toda su vida, de  $\approx 4.5\times 10^{9}$ años (ver Sección \ref{EdadTierra}).

Sin embargo, tan alta longevidad de los cometas parece incompatible con  el hecho de que los cometas son susceptibles  a disgregarse  por pérdidas de materia del núcleo, y con el hecho de que las perturbaciones gravitatorias producidas por los encuentros cercanos  con estrellas y nubes moleculares modifican las órbitas cometarias y muchos cometas son expulsados al medio interestelar. Por ello, algunos investigadores piensan que gran parte de los cometas primigenios de la nube de Oort\index{Oort, nube de} se habría perdido  y consecuentemente  algún mecanismo de reposición de  cometas operaría. Incluso, tal como lo muestra el epígrafe de esta sección, algunos investigadores plantean la posibilidad de que la nube de Oort\index{Oort, nube de} es el resultado de una captura reciente de cometas interestelares.

Un mecanismo de destrucción de la nube de Oort\index{Oort, nube de}  es el choque de los cometas de la nube de Oort\index{Oort, nube de} con el flujo de cometas interestelares. Con propósitos ilustrativos,  trataremos  primero  dicho simple mecanismo en forma independiente de los restantes factores de pérdida de cometas.  Si, en el tiempo $t_{0}$, $n(t_{0})$ es el número total de cometas de la nube de Oort, el área total que exponen  esos cometas a la corriente  de cometas interestelares es $n(t_{0})\, A$, donde $A$ es la sección de área de cada cometa perpendicular a la corriente cometaria. Suponemos que todos los cometas tienen un mismo radio $a$ y en consecuencia $A=\pi a^{2}$. Como  el Sistema Solar se mueve con una  velocidad peculiar $v_{\odot}$ con respecto al medio interestelar, la corriente de cometas interestelares incide en promedio sobre la nube de Oort con la velocidad $-v_{\odot}$. Si $\Delta t_{1}=t_{1}-t_{0}$ es el tiempo que debe transcurrir para que se produzca el primer choque entre un cometa solar y uno interestelar, en el volumen $dV= n(t_{0})\,  A \,  \mid v_{\odot} \mid \,\Delta t_{1}$, debe haber un cometa interestelar. En consecuencia, $ n_{ci} \, dV =  n_{ci} \, n(t_{0})\,  A \,  \mid v_{\odot} \mid \,\Delta t_{1}=1$, donde $ n_{ci}$ es la densidad numérica de cometas interestelares (ver Subsección \ref{Diseminación}). Despejando $\Delta t_{1}$ de la última ecuación, obtenemos
\begin{equation}
\Delta t_{1}=  \frac{1}{ \gamma\, n(t_{0})},
\end{equation}
donde $\gamma= n_{ci} \, A \,   \mid v_{\odot} \mid$.

Suponemos que en el choque se destruyen los dos cometas y que  una parte de los escombros queda atrapada en la nube de Oort\index{Oort, nube de} y el resto escapa al medio interestelar. Entonces, en el tiempo $t_{1}$, $n(t_{1})=n(t_{0})-1$, dado que se eliminó un cometa de la nube de Oort\index{Oort, nube de}, el número de blancos disminuyó en una unidad. Llamando $\Delta t_{2}=t_{2}-t_{1}$ al tiempo transcurrido para producir el segundo choque, 
 podemos escribir: $n_{ci} \, n(t_{1})\,  A \,  \mid v_{\odot} \mid \,\Delta t_{2}=1$, de donde obtenemos 
 \begin{eqnarray}
   n(t_{1}) & = & n(t_{0})-1,  \nonumber\\
 \Delta t_{2}& = & \frac{1}{ \gamma\, n(t_{1})}.
  \end{eqnarray}
Similarmente, 
\begin{eqnarray}
 n(t_{2}) & = &  n(t_{1})-1=n(t_{0}) -2, \nonumber\\
 \Delta t_{3}& = &  \frac{1}{ \gamma \, n(t_{2})}. 
\end{eqnarray}
Generalizando, 
\begin{eqnarray}
n(t_{i}) & = & n(t_{0}) -i, \nonumber\\
 \Delta t_{i+1}& = & \frac{1}{ \gamma  \, n(t_{i})},
 \label{DestruccionCometas}
  \end{eqnarray}
donde $i=0,1,2,3,... N_{d}$, cuenta los cometas destruidos.

Reemplazando $n(t_{i})$ por $n(t_{0}) -i$ en el denominador de la ecuación (\ref{DestruccionCometas}), obtenemos $\Delta t_{i+1} =   \frac{1}{\gamma\,(n(t_{0}) -i)}$. El intervalo de tiempo $t-t_{0}$ que debe transcurrir para destruir el número $N_{d}$ de cometas de la nube de Oort\index{Oort, nube de} es dado por  $t-t_{0}=\sum_{i=0}^{N_{d}} \Delta t_{i+1} =\sum_{i=0}^{N_{d}}  \frac{1}{\gamma \, (n(t_{0}) -i)}$. Esta última ecuación podemos escribirla del siguiente modo:

\begin{equation}
t-t_{0}=\frac{1}{\gamma}  \sum_{i=0}^{N_{d}}  \frac{1}{(1 -\frac{i}{n(t_{0})})} \frac{\Delta i}{n(t_{0})},
\label{DestruccionCometas2}
\end{equation}
donde $\Delta i=1$. Si definimos la variable $u=\frac{i}{n(t_{0})}$, $\Delta u= \frac{\Delta i}{n(t_{0})}$ y la ecuación (\ref{DestruccionCometas2}) se transforma en

\begin{equation}
t-t_{0}= \frac{1}{\gamma}  \sum_{u=0}^{\frac{N_{d}}{n(t_{0})}} \frac{1}{(1-u)}\Delta u.
\label{DestruccionCometas3}
\end{equation}
En el cálculo diferencial e integral, la suma de la ecuación (\ref{DestruccionCometas3}) se puede igualar a $ \frac{1}{\gamma} \int_{0}^{\frac{N_{d}}{n(t_{0})}}
 \frac{1}{(1-u)} du$. Esta integral tiene una solución analítica dada por $\int
 \frac{1}{(1-u)} du=-ln(1-u)$, con lo cual la  expresión (\ref{DestruccionCometas3}) resulta $t-t_{0}= -\frac{1}{\gamma}\, ln(1-\frac{N_{d}}{n(t_{0})})$ o usando la definición de logaritmo natural ($ln(x)=y \Leftrightarrow e^{y}=x$), $1-\frac{N_{d}}{n(t_{0})}= e^{-\gamma (t-t_{0})}$. Por lo tanto
 
 \begin{equation}
 N_{d}= n(t_{0})\, ( 1- e^{- \gamma (t-t_{0})}),
 \label{PorcentageDestruccion1}
 \end{equation}
 y teniendo en cuenta que $n (t)= n(t_{0})-N_{d}$, 
 
 \begin{equation}
 n(t) = n(t_{0})\, e^{- \gamma (t-t_{0})}.
  \label{PorcentageDestruccion2}
 \end{equation}
 
 Si designamos con $t_{0}$ el tiempo de origen del Sistema Solar y con $t$ el tiempo presente, $t-t_{0}=4.5 \times 10^{9}$ años, es decir,  la edad  del Sistema Solar.
 A fin de  calcular el valor de $\gamma=n_{ci} \, A \,   \mid v_{\odot} \mid$, adoptamos $a=4$ km para el radio de los cometas, con lo cual obtenemos el valor de $A (=\pi a^{2})$.  Para  los parámetros restantes de $\gamma$, adoptamos   $v_{\odot}=40$ km s$^{-1}$ y $n_{ci} =7 \times 10^{14}$ cometas por pc$^{3}$ (ver Subsección \ref{Diseminación}). Si bien la velocidad peculiar actual del Sol es de 20 km s$^{-1}$, se piensa que en el pasado fue mucho mayor (ver Sección \ref{TioVivo}). Con todo ello, obtenemos que $\gamma=1.7  \times 10^{-15}$ (años$)^{-1}$. Al tiempo que se necesita para que  $n(t)$ decaiga a la mitad de su valor inicial, $n(t_{0})$, se le llama  vida media y es dado por $t_{\frac{1}{2}}=\frac{ln 2}{\gamma}$ (ver la fórmula \ref{Half-time}). Con el valor de $\gamma$ que obtuvimos, $t_{\frac{1}{2}}= 6 \times 10^{14}$ años, es decir,  mucho mayor que edad del Sistema Solar. Por lo tanto, la destrucción de cometas por choques con cometas interestelares es relativamente muy pequeña y no ha afectado  significativamente a la nube de Oort. Nosotros utilizaremos ese hecho para obtener más abajo una mejor estimación de $N_{d}$ que la que se obtendría mediante la fórmula (\ref{PorcentageDestruccion1}).
 
 Se piensa que la nube de Oort fue fuertemente  despojada de cometas en los encuentros del Sistema Solar con estrellas y NMGs, y que ello sería el principal factor de desintegración de la nube de Oort\index{Oort, nube de}. Podemos suponer que la disminución de la población de cometas sigue la ley exponencial expresada por la fórmula (\ref{PorcentageDestruccion2}), pero con un $\gamma$ diferente que denotamos $\gamma^{\star}$:
 \begin{equation}
 n^{\star}(t) = n(t_{0})\, e^{-\gamma^{\star} (t-t_{0})}.
  \label{PorcentageDestruccion3}
 \end{equation}
 Como hemos visto $\gamma^{\star}$ está relacionada con la vida media $t^{\star}_{\frac{1}{2}}$ de la nube de Oort\index{Oort, nube de}, mediante  $\gamma^{\star}=(ln 2)/(t^{\star}_{\frac{1}{2}})$. Se estima que $t^{\star}_{\frac{1}{2}}\approx 10^{9}$ años \cite{Bailey}, es decir mucho menor que la edad del Sistema Solar. Dado que en principio conocemos el número de cometas en el presente, es decir $n^{\star}(t)$, podemos  a partir de la ecuación (\ref{PorcentageDestruccion3}) despejar $n(t_{0})$ y determinar su valor. Las estimaciones para el valor de $n^{\star}(t)$ son muy inciertas y  varían entre $2 \times 10^{12}$ cometas, valor clásico obtenido por Oort, y $10^{14}$ cometas si la nube de Oort\index{Oort, nube de} posee un masivo núcleo interior \footnote{Cabe destacar que la estimación del número de cometas de la nube de Oort se obtiene a partir del flujo de cometas de largo periodo que llegan a la zona de visibilidad gracias a las perturbaciones gravitatorias de estrellas y NMGs que impulsan a los cometas hacia el interior del Sistema Solar. De otro modo, ignoraríamos la existencia de la nube de Oort. Una fracción de los cometas de largo periodo, los cuales provienen de capas exteriores de la nube de Oort, se convierten en cometas de corto período a causa de las perturbaciones de los grandes planetas. Sin embargo, ello no explicaría la magnitud del flujo de cometas de corto período que se acercan al Sol y, por lo tanto, algunos investigadores postulan la existencia de un masivo núcleo interior de la nube de Oort como la fuente de dichos cometas.}. Adoptando conservadoramente $n^{\star}(t)=2 \times 10^{12}$,  obtenemos $n(t_{0})\approx 5 \times 10^{13}$ cometas. 
 
Ahora estamos en condiciones de hacer una mejor evaluación del número de cometas destruidos  $N_{d}$ debido a choques entre cometas interestelares y cometas de la nube de Oort\index{Oort, nube de}. Dado que la pérdida de cometas  por este mecanismo  es relativamente pequeño, el número total de cometas en función del tiempo es determinado esencialmente por $n_{\star}(t)$. Por lo tanto, podemos escribir $dN_{d}=n_{ci} \, A \,   \mid v_{\odot} \mid \, n_{\star}(t) \, dt= \gamma \,  n_{\star}(t)\, dt$, de lo cual resulta que
 \begin{equation}
N_{d}= \gamma \int_{t_{0}}^{t} n_{\star}(t) dt.
\label{PorcentageDestruccion4}
 \end{equation}
 La integración de la ecuación (\ref{PorcentageDestruccion4})  con la expresión de $n_{\star}(t)$ dada por la fórmula (\ref{PorcentageDestruccion3}) resulta
 \begin{equation}
 N_{d}=n(t_{0}) \frac{ \gamma }{\gamma^{\star}} (1- e^{-\gamma^{\star} (t-t_{0})}),
 \end{equation}
 con lo cual obtenemos $N_{d}=8 \times 10^{7}$ cometas. Suponemos que por cada cometa de la nube de Oort\index{Oort, nube de} que se destruye, se destruye uno interestelar. Si además suponemos que solo la mitad de la masa total de los escombros cometarios queda ligada gravitacionalmente a  la nube de Oort\index{Oort, nube de} y que   $m_{c}=8 \times 10^{14}$ Kg  es la masa media de cada cometa,  la masa de escombros esparcidos dentro de la nube de Oort es $N_{d} m_{c}=0.01 M_{\oplus}$. A dicha masa, equivalente al uno por ciento de la masa de la Tierra, habría que sumarle los escombros producidos durante la vida del Sistema Solar por los choques entre los mismos cometas de la nube de Oort\index{Oort, nube de}.
 
 Un punto débil de nuestra argumentación desarrollada arriba es que solo  hemos considerado las pérdidas de cometas y, implícitamente,  despreciado  la posibilidad de reposición de cometas. El Sistema Solar puede capturar cometas interestelares al atravesar densas nubes moleculares interestelares, donde el contenido de cometas es probablemente alto. Sin embargo, dada la relativamente alta velocidad peculiar del Sol y, por lo tanto, de la velocidad relativa de los encuentros con las nubes moleculares, la probabilidad de captura de cometas por parte del Sol es muy baja. Entonces, en los encuentros con las nubes moleculares, la pérdida de cometas de la nube de Oort sería en general mucho mayor que la ganancia.

Los cometas, o asteroides derivados de ellos, fueron seguramente los principales cuerpos cósmicos que originaron los cráteres  de impacto de la Tierra y de la Luna. Por otro lado, hemos visto que el número  de cometas de la nube de Oort decrecería exponencialmente a medida que nos acercamos al presente y, por lo tanto, el número de cráteres jóvenes  debería ser menor que el número de cráteres más viejos. Sin embargo, la distribución de edades  de cráteres lunares y terrestres parece contradecir en parte dicha predicción. En efecto, en el estudio de un equipo de investigadores japoneses: Dres. Kentaro Terada, Tomokatsu Morota y Mami Kato, se encuentra que la tasa de formación de cráteres aumentó abruptamente durante los últimos 800 Millones de años, y previamente se mantuvo  bajo y casi constante hasta los 2500 millones de años atrás (ver Fig. 3 de \cite{Terada}).

El comienzo de la ráfaga de cometas sobre la Tierra y la Luna coincide aproximadamente con el inicio del período Criogénico\index{Criogénico} y también coincide con la captura del Sol por parte de la supernube interestelar local (ver la siguiente sección), propuesta por este autor \cite{Olano1}  \cite{Olano8} \cite{Olano2b} \cite{Olano4} \cite{Olano7}. Por lo tanto, nosotros asociamos dichos hechos con un episodio singular de captura masiva de cometas interestelares por parte del Sistema Solar. En este contexto, es muy probable que el Sol se haya movido con muy baja velocidad a través de  un extenso complejo de densas nubes moleculares repletas de cometas interestelares. En estas condiciones, una significativa fracción de los cometas interestelares que ingresaban a la esfera de influencia gravitatoria del Sol  quedó ligada gravitacionalmente  al Sistema Solar. En conclusión, antes del período Criogénico\index{Criogénico}, la nube de Oort\index{Oort, nube de} original se habría vaciado en su  mayor parte y se volvió a llenar con cometas interestelares capturados  500-700 millones de años atrás. La cita de Bobrovnikoff\index{Bobrovnikoff, N.T.} al comienzo de esta sección muestra la seria atención prestada a la posibilidad de que la mayoría de los cometas de la nube de Oort\index{Oort, nube de} no fuera primordial sino adquirida recientemente. 

El hecho singular es que el Sol cambió de hábitat, permaneciendo cautivo de una nube interestelar gigante con extensas nubes moleculares en su interior y regiones de formación de estrellas, y concatenado con ello el Sol se rodeó de ``nuevos acompañantes'', un enjambre de cometas interestelares que formaron la nueva nube de Oort\index{Oort, nube de}. En el proceso de captura de cometas que habría ocurrido entre 500 y 700 Ma atrás, probablemente andanadas de cometas cayeron sobre el Sistema Solar y particularmente sobre la Tierra y la Luna, tal como lo documentan los cráteres de impacto que datan de ese período, y originaron las grandes glaciaciones del período Criogénico\index{Criogénico}. Desde el período Criogénico\index{Criogénico} hasta el presente, la nube de Oort\index{Oort, nube de} recargada de cometas tuvo  consecuencias positivas como posiblemente la gestación de la explosión de vida del Cámbrico\index{Cámbrico} y negativas como las recurrentes catástrofes climáticas que provocaron extinciones masivas de especies (ver la sección siguiente).
\newpage
\section{Impactos cósmicos periódicos y extinciones masivas de las especies\label{TioVivo}}

\begin{quote}
\small \it{My own view is that there is not longer any controversy except over the possible periodicity of the impacts, in which I am one of the few believers.}\rm

Luiz W. Alvarez\index{Alvarez, L.W.} (1911-1988) \footnote{Descubridor del tritio y  premio Nobel de física en 1968. También descubrió que la capa geológica correspondiente al límite K-T posee niveles anormalmente altos de iridio y que ello implicaría que un gran asteroide o cometa chocó contra la Tierra ocasionando la extinción masiva de fines del Cretácico.}
\end{quote}

Hemos recorrido a vuelo de pájaro los primeros 4000 Ma de la historia geológica de la Tierra y nos quedan por recorrer los últimos 550 Ma, que son los más ricos desde un  punto de vista biológico. Si bien ya nos hemos referido brevemente a la era Cenozoica, en relación con las glaciaciones del Cuaternario, restan las eras Mesozoica (245-66.4 Ma) y Paleozoica (550-245 Ma), donde el primer período es el Cámbrico.

 Al comienzo del período del Cámbrico, es decir $\approx 550$ Ma atrás, se produjo un hecho singular: una explosión de vida animal, llamada ``Explosión del Cámbrico''\index{Cámbrico, la explosión del}, o parangonando con el origen del Universo, ``el big bang'' de la biología.  Una asombrosa variedad  de  organismos pluricelulares complejos surge de pronto y como si fuese de la nada. Las evidencias de los registros fósiles disponibles son contundentes y no corroboran, para este caso singular, la idea de Charles Darwin\index{Darwin, C.} de  la evolución gradual por selección natural. El propio  Darwin  reconoció que su teoría de la evolución no podía explicar el repentino estallido del Cámbrico, por ello también llamado ``el dilema de Darwin''. En efecto, los organismos  se formaron en un intervalo de tiempo  geológico muy corto y  sin  antepasados a partir de los cuales  evolucionar \footnote{En un corto período que antecede inmediatamente al Cámbrico, llamado Ediacárico, existieron misteriosos organismos de cuerpos blandos que se extinguieron antes de la explosión del Cámbrico y que aparentemente no tuvieron vínculos  evolutivos con la vida del Cámbrico.}. Una rica representación de la exótica fauna de la explosión cámbrica, entre ellos la fascinante Hallucigenia\index{Hallucigenia, criatura del Cámbrico} (de allí su nombre)  y los emblemáticos trilobites\index{trilobites, criatura del Cámbrico }, se encontró en Burgues Shale (Canadá)\index{Burgues Shale, Canadá}, una formación geológica célebre por el buen estado de preservación de los fósiles más antiguos.
 
 Un hecho que llamó la atención a los primeros paleontólogos es que la mayoría de las especies aparecen de pronto en los registros geológicos y perduran sin cambios por millones de años hasta su extinción por sucesos catastróficos. Por ello, los más eminentes paleontólogos, como Cuvier\index{Cuvier, G.} y Agassiz\index{Agassiz, L.} entre otros,   defendían  el creacionismo y  la inmutabilidad de las especies\footnote{El gran biólogo francés Jean Baptiste Lamarck (1744-1829)\index{Lamarck, J.B.},   contemporáneo de  Cuvier, expuso sus ideas sobre la evolución de las especies en 1809, 50 años antes que  Darwin publicara su famoso libro. Cuvier fue naturalmente crítico de las ideas de  Lamarck. En  el Lamarckismo, el individuo juega un papel activo en la evolución de su especie; en cambio en el Darwinismo, su papel es pasivo: solo el azar y la selección natural determinan la evolución. En la teoría moderna de la evolución, la  apariencia de  inmutabilidad de las especies puede asimilarse a un ritmo evolutivo lento y la idea metafísica de creacionismo  puede asimilarse a la idea de evolución explosiva o radiación evolutiva.}. En la teoría moderna de la evolución, se acepta  que una nueva especie o subespecie puede originarse en un pequeño grupo de individuos de una especie, grupo  que se encuentre, por factores geográficos o ecológicos,  aislados  de la población madre. Al no tener contacto con la población restante de la especie, las mutaciones  genéticas del pequeño grupo se conservan y refuerzan  dando lugar a una nueva especie o subespecie. Al finalizar las grandes glaciaciones y catástrofes medioambientales, como la de fines del Cretácico,  se produce la aparición de muchas nuevas especies, fenómeno llamado radiación evolutiva  o equilibrio puntuado \footnote{Estudiado en detalle por los  paleontólogos norteamericanos N. Eldredge\index{Eldredge, N.}   y S. J. Gould. Stephen J. Gould\index{Gould, S.J.} (1941-2002) escribió el voluminoso libro ``La estructura de la teoría de la evolución'', donde se describe amenamente  el desarrollo histórico de la teoría de la evolución.}. Ello podría explicarse porque las extremas condiciones climáticas  obligaron a grupos de  las distintas especies sobrevivientes a   vivir  en refugios o territorios aislados unos de otros, favoreciendo así la gestación de nuevas especies. A veces se menciona  al fenómeno de radiación evolutiva, o equilibrio puntuado, como explicación de  la explosión Cámbrica. Sin embargo, la explosión del Cámbrico fue un caso único en la historia de la Tierra en que no hubo aparentemente especies progenitoras. De modo que el dilema de Darwin\index{Darwin, dilema} aún persiste entre los evolucionistas.
 
 Los geólogos del siglo XIX elaboraron una escala de tiempo geológico sobre la base de la aparición y desaparición de especies en los registros fósiles. De allí los nombres de Era Paleozoica (era de la vida antigua), Era Mesozoica (era de la vida media) y Era Cenozoica (era de la vida nueva), las cuales a su vez se dividen en Períodos.
 Por lo tanto,  los límites de los Períodos se corresponden generalmente con eventos de extinciones significativas de animales y plantas. Las cinco más traumáticas extinciones masivas, en las que desaparecieron del 70 al 90 por ciento de las especies de toda la Tierra, ocurrieron en los  límites: Ordovícico-Silúrico (440 Ma)\index{Ordovícico-Silúrico }, Devónico-Carbonífero (365 Ma)\index{Devónico-Carbonífero}, Pérmico-Triásico (240 Ma)\index{Pérmico-Triásico}, Triásico-Jurásico (210 Ma)\index{Triásico-Jurásico} y Cretácico-Terciario (65 Ma)\index{Cretácico-Terciario}.
 
 A partir de un estudio de las extinciones de organismos marinos  en el pasado geológico, los paleontólogos norteamericanos David M. Raup (1933-2015)\index{Raup, D.M.} y  Jack Sepkoski (1948-1999)\index{Sepkoski, J.} encontraron que además de las ``cinco grandes'' hubo muchas otras extinciones menores, pero significativas, que se sucedieron con una periodicidad de $\approx 26$ Ma \footnote{Las primeras evidencias de ritmos geológicos fueron presentadas  en el año 1927 por el geólogo británico Arthur Holmes\index{Holmes, A.} (1890-1965). El ciclo tectónico de Holmes es de $\approx$ 30 Ma. Un ciclo similar para las extinciones masivas de especies fue encontrado  por  Fischer\index{Fischer, A.G.} y Arthur\index{Arthur, M.A.} en el año 1977.}. Este notable descubrimiento llevó a sospechar que una  causa astronómica desencadenó  regularmente  las extinciones masivas. De hecho, en varias estractos geológicos correspondientes a las extinciones masivas, se encontraron vestigios de grandes impactos cósmicos (ver la Fig. 2 de \cite{Rampino1}). En un estudio de la distribución de las edades de cráteres de impacto en  los últimos 260 Ma,  se encontró una periodicidad de las edades de los cráteres, de  $\approx 26$ Ma, en fase con las edades correspondientes a las extinciones masivas, lo cual sugiere  una relación causal \cite{Rampino3} (ver también  \cite{Rampino12}).
 
 Conocemos los proyectiles, es decir cometas o asteroides, que impactaron contra la Tierra, pero no el arma que los disparó automáticamente  una vez cada 26 Ma. Sabemos también que  el Sistema Solar está inmerso en una extensa nube de cometas, la nube de Oort\index{Oort, nube de}, cuyo borde externo se interna en el medio interestelar y se acerca a las estrellas más cercanas al Sol. De modo que la nube de Oort\index{Oort, nube de} es una suerte de cuerda o puente que une al Sistema Solar con la Galaxia. Por lo tanto, cabe esperar que fuerzas galácticas puedan tensionar la nube de Oort\index{Oort, nube de} y producir un alud de cometas hacia el interior del Sistema Solar y bombardear la Tierra. El plano Galáctico está poblado por  relativamente densas nubes interestelares (ver Sección \ref{MedioInterestelar}) y, en particular, por  nubes moleculares gigantes, las cuales debido a sus altas masas (de miles hasta millones de $M_{\odot}$) pueden perturbar gravitacionalmente a la nube de Oort\index{Oort, nube de} en un   encuentro cercano con el Sistema Solar (ver Sección \ref{Penetrante1}). 
 
 Tenemos entonces a los  objetos galácticos que potencialmente pueden inducir una lluvia de cometas hacia los planetas del Sistema Solar. Sin embargo, ¿a qué se debió que  el Sistema Solar se  haya encontrado periódicamente con densas nubes interestelares? El Sol, y por supuesto su séquito de planetas, orbita alrededor del centro Galáctico sobre el plano medio del disco Galáctico y a la vez oscila en torno al plano galáctico. El movimiento galáctico del Sol es similar al del caballito de carrusel que sube y baja, mientras gira con la plataforma del carrusel. El período de oscilación vertical del Sol es de $\approx 60$ Ma. Por lo tanto, cada $\approx 30$ Ma, el Sol   cruza al plano Galáctico, donde se concentran las nubes interestelares y donde se producirían los encuentros con el Sol. Este mecanismo para explicar la periodicidad de las colisiones cósmicas y las extinciones masivas asociadas fue propuesto en el año 1984 por el geólogo Michael R. Rampino\index{Rampino, R.M.} y el astrónomo Richard B. Stothers (1939-2011)\index{Stothers, R.B.}\cite{Rampino2}.

 En el mecanismo de Rampino\index{Rampino, R.M.} y Stothers\index{Stothers, R.B.}, se supone al Sol como un cuerpo aislado que oscila por la componente vertical de la fuerza gravitatoria que ejerce el disco galáctico en su conjunto. Sin embargo, el Sol forma parte de un sistema local de gas y estrellas. Una conspicua parte de esta estructura local se observa en el cielo como  una banda de estrellas brillantes, llamada cinturón de Gould en honor al astrónomo norteamericano Benjamín A. Gould (1824-1896)\index{Gould, Benjamín A.} \footnote{El cinturón de Gould fue advertido por primera vez por sir John Herschel\index{Herschel, J.} en 1847. Tres décadas después, desde Argentina, Gould estudió el fenómeno más a fondo. El presidente argentino Domingo Faustino Sarmiento\index{Sarmiento, D.F.} fundó el  observatorio astronómico de Córdoba (Argentina) y nombró a Gould como su primer director.}, que se destaca porque se encuentra inclinada en $\approx 20^{\circ}$ con respecto al plano de la Vía Láctea. En un párrafo de la Uranometría Argentina que transcribimos a continuación, Gould describe el mencionado fenómeno: ``Hay una faja, o corriente, de estrellas brillantes que parece ceñir el cielo casi en un gran círculo máximo, que se intersecta con la Vía Láctea...''  En otro pasaje de su libro, Gould señala el carácter local del Cinturón: ``es un pequeño cúmulo de menos de 500 estrellas de aspecto chato y distinto de la vasta organización de la Vía Láctea''. También, Gould  reconocía que la ubicación del Sol dentro del sistema es algo excéntrica: ``puede inferirse  que nuestra posición no está lejos del borde...''.
 
 El astrónomo sueco Per Olof Lindblad\index{Lindblad, P.O.}, en  un análisis de extensas observaciones de la línea de 21 cm  del  HI galáctico \footnote{El átomo de hidrógeno neutro en las condiciones generales del medio interestelar se encuentra en el nivel energético fundamental, el cual se desdobla en dos niveles cuánticos  hiperfinos. En estas condiciones, el electrón de un átomo de HI se desexcita mediante un salto cuántico del nivel hiperfino superior al inferior, y en el proceso emite un fotón de 21 cm de longitud de onda o 1420 MHz de frecuencia. Esta radiación de 21 cm se puede observar a través de radiotelescopios}, encontró  que una componente  de velocidad del HI, por su distribución y cinemática, podía asociarse al cinturón de Gould e interpretarse como un elongado anillo en expansión. El modelo original de Lindblad fue extendido  por el autor de esta monografía \cite{Olano6}. El cinturón de Gould se encuentra en la región central  de una estructura alargada  de gas, polvo  y estrellas jóvenes, llamada brazo local o brazo de Orión. Probablemente, el cinturón de Gould y el brazo de Orion tuvieron hace 60 Ma  un origen común, a juzgar por las edades de sus estrellas,  y ellos se ven en el presente,  dentro del gran diseño de la Galaxia,  como una isla  entre el  brazo espiral de Sagitario y el de Perseo \cite{Olano7} \cite{Olano8}.  La región local contiene además un grupo de estrellas, el supercúmulo de Sirio, con una edad (de $\approx 500$ Ma) y   una cinemática  diferentes de las del cinturón de Gould y del brazo de Orión. En cierto modo, las estrellas del supercúmulo de Sirio  tienden a mantener su pasado dinámico, dado que, a diferencia del gas, las estrellas casi no son frenadas al atravesar el medio interestelar. Por lo tanto es probable que  el supercúmulo de Sirio y los precursores del cinturón de Gould y del brazo de Orión constituyeron una supernube, donde  compartieron  una cinemática similar en tiempos pasados mayores a $\approx 60$ Ma. Entonces, el progenitor del sistema local fue  probablemente  una supernube con diámetro y masa del orden de 600 parsec y $2 \times 10^{7}$ masas solares, valores típicos de supernubes encontradas en nuestra Galaxia y en galaxias externas. En resumen, el supercúmulo de Sirio,  supuesto casi coetáneo de la supernube gaseosa, tiende a conservar  la cinemática que compartió con el gas de la supernube  antes de la formación del cinturón de Gould y del brazo de Orión. Ello permite reconstruir la trayectoría  galáctica pasada de la supernube.  \cite{Olano7} \cite{Olano8}.

 Poco después de la formación de la supernube gaseosa, se extendió por toda la supernube  la formación de estrellas de edades similares, dando origen al supercúmulo de Sirio. Por lo tanto, la edad de la supernube sería del orden o un poco mayor que la edad del supercúmulo, digamos entre $500$ y $700$ Ma, y la velocidad inicial de la supernube sería similar a la velocidad del supercúmulo de Sirio en el momento de su formación. Nuestros cálculos indican que la supernube  tuvo  en su origen una relativamente alta velocidad peculiar del orden de 30-40 km s$^{-1}$. Ello indicaría que un frente de choque de gran escala, como por ejemplo al asociado con la onda espiral galáctica, acumuló y aceleró gas interestelar hasta formar una supernube con una velocidad similar a la velocidad de dispersión de las estrellas de campo. En el proceso de aceleración de la supernube en formación, un porcentaje significativo de las estrellas de campo que se encontraban durante el proceso dentro de la supernube quedó ligado gravitacionalmente a la supernube \cite{Olano4}. Por lo tanto, es posible que el Sol mismo haya sido capturado por la supernube, con lo cual se explicaría su actual posición dentro del sistema local y su relativamente baja velocidad peculiar.
 
A semejanza con el modelo de  Rampino\index{Rampino, R.M.} y Stothers \index{Stothers, R.B.}\cite{Rampino2}, en nuestro modelo de la supernube suponemos que el movimiento en $z$ (perpendicular al plano Galáctico) del Sol determina la cuasi-periodicidad de los encuentros cercanos del Sol con nubes interestelares y en consecuencia la cuasi-periodicidad de las extinciones masivas de especies. La diferencia con el modelo de Rampino\index{Rampino, R.M.} y Stothers \index{Stothers, R.B.} radica en que en nuestro caso la componente $z$ de la atracción gravitatoria que actúa sobre el Sol corresponde  a ambas partes, al disco Galáctico y a la supernube, con lo cual el semiperiodo  de la oscilación  en $z$ del Sol es de 26 millones de años \citep{Olano2b}. La otra diferencia es que el Sol se encuentra con nubes de tamaño medio que pertenecen a la misma Supernube y por lo tanto lo acompañan en conjunto. En la Fig. 4 de 
\citep{Olano2b} se representa la altura del Sol con respecto al plano medio de la Supernube, durante los últimos 100 Ma, lo cual muestra que   los extremos de la oscilación del Sol coinciden aproximadamente con los tiempos de las extinciones masivas, es decir cada $\approx 26$ Ma. Ello se explicaría por el hecho de  que las más intensas perturbaciones de la nube de Oort\index{Oort, nube de} ocurren  cerca de 
los extremos  del movimiento vertical del Sol donde la velocidad de los encuentros es mínima. En efecto, la fórmula (\ref{ImpulsiveV})  índica que  la magnitud de la fuerza de marea, e intensidad de la lluvia de cometas que esta desata,  es  inversamente proporcional a la velocidad relativa $V_{r}$ del encuentro.

\section{¿Se adquirió la nube de Oort\index{Oort, nube de} durante el período Criogénico\index{Criogénico} por una masiva captura de cometas interestelares por parte del Sol?\label{CapturaCometas}}
En las Secciones \ref{criogenico1} y \ref{Reposición}, hemos mencionado nuestra tesis de que, durante el período Criogénico\index{Criogénico},  el Sol se sumergió parsimoniosamente en un extenso complejo de densas nubes interestelares de gas, polvo,  cometas y estrellas en formación. Como consecuencia de ello, el Sol quedó  rodeado de una numerosa población de cometas interestelares y  fue intermitentemente envuelto y oscurecido por densas concentraciones de polvo que originaron eras de hielo en la Tierra. El autor de esta monografía sostiene la hipótesis de que el Sol fue capturado por una supernube, en coincidencia con el período Criogénico\index{Criogénico} (ver sección \ref{TioVivo}, y referencias \cite{Olano2b}, \cite{Olano4} y \cite{Olano7}). En ese escenario, es posible imaginar que circunstancialmente  el Sol  se haya  deslizado lentamente entre densas y frías nubes moleculares con una velocidad relativa pequeña y similar a la dispersión de velocidades de las distintas componentes y subestructuras del complejo molecular. Aquí mostraremos que en esas condiciones el Sol puede capturar eficientemente cometas de las nubes interestelares.
 
 Si designamos con el símbolo $\sigma$ la dispersión de velocidades del complejo   de nubes moleculares que atravesó el Sol en el período Criogénico\index{Criogénico}, y si $v_{\odot}$ fue la velocidad de Sol  relativa a la velocidad media del complejo molecular, adoptamos, de acuerdo  con nuestro modelo,
 \begin{equation}
 v_{\odot}=\sigma.
 \label{vsolcriogenico}
\end{equation} 
  La dispersión de velocidades en las nubes interestelares es indicada por el ensanchamiento de las líneas espectrales de sus elementos químicos. En las densas y frías nubes moleculares, el ensanchamiento por turbulencia predomina sobre el térmico y es en general de unas pocas décimas de km\, s$^{-1}$.

 El problema entre manos incluye la interacción gravitatoria de múltiples cuerpos: el Sol, cometas y nubes moleculares. Como bien sabemos, el problema de $n$ cuerpos no tiene una solución analítica y debe resolverse mediante arduos cálculos numéricos. Sin embargo, el concepto de esfera de influencia gravitatoria del Sol nos permite  simplificar dicho  problema complejo a uno de solo  dos cuerpos, en nuestro caso  el Sol y un cometa. La esfera de influencia  del Sol es aquella esfera  de radio $r_{ei}$ con centro en el Sol dentro de la cual la fuerza gravitatoria del Sol predomina con respecto a las fuerzas  del entorno. A su vez,  el valor de $r_{ei}$ depende de cómo  se distribuye el material interestelar circundante al Sol, con lo cual en cierto modo entramos en un circulo vicioso. De todos modos, podemos trabajar con valores tentativos para $r_{ei}$.

La esfera de influencia del Sol barre con la velocidad $v_{\odot}$ (ecuación \ref{vsolcriogenico}) el complejo molecular, y nuestro propósito es determinar las órbitas de los cometas interestelares, asociados con el complejo molecular, que ingresan a la esfera de influencia solar. Para ello, definimos un sistema Cartesiano de coordenadas $(XYZ)$ solidario con el Sol cuyo eje $X$ coincide con la dirección y sentido de $\vec{v}_{\odot}$. Por lo tanto, un objeto en reposo en el complejo molecular ingresa a la esfera de influencia solar con la velocidad  $ -\vec{v}_{\odot}$.  El plano perpendicular a $\vec{v}_{\odot}$ que pasa por el origen de coordenadas divide el espacio y determina la semiesfera de influencia solar por donde ingresan los cometas interestelares (es decir, $X\geq0$).

\begin{figure}
\includegraphics[scale=0.95]{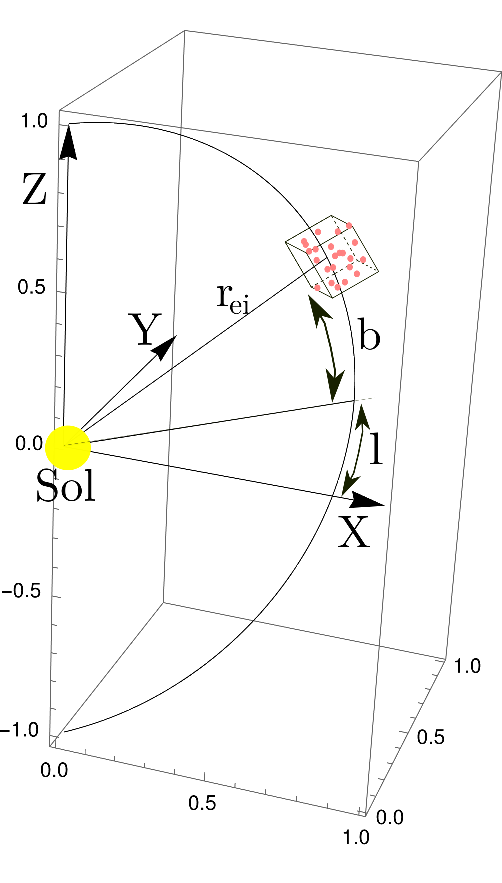} 
\caption{Esquema de un elemento de volumen con cometas, cubo con puntos rojos, que ingresa a la esfera de influencia solar de radio $r_{ei}$. Las coordenadas esféricas $(l,b)$ y cartesianas (X,Y,Z) de la posición del elemento de volumen se refirieren a los ejes cartesianos X,Y,Z con origen en el Sol.}
\label{EsferaInfluencia} 
\end{figure}

Para identificar un punto $P$ sobre una esfera, en nuestro caso sobre una semiesfera, es conveniente emplear un sistema de coordenadas esférico $(l,b)$, el cual tiene como eje polar el eje $Z$ y su plano ecuatorial el plano $X-Y$. Definimos la longitud $l$ del punto $P$ como el ángulo medido sobre el ecuador entre el meridiano que pasa por $P$ y el meridiano que contiene el eje $X$. Mientras, la latitud $b$ del punto $P$ es el ángulo entre el ecuador y el punto medido a lo largo del meridiano del punto $P$ (ver Fig. \ref{EsferaInfluencia}). 
Por lo tanto, un  punto de  coordenadas $(l,b)$  sobre la semiesfera de radio 
$r_{ei}$ tiene las siguientes coordenadas Cartesianas:
 
\begin{eqnarray}
X & = & r_{ei}\, cos \, b \,\,cos \, l, \nonumber \\
Y & = & r_{ei}\, cos \, b \,\, sen  \, l, \nonumber \\
Z & = & r_{ei}\, sen \, b.
\label{RefXYZ}
\end{eqnarray}
A fines de la representación de la Fig. \ref{EsferaInfluencia}, usamos como unidad de escala $r_{ei}=1$.

Consideremos un diferencial de área $dA$ centrada en $(l,b)$ sobre la semiesfera de influencia, con el propósito de analizar las órbitas de los cometas que ingresan a través de $dA$. Si expresamos $dA$ como un cuadrado en función de las coordenadas esféricas $(l,b)$, y dado que los dos lados paralelos al ecuador son iguales a $ r_{ei}\, \Delta\, l \, cos\, b$ y los dos perpendiculares $r_{ei} \Delta\, b$, $\Delta\, l \,cos\, b=\Delta\, b$. Fijaremos $\Delta\, b=5^{\circ}$  y por lo tanto $dA=( r_{ei}\, sen \, 5^{\circ} )^{2}$. Note que $\Delta\, l $ depende de $b$, pues al aumentar $b$, el segmento que une perpendicularmente  dos meridianos se acorta al acercarse al polo. En la Fig. \ref{EsferaInfluencia}, representamos el $dA$ como parte de un cubo de profundidad $\Delta r_{ei}= r_{ei}\, sen \, 5^{\circ}=8716 $ unidades astronómicas (UA), si adoptamos $r_{ei}=100000$ UA. En otras palabras, adoptamos como  cubo elemental o diferencial de volumen,  $dV=dA\, \Delta r_{ei}=$ (8716)$^{3}$ UA$^{3}$.

En el sistema de referencia $(XYZ)$, el material interestelar y los cometas (simbolizados por puntos violetas en la Fig. \ref{EsferaInfluencia}) contenidos en el cubo elemental  ingresan como un todo a la región de influencia con las componentes de velocidad $V_{X}=-\sigma$ (ver ecuación \ref{vsolcriogenico}), $V_{Y}=0$ y $V_{Z}=0$, la cual corresponde a la velocidad media del contenido del cubo. Sin embargo, los cometas allí contenidos tienen velocidades peculiares que diferencian sus órbitas dentro de la esfera de influencia solar. Por lo tanto, es conveniente definir un sistema de referencia $(xyz)$ centrado y fijo al cubo al cual referir las velocidades peculiares de los cometas contenidos en el cubo. Las direcciones y sentidos de los ejes coordenados del sistema  $(xyz)$ son coincidentes con los correspondientes al sistema $(XYZ)$. Las componentes coordenadas $(X_{0},Y_{0},Z_{0})$  de la posición   del origen  del sistema $(xyz)$ con respecto al sistema $(XYZ)$ se calculan  mediante las ecuaciones (\ref{RefXYZ}) con las coordenadas $(l,b)$ del centro de  $dA$. Las componentes de  velocidad del sistema $(xyz)$ con respecto al sistema $(XYZ)$ son   $V_{X_{0}}=-\sigma$ , $V_{Y_{0}}=0$ y $V_{Z_{0}}=0$, como se infiere por lo dicho antes.

 A semejanza de los movimientos de las moléculas dentro de un volumen de gas, suponemos que las velocidades peculiares $v$ de los cometas dentro de $dV$ se distribuyen isotrópicamente y siguen  una distribución de Maxwell-Boltzmann de la forma:
 
 \begin{equation}
 f(v)= \frac{2}{\sqrt{2 \pi} \sigma^{3}} v^{2} e^{-\frac{(v-v_{0})^{2}}{2 \sigma^{2}}},
 \label{Maxwell-Boltzmann}
 \end{equation}
 donde $v$ es la magnitud o módulo del vector velocidad, $\mid \vec{v} \mid$, y $\sigma$ es la dispersión de las velocidades peculiares de los cometas en el diferencial de volumen $dV$ y es coincidente con la dispersión  expresada en la ecuación (\ref{vsolcriogenico}). De aquí en adelante, adoptamos $\sigma= 0.2$ km s$^{-1}$. Dado que las velocidades $v$ son  referidas al sistema $(xyz)$, la velocidad media de los cometas es $v_{0}=0$.
 
 A fin de especificar la dirección de un vector de velocidad,  usamos un par de ángulos, $\theta$ y $\phi$, los cuales equivalen a las coordenadas  esféricas  $l$ y $b$, respectivamente. Por lo tanto, las componentes de un vector velocidad $v$ en el sistema $(xyz)$ son
 
 \begin{eqnarray}
v_{x} & = & v \, cos \, \phi  \,\,cos \, \theta,\nonumber \\
v_{y} & = & v \, cos \, \phi \,\, sen  \, \theta, \nonumber \\
v_{z} & = & v \, sen \, \phi.
\label{vxyz}
\end{eqnarray}

Para calcular las trayectorias de los cometas con respecto al Sol,  las velocidades de los cometas que se encuentran en el volumen $dV$ deben referirse al
sistema $(XYZ)$. Para ello, a las velocidades peculiares, referidas al sistema $(xyz)$, debemos sumarles la velocidad del sistema $(xyz)$ con respecto al sistema $(XYZ)$. Por lo tanto, $V_{X}=V_{X_{0}}+v_{x}$,$V_{Y}=V_{Y_{0}}+v_{x}$ y $V_{Z}=V_{Z_{0}}+v_{x}$, y teniendo en cuenta las relaciones (\ref{vxyz}) obtenemos

\begin{eqnarray}
V_{X} & = & -\sigma +  v \, cos \, \phi  \,\,cos \, \theta, \nonumber \\
V_{Y} & = & v \, cos \, \phi \,\, sen  \, \theta, \nonumber \\
V_{Z} & = & v \, sen \, \phi.
\label{VXYZ}
\end{eqnarray}
Las velocidades dadas por (\ref{VXYZ}) corresponden al momento en que el centro del cubo posicionado en $(X_{0},Y_{0}, Z_{0})$ ingresa al interior de la esfera de influencia del Sol (ver Fig. \ref{EsferaInfluencia}). Nosotros adoptaremos un conjunto de valores para $v$ y $(\theta,  \phi)$ que representen la distribución de los cometas dentro del cubo, cuya posición central es definida dando los valores de sus coordenadas esféricas $(l,b)$.
 
 De la dinámica de dos cuerpos en interacción gravitatoria, en nuestro caso entre el Sol y un cometa, se deriva la siguiente importante fórmula sobre la velocidad total $V$ del cometa, relacionada con la conservación de la energía del sistema:
 \begin{equation}
  V^{2}= \mu \left( \frac{2}{r}-\frac{1}{a} \right),
  \label{Vtotal}
 \end{equation}
 donde $\mu= G M_{\odot}$, $G$ es la constante gravitatoria y $M_{\odot}$ la masa del Sol. En el caso de que el cometa esté ligado al Sol, $a$ es el semieje mayor de la órbita elíptica y $r$ es el módulo del  vector posición, o radio vector, del cometa (ver Fig. \ref{Elipse}). La constante $h$ relacionada con la segunda ley de Kepler (ver ecuación \ref{LeyKepler2}) satisface  la siguiente expresión: 
 \begin{equation}
 h^{2}=\mu \, a (1-e^{2}),
 \label{hconstante}
\end{equation}
donde $e$ es la excentricidad de la órbita del cometa. 

Dadas la posición y velocidad  de un cometa dentro del cubo elemental en el momento de ingreso a la esfera de influencia solar podemos obtener $r$, $V$ y $h$ para el cometa y con ello  determinar sus parámetros orbitales $a$ y $e$ resolviendo las ecuaciones (\ref{Vtotal}) y (\ref{hconstante}). En efecto, 
\begin{equation}
a = \left( \frac{2}{r}-\frac{V^{2}}{\mu} \right)^{-1}, 
\label{a}
\end{equation} 
y despejando $e$ de (\ref{hconstante}) obtenemos

\begin{equation}
e  = \sqrt{1-\frac{h^{2}}{\mu a}}. 
\label{e}
\end{equation}
Reemplazando el valor de $a$ obtenido con (\ref{a}) en (\ref{e}) determinamos el valor de $e$.

Dado que el volumen del cubo elemental es relativamente pequeño, consideramos que todos los cometas del cubo tienen las mismas posiciones, coincidentes con la posición del centro del cubo. En consecuencia, $r=\mid \vec{r} \mid=\sqrt{X_{0}^{2}+Y_{0}^{2}+Z_{0}^{2} }=r_{ei}$ (ver ecuaciones \ref{RefXYZ}). Por lo tanto, consideramos que todos los cometas ingresan a la esfera de influencia del Sol al mismo tiempo. El vector posición puede  escribirse como $\vec{r}=(X_{0},Y_{0},Z_{0}) = (r_{ei} \, cos \, b \,\,cos \, l, 
 r_{ei} \, cos \, b \,\, sen  \, l , r_{ei} \, sen \, b )$, donde $l$ y $b$ son la longitud y latitud del centro del cubo elemental, de acuerdo con las relaciones (\ref{RefXYZ}). Por otra parte, el vector de la velocidad total $V$ de un cometa es dado por $\vec{V}=(V_{X}, V_{Y}, V_{Z})$ y el módulo de $V$ por 
$V=\mid \vec{V} \mid=\sqrt{V_{X}^{2}+V_{Y}^{2}+V_{Z}^{2}} $, donde las componentes de velocidad son dadas por las fórmulas (\ref{VXYZ}).

 Ahora, solo nos resta determinar el valor de $h$, para ello empleamos la relación $h= V_{T}\,\, r$, donde $V_{T}$ es la velocidad tangencial orbital (ver Sección \ref{Cuaternario}). Como conocemos $V$, podemos determinar $V_{T}$ proyectando $\vec{V}$ perpendicularmente a $\vec{r}$. En efecto, $V_{T}= V \,sen \,\alpha$, donde $\alpha$ es el ángulo entre  $\vec{r}$ y $\vec{V}$. El producto escalar de dos vectores es igual al coseno del ángulo entre ellos, en nuestro caso 
 $cos\, \alpha= \frac{\vec{r}.\vec{V}}{r \,V }$. Dado que $sen \alpha=\sqrt{1-(cos\, \alpha) ^{2}}$, $V_{T}= V \sqrt{1-(cos\, \alpha) ^{2}} $ y  $h=r V \sqrt{1-(cos\, \alpha) ^{2}}$.

\begin{figure}
\includegraphics[scale=1.5]{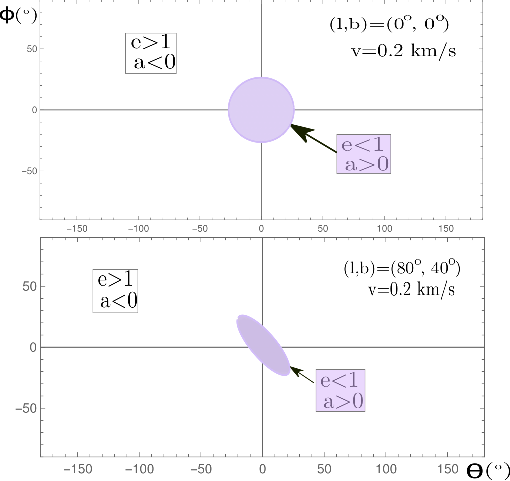} 
\caption{Posibles direcciones  $(\theta, \phi)$ de las  velocidades de magnitud   $v=0.2$ km s$^{-1}$ de los cometas contenidos en el volumen elemental centrado en la posición celeste $(l,b)=(0^{0}, 0^{0})$ (gráfico superior) y en $(l,b)=(80^{0}, 40^{0})$ (gráfico inferior). Las coordenadas  $(\theta$ y $\phi)$ son expresadas en grados de arco. El área circular en la parte superior y el área elipsoidal en la parte inferior del gráfico, ambas coloreadas, contienen las coordenadas $(\theta, \phi)$ de los cometas capturados, es decir con los parámetros $e<1$ y $a>0$ recuadrados.}
\label{AnguloSolido}
\end{figure}

 Fijada la posición celeste $(l,b)$ del cubo elemental considerado, los elementos orbitales $a$ y $e$ de los cometas que de allí provienen son funciones de la  dirección ($\theta$, $\phi$) y magnitud  $v$  de sus respectivas velocidades. Es decir,  $a(v,\theta, \phi)$ y $e (v,\theta, \phi)$, donde los posibles valores de $v$ yacen en  el rango de $0$ a $2 \sigma$ km s$^{-1}$, los de $\theta$ en el rango de $-180^{\circ}$ a $180^{\circ}$ y los de $\phi$ en el rango de  $-90^{\circ}$ a $90^{\circ}$. A fin de que  un cometa pueda quedar ligado al Sol, la distancia $r_{a}$ del afelio de su órbita en torno al Sol debe ser menor o del orden de $r_{ei}$. Con los valores  $a$ y $e$ obtenidos para  un cometa, y dado que  $r_{a}=a (1+e)$ (ver Sección \ref{Penetrante1} y fórmula \ref{Rafelio}),  el   criterio de que $r_{a} \le r_{ei}+ \frac{\Delta r_{ei}}{2}=104360$ UA nos permite determinar el número de cometas capturados por el Sol. Considerado un conjunto de vectores  $v$ de la misma   magnitud pero que apuntan en todas direcciones, podemos determinar la región  de espacio, especificadas por $(\theta, \phi)$, hacia donde apuntan las velocidades  
 de los cometas con órbitas elípticas ligadas al Sol. Como ejemplo, en la Fig. \ref{AnguloSolido}, se indican las direcciones de las velocidades de cometas, inscritos en las áreas coloreadas, que satisfacen el criterio de captura. El gráfico superior e inferior de la figura de marras corresponden a la posición celeste $(l,b)=(0^{0}, 0^{0})$ y  $(l,b)=(80^{0}, 40^{0})$, respectivamente,  en ambos gráficos $v=0.2$ km s$^{-1}$. Las direcciones de $v$ que se encuentran fuera de las áreas coloreadas, en esos dos casos una circular y otra elíptica, corresponden a cometas con órbitas hiperbólicas,
 
\begin{figure} 
\includegraphics[scale=0.65]{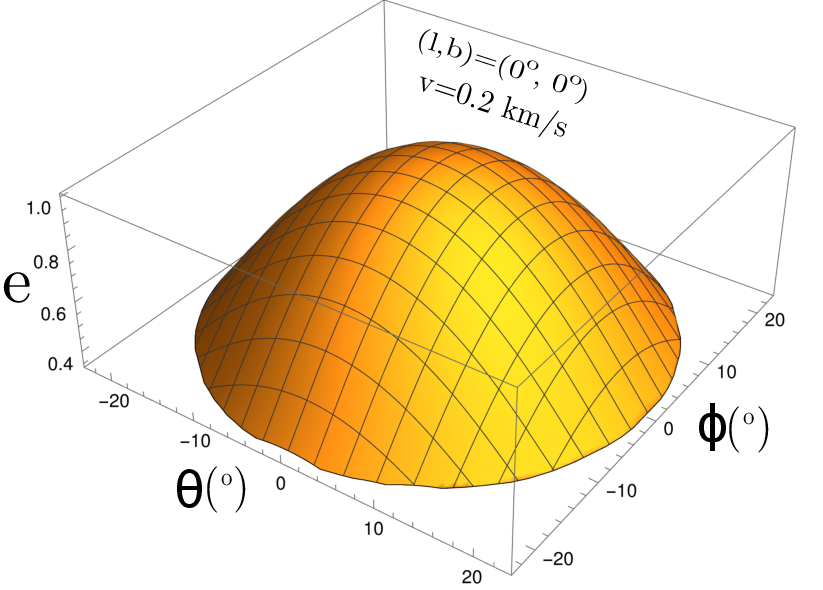} 
\caption{El valor de la excentricidad $e$  en función $(\theta$, $\phi)$ para los cometas capturados que están dentro del diferencial de volumen  de posición inicial $(l,b)=(0^{0}, 0^{0})$ y tienen un módulo de  velocidad $v=0.2$ km s$^{-1}$.}
\label{DistribucionExcentricidad}
\end{figure}

\begin{figure}
\includegraphics[scale=0.65]{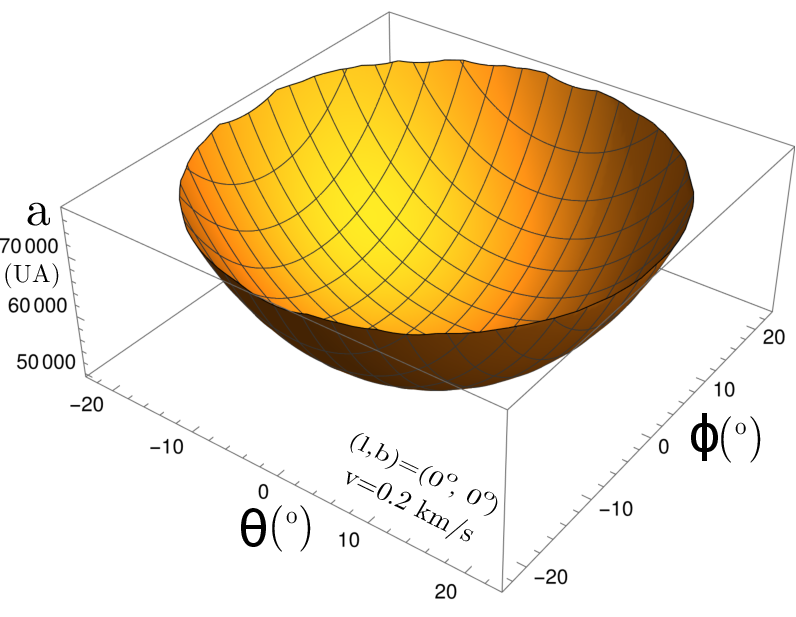} 
\caption{Lo mismo que la Fig. \ref{DistribucionExcentricidad}, pero para el semieje mayor $a$.}
\label{DistribucionEjeA}
\end{figure}

\begin{figure}
\includegraphics[scale=0.65]{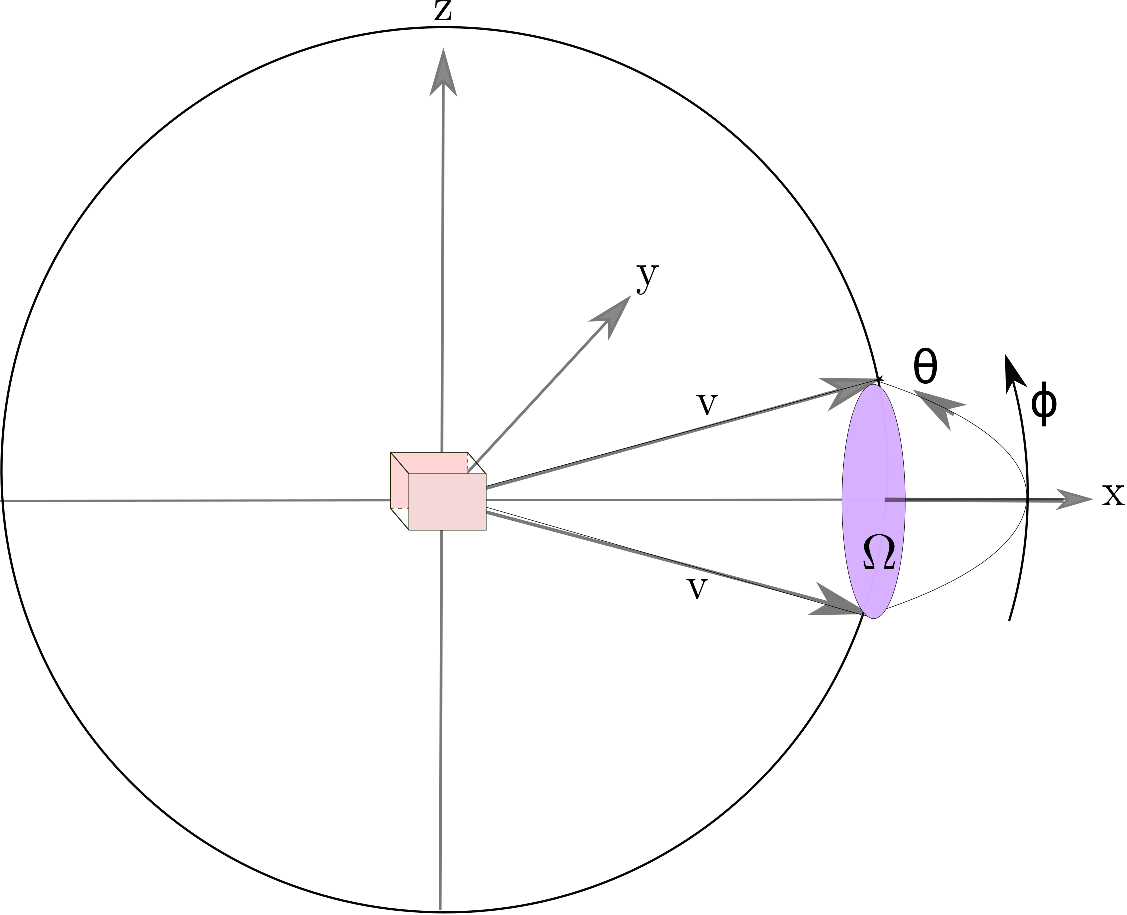} 
\caption{Esquema que representa  en 3 dimensiones los vectores de velocidad de los cometas  que se encuentran dentro del pequeño cubo, centrado en el origen del sistema de referencia x-y-z, y que pueden ser capturados por el Sol. Dichos vectores de velocidad llenan un cono, o más precisamente un sector esférico, que subtiende un ángulo sólido $\Omega$, cuya  relación con los ángulos  $(\theta$ y $\phi)$ se muestra.}
\label{cono}
\end{figure}

La Fig. \ref{DistribucionExcentricidad} muestra la distribución de la excentricidad $e$ en el dominio $(\theta, \phi)$ en el cual $e<1$ y $a>0$, representado por la superficie coloreada en el gráfico superior de la Fig. \ref{AnguloSolido}. En forma similar a la Fig. \ref{DistribucionExcentricidad}, la Fig. \ref{DistribucionEjeA} muestra la distribución de $a$ expresado en UA. A fin de visualizar en tres dimensiones  los gráficos de la Fig. \ref{AnguloSolido}, representamos,  en la Fig. \ref{cono}, un elemento de volumen centrado en el sistema x-y-z y el cono de velocidades de módulo $v$ de los cometas contenidos en el elemento de volumen que poseen  las condiciones de ser capturados por el Sol. El eje del cono coincide con el eje x positivo. Si $S$ es la superficie que resulta de la intersección del cono de velocidades con la esfera de radio $v$, el ángulo sólido que subtiende el cono es 
$\Omega= \frac{S}{v^{2}}$. Teniendo en cuenta que $S=\sum dS$ y $dS=v^{2}\, d\theta\,
 d\phi\, cos\,\phi$, $\Omega= \sum d\theta\, d\phi\, cos\,\phi$. Si dividimos el área circular o elíptica de la Fig. \ref{AnguloSolido} en cuadrículas de $n$  cuadros, donde el área de cada  cuadro es  $d\theta \, d\phi\, cos\,\phi$,  $\Omega= n\, d\theta\, d\phi\, cos\,\phi \,\,  (cos\,\phi \approx 1)$.
 
\begin{figure}
\includegraphics[scale=1.0]{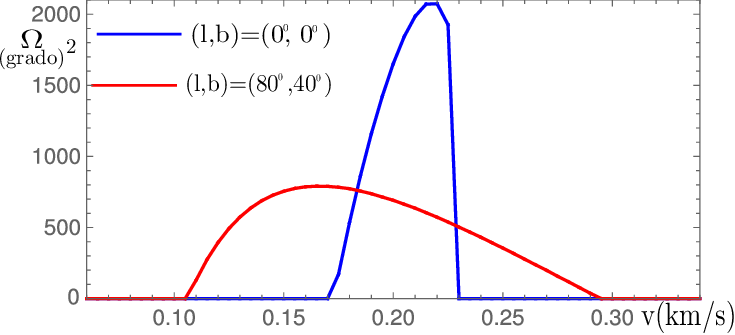} 
\caption{El ángulo sólido $\Omega$ en función de la magnitud de la velocidad $v$ de los cometas que pueden ser capturados por el Sol.}
\label{AngSolLBV}
\end{figure}

Todos los vectores de velocidad contenidos dentro del  ángulo sólido $\Omega$ tienen la misma magnitud, $v$, pero apuntan en distintas direcciones dentro de la apertura del ángulo sólido  $\Omega$. El valor del ángulo sólido $\Omega$ depende de la posición $(l,b)$ del elemento de volumen, como lo muestra la Fig. \ref{AnguloSolido}, como así también de $v$ (ver Fig. \ref{AngSolLBV}),  es decir $\Omega(l,b,v)$. Toda la esfera subtiende un ángulo sólido de aproximadamente  41253 grados cuadrados $(4 \pi (\frac{180}{\pi})^{2})$. Si expresamos $\Omega(l,b,v)$ en grados cuadrados, la fracción $f_{cc}$ de cometas del elemento de volumen con las condiciones de ser capturados por el Sol es 
\begin{equation}
f_{cc} (l,b)=\int_{0}^{2 \sigma} f(v) \frac{\Omega(l,b,v)}{41253} dv,
\label{fraccionCometas}
\end{equation}
 donde $f(v)$ es dada por la fórmula (\ref{Maxwell-Boltzmann}). La integral de la ecuación (\ref{fraccionCometas}) se resuelve numéricamente y se encuentra que la variación de $f_{cc} (l,b)$ con la posición $(l,b)$ del elemento de volumen sobre la superficie de la semiesfera de influencia solar es insignificante. En efecto,   $\overline{f_{cc} (l,b)}=0.00489 \pm  0.00002 $, valor que adoptaremos.

 Expresando la densidad $n_{ci}$ de cometas interestelares por UA cúbicas, el número total de cometas que contiene cada elemento de volumen, $dV= (8716)^{3}$ UA$^{3}$, es  $n_{ci} dV= n_{ci} (8716)^{3}$ y en consecuencia el número total de cometas capturados de cada elemento de volumen que ingresa a la esfera de influencia solar es  
 \begin{equation}
 dN_{cc} = \overline{f_{cc} (l,b)} n_{ci} (8716)^{3}. 
 \label{difNcc}
 \end{equation}
 Por lo tanto,
 \begin{equation}
 dN_{cc}= 3.24\times 10^{9}\, n_{ci}\, \, \, \, \, \rm (cometas).
 \label{CometasCapturados}
 \end{equation}

 Según lo adoptado, el volumen elemental ocupa un área $dA$, sobre la superficie de la semiesfera de influencia solar, de $25$ grados cuadrados ($5^{\circ} \times 5^{\circ}$). Puesto que la semiesfera tiene $20626.5$  grados cuadrados ($=\frac{41253}{2}$), el número $n_{ve}$ de volúmenes elementales que se encuentran sobre la superficie de  la semiesfera de influencia gravitatoria solar es $n_{ve}=\frac{20626.5}{25}=825$. En otras palabras, la capa o el cascarón de la nube interestelar  contenido entre el radio interior $r_{ei}$ y radio exterior $r_{ei}+\Delta r_{ei}$ de la semiesfera fue dividido en 825 celdas cuadradas para realizar los cálculos. Por lo tanto, el número total de cometas interestelares que ingresan a través de la semiesfera de influencia solar en un determinado tiempo y que son potencialmente capturados por el Sol es $N_{cc}=n_{ve} dN_{cc}= 825\, dN_{cc}$. Teniendo en cuenta la expresión (\ref{CometasCapturados}), obtenemos
 \begin{equation}
N_{cc}= 2,67 \times 10^{12} \, n_{ci} \, \, \, \, \, \rm (cometas).
\label{CometasCapturadosTotal}
 \end{equation}

\begin{figure}
\includegraphics[scale=1.0]{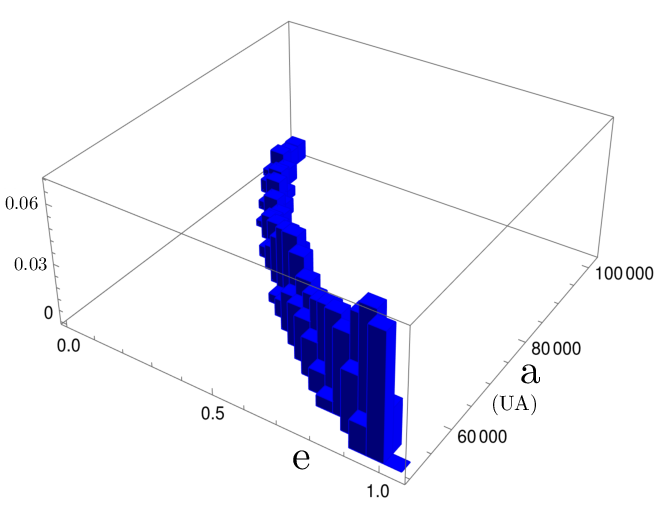} 
\caption{Las alturas de las barras indican la frecuencia o cantidad relativa al total  de cometas capturados con $(e,a)$ que yacen dentro de las respectivas bases de las barras.}
\label{Histograma}
\end{figure}
 En nuestra simulación numérica, se calculó un conjunto de pares $(e,a)$, donde cada par representa la excentricidad y semieje mayor de la órbita de un cometa capturado. La Fig. \ref{Histograma} muestra la distribución de  $(e,a)$ de los cometas capturados del elemento de volumen $(l,b)=(0^{\circ},0^{\circ})$.  El resto de  los elementos de volumen que cubren la semiesfera presenta distribuciones $(e,a)$ similares a la representada por la Fig. \ref{Histograma}, las cuales tienen la forma de una cordillera con los picos más altos cerca   
del extremo cercano a $e=1$. La forma curva de dicha cordillera indica una relación funcional entre $e$ y $a$, lo cual es natural dado que $r_{a}= a (1+e)\approx 100000$ UA. Los picos de la cordillera que se aproximan a $e=1$, y por lo tanto a $a=50000$ UA, indican que la mayor cantidad de cometas capturados ``cae'' hacia el interior del Sistema Solar.

De acuerdo con la igualdad (\ref{vsolcriogenico}), el Sol se traslada en  dirección al centro de la semiesfera con  $v_{\odot}=0.2$ km s$^{-1}$ y, por lo tanto en el  intervalo de tiempo de deriva $  \Delta t_{d}=\frac{\Delta r_{ei}}{v_{\odot}}$, el cascarón contiguo, entre $r_{ei}+\Delta r_{ei}$ y  $r_{ei}+2\, \Delta r_{ei}$, se incorpora a la zona de influencia gravitacional solar. En consecuencia, la población $P_{cc}$ de la nube de cometas interestelares capturados crece según la fórmula:
\begin{equation}
\frac{dP_{cc}}{dt}=\frac{N_{cc}}{\Delta t_{d}}.
\label{PoblacionCometas}
\end{equation}
Recordando que $\Delta r_{ei}=8716$ UA, $\Delta t_{d}= 207287$ años y reemplazando 
el valor obtenido para $\Delta t_{d}$ y $N_{cc}$ dada por (\ref{CometasCapturadosTotal}) en (\ref{PoblacionCometas}) obtenemos que $\frac{dP_{cc}}{dt}= 1.29 \times 10^{7}\, n_{ci}$ cometas por año. Si suponemos que este proceso transcurrió durante  todo el período Criogénico\index{Criogénico}, 
\begin{equation}
P_{cc}= 1.29 \times 10^{7}\, n_{ci}\, t_{PC} \,\,\, \rm (cometas),
\label{PoblacionCometas2}
\end{equation}
donde $t_{PC}$ es la  duración en años del período Criogénico. Expresando la densidad de cometas interestelares por parsec cúbico dada por la igualdad (\ref{MateriaOscura}) en unidades astronómicas cúbicas, $n_{ci}\simeq 2.2\times 10^{-2}$ cometas por UA$^{3}$ y teniendo en cuenta que  $t_{PC} \simeq 2\times 10^{8}$ años, de la fórmula (\ref{PoblacionCometas2}) obtenemos

\begin{equation}
P_{cc}\simeq 5.68\times 10^{13}      \,\,\,(\rm cometas).
\label{PoblacionCometas3}
\end{equation}
Ese valor de $P_{cc}$ derivado de nuestro modelo se encuentra dentro del rango de $10^{12}$ a $10^{14}$ de cometas estimados estadísticamente para la población de la nube de Oort, con lo cual la hipótesis de una captura relativamente reciente de la nube de Oort\index{Oort, nube de} o de una parte de ella es plausible.

En nuestra hipótesis, el Sol permaneció durante todo el período Criogénico dentro de una densa nube interestelar de gas molecular, polvo y cometas y posiblemente se haya movido a la deriva dentro de ella, con lo cual la orientación orbital de los cometas capturados por el Sol tendería  a ser al azar.  Si suponemos, sin embargo, que la dirección de la velocidad del Sol  $v_{\odot}$ ($=0.2$ km s$^{-1})$  no se desvió durante el periodo Criogénico\index{Criogénico},  la dimensión $L$ de la nube interestelar en la cual  el Sol estuvo inmerso fue $L \sim v_{\odot} t_{PC}= 40$ parsec.

Las excentricidades de los cometas capturados se concentran en  valores que se acercan a $e=1$ (ver histograma de la Fig. \ref{Histograma}) y, en consecuencia, la ``caída'' de cometas hacia regiones cercanas al Sol, en particular hacía la órbita terrestre, debió ser intensa. Por ello, nos interesa  estimar el número de impactos cometarios que sufrió la Tierra durante el periodo Criogénico\index{Criogénico}. Los cometas que tienen posibilidades de colisionar con la Tierra son aquellos cuyas distancias al perihelio $r_{p}=a (1-e)$ cumplen con la condición $r_{p}\le 1 $ UA. Empleando los mismos procedimientos y criterios por lo cuales obtuvimos $f_{cc}(l,b)$ mediante la ecuación (\ref{fraccionCometas}), pero agregando la condición $r_{p}\le 1 $ UA, encontramos  $\overline{f_{cc} (l,b)}=1.45 \times 10^{-7}$. Entonces, con la ecuación (\ref{difNcc}) y el hecho  de que $N_{cc}=825\,\, dN_{cc} (r_{p}\le 1$ UA),

\begin{equation}
N_{cc} = 1.748 \times 10^{6} \, \, \rm (nro. \, de\, cometas\,\, con \,\,r_{p}\le 1 UA),
\label{Tanda}
\end{equation}
donde $N_{cc}$ representa una tanda de cometas capturados con $r_{p}\le 1 UA$,  que se repite cada $\Delta t_{d}\, (= 207287$ años).

El número $n_{t}$ de tandas de cometas capturados durante el período Criogénico \index{Criogénico} es $n_{t}=\frac{t_{PC}}{\Delta t_{d}}$. A fin de estimar el número ${\cal N}_{tc}$ de tránsitos de cometas a través de una esfera de 1 UA de radio centrada en Sol, la cual naturalmente envuelve la orbita terrestre, debemos calcular el periodo $T$ de revolución de las órbitas cometarias con $0<r_{p}\le 1 UA$. De acuerdo con la fórmula (\ref{LeyKepler3}) y adoptando $a=50000$ UA (ver el histograma de la Fig. \ref{Histograma}), $T=1.1\times 10^{7}$ años. Por lo tanto, el número de revoluciones que dieron los cometas de la primera tanda (i=1) en todo el período Criogénico\index{Criogénico} es $\frac{t_{PC}}{T}$ y el  de la segunda tanda $(i=2)$ $\frac{t_{PC}-\Delta t_{d}}{T}$ y generalizando  $\frac{t_{PC}- (i-1)\Delta t_{d}}{T}$. Con lo cual, obtenemos la siguiente fórmula para ${\cal N}_{tc}$:
\begin{equation}
{\cal N}_{tc}= N_{cc} \sum_{i=1}^{n_{t}} \frac{(t_{PC}-(i-1) \Delta t_{d}) }{T}=1.6 \times 10^{10} \,\, \rm (nro.\, de\, tránsitos), 
\end{equation}
donde  $N_{cc}$ es dado por la expresión (\ref{Tanda}).

Ignoramos la inclinación de la orbita terrestre, en torno al Sol, que tuvo en  nuestro sistema de coordenadas XYZ, pero sabemos que la Tierra siempre se encontraba sobre alguna posición sobre la esfera de aproximadamente 1 UA de radio . Como los cometas de $r_{p}\le 1 UA$ penetraron la superficie de la esfera desde distintas direcciones, la probabilidad $p$ del choque de un cometa contra la Tierra resulta del área que expone la Tierra ($\pi R_{T}^{2}$) dividido por el área de la superficie de la esfera $4 \pi d_{TS}^{2}$, donde $R_{T}$ es el radio de la Tierra y $d_{TS}=1$ UA es la distancia Tierra-Sol. Es decir,

\begin{equation}
p= 2 \frac{\pi R_{T}^{2}}{4 \pi d_{TS}^{2}}=8 \times 10^{-10}.
\label{probabilidad}
\end{equation}
El factor 2 de $p$  (ecuación \ref{probabilidad}) se debe a que cada cometa generalmente atravesó la superficie de la esfera dos veces, al ingresar y al emerger de ella. En consecuencia, el número ${\cal N}_{i}$ de impactos que recibió la Tierra debido a la captura de cometas interestelares durante todo el período Criogénico\index{Criogénico} es 

\begin{equation}
{\cal N}_{i}=p \, {\cal N}_{tc} \sim 13 \,\, \rm (choques\, de\, cometas\, contra\, la\, Tierra),
\end{equation}
un número considerable, teniendo en cuenta el daño climatológico que produce el choque de un gran cometa. El flujo de cometas que atravesó el entorno solar con distancias del perihelio menores que una UA produjo también muy probablemente impactos en la Luna, Venus y Mercurio.

Esos números  son naturalmente estimaciones basadas en simplificaciones, como por ejemplo  el hecho de ignorar que,  al pasar por las cercanías del Sol, los cometas están sujetos a  intensas fuerzas de mareas ejercidas por Sol y como consecuencia los cometas  se estiran y se despedazan. Con lo cual, como las bombas de racimo, aumenta el número de fragmentos cometarios y  el área de impacto. Por otro lado, al acercarse al Sol, los cometas pierden masa por evaporación. Otro posible hecho que ignoramos es que no solo precipitaron  cometas sobre el Sol, sino que también nubes de gas molecular y polvo interestelar que embebieron al interior del Sistema Solar en una densa atmósfera  en cuasi equilibrio hidrostático. Por el reiterado roce de los cometas con dicha atmósfera, los cometas se fueron lentamente disgregando y perdiendo energía. Espesas nubes de la mencionada atmósfera,  asentadas sobre el disco planetario, absorbieron la luz solar y fueron  probablemente responsables  de tan largos períodos  de congelamiento de la Tierra (ver Sección \ref{Penetrante1}). Al concluir  la gran captura de cometas a fines del período Criogénico,  los cometas que circularon repetidamente por las cercanías del Sol fueron desintegrándose  paulatinamente  y, por lo tanto, la tasa de colisiones cometarias sobre los planetas interiores debe haber decaído significativamente después del período Criogénico.

El hipotético escenario que hemos aquí explorado  nos presenta el período Criogénico como un punto de inflexión en la historia geológica y biológica de Tierra. El  período Criogénico abarcó 200 millones de años en el cual la Tierra estuvo envuelta en hielo. Este extraordinario fenómeno podría encontrar una explicación  en el hecho, también singular, de que  la Tierra habitó dentro de una  densa nube interestelar durante todo el período Criogénico.  En la Sección \ref{Penetrante1}, encontramos que  la Tierra sumergida en una pequeña nube interestelar de  densidad mayor a $3 \times 10^{7}$ moléculas por cm$^{3}$ sufre consecuencias climáticas severas. Además, en esta Sección, hemos mostrado que una larga permanencia del Sistema Solar  en el seno de una densa nube de gas, polvo y cometas interestelares hace posible que el Sol capture cometas de la nube molecular,  en una cantidad similar a la cantidad de cometas  que posee la nube de Oort (ver expresión \ref{PoblacionCometas3}). De modo que no es una casualidad  que la captura relativamente reciente de  cometas coincida con el período Criogénico. Este fenómeno astronómico es empíricamente avalado por el aumento de la tasa de formación de cráteres lunares durante período Criogénico \cite{Terada} y, aún más, estaría conectado con el asombroso hecho de la abrupta aparición de la vida compleja al comienzo del Cámbrico (ver Secciones \ref{TioVivo} y \ref{panspermia}). 

\begin{figure}
\includegraphics[scale=0.6]{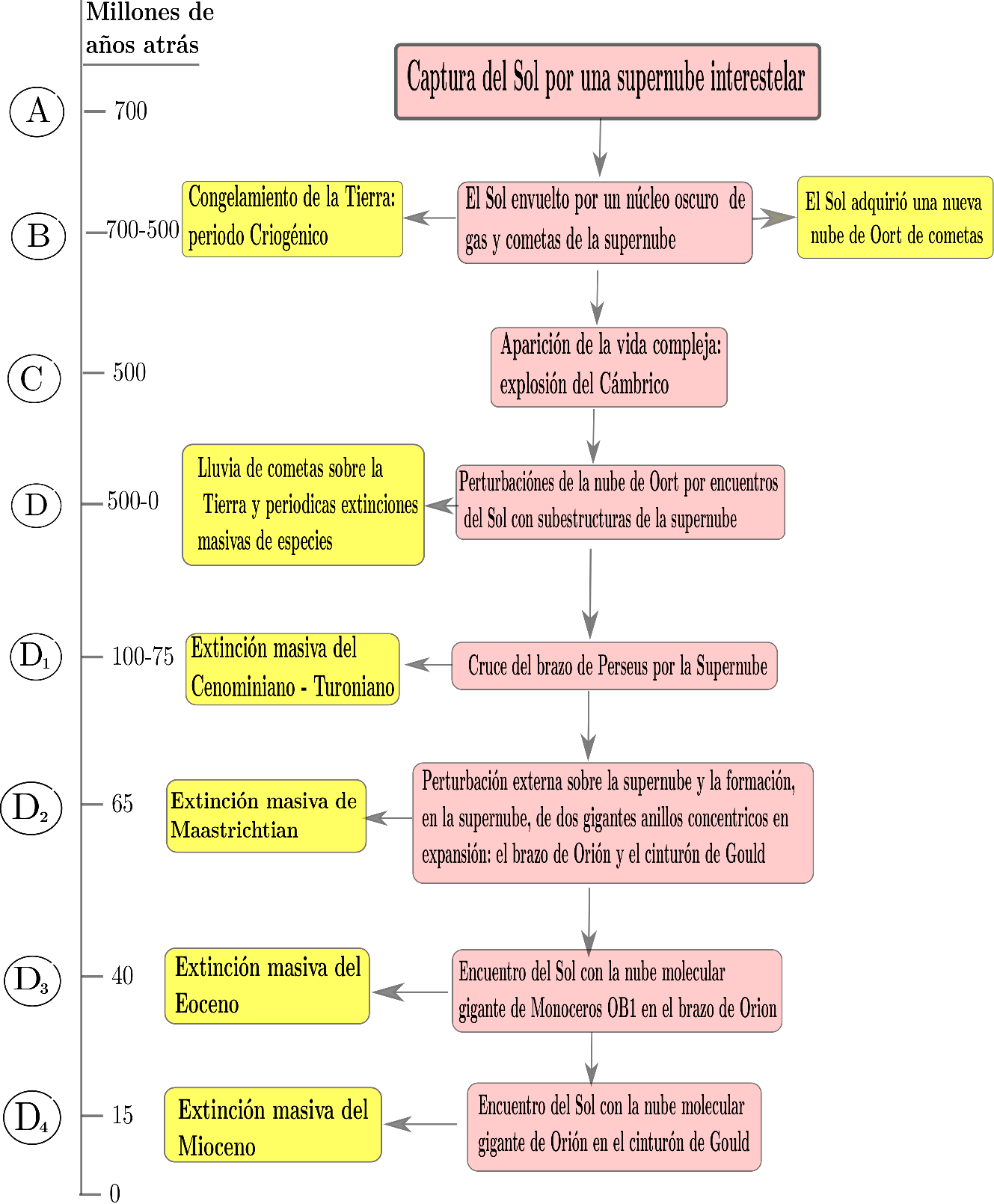} 
\caption{Secuencia de los principales sucesos  que se habrían desencadenado sobre la Tierra y el Sistema Solar como consecuencia  del cautiverio  del Sol por parte  de una supernube interestelar.}
\label{FlechaEvolutiva}
\end{figure}

Dichos procesos pueden entenderse dentro del marco de nuestra teoría de que el Sol fue capturado por una supernube interestelar poco antes del período Criogénico. El esquema de la Fig. \ref{FlechaEvolutiva} resume el papel central que jugó la relación dinámica y física del Sol con la supernube en el fenómeno de la extinciones masivas de las especies con una cuasi-periodicidad de 26 millones de años. En la figura de marras designamos con la letra D al conjunto de las extinciones masivas, las cuales incluyen las extinciones designadas con los subíndices 1, 2, 3 y 4 de D  que se vinculan a eventos específicos investigados por este autor (ver Figuras 5 y 12  de \cite{Olano7} y referencias  \cite{Olano10} \cite{Olano1} \cite{Olano9}).

\newpage

\section{El origen de la vida: ¿la vida  provino  del medio interestelar?\label{panspermia}}

\begin{quote}
\small \it{La inteligencia humana es un azar--no está en nuestra mano. Tiene un carácter de inspiración, de insuflamiento casual y discontinuo. No sabemos nunca si, en un caso dado, seremos inteligentes ni si el problema que nos urge resolver será soluble para la inteligencia.} \rm 

José Ortega y Gasset \index{Ortega y Gasset, J. } (1883-1955) en el libro ``Apuntes sobre el pensamiento'' 
\end{quote}

\begin{quote}
\small \it{... a partir de todo lo que hemos aprendido sobre la estructura de la materia viva, debemos estar dispuestos a encontrar que funciona de una manera que no puede reducirse a las leyes ordinarias de la Física...La sucesión de acontecimientos en el ciclo vital de un organismo exhibe una regularidad y un orden admirables, no rivalizados por nada de lo que observamos en la materia inanimada.} \rm 

Erwin Schrödinger\index{Schrödinger,  E.} \footnote{Físico austríaco a quien en 1933 le otorgaron el premio Nobel de Física, junto  a  Paul Dirac\index{Dirac,  P.}, por  la formulación matemática de la mecánica cuántica, en la cual una ecuación de onda fundamental lleva su nombre.} en el libro ``¿Qué es la vida?'', escrito en 1943 con el título ``What is life? The physical aspect of living cell''
\end{quote}

Se puede decir quizá en tono de crítica  que para referirse a los orígenes hay que ser un poco poeta y astrónomo. Sin embargo, con el descubrimiento de miles de exoplanetas, la astronomía se ganó el derecho a investigar el origen de la vida y la vida en el Universo como un imperativo cósmico. De hecho, la Astrobiología es una nueva disciplina científica. Como sugerimos en esta breve sección, ``no somos sólo de este mundo'', parafraseando la expresión ``Wir sind nicht nur von dieser Welt'', con la cual el  escritor científico alemán Hoimar von Ditfurth \index{von Ditfurth, H.} (1921-1989) tituló uno de sus libros.

 Las ideas de que la vida se originó en la Tierra primitiva o de que la vida provino desde el espacio exterior fueron planteadas por los filósofos presocráticos, Anaximandro\index{Anaximandro} y  Anaxágoras\index{Anaxágoras}, hace más de 2500 años.  Anaximandro postuló que la vida se engendró en los mares antiguos y que todos los seres vivos actuales evolucionaron a partir  de los primeros peces, adelantándose a las ideas evolutivas de Charles Darwin\index{Darwin, C.}.  El mismo Darwin imagina el origen de  la vida en una charca de agua templada.  
 
 A comienzos del siglo veinte, el bioquímico ruso Alexandr Ivánovich Oparin\index{Oparin, A.I.} retoma la idea de la charca de Darwin. La teoría de Oparín,  que se hizo  popular con el nombre de ``caldo primigenio'', propone que la vida surgió de  una sopa de elementos químicos complejos en la Tierra primitiva. La teoría de Oparin recibió un fuerte apoyo con el experimento realizado en el año 1953 por Stanley Miller\index{Miller, S.} y Harold Clayton Urey\index{Urey, H.C.} en la Universidad de Chicago\index{Chicago, Universidad}. Miller y Urey diseñaron un aparato que producía descargas eléctricas,  simulando los rayos de una tormenta,  sobre una mezcla de elementos (agua, metano, amoníaco e hidrógeno), simulando la atmósfera de la Tierra primitiva,  y encontraron que se formaba materia orgánica como  aminoácidos.
 
 Los aminoácidos son moléculas que forman las proteínas, las cuales son esenciales para la vida. El descubrimiento  de Miller y Urey parece respaldar la teoría de que la vida se originó en la Tierra. Sin embargo,  los aminoácidos han sido también hallados  en meteoritos y cometas \citep{Elsila}, y por lo tanto es también posible que la vida haya llegado a la Tierra  a través de un meteorito o cometa, tal como lo intuyó  2500 años atrás el filósofo griego Anaxágoras. Uno de los primeros defensores del origen exógeno de la vida terrestre  fue Svante Arrhenius\index{Arrhenius, S.}, quien recuperó esta teoría denominándola: panspermia. Científicos ilustres, además de Arrhenius, como Lord Kelvin\index{Kelvin Lord}, Hermann von Helmholtz\index{von Helmholtz, H.}, Fred Hoyle\index{Hoyle, F.} y Francis Crick\index{Crick, F.} \footnote{ Junto a  Leslie Orgel\index{Orgel, L.}  propuso la teoría de la panspermia dirigida. Por otra parte,  Francis Crick, James Dewey Watson \index{Watson, J.D.} y Maurice Wilkins \index{Wilkins, M.} descubrieron  la estructura en doble hélice de la molécula de ADN y por ello recibieron el Premio Nobel de Medicina en 1962.} argumentaron a favor de la panspermia.
 
 Las densas y oscuras nubes interestelares como la emblemática nube molecular de Taurus 1 (TMC1 en inglés por Taurus Molecular Cloud 1) son laboratorios naturales donde se sintetizan moléculas complejas \citep{Ehrenfreund} \citep{OlanoOtros} y por lo tanto pueden proveer de los materiales prebióticos necesarios para generar la vida. Sin embargo, como lo señala Robert Shapiro\index{Shapiro, R.} en su libro de divulgación Orígenes, ``una mezcla de compuestos químicos simples, aun cuando esté enriquecida con aminoácidos, se parece tanto a una bacteria como un montoncito de palabras sin sentido, escrita cada cual en un pedazo de papel, a las obras completas de Shakespeare\index{Shakespeare}''. Debemos estar dispuestos a aceptar que la materia viva puede estar gobernada por leyes que apenas  sospechamos y que difícilmente puedan reducirse a las leyes ordinarias que rigen la materia inorgánica, tal como sugiere Erwin Schrödinger\index{Schrödinger,  E.}\ en la cita del  epígrafe.
 
 El Sol se encuentra dentro de un extenso complejo de gas y estrellas, que incluyen el brazo de Orión\index{Orión, brazo de} y el cinturón de Gould\index{Gould, cinturón de} y que en conjunto formarían en el presente una supernube en estado de desintegración \citep{Olano8} \citep{Olano7} (ver también  Sección \ref{TioVivo}). El autor de esta monografía ha mostrado  que el Sol pudo ser  capturado por la supernube cuando esta se formaba, $\approx 700$ millones de años atrás \cite{Olano4}, un hecho astronómico con profundas implicancias  para el Sistema Solar y para la Tierra en particular. El hecho sugestivo es que la captura del Sol por parte de la supernube ocurrió  en coincidencia con el período Criogénico\index{Criogénico}. Ello podría indicar que, en el proceso de captura, el Sol penetró a baja velocidad una subestructura de la supernube: una densa nube de gas, polvo y cometas interestelares. Como consecuencia, se produjo una masiva captura de cometas interestelares que rellenaron la nube de Oort (ver Secciónes \ref{Reposición} y \ref{CapturaCometas}). La profusa caída de polvo, gas molecular y cometas interestelares sobre la atmósfera terrestre durante 200 millones de años condujo al período Criogénico \index{Criogénico} convirtiendo a la Tierra  en una bola de hielo (ver Secciones \ref{criogenico1} y \ref{Penetrante1}). Conjeturamos que como consecuencia de la lluvia de dichos materiales cósmicos sobre la Tierra sobrevino un hecho asombroso: un evento de panspermia de vida compleja (multicelular) que originó la explosión de vida del Cámbrico\index{Cámbrico}.

La temprana aparición de la vida microbiana en la Tierra primitiva indicaría que los cometas interestelares ligados a la nube molecular que formó el disco protoplanetario  transportaban vida microbiana, lo cual significaría que la propagación de la vida microbiana a través de la Galaxia sería relativamente eficiente. La Galaxia se hallaría preñada de vida bacteriana. En cambio, la vida compleja, que por un lado tardaría mucho en desarrollarse y que por otro lado puede ser destruida  más fácilmente que las bacterias en los largos viajes interestelares a bordo de cometas, sería de relativamente escasa distribución en la Galaxia. Ello llama a nuestra responsabilidad en la protección de la vida, que ya  no sería  sola planetaria sino también galáctica.
 
\newpage
 
Así como la teoría de la evolución no explica el origen de la vida, si bien ese es el título del famoso libro de Darwin\index{Darwin, C.}, la teoría de la panspermia tampoco lo explica. Se dice que ``de la ameba al hombre hay un solo paso'', pero de la materia  inorgánica a la vida parece haber un abismo. La teoría de la panspermia tiene la ``ventaja'' de que enmarca el fenómeno de la vida  en un escenario de tiempo y espacio inconmensurables que hace posible la realización de un suceso poco probable \footnote{El biólogo y bioquímico francés Jacques Monod\index{Monod, J.} en su libro \it{El azar y la necesidad} \rm hace una sugerencia sorprendente: la vida surgió de la materia inanimada por una conjunción altamente improbable de circunstancias fortuitas; este suceso no fue meramente de baja probabilidad, sino de probabilidad cero, es decir un evento único e irrepetible.  Monod ganó del Premio Nobel de Fisiología o Medicina en 1965.}. Permítase concluir con la última frase del libro \it{Polvo Vital} \rm de Christian de Duve \index{de Duve, C.}\footnote{\rm Christian de Duve (1917-2013) recibió el premio Nobel de Biología o Medicina en el año  1974. El título completo del libro mencionado es ``Polvo Vital: El origen y evolución de la vida en la Tierra''.} ``debemos inclinarnos ante el misterio''.\rm 

\newpage

\section*{Agradecimientos}
Agradezco al Dr. Andrés Colubri la lectura del manuscrito y sus comentarios constructivos. Mis agradecimientos al Dr. Fernando Espí que difundió la primera parte de esta monografía con  el podcast ``Amenazas astronómicas del planeta Tierra'' en El último humanista. Muchas gracias a mi esposa, Susana,  mis hijos, Sebastián y José María, mi nuera Gabriela y mi nieto Máximo, sin el apoyo y estimulo de ellos hubiera sido imposible este artículo. Quiero también agradecerles a mis amigos: Pablo Alvarez, Estefanía y Thiago Poratti, quienes con sus visitas a mi casa de campo Puesto Viejo, donde elaboré la mayor parte de este trabajo, me alentaban en mi tarea.

\renewcommand{\refname}{Bibliografía}

\,

\addcontentsline{toc}{}{Índice alfabético de nombres}
\printindex 

\end{document}